\documentclass[aps,prb,showpacs,onecolumn,notitlepage,superscriptaddress,nofootinbib]{revtex4-2}

\usepackage{stix}
\usepackage{bm,color,amsmath,graphicx,psfrag,accents,float}
\usepackage{multirow}
\usepackage{dcolumn}
\usepackage{xcolor}
\usepackage{comment}
\usepackage{tcolorbox}
\usepackage{ulem,tikz}

\newcolumntype{L}[1]{>{\raggedright\arraybackslash}p{#1}}
\newcolumntype{C}[1]{>{\centering\arraybackslash}p{#1}}
\newcolumntype{R}[1]{>{\raggedleft\arraybackslash}p{#1}}

        \definecolor{AAcolor}{rgb}{0.7,0.1,0.4}

  \newcommand{\noi}[1]{\noindent (#1)}	
	
			\newcommand{\e}[1]{\begin{align}{#1}\end{align}}	
			\newcommand{\lin}{\notag \\}
		
		\newcommand{\f}[2]{\frac{#1}{#2}}
		\newcommand{\tf}[2]{\tfrac{#1}{#2}}
		
		\newcommand{\p}[2]{\frac{\partial #1}{\partial #2}}

				\newcommand{\pe}{\partial_{\be}}

		\newcommand{\la}[1]{\label{#1}}

		\newcommand{\q}[1]{Eq.\ (\ref{#1})}
		
		\newcommand{\qq}[2]{Eqs.\ (\ref{#1})-(\ref{#2})}
		\newcommand{\s}[1]{Sec.\ \ref{#1}}
		\newcommand{\fig}[1]{Fig.\ \ref{#1}}		
		\newcommand{\app}[1]{App.\ \ref{#1}}				
		\newcommand{\tab}[1]{Tab.\ \ref{#1}}
		\newcommand{\ocite}[1]{Ref.\ \onlinecite{#1}}

\newcommand{\sma}[1]{\scriptscriptstyle{#1}}
\newcommand{\mone}{^{\scriptscriptstyle{-1}}}

\newcommand{\mo}{{\scriptscriptstyle{-1}}}
\newcommand{\gmo}{g^{\scriptscriptstyle{-1}}}
\newcommand{\hgmo}{\hat{g}^{\scriptscriptstyle{-1}}}

		\newcommand{\ri}{\rightarrow}
		
		\newcommand{\lea}{\leftarrow}
				
		\newcommand{\iand}{\ins{and}}

		\newcommand{\sgn}{\text{sgn}}

		\newcommand{\imp}{\;\;\Rightarrow\;\;}
		\newcommand{\eq}{=&\;}
		\newcommand{\condeq}[1]{\as \substack{\sma{#1}\\=}\as}
        	\newcommand{\condapprox}[1]{\as \substack{\sma{#1}\\ \approx}\as}

        \newcommand{\refeq}[1]{\as \substack{\sma{\text{Eq. } (\ref{#1}})\\=}\as}
        \newcommand{\reftwoeq}[2]{\as\substack{\sma{\text{Eq. }(\ref{#1},\ref{#2})}\\\sma{=}}\as}
        
		\newcommand{\refimp}[1]{\as \substack{\sma{\text{Eq. } (\ref{#1}})\\ \Rightarrow}\as}
		\newcommand{\limit}[1]{\substack{\text{lim}\\#1}\;}		
		 
		\newcommand{\appr}{\approx &\;}

		\newcommand{\R}{\mathbb{R}}
		\newcommand{\Z}{\mathbb{Z}}
		
		\newcommand{\C}{\mathbb{C}}
		\newcommand{\N}{\mathbb{N}}

\newcommand{\mathsout}[1]
{\bgroup\mathchoice
  {\sbox0{$\displaystyle{#1}$}%
    \usebox0\hspace{-\wd0}%
    \rule[0.5\ht0-0.5\dp0-.5pt]{\wd0}{1pt}}%
  {\sbox0{$\textstyle{#1}$}%
    \usebox0\hspace{-\wd0}%
    \rule[0.5\ht0-0.5\dp0-.5pt]{\wd0}{1pt}}%
  {\sbox0{$\scriptstyle{#1}$}%
    \usebox0\hspace{-\wd0}%
    \rule[0.5\ht0-0.5\dp0-.5pt]{\wd0}{1pt}}%
  {\sbox0{$\scriptscriptstyle{#1}$}%
    \usebox0\hspace{-\wd0}%
    \rule[0.5\ht0-0.5\dp0-.5pt]{\wd0}{1pt}}%
\egroup}

	\newcommand{\eikr}{e^{i\bk \cdot \br}}

	\newcommand{\emikr}{e^{-i\bk \cdot \br}}

\newcommand{\Nimp}{N_{\text{imp}}}
\newcommand{\mD}{m_{\sma{\text{D}}}}

\newcommand{\curl}{\nabla \times}

\newcommand{\nabk}{\boldsymbol{\nabla_{k}}}

\newcommand{\nabr}{\boldsymbol{\nabla_{r}}}

\newcommand{\nab}{\boldsymbol{\nabla}}

\newcommand{\dekkp}{\delta_{\bk,\bk'}}

\newcommand{\para}{\shortparallel}
\newcommand{\per}{\perp}

\newcommand{\bkper}{\boldsymbol{k}_{{\perp}}}

\newcommand{\kper}{k_{\perp}}

\newcommand{\bGper}{\bG_{\perp}}
\newcommand{\bpper}{\bp_{\perp}}
\newcommand{\bqper}{\bq_{\perp}}
\newcommand{\brper}{\br_{\perp}}
\newcommand{\hbrper}{\hbr_{\perp}}

\newcommand{\bkp}{\bk'}

\newcommand{\vcell}{{\cal V}_{\scriptscriptstyle{\text{cell}}}}

\newcommand{\ncell}{{\cal N}_{\scriptscriptstyle{\text{cell}}}}
\newcommand{\rootv}{\sma{\sqrt{\calv}}}

\newcommand{\Vim}{V^{\text{im}}}
\newcommand{\Vex}{V^{\text{ex}}}

\newcommand{\Eex}{E_{\sma{\text{ex}}}}
\newcommand{\Ek}{E_{\bk}}

\newcommand\as{\;\;\;\;}

\newcommand{\hd}{\hat{d}}
\newcommand{\hg}{\hat{g}}

\newcommand{\hz}{\hat{z}}

\newcommand{\hH}{\hat{H}}

\newcommand{\hU}{\hat{U}}

\newcommand{\iden}{\mathbb{1}}

\newcommand{\hbd}{\boldsymbol{\hat{d}}}

\newcommand{\hbr}{\hat{\br}}

\newcommand{\hbv}{\hat{\bv}}
\newcommand{\hbx}{\hat{\bx}}
\newcommand{\hby}{\hat{\by}}
\newcommand{\hbz}{\hat{\bz}}

\newcommand{\overlr}[1]{\overleftrightarrow{{#1}}}

\newcommand{\overg}{\overleftrightarrow{\bg}}

\newcommand{\overT}{\overleftrightarrow{T}}

\newcommand{\ba}{\boldsymbol{a}}

\newcommand{\bb}{\boldsymbol{b}}
\newcommand{\bc}{\boldsymbol{c}}
\newcommand{\bd}{\boldsymbol{d}}
\newcommand{\be}{\boldsymbol{e}}

\newcommand{\fbk}{f_{\bk}}
\newcommand{\fbkp}{f_{\bk'}}

\newcommand{\bg}{\boldsymbol{g}}

\newcommand{\bj}{\boldsymbol{j}}
\newcommand{\bk}{\boldsymbol{k}}
\newcommand{\bl}{\boldsymbol{l}}

\newcommand{\bp}{\boldsymbol{p}}
\newcommand{\bq}{\boldsymbol{q}}
\newcommand{\br}{\boldsymbol{r}}

\newcommand{\bv}{\boldsymbol{v}}

\newcommand{\bx}{\boldsymbol{x}}
\newcommand{\by}{\boldsymbol{y}}
\newcommand{\bz}{\boldsymbol{z}}
\newcommand{\bA}{\boldsymbol{A}}
\newcommand{\bB}{\boldsymbol{B}}

\newcommand{\bD}{\boldsymbol{D}}
\newcommand{\bE}{\boldsymbol{E}}

\newcommand{\bG}{\boldsymbol{G}}

\newcommand{\bI}{\boldsymbol{I}}

\newcommand{\bR}{\boldsymbol{R}}
\newcommand{\bS}{\boldsymbol{S}}

\newcommand{\bze}{\boldsymbol{0}}

\newcommand{\bOmega}{\boldsymbol{\Omega}}

\newcommand{\bsigma}{\boldsymbol{\sigma}}
\newcommand{\bSigma}{\boldsymbol{\Sigma}}
\newcommand{\btau}{\boldsymbol{\tau}}

\newcommand{\jball}{\bj_{\sma{\text{ballistic}}}}
\newcommand{\jshift}{\bj_{\sma{\text{shift}}}}

\usetikzlibrary{arrows,scopes} 

\newcommand{\widephi}{\widetilde{\phi}}

\newcommand{\cheg}{\check{g}}
\newcommand{\chegmo}{\check{g}^{\mo}}
\newcommand{\cher}{\check{r}}
\newcommand{\chebr}{\check{\br}}
\newcommand{\cheH}{\check{H}}

\newcommand{\gk}{g\cdot\bk}

\newcommand{\matrixtwo}[4]{\begin{pmatrix} #1 & #2 \\ #3 & #4 \end{pmatrix}}
\newcommand{\diagmatrixtwo}[2]{\begin{pmatrix} #1 & 0 \\ 0 & #2 \end{pmatrix}}
\newcommand{\diagmatrixthree}[3]{\begin{pmatrix} #1 & 0 &0 \\ 0 & #2 & 0 \\ 0 & 0 & #3 \end{pmatrix}}
\newcommand{\matrixthree}[9]{\begin{pmatrix} #1 & #2 & #3 \\ #4 & #5 & #6 \\ #7 & #8 & #9 \end{pmatrix}}

\newcommand{\sx}{\sigma_{\sma{1}}}
\newcommand{\sy}{\sigma_{\sma{2}}}
\newcommand{\sz}{\sigma_{\sma{3}}}

\newcommand{\ins}[1]{\;\;\;\;\text{#1}\;\;\;\;}

\newcommand{\imag}{{\text{Im}\;}}
\newcommand{\real}{{\text{Re}\;}}

\newcommand{\cale}{{\cal E}}
\newcommand{\bcale}{\boldsymbol{\cal E}}

\newcommand{\calv}{{\cal V}}

\newcommand{\scrb}{{\mathscr B}}

\newcommand{\scrd}{{\mathscr D}}
\newcommand{\bscrd}{\boldsymbol{\scrd}}

\newcommand{\scrf}{{\mathscr F}}
\newcommand{\bscrf}{\boldsymbol{\mathscr F}}

\newcommand{\scri}{{\mathscr I}}

\newcommand{\scrr}{{\mathscr R}}

\newcommand{\braket}[2]{\langle #1 \,|\, #2 \rangle}
\newcommand{\ketbra}[2]{|\,  #1  \rangle \langle #2 \,| }
\newcommand{\braopket}[3]{\langle #1 \,|\, #2 \,|\, #3 \rangle}
 		
\newcommand{\bra}[1]{\langle\,#1\,|}
\newcommand{\ket}[1]{|\,#1\,\rangle}

\newcommand{\cbraket}[2]{( #1 \,|\, #2 )}

\newcommand{\cbraopket}[3]{( #1 \,|\, #2 \,|\, #3 )}	
\newcommand{\cbra}[1]{(\,#1\,|}
\newcommand{\cket}[1]{|\,#1\,)}

\newcommand{\half}{\frac{1}{2} }
\newcommand{\thalf}{\tfrac{1}{2} }

\newcommand{\im}{\mathrm{Im}}

\newcommand{\ab}{\alpha\beta}

\newcommand{\bpm}{\begin{pmatrix}}
\newcommand{\epm}{\end{pmatrix}}

\newcommand{\bal}{\begin{align}}

\newcommand{\eps}{\epsilon}

\usepackage{stix}
\usepackage{bm,color,amsmath,graphicx,psfrag,accents,float}
\usepackage{multirow}
\usepackage{dcolumn}
\usepackage{xcolor}
\usepackage{comment}
\usepackage{tcolorbox}
\tcbuselibrary{skins, breakable}
\usepackage{tikz}
\usepackage{cancel}
\usepackage{longtable}

\usepackage[caption=false]{subfig}

\usepackage[colorlinks=true,urlcolor=blue,citecolor=blue,linkcolor=blue,breaklinks=true]{hyperref}

\newcommand{\DV}{D_{\sma{V}}}
\newcommand{\DVmr}{D_{\sma{V}}^{\sma{\text{mr}}}}
\newcommand{\dfield}{$\bscrd$\text{irector field} }

\newcommand{\HD}{H^{\sma{\text{D}}}}

\newcommand{\df}[1]{\,\delta{\left(#1\right)}}

\newcommand{\cA}{\bm{\mathcal{A}}}
\newcommand{\Bde}{B^{\text{de}}}

\newcommand{\Sigmade}{B^{\text{de}}}
\newcommand{\Sigmadeabc}{B^{\text{de}}_{abc}}

\begin{document}

\title{Vortex-enhanced photovoltaic current in disordered topological materials}

\author{Pavlo Sukhachov}
\thanks{These authors contributed equally to this work}
\affiliation{Department of Physics and Astronomy, University of Missouri, Columbia, Missouri, 65211, USA}
\affiliation{MU Materials Science \& Engineering Institute, University of Missouri, Columbia, Missouri, 65211, USA}

\author{Penghao Zhu}
\thanks{These authors contributed equally to this work}
\affiliation{Department of Physics, The Ohio State University, Columbus, OH 43210, USA}

\author{Ella Banyas}
\affiliation{Molecular Foundry, Lawrence Berkeley National Laboratory, Berkeley, California 94720, USA}

\author{Liang Z. Tan}
\affiliation{Molecular Foundry, Lawrence Berkeley National Laboratory, Berkeley, California 94720, USA}

\author{A. Alexandradinata} \affiliation{Department of Physics and Santa Cruz Materials Center, University of California Santa Cruz, Santa Cruz, CA 95064, USA}


\begin{abstract}
In disordered topological materials, real-space crystalline defects interplay with
momentum-space wave function singularities to \textit{enhance} the bulk photovoltaic current. What's singular is the interband Berry phase, or equivalently the phase of the \textit{optical} dipole matrix element, which has a \textit{vortex} structure in momentum space. Such \textit{optical vorticity} is guaranteed to exist in all topological materials associated with nontrivial Chern numbers. These vortices enhance electron-impurity skew scattering, which manifests as a ballistic photovoltaic current that is sensitive to (a) the topological material class, (b) the symmetry class of crystalline defects, and (c) the light polarization. This sensitivity manifests in two ways: firstly, by (a-c)-dependent frequency exponents for the photovoltaic current $\propto \omega^{\text{exponent}}$ in topological semimetals, with $\omega$ the frequency of the light source. Secondly, by (a-c)-dependent constraints of the bulk photovoltaic tensor, which are explainable only by emergent, \textit{magnetic} symmetries of \textit{time-reversal-invariant} topological materials. These ideas are concretized by case studies on multifold fermions, 3D $m$-order Weyl semimetals and 2D $n$-order Dirac systems, which include $n$-layer rhombohedral graphene, transition metal dichalcogenides, and topological surface states. Theoretical guidance is provided for a tri-pronged experimental program that combines frequency-tuned photoconductivity measurements, defect characterization and defect engineering.
\end{abstract}
\date{\today}

\maketitle

{\tableofcontents \par}

\section{Introduction and outline}

Irrespective of a material's symmetry, the longitudinal linear conductivity is usually degraded by an increasing concentration of crystalline defects.\cite{Drude1900a,ashcroft_mermin} Nonlinear optical responses can be qualitatively different: only in noncentrosymmetric materials can a dynamic electric field (associated to a light wave) drive a \textit{bulk photovoltaic }current in quadratic response,\cite{kraut_anomalousbulkPV} and this current can be enhanced by crystalline defects.\cite{belinicher_kinetictheory,belinicher_ballistic} How is this possible? It is not as simple as saying that noncentrosymmetric defects (e.g., electric dipolar impurities) reduce the symmetry of the material, because  centrosymmetric defects (e.g., electric monopolar impurities) can also mediate a {bulk photovoltaic} current, as a matter of principle.\cite{isobe_rectification} Evidently, centrosymmetry-breaking of the material's band structure must also play a role in activating the bulk photovoltaic current. But there is no centrosymmetry-breaking of the band energy-momentum relation in non-magnetic materials, because $\Ek=E_{-\bk}$ is guaranteed by time-reversal symmetry. Apparently, it must be a centrosymmetry-breaking of the electron wave function: $|\psi_{\bk}) \neq \text{Inversion} |\psi_{-\bk})$.  \\

\begin{figure}[H]
\centering
\includegraphics[width=0.7\textwidth]{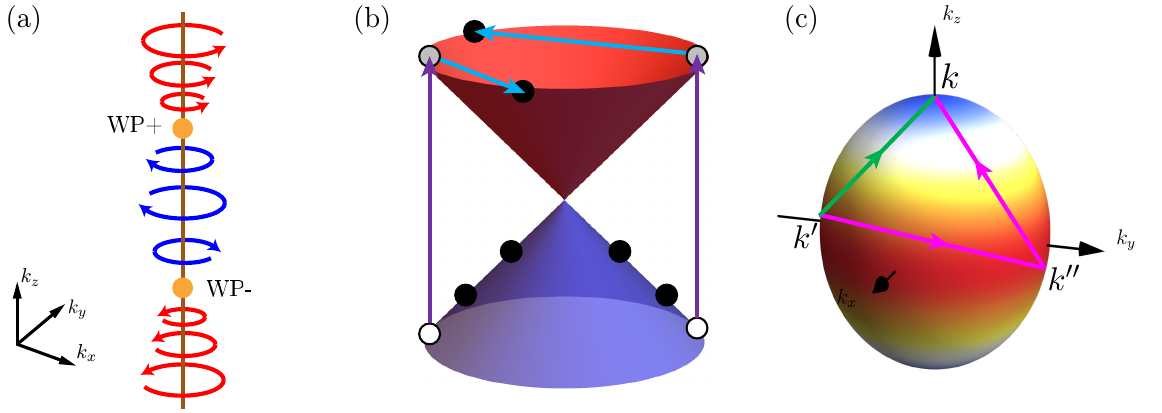}
\caption{ Topological band-touching $\bk$-points (orange dots) are not just sources/sinks of the intraband Berry curvature but also sources/sinks of vorticity in the interband Berry phase. In panel (a), a brown-colored
optical vortex line threads two band-touching Weyl points (of opposite Weyl charge) and is wrapped by the circulating vector field: $\nabk \arg \be\cdot \bA_{vc\bk}$.
(b) For a topological semimetal, asymmetric electron-defect scattering results in the electron distribution  $f_{\bk}\neq f_{-\bk}$. 
(c) An optical vortex makes the photo-excitation rate anisotropic over $\bk$-space: largest at the equator, zero at the poles, as indicated by smoothly varying colors. An electron photo-excited at $\bk'$ (on the equator) can subsequently collide with a defect and scatter directly (green arrow) to $\bk'$ (at the north pole), or scatter indirectly (magenta arrow) as $\bk'\lea \bk''\lea \bk$. The quantum interference between both scattering processes results in the asymmetric scattering of panel (b), and is fundamentally characterized by the Bargmann invariant: $\braket{u_{\bk'}}{u_{\bk}}\braket{u_{\bk}}{u_{\bk''}}\braket{u_{\bk''}}{u_{\bk'}}$, a wave function geometric characteristic that is generally non-reducible to the quantum metric or Berry curvature. 
}
\label{fig:ballistic}
\end{figure}

What geometric or topological properties of the electron wave function are relevant to the component of the bulk photovoltaic current which grows in  proportionality to  the concentration of crystalline defects? This paper proposes that in disordered topological materials, momentum-space wave function singularities interplay with real-space crystalline defects 
 to {enhance} the bulk photovoltaic current. These are singularities of  the interband Berry connection: $\bA_{vc\bk}=\braket{u_{v\bk}}{i\nabk u_{c\bk}}$ between the low-energy valence band and the high-energy conduction band. The manifestation of a singularity is a vortex structure in the vector field $\nabk \arg \be\cdot \bA_{vc\bk}$ for any  vector $\be$, as illustrated in \fig{fig:ballistic}(a). Because $\be$ can be interpreted as the polarization vector of a source-driven light wave, and $i\nabk$ corresponds in quantum mechanics to the real-space position, the aforementioned vortex structure is equivalently a \textit{vortex} structure in the phase of the \textit{optical} dipole matrix element.  \\

Such \textit{optical vortices} are guaranteed to exist in all topological materials associated with nontrivial Chern numbers, including (but not exhausted by) Chern insulators, $n$-order Weyl semimetals,\cite{chen_multiweyl} multifold fermions with pseudospin-1 (spin-one Weyl fermion) and -3/2 (Rarita-Schwinger fermions).\cite{bradlyn_multifold} This topological guarantee of optical vorticity follows from a theorem~\cite{zhuAA_anomalousshift} relating the vorticity of the \textbf{inter}band Berry phase to the quantized \textbf{intra}band Berry phase, which is what a Chern number is.
Historically, topological materials have been characterized by their intraband Berry phase, which is the holonomy associated with the parallel transport of an electron within a band.\cite{TKNN,Haldane1988,wan_weylsemimetal} One main message of this paper is that an alternative history could have happened in which topological materials were first characterized by their interband Berry phase.\\

The second main  message of this paper is worth reiterating:
\e{
\substack{\text{Optical vortex in}\\ \text{disordered topological materials}} \as \ri\as  \substack{\text{Asymmetric electron-defect} \\ \text{scattering}} \as\ri\as \substack{\text{Ballistic}\\ \text{photovoltaic current}} \la{dictum0}
}
By `asymmetric', we mean that electron-defect scattering creates an electron distribution $f_{b\bk}$ that is non-symmetric in inverting $\bk$, with $b$ being a band index, and with the difference $(f_{b\bk}- f_{b,-\bk})$ being proportional to the light intensity, as illustrated in \fig{fig:ballistic}(b). The resultant photocurrent is called the 
\e{
\substack{\text{Defect-mediated ballistic}\\ \text{photovoltaic current density}} \as \mathbf{j}_{\text{ballistic}}=-\frac{|e|}{2\text{Volume}}\sum_{b\bk}\bv_{b\bk}(f_{b\bk}-f_{b,-\bk}),
}
with the group velocity satisfying $\bv_{b\bk} = -\bv_{b,-\bk}$ owing to time-reversal symmetry. The qualifier of `defect-mediated' specifies the scattering mechanism behind $\jball$; there are other mechanisms\cite{belinicher_ballistic,belinicher_phononmechanism,zhenbang_phononballistic,zhenbang_electronholeballistic,Tan-Ogitsu-DynamicsBallisticPhotocurrents-2026} of the bulk photovoltaic current which are not the focus of this work. \\

As a general property of bulk photovoltaic currents, $\jball$ is distinguished phenomenologically (from conventional photovoltaic currents\cite{wurfel_solarcells}) by its sensitivity to the polarization of the driving light wave, which is encoded in the photovoltaic tensor, $B^{\text{de}}$, defined by
\e{
j_{\text{ballistic}}^a=\sum_{bc}\Sigmadeabc \cale^b_{\bq\omega}\overline{\cale^c_{\bq\omega}}; \as a,b,c \in \{x,y,z\}, \la{defineBde}
}
with $\bcale_{\bq\omega}$ being an electric wave of wavevector $\bq$ and frequency $\omega$.
A bulk photovoltaic theory of  disordered topological materials must therefore \textbf{interconnect (a) the polarization sensitivity of $\jball$ to (b) the symmetry class of defects and (c) the topological class of the underlying pristine material}.  This interconnection forms the \textbf{central theme }for the rest of the paper, and  is explored in two related ways:\\

\begin{figure}[H]
\centering
\includegraphics[width=9 cm]{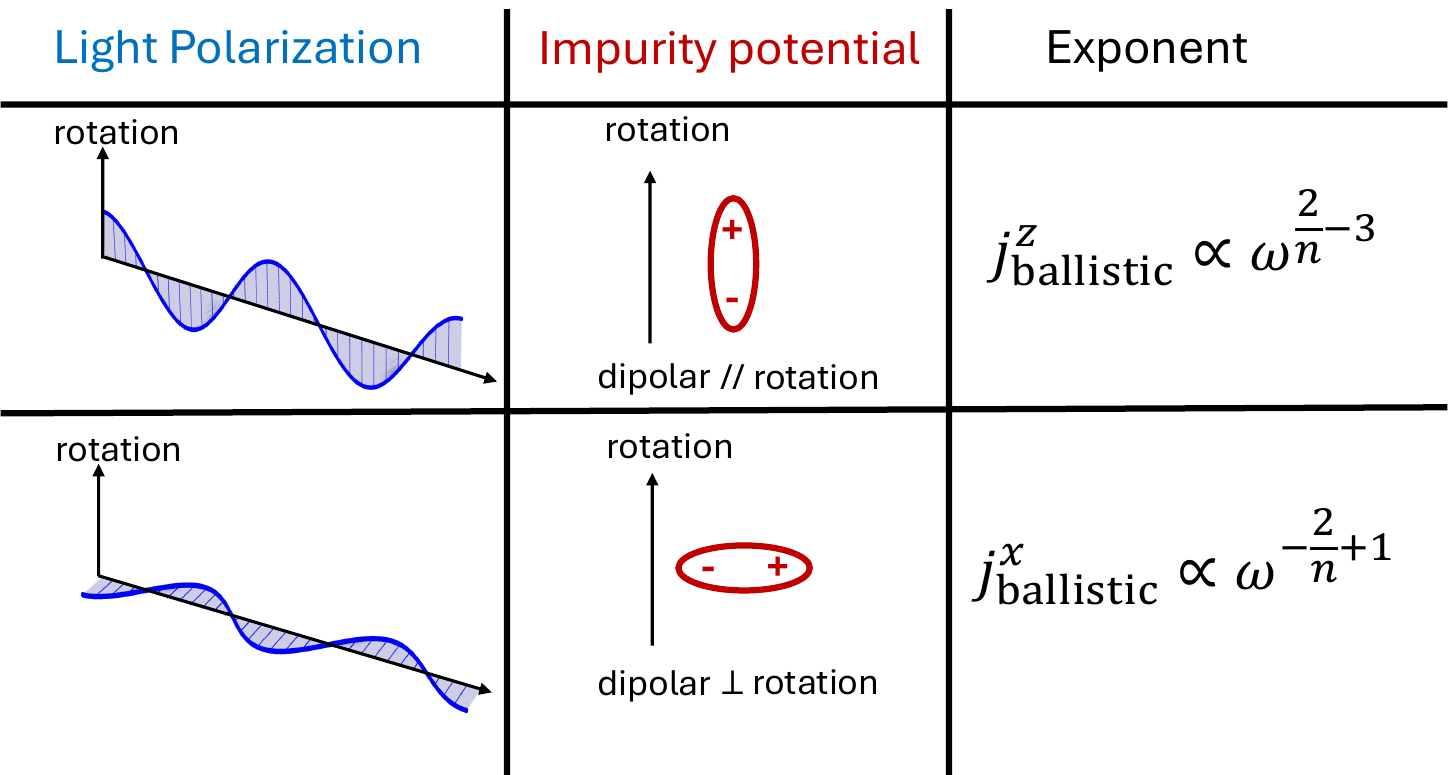}
\caption{The frequency exponent   depends on the order $n$ of $n$-Weyl fermions, the  light polarization (linear polarization vector parallel or orthogonal to the rotation axis), and the impurity type (electric dipole moment  parallel or orthogonal to the rotation axis).  }
\label{fig:nweylexponent}
\end{figure}

\noi{i} Firstly, because topological semimetals have a topologically-enforced zero band gap, they are uniquely fingerprinted by  the low-frequency bulk photovoltaic response which has a power law form: $\jball \propto \omega^{\text{exponent}}$. It will turn out that this frequency exponent depends on the factors (a-c) of our \textbf{central theme}.  A representative result for $n$-order Weyl semimetals is given in \fig{fig:nweylexponent}  and will be greatly generalized.\\

\noi{ii} Secondly, the low-frequency bulk photovoltaic response (of disordered topological materials) is determined by low-energy effective Hamiltonians whose point-group symmetries are determined by  the  \textbf{(b) the symmetry class of defects and (c) the topological class of the underlying pristine material}. We have systematically determined what these symmetries are for Dirac-Weyl topological systems and for point defects ($\equiv$ \textit{impurities}) with electric monopole and/or dipole moments.  These symmetries are encoded as constraints on the photovoltaic tensor (e.g., $\Sigmadeabc=0$ or $\Sigmadeabc=\pm B^{\text{de}}_{a'b'c'}$) which determines the \textbf{(a) polarization-dependence of $\jball$}.\\

As a representative illustration, consider $n$-order Dirac fermions characterized by a Berry phase of $n\pi$ and materialized by $n$-layer rhombohedral graphene, or $n$-\textit{graphene} in short.\cite{Koshino-McCann-TrigonalWarpingBerrys-2009} The symmetry of the low-energy effective Hamiltonian\cite{Koshino-McCann-TrigonalWarpingBerrys-2009} (of pristine $n$-graphene) is the magnetic point group $\bar{3}'m$, which contains $C_{3v}$ as a subgroup and additionally magnetic symmetries like $TS_6 \equiv$ (time reversal) composed with (improper six-fold reflection). Magnetic symmetries like $TS_6$ impose symmetry constraints on the photovoltaic tensor ($\Sigmade$) that we formalize into a symmetry theorem, newly presented here. The  symmetry of disordered $n$-graphene can be lower than that of $\bar{3}'m$ depending on the symmetry class of crystalline defects (e.g., monopolar vs dipolar impurity), and depending on whether a Dirac mass is generated externally by an out-of-plane displacement field or generated internally by a correlation-induced, band-energy gap. A lower symmetry than $\bar{3}'m$ relaxes some of the symmetry constraints on $\Sigmade$, which thus affects the polarization dependence of $\jball$. We see (in $n$-graphene) the tight interconnection of ideas encapsulated in our \textbf{central theme}, which we will elaborate on and generalize to other classes of topological materials.\\

Here is an outline of our paper.\\

\centerline{\underline{\s{sec:vorticesinTSM}: Optical vortices in topological semimetals and Dirac-Weyl materials}}

\begin{itemize}
    \item \ref{sec:chernvortex}: A Chern-vorticity theorem relates the quantized {intra}band Berry phase  to the vorticity of the {inter}band Berry phase. 
\item \ref{sec:vorticity3D}:  This  theorem is applied to uncover  the structure of optical vorticity in 3D topological semimetals associated with nontrivial Chern numbers, including isotropic Weyl fermions, anisotropic $n$-Weyl fermions and multi-fold fermions.
\item \ref{sec:vorticity2D}: The structure of optical vorticity is also uncovered for 2D Dirac systems associated with nontrivial Chern numbers, including Chern insulators, transition metal chalcogenides and topological surface states.
\item \ref{sec:directorfield}: To  compactly represent the $\bk$-location of optical vortices, we introduce the \textit{director field}, a wave-function-geometric vector field over $\bk$-space. An optical vortex is shown to be dual to a skyrmion in the director field. 
\end{itemize}

\centerline{\underline{\s{sec:current}: Ballistic photovoltaic current in disordered Dirac-Weyl materials} }

\begin{itemize}
    \item \ref{sec:interplay}: An argument is presented for the enhancement of the ballistic photovoltaic current by an interplay between momentum-space optical vortices and real-space crystalline defects.
\item \ref{sec:powerlaws}: Focusing on impurities with  an electric dipolar component, we formulate impurity-dependent power laws which determine how the frequency exponent [in $\jball \propto \omega^{\text{exponent}}$] depends on the dipole orientation, light polarization, and topological class of the Dirac-Weyl material. 

\item \ref{sec:symmetryreduction}: Because dipolar impurities are lower in symmetry than electric monopolar impurities,  the monopolar-impurity-mediated $\jball$ is nontrivial only in a smaller class of symmetry-reduced Dirac-Weyl materials. To determine the boundary of this `smaller class', we perform a symmetry classification of the low-energy effective Hamiltonians (of Dirac-Weyl materials); knowing their symmetry is tantamount to knowing the symmetry constraints on the photovoltaic tensor, by application of symmetry theorems formulated in the next section.
\end{itemize}

\centerline{\underline{\s{sec:symmetry}: Symmetry theorems for bulk photovoltaic response tensors} }

\begin{itemize}
    \item \ref{sec:generalsymmetryconstraints}: Theorems are established which constrain the mechanism-agnostic bulk photovoltaic tensor (inclusive of shift and ballistic contributions) based on the point-group symmetries of general Hamiltonians -- effective or not. Emphasis is placed on magnetic point-group symmetries of the form: (time reversal) composed with (spatial transformation), which emerge in the low-energy effective Hamiltonians of the previous subsection. 
\item \ref{sec:Hspecificsymmetry}: Additional theorems are established which constrain the mechanism-specific bulk photovoltaic tensor associated to defect-mediated skew scattering.
\end{itemize}

\centerline{\underline{\s{sec:discussion}: {Discussion, generalization, outlook}}}

\begin{itemize}
    \item A high-level, broad-perspective summary of results is presented. Generalizations are discussed, with an outlook for future theoretical and experimental progress. In particular, the notion of optical vorticity is extended to broader classes of topological materials and even to non-topological materials.
    
    \item Additional vortex-enhanced photovoltaic effects are discussed, including more mechanisms of the bulk photovoltaic effect and also the surface photovoltaic effect.
    
    \item A case is made for bulk photovoltaics as an experimental probe of correlated insulating phases, with an application to rhombohedral $n$-graphene.
    
    \item Theoretical guidance is provided for a tri-pronged experimental program that combines frequency-tuned photoconductivity measurements, defect characterization and defect engineering.
\end{itemize}

The rest of the paper are appendices targeted to the technically-oriented audience and separately outlined above \app{app:preliminaries}.\\

\s{sec:directorfield} and \s{sec:symmetry} are mathematically-oriented and written to be self-contained; a physically-oriented reader can skip these subsections without losing sight of the logical thread running through the paper.

\section{Optical vortices in topological semimetals and Dirac-Weyl materials}\la{sec:vorticesinTSM}

\subsection{Optical vortices and the Chern-vorticity theorem}\la{sec:chernvortex}

Because the interband Berry connection $\be \cdot\bA_{bb'\bk}$ is a complex number with two real parameters, the zeros of $\be \cdot\bA_{bb'\bk}$  have co-dimension two, meaning they generically form lines in the 3D Brillouin  zone and points in the 2D Brillouin zone; such lines and points are collectively referred as \textit{vortex centers}. The interband Berry phase $(\arg \be \cdot\bA_{bb'\bk})$ winds around any infinitesimally small loop ($\partial \text{vortex}$) encircling a vortex center; this winding number is referred to as the \textit{vorticity}, and it  can be calculated as the circulation of the photonic shift vector:
\e{
\text{Vorticity}\; = \oint_{\partial \text{vortex}} \bS^{\be}_{bb'\bk}\cdot \tf{d\bk}{2\pi}; \as
\text{Photonic shift vector}\; \bS^{\be}_{bb'\bk}\eq -\nabk \arg \be \cdot\bA_{bb'\bk}+\bA_{bb\bk}-\bA_{b'b'\bk}. \la{photonicshift}
}
The last two terms involving the intraband Berry connection have been added so that the shift vector is gauge invariant.\footnote{If a gauge is chosen so that $\bA_{cc}-\bA_{vv}$ is smoothly defined in the neighborhood of the vortex point, then the line integral of  $\bA_{cc}-\bA_{vv}$ vanishes $\partial vortex$ is infinitesimally small, and what remains is the winding number of the interband Berry phase. }\\

Throughout this work, we focus on a subcategory of topological semimetals whose energy bands are either energy-nondegenerate at generic $\bk$, or doubly-degenerate at generic $\bk$ due to the spin-orbit interaction being negligible.\footnote{This subcategory excludes the 3D Dirac semimetals.} For such semimetals, the existence of optical vortices is guaranteed by the Chern-vorticity theorem\cite{zhuAA_anomalousshift}, which colloquially states: wherever there is a quantized intraband Berry phase, there is also a quantized interband Berry phase. More precisely, for any pair of bands indexed by $b$ and $b'$, the
\e{\text{Chern-vorticity theorem:}\as C_b-C_{b'}= \text{Vor}^{\be}_{b'b} =\sum_{vortex}\oint_{\partial \text{vortex}} \bS^{\be}_{b'b\bk}\cdot \tf{d\bk}{2\pi}, \label{chernvortex}
}
relates the \textit{interband Chern number} ($C_b-C_{b'}$)  to the\textit{ interband optical vorticity} $(\text{Vor}^{\be}_{b'b})$, which is  
 the net circulation (of the shift vector) over all $\be$-vortex points in a closed 2D $\bk$-manifold, which can be a 2D Brillouin zone, or a two-toroidal/spherical cut of the 3D Brillouin zone. This theorem is
  proven in \app{app:chernvortex2sphere}. In the application to 3D topological semimetals, the closed 2D $\bk$-manifold is a sphere (enclosing the topological band touching) characterized by a nontrivial interband Chern number, thus guaranteeing optical vorticity by the theorem, as the next Section [\s{sec:vorticity3D}] elucidates.

\subsection{Optical vorticity of 3D topological semimetals} \la{sec:vorticity3D}

\subsubsection{Isotropic Weyl fermions} \la{sec:isotropicweyl}

An isotropic Weyl fermion has a two-band effective Hamiltonian:
\e{
H^{i,w}_{\bk}= d_{\bk}\cdot \bsigma; \as d_{\bk}=v\bk \as \bsigma=(\sigma_1,\sigma_2,\sigma_3), \la{hamiltonianisotropicweyl}
}
which is the dot product of a \textit{Hamiltonian vector} ($\bd \in \R^3$) with the vector $\bsigma$ of Pauli matrices. Throughout, we set $\hbar=1$.\\

For a $\bk$-surface surrounding a Weyl band-touching point, the Chern numbers of the conduction and valence bands satisfy: $C_c=-1=-C_v$,
implying (via the Chern-vorticity theorem) that $\text{Vor}^{\be}_{cv}=2$ for any light polarization $\be$, complex or not. Because this conclusion holds for an arbitrarily small radius of the $\bk$-surface, it must be that at least one vortex line connects to (or emerges from) the Weyl point. \\

Where is this line exactly? The $\bk$-location of this line must also be a zero of the \textit{optical affinity} defined as
\e{
\text{Optical affinity}:\as \text{Aff}_{cv\bk}^{\be}\equiv |\be\cdot \bA_{cv\bk}|^2.\la{defineopticalaffinity}
}
If $\be$ is viewed as the light polarization vector, then the optical affinity is essentially the absolute-valued square of the dipole matrix element and is a measure of the optical absorption.\footnote{This is generally true for  bulk 3D materials, and  remains true for 2D materials if $\be$ lies in the plane of the 2D material. Conversely, if $\be$ has an out-of-plane  component (say, $e_z$), then the quantum-mechanical correspondence between the position $z$ and $i\nabla_{k_z}$ breaks down.  } 
For any two-band, effective Hamiltonian of the form: 
\e{
H_{\bk}=\bd_{\bk}\cdot \bsigma + \delta E_{\bk} \iden_2; \as \bsigma=(\sx,\sy,\sz); \as \iden_2=\diagmatrixtwo{1}{1}, \la{twobandham}
} 
the affinity is expressible [\app{app:affinity2band}] as
\e{
\text{Aff}_{cv\bk}^{\be}= \sum_{ab} 
\left(g_{ab,\bk}\real (e_a \overline{e_b})-\frac{1}{2}\Omega_{ab,v\bk}\imag(e_a\overline{e_{b}})\right)
\la{eq:aff_2band}
}
in terms of the
\begin{equation} 
\label{eq:metric_Berrycuvature}
    \text{Quantum metric tensor}\as g_{ab,v\bk} \equiv \frac{1}
    {4}\partial_a\hat{\mathbf{d}}_{\bk}\cdot\partial_b\hat{\mathbf{d}}_{\bk}, \as\text{and Berry curvature tensor} \as \Omega_{ab,v\bk}\equiv \frac{1}{2}\hat{\mathbf{d}}_{\bk}\cdot(\partial_a\hat{\mathbf{d}}_{\bk}\times\partial_b\hat{\mathbf{d}}_{\bk}),
\end{equation} 
of the valence band, with $\hat{\mathbf{d}}_{\mathbf{k}}\equiv \bd_{\bk}/|\bd_{\bk}|$. 
In particular, the optical affinity of the isotropic Weyl fermion [\q{hamiltonianisotropicweyl}] is obtained by plugging $\bd=v\bk$ into \q{defineopticalaffinity}, which we write for real-valued $\be$ as
\e{
\text{Optical affinity of Weyl fermion}:\as \text{Aff}_{cv\bk}^{\be \in \R}=  \tf{1-(\hbd\cdot \be)^2}{4k^2}; \as \hbd=\tf{\bk}{k}; \as k=|\bk|. \la{affinityisotropicweyl}
}
This tells us the $\be$-vortex lies on a $\bk$-line that is parallel to $\be$ and intersects the Weyl point.\\

\begin{figure}[th!]
\centering
\includegraphics[width=0.9\textwidth]{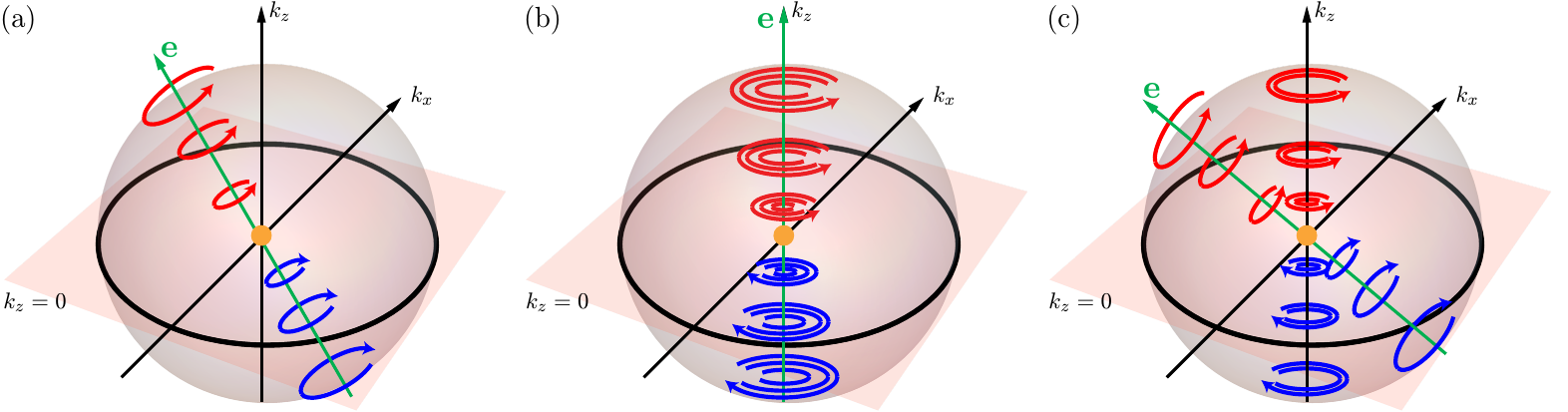}
\caption{
(a) Vortex-antivortex line of an isotropic Weyl fermion for the light polarized along the $\mathbf{e}$ direction. 
The shift vector for the light polarized along the $\mathbf{e}$ direction is $\bS^{\mathbf{e}}_{cv,\bk} = \frac{v}{k} \frac{(\mathbf{e}\cdot \hbd) \left(\mathbf{e}\times \hbd\right)}{1-(\mathbf{e} \cdot \hbd)^2}$.
(b) and (c) Vortex-antivortex line of a three-fold Weyl fermion for the light polarized along the $\mathbf{z}$ direction (b) and along a general direction $\mathbf{e}\neq \mathbf{z}$  (c). 
In all panels, the shift vector vanishes on the equatorial plane.
In both panels, red and blue arrows represent the circulating shift vectors.
}
\label{fig:shiftweyl}
\end{figure}

This line intersects the $\bk$-surface at two points, and each vortex point contributes $-1$ to the net vorticity ($\text{Vor}^{\be}_{cv}$) of the $\bk$-surface, as is consistent with the Chern-vorticity theorem. When calculating the vorticity of a $\bk$-surface, the orientation of the line integral in \q{photonicshift} is right-handed with the thumb pointing to the outward surface normal direction. However, if one orients the vorticity with a fixed thumb, say, pointing always in the positive z direction, then the vortex orientation flips discontinuously across the Weyl point as one moves along the $k_z$ axis, as illustrated in \fig{fig:shiftweyl}. We may say that the Weyl point is an optical singularity: a connection point between a vortex and antivortex line.\footnote{This is compatible with time-reversal symmetry, which constrains the shift vector field as [cf.\ \q{symmetryshift}]
\e{
\text{Time-reversal symmetry}:\as\bS_{cv, \bk}^{\be} = \bS_{cv,-\bk}^{\overline{\be}},
}
implying that a vortex point at $\bk$ has a time-reversed anti-vortex point at $-\bk$. This holds for  Kramers-Weyl fermions in a nonmagnetic, noncentrosymmetric crystal with nontrivial spin-orbit coupling.\cite{nielsen_ninomiya}}\\

\begin{table}[th!]
\centering
\scriptsize
\renewcommand{\arraystretch}{1.25}
\resizebox{\textwidth}{!}{%
\begin{tabular}{|c|c|c|c|c|c|c|c|} 
\hline
\text{Fermion}  & $n$-Weyl $(\mathbf{e} \neq \mathbf{z})$  & \multicolumn{2}{c|}{Pseudospin-1 $(\mathbf{e}=\mathbf{z})$} & \multicolumn{4}{c|}{Pseudospin-$3/2$ $(\mathbf{e}=\mathbf{z})$} \\ \hline  
$(b,b')$ & $(1,-1)$ & $(1,0)$ & $(1,-1)$ & $(3/2,1/2)$ & $(1/2,-1/2)$ & $(3/2,-1/2)$ & $(3/2,-3/2)$ \\ \hline
$\text{Vor}^{\mathbf{e}}_{bb'}$  & $2(n-1)+2$ & $2$ & $4$ & $2$ & $2$ & $4$ & $6$ \\ \hline
$C_b-C_{b'}$  & $-2n$ & $-2$ & $-4$ & $-2$ & $-2$ & $-4$ & $-6$ \\ \hline
\raisebox{5\height}{\text{Illustration}} & \includegraphics[width=0.08\textwidth]{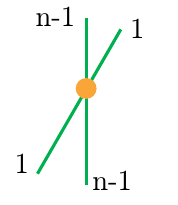} & \includegraphics[width=0.08\textwidth]{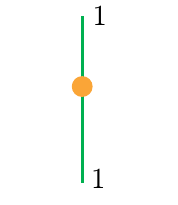} & \includegraphics[width=0.08\textwidth]{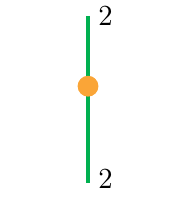} & \includegraphics[width=0.08\textwidth]{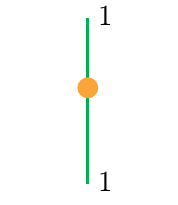} & \includegraphics[width=0.08\textwidth]{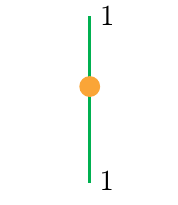} & \includegraphics[width=0.08\textwidth]{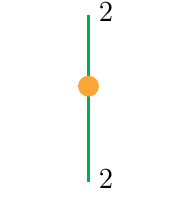} & \includegraphics[width=0.08\textwidth]{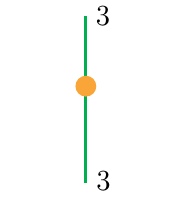} \\ \hline
\end{tabular}%
}
\caption{ 
The structure of vortex lines and the vorticity for isotropic Weyl, $n$-Weyl, pseudospin-1, and pseudospin-3/2 fermions. A number on the green-colored vortex half-line indicates its contribution to the interband vorticity ($\text{Vor}^{\mathbf{e}}_{bb'}$) of a surface enclosing the topological band-touching point (orange colored).
Owing to both effective Hamiltonians in \q{multifold-3D-H} having a symmetry that relates positive-energy eigenstates to negative-energy eigenstates, the interband Chern numbers $C_{b}-C_{b'}$ and interband vorticities are identical for $(b,b')$ and $(-b',-b)$. To resolve some of the above vortex lines, point-group-symmetric perturbations have been added to both effective Hamiltonians in \q{multifold-3D-H}, as detailed in App.~\ref{app:models-multi}. 
}
\label{fig:shift-multi-weyl}
\end{table}

\subsubsection{$n$-Weyl fermions and higher-order vortices}\la{sec:nweylvortex}

Optical vorticity is qualitatively different for the $(n{>}1)$-Weyl Hamiltonian:\cite{chen_multiweyl}
\e{
H^{n,w}\eq \bd\cdot \bsigma; \as \bd = (d_1,d_2,d_3)=\big(f_{n}\real  (k_+)^n,  f_{n} \imag(k_+)^n, v k_z\big); \as k_{\pm}=k_x\pm ik_y.
\la{multiWeyl-H}
}
(This Hamiltonian reduces to the isotropic Weyl Hamiltonian \q{hamiltonianisotropicweyl} for $n=1$ {and $f_1=v$}, but in this section, we focus on $n>1$.) For an \textit{enclosing $\bk$-surface} surrounding an $n$-fold Weyl band touching point, the Chern numbers are $C_c=-C_v=-n$.  Such a band touching is stable, meaning unsplittable into $n$ 1-fold Weyl points, assuming the crossing is between certain representations of an $M$-fold rotational symmetry, with  $n$ nontrivially dividing $M$.\cite{chen_multiweyl}\\

\noi{i} For $\be=\bz$, \fig{fig:shiftweyl}(b) illustrates a single vortex-antivortex line along the rotation-symmetric $k_z$ axis. This line intersects the  enclosing $\bk$-surface at two vortex points, which each point contributing $n$ to the vorticity of the surface, in accordance with the 
Chern-vorticity theorem [\q{chernvortex}]. The vortex-antivortex line will be described as \textit{higher-order}; more precisely, it is $(n{>}1)$-order, meaning that the shift vector winds by $\pm n$ for an infinitesimal loop linking the line [\q{chernvortex}]. It is not possible to split the $n$-order vortex-antivortex line into $n$ 1-order vortex-antivortex line without violating the same $M$-fold symmetry that prevents the splitting of the $n$-order Weyl. This is because the photonic shift vector field is
\e{
\text{$M$-fold rotation symmetric} \as \bS^{\bz}_{cv,\bk} \eq \{R_{z}^{\sma{2\pi/M}}\}^{\mo}\cdot \bS^{\bz}_{cv,R_{z}^{\sma{2\pi/M}}\cdot \bk}; \as R_z^{\theta}=\matrixthree{\cos\theta}{\sin\theta}{0}{-\sin\theta}{\cos\theta}{0}{0}{0}{1}, \la{shiftCinf}
}
this being a particular case of  a more general constraint [proven in \app{app:symmetryshift}]
\e{
\text{Point-group-symmetric}\as \bS^{\be}_{cv,\bk} 
\eq g^{s,\mo}\cdot \bS^{g^s\cdot\overline{\be}^g}_{cv,\gk}; \as \gk= i_g g^s\cdot \bk,\la{symmetryshift}
}
arising from the crystalline Hamiltonian having  a point-group symmetry $g$ that maps spacetime as 
\e{
\text{Point-group symmetry} \as g\circ (\br,t) \ri (g^s\cdot \br,i_g t); \as g^s \in O(3)\;\text{matrix}; \as i_g=\pm 1.
}
In \q{symmetryshift},  $\overline{A}^g=\overline{A}$ = (complex conjugate of $A$) if $g$ inverts time, and otherwise $\overline{A}^g=A$. \\

\noi{ii} For $\be\neq \bz$, there is an $(n-1)$-order vortex-antivortex line (lying along the rotation-symmetric $k_z$ axis and contributing $2n-2$ to the vorticity of the enclosing surface), and an additional 1-order vortex-antivortex line (non-parallel to the $k_z$ axis and contributing $2$ to the vorticity), as illustrated in Fig.~\ref{fig:shiftweyl}(c).\\

\subsubsection{Multifold fermions: Pseudospin-1 and -3/2}

The Chern-vorticity theorem [\q{chernvortex}] guarantees optical vorticity in all 3D topological semimetals associated with nontrivial Chern numbers.\footnote{3D topological semimetals (with nontrivial Chern numbers) fall into two categories: (a) having nontrivial spin-orbit coupling with spin-split energy bands, and (b) having negligible spin-orbit coupling with spin-degenerate energy bands. In category (a), bands are energy-nondegenerate at generic $\bk$; in category (b),  bands (restricted to one spin sector) are energy-nondegenerate at generic $\bk$.} These include also higher-angular-momentum generalizations of the 1-Weyl fermion [\q{hamiltonianisotropicweyl}]: while the 1-Weyl fermion has pseudospin-half, multifold fermions can have pseudospin-1 and pseudospin-3/2. \\

Pseudospin-1 fermions (sometimes called `spin-1 Weyl/chiral fermions')~\cite{bradlyn_multifold} are materialized in the family RhSi, CoSi, and AlPt, if the spin-orbit coupling is neglected and in a magnetic phase of Mn$_3$IrSi~\cite{Cano-Vergniory-MultifoldNodalPoints-2019}; pseudospin-3/2 fermions (sometimes called `Rarita-Schwinger(-Weyl) fermions') are materialized in RhSi, CoSi, and AlPt, in the presence of the spin-orbit coupling.\cite{bradlyn_multifold, Liang-Yu-SemimetalBothRaritaSchwingerWeyl-2016, Schroter-Felser-ObservationControlMaximal-2020}. Their respective effective Hamiltonians (at linear order in $\bk$) are \cite{bradlyn_multifold, Boettcher-InterplayTopologyElectronElectron-2020, Link-Herbut:2020}
\begin{equation}
\label{multifold-3D-H}
    \text{Pseudospin-1}\;\; \mathcal{H}^{s1} = 
    v(\hat{\mathbf{L}}\cdot\mathbf{k}) \as\text{and Pseudospin-3/2}\;\;  \mathcal{H}^{s3/2} = v(\hat{\mathbf{J}}\cdot\mathbf{k}),
\end{equation}
with  $\hat{L}_{i=1,2,3}$ (resp. $\hat{J}_{i=1,2,3}$) being spin-1 (resp. spin-3/2) matrices detailed in App.~\ref{app:models-multi}.  \\

$\mathcal{H}^{s1}$ describes a touching of three bands which we index by their angular momenta: $1,0$ and $-1$. There are then three possible pairings of band indices: $(1,0), (1,-1), (0,-1)$, and the Chern-vorticity theorem [\q{chernvortex}] applies separately to each pairing. Likewise, $\mathcal{H}^{s3/2}$ describes a touching of four bands (indexed by $3/2,1/2,-1/2,-3/2$) and the theorem applies separately to each of the six pairings. For each pairing, interband Chern numbers and optical vortex lines are depicted in Tab.~\ref{fig:shift-multi-weyl}, demonstrating full consistency with the theorem.

\subsection{Optical vorticity of 2D Dirac systems}\la{sec:vorticity2D}

A number of prominent 2D Dirac systems can be described by a
\e{
\text{Mass-twisted Dirac Hamiltonian}\as H^{\sma{\text{D}}}_{k_x,k_y}= v(k_x\sigma_1 +k_y\sigma_2) + \big(\mD v^2 + \text{Twist}_{\bk}\big) \sigma_3,\la{masstwistH}
}
with $(k_x,k_y)$ parametrizing the 2D Brillouin zone (of a 2D material) or a reduced Brillouin zone (associated with a surface facet of a 3D crystal). The name `mass-twisted' emphasizes a $\bk$-dependent Twist term which adds to the Dirac-massive $\mD v^2$; what Twist to add depends on the model/material application, as exemplified in \tab{tab:twist}, which also lists their compatible rotational symmetry: $C_{2z}, SO(2)$ and $C_{3z}$.\\

\begin{table}[H]
\centering
\begin{tabular} {|c|c|c|} \hline
			
\text{Twist}  & \text{  Symmetry  } & \text{Application}   \\  \hline  
		  $v_{\sma{T}} \,k_x$ & $C_{2y}T$ & $\bk$-plane of BiTeI [Ref. \cite{zhuAA_anomalousshift}] \\ 
          {$(k_x^2+k_y^2)/(2m)$} & $SO(2)$ & \text{Chern insulator} [Ref.~\cite{qi2005} ]; \;\text{TMD} [Ref.~\cite{Rostami-Asgari-EffectiveLatticeHamiltonian-2013}]\\
$\text{Tw}_{3}\, k_x(k_x^2-3k_y^2)\as $ & $C_{3z}$		& \text{Topological surface states} [Ref.~\cite{Fu-HexagonalWarpingEffects-2009}]\\ \hline
\end{tabular}
\caption{  Twists of the Dirac Hamiltonian $\HD$ in \q{masstwistH}. The first column gives the $\bk$-dependence of $\text{Twist}_{\bk}$, the second column states the rotational symmetry of  $\HD$ with a given Twist, and the third column lists several applications of $\HD$, with TMD $\equiv$ transition metal dichalcogenides.}\la{tab:twist}
\end{table}

\begin{figure}[th!]
\centering
\includegraphics[width=0.9\textwidth]{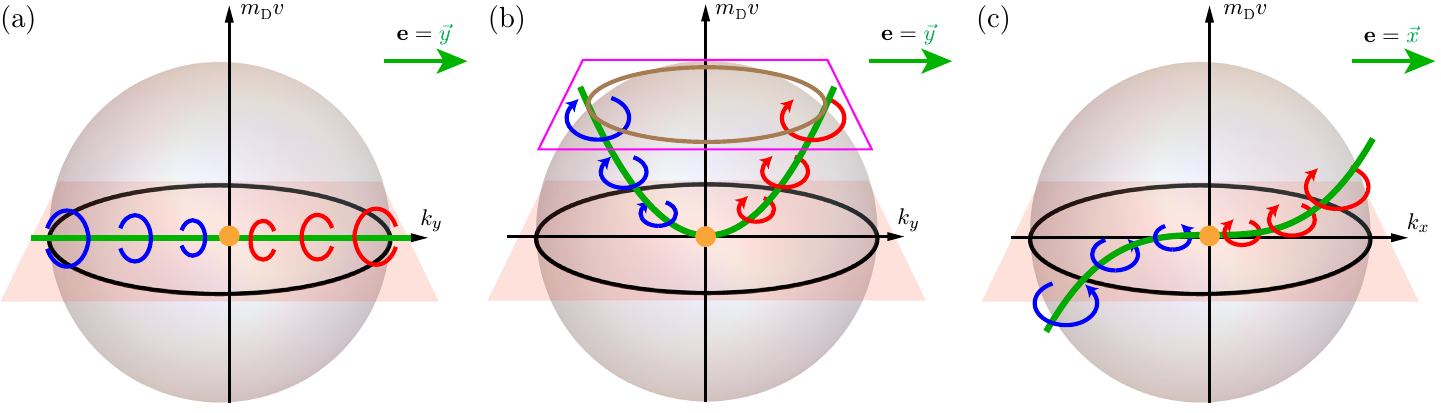}
\caption{ 
(a) $\by$-vortex line (green) of a massive Dirac Hamiltonian [\q{masstwistH}] with Twist $=0$. (b) The same $\by$-vortex line curves into a quadratic parabola when Twist $=(k_x^2+k_y^2)/2m$.
(c) The $\bx$-vortex   line curves into a cubic parabola when Twist $=k_x(k_x^2 -3k_y^2)$.
}
\label{fig:massivedirac}
\end{figure}

Absent the Twist, the optical affinity vanishes  only if $\mD=0$, and this vanishing occurs along a straight $\bk$-line parallel to $\be$ and intersecting $\bk=\bze$, as deducible from \q{hamiltonianisotropicweyl} [with $vk_z$ re-interpreted as $(\mD v^2 + \text{Twist}_{\bk}$)] and illustrated in \fig{fig:massivedirac}(a) for $\be=\by$. If a Twist term is added to the Hamiltonian, the zero-affinity line curves. How it curves depends on the choice of Twist.\\

Suppose our choice is the $SO(2)$-symmetric Twist from \tab{tab:twist}.   By `$SO(2)$', we mean a $U(1)$ rotational symmetry about the z axis, with $\sz$ being the matrix-represented generator of rotation.
The resultant Hamiltonian is the small-momentum effective Hamiltonian of a representative Chern insulator,\cite{bernevig_book} such that $\mD=0$ marks a topological phase transition associated with a unit change of the Chern number.\footnote{
{In the untwisted case, the integrated Berry flux is $\int d^2 k\, \Omega_{xy}^{v}/(2\pi) = \mD/4 \int_0^{\infty} dx/\left(\mD^2 +x\right)^{3/2} = \sgn{\,\mD}/2$~\cite{bernevig_book}.}
}

\fig{fig:massivedirac}(b) shows how the $SO(2)$-symmetric Twist curves the zero-$|A^y_{cv}|^2$ line  [in $(k_x,k_y,\mD v)$-space] into a quadratic parabola, such that the parabola  intersects a constant-$(\mD v>0)$ plane at two isolated $\bk$-points, which are then the $\by$-vortex points for the Hamiltonian with that value of $\mD$.   This is consistent with a previous observation in \s{sec:chernvortex}, namely that optical vortex centers are generically points in a 2D $\bk$-space; after all, Twist terms of one form or another generically exist as higher-order-in-momentum, Taylor-series corrections to the massive Dirac Hamiltonian. \\

To appreciate the qualitative features of \fig{fig:massivedirac}(b), we highlight the role of 
 \textit{magnetic reflection symmetry}  ($TR_x=$ composition of time-reversal  and x-reflection) in the Hamiltonian and the shift vector field:
 \e{
\widehat{TR_x}H^{\sma{\text{D}}}_{k_xk_y}\widehat{TR_x}^{\mone}=H^{\sma{\text{D}}}_{k_x,{-}k_y}; \as \widehat{TR_x}=K=\,\text{complex conjugation};
}
\e{
\bS^{\be}_{cv,k_x,k_y} \eq \overleftrightarrow{R_x}\cdot \bS^{\be}_{cv,k_x,-k_y}; \as \overleftrightarrow{R_x}=\matrixtwo{-1}{0}{0}{1}; \as \be=\vec{x} \;\text{or}\; \vec{y}.\la{shiftTRx}
}
The last equation is a particular case of the general symmetry constraint [\q{symmetryshift}].\footnote{Bear in mind the shift vector is invariant under scaling $\be$ by a real multiplicative constant:  $\bS^{\be}_{cv\bk}=\bS^{\text{constant}\cdot \be}_{cv\bk}$. After all, $\bS$ depends on $\be$ only through $\nabk \arg \{\be \cdot \bA_{cv}\}$, which is scale-invariant.  Moreover, for the shift vector $\bS^{\be}_{cv\bk}$ of 2D electrons in the $xy$-plane, we may as well restrict $\be$ to lie in the $xy$-plane, because $\bS^{\be}_{cv\bk}=\bS^{\be'}_{cv\bk}$ if $\be$ and $\be'$ differ only in their z components. This is because of the aforementioned scale invariance and the $k_z$-independence of energy eigenstates $\ket{u_{b,k_x,k_y}}$  in a 2D material.}  Because a vortex is a circulation of the shift vector [\q{photonicshift}], \q{shiftTRx} implies that $TR_x$ maps a $\by$-vortex at generic $k_y$ to a $\by$-vortex at $-k_y$, and not to an $\by$-anti-vortex at $-k_y$; in other words, $TR_x$-related $\by$-vortex pairs have the same circulation, as illustrated in \fig{fig:massivedirac}(b).  Because the $\by$-vortex line bends upward (toward $\mD v=+\infty$), the interband vorticity changes by $\Delta \mathrm{Vor}^{\mathbf{y}}_{cv}=-2$ as  $\mD$ is tuned from negative to positive. 
This is consistent with the Chern-vorticity theorem [\q{chernvortex}], since the interband Chern number, $C_c-C_v$, also changes by $-2$ across the transition.\footnote{For the $k\cdot p$ Dirac model, when examining the Chern-vorticity theorem, it is only meaningful to consider the changes in the vorticity and Chern number across the transition, rather than their absolute values, because the continuum description is valid only locally near the Dirac point and the absolute topological invariants generally depend on the ultraviolet regularization.}

Due to the $SO(2)$ rotational symmetry,  $\Delta\text{Vor}^{\vec{y}}_{cv}=-2$ across the transition leads us immediately to  $\Delta\text{Vor}^{\be}_{cv}=-2$ for any $\be$ in the xy-plane; this means that if we embed the parabolic vortex line in a plane (which is the $\mD v-k_y$ plane for $\be=\vec{y}$), then the plane simply rotates continuously as $\be$ is rotated, this being a  rotation-symmetric constraint on the shift vector field [\q{shiftCinf} with $M$ being arbitrarily large].\\

Qualitatively different vortex lines are possible. \fig{fig:massivedirac}(c) shows a cubic-parabolic $\bx$-vortex line for the mass-twisted Dirac Hamiltonian [\q{masstwistH}] with a $C_{3z}$-symmetric Twist [\tab{tab:twist}], which apply to the surface states of topological insulators.\cite{Fu-HexagonalWarpingEffects-2009}

\subsection{Interlude: Optical vortices from the director field}\la{sec:directorfield}

\subsubsection{Vortex-director relation for 3D $\bk$-space}

To represent where the vortex lines are in $\bk$-space, we could pick some representative polarization vectors $\{\be_1,\be_2,\ldots,\be_N\}$ and draw $N$ corresponding vortex lines, labeling each line by the corresponding $\be_j$.  Such a drawing quickly becomes convoluted as $N\ri \infty$, i.e., as we attempt to capture the where-information for all polarizations. \\

The challenge is to compactly represent the where-information of all vortices. Our proposed solution is to represent this infinite family of vortex lines  by a  real vector field $\bscrd_{cv\bk}$,  such that the condition $\bscrd_{cv\bk}=\be$ is generically satisfied by  lines in a 3D Brillouin zone, which hypothetically coincide with the $\be$-vortex lines.  In other words, for any fixed point $\bk$, $\bscrd_{cv\bk}$ is the direction of the  real polarization vector  $\be$ for which an $\be$-vortex hypothetically exists. For this reason, we call $\bscrd_{cv\bk}$ the $\bscrd$\textit{irector field}.\\

For any pair of bands labeled $c$ and $v$, we propose that the \dfield is  the normalized cross product  of the real and imaginary components of the  interband Berry connection vector field:
\e{\bscrd\text{irector field} \as \bscrd_{cv\bk} \eq  \f{\bD_{cv\bk}}{|\bD_{cv\bk}|};\as \text{Director field}\as \bD_{cv\bk}=\bR_{cv\bk} \times \bI_{cv\bk}; \as \bR_{cv\bk}=\real \bA_{cv\bk}; \as \bI_{cv\bk}=\imag \bA_{cv\bk}.\la{directorfield}
}
The above cross product is invariant\footnote{
This transformation sends $\bA_{cv\bk}\ri\; \bA_{cv\bk} e^{-i\phi_{cv\bk}}$ with $\phi_{cv\bk}=\phi_{c\bk}-\phi_{v\bk}$. Because $\bR \ri \bR \cos \phi_{cv} +\bI \sin \phi_{cv}$ and $\bI \ri -\bR \sin \phi_{cv} +\bI \cos \phi_{cv}$, a gauge transformation may be viewed as a (generalized) rotation of (non)orthogonal vectors within the plane spanned by $\bR$ and $\bI$;  such a rotation does not affect the cross product which points out of plane.}
under the gauge transformation $\ket{u_{b\bk}}\ri e^{i\phi_{b\bk}}\ket{u_{b\bk}}$, with $b=v$ or $c$.   Crucially,   $\bD_{cv\bk}\cdot \bA_{cv\bk}=0$,
\footnote{
Because $\bD_{cv}$ is real while $\bA_{cv}$ is complex,
\e{|\bD\cdot \bA_{cv}|^2 \eq |\bD\cdot \bR|^2 + |\bD\cdot \bI|^2= |\bR \times \bI\cdot \bR|^2 + |\bR \times \bI\cdot \bI|^2
 =0.}
At a generic wavevector, we assume that both $\bD_{cv}$ and $\bA_{cv}$ are nonzero, finite vectors; then  $\bD_{cv}\cdot \bA_{cv}=0$ if and only if $\bscrd_{cv}\cdot \bA_{cv}=0$.
} 
which implies that $\bscrd_{cv\bk}$ is the direction along which $\bA_{cv\bk}$ vanishes:\footnote{ 
Moreover, there can be no other direction  along which  $\bA_{cv\bk}$  vanishes:
\e{
\text{No-misdirection lemma}: \as \be\cdot \bA_{cv\bk} =0\as  \text{is incompatible with} \as \bscrd_{cv\bk} =\be' \neq \pm \be,\la{nomisdirection}
}
  Proof by contradiction: supposing the two statements are compatible, then $\be\cdot \bR=\be\cdot \bI=\be'\cdot \bR=\be'\cdot \bI=0$, implying that $\bR // \bI // \be\times \be'$, with `$//$' meaning `collinear to'. This implies $\bD_{cv\bk}=\bze$, contradicting $\bscrd_{cv\bk} =\be'$.
}
\e{
\bD_{cv\bk_0}\neq 0 \iand \bscrd_{cv\bk_0}=\be \imp \be\cdot \bA_{cv\bk_0}=0 \as \text{(optical zero)}.
}
The last equality defines an \textit{optical zero} at $\bk_0$. An optical vortex is more than an optical zero: in addition to the defining condition for an optical zero, a vortex  requires that the interband Berry phase (arg$\be\cdot \bA_{cv\bk}$) winds as $\bk$ is varied in a small circle around $\bk_0$ [\q{photonicshift}]. Optical zeros without interband Berry phase winding exist.\cite{aa_quantizationintrainter} If $\bscrd_{cv\bk_0}=\be$, can one be sure that there is an $\be$-vortex at $\bk_0$ and not a winding-less optical zero? Yes, if the \dfield has a nonzero 
\e{
\text{Skyrmion density}\as \bscrf_{cv\bk} \eq \eps_{abc}\scrd^a_{cv} \nabk \scrd^b_{cv}\times \nabk \scrd^c_{cv}\bigg|_{{\bk}} \la{skyrmiondensity}
}
at $\bk_0$, as representatively illustrated in \fig{fig:localskyrmion}.
For order-one vortices (defined to have  interband Berry phase winding equaling $\pm 1$), we propose a duality [proven  in \app{app:skyrmionvortex}] between vortices of the interband Berry phase and skyrmions of the $\bscrd$\text{irector field}:
\e{
\text{Skyrmion-vortex duality:}  \as \bscrd_{cv\bk_0}=\be \as\text{and}\as  \be\cdot \bscrf_{cv\bk_0} \neq &\; 0 \iff \;\be\text{-zero}\;= \;\text{order-1}\;\be\text{-vortex at }\bk_0.  \la{skyrmionvortexduality}
}\\

For illustration, we present the 
\dfield and skyrmion density
of the isotropic Weyl Hamiltonian [\q{hamiltonianisotropicweyl}]:
\e{
\text{Isotropic Weyl:} \as \bscrd_{cv\bk}=-\hbd=-\bk/k; \as \bscrf=-2\hbd/k^2.
}
Thus for $\be=-\hbd$, both conditions on the left-hand side of the skyrmion-vortex duality [\q{skyrmionvortexduality}] apply, implying that there is an $\hbd$-vortex at $\bk$.  
 As illustrated in Fig 4(a), the SO(3)-symmetric, hedgehog-like \dfield implies that vortex lines are straight lines intersecting the Weyl point. \\

 In contrast, the director field of  the $(n{>}1)$-Weyl Hamiltonian [\q{multiWeyl-H}] is anisotropic with curved vortex lines intersecting the $n$-Weyl point:
\e{
 \text{$n$-Weyl director field}
\as \bD_{cv\bk} = -\f{n}{4}\f{ {f_{n}^2 v}\, (k_x^2+k_y^2)^{n-1}}{|\bd_{\bk}|^3}\bq; \as \bq=(k_x,k_y,nk_z), \la{directormultiWeyl}
}
with $|\bd_{\bk}|$ determined from \q{multiWeyl-H}. Observe that  the $n$-Weyl director field vanishes everywhere along the rotation-symmetric $k_z$ axis, which is the same axis where  an $\be$-vortex line exists for any $\be$, as previously ascertained in \s{sec:nweylvortex}.  This rotation-symmetric vortex line  may be said to be \textit{directionless}, in the sense that the director field just vanishes. To recapitulate,  an $\be$-vortex is either directed by the \dfield (meaning $\bscrd // \be$), 
 or directionless  (if $\bD=\bze$), and admits no other possibility; in particular, a misdirected vortex does not exist [\q{nomisdirection}].\\

\begin{figure}[th!]
\centering
\includegraphics[width=0.9\textwidth]{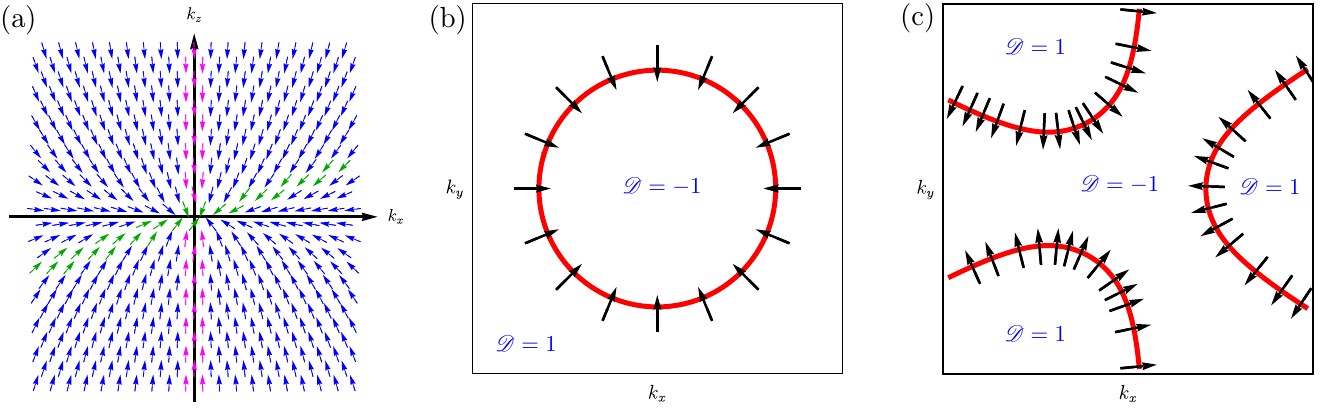}
\caption{
(a) \dfield of the $3$-Weyl fermion, restricted to the $k_x-k_z$ plane. The interpolation of all arrows collinear with a directional vector $\be$ is 
the $\bm{e}$-vortex line; representative examples are provided  with $\bm{e}= (\bx+\bz)/\sqrt{2}$ [green arrows] and   $\be=\bm{z}$ [magenta arrows].
For the massive Dirac Hamiltonian with an $SO(2)$-symmetric twist [panel (b)] and $C_{3z}$-symmetric twist [panel (c)], the \dfield  is indicated by blue values, and a black arrow (at $\bk$ on the red director-zero curve) indicates the direction $\be$ of an $\be$-vortex point at $\bk$. 
} 
\label{fig:directorchern}
\end{figure}

\subsubsection{Vortex-director relation for 2D $\bk$-space}

For the $(k_x,k_y)$-space of a 2D material/surface, the director field [\q{directorfield}] can only be collinear to $\vec{z}$: $\bD_{cv\bk}=D_{cv\bk}\vec{z}$, hence it is reducible to a real-valued
\e{
\scrd\text{irector scalar field} \as \scrd_{cv\bk}=\tf{D_{cv\bk}}{|D_{cv\bk}|}; \as D_{cv\bk}=R^x_{cv\bk}I^y_{cv\bk}-R^y_{cv\bk}I^x_{cv\bk},
}
with $R_{cv\bk}^x$ (resp. $I_{cv\bk}^y$) meaning the x-component of the real part (resp. y-component of the imaginary part) of the interband Berry connection [\q{directorfield}].  Because it takes one real parameter to tune a real scalar number to zero, the zeros of the director field are co-dimension one, meaning they form $\bk$-lines (in our 2D $\bk$-space) separating domains with opposite signs for $\scrd_{cv\bk}=\pm 1$. We will refer to these lines as \textit{director-zero lines}. Because of the no-misdirection lemma [\q{nomisdirection}], an $\be$-vortex point (with $\be$ in the xy-plane) can only lie in a director-zero line.  In other words, all $\be$-vortex points (in 2D materials/surfaces) are directionless, like the higher-order vortex of $n$-fold Weyl fermions in \s{sec:isotropicweyl}.\\

\begin{figure}[th!]
\centering
\includegraphics[width= 0.9\textwidth]{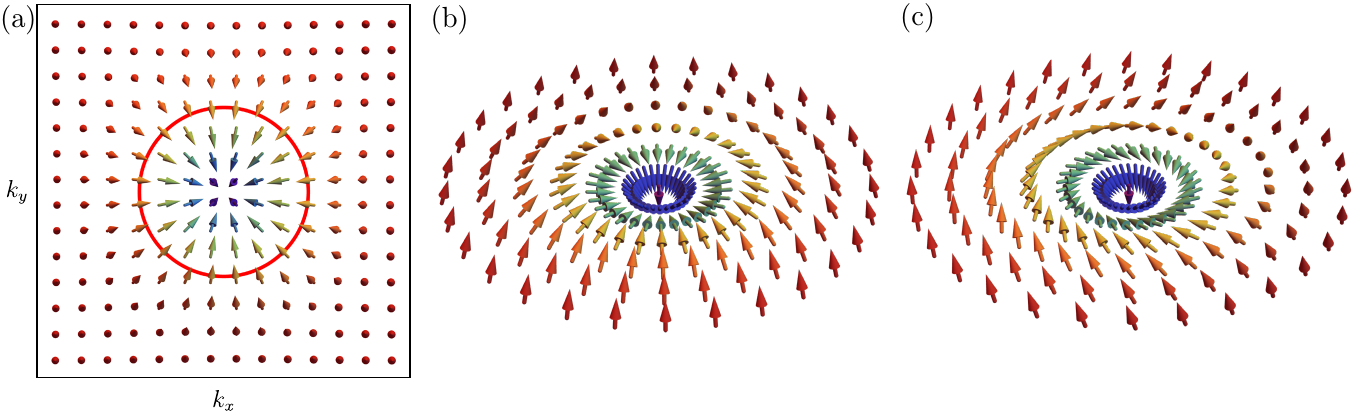}
\caption{
(a) Skyrmionic director field of quasi-2D Chern insulator obtained by adding a Hamiltonian perturbation, which nontrivially depends on $k_z$. Such a perturbation transforms the scalar field of $\Z_2$ domains to a skyrmionic director field, as is consistent with the skyrmion-vortex duality in \q{skyrmionvortexduality}. 
(b) and (c) The director field in a neighborhood of $\bk_0$ determines vorticity: as long as the director field is locally tilting into a hedgehog skyrmion (b) or a helical skyrmion (c) or some linear combination of the two, one can be sure an optical zero is an optical vortex.
} 
\label{fig:localskyrmion}
\end{figure}

An illustrative example is the massive Dirac Hamiltonian [\q{masstwistH}] with an $SO(2)$-symmetric Twist [\tab{tab:twist}]. As observed in \s{sec:vorticity2D}, the set of vortex points (for all real $\be$) fills out a circle of radius $k_r=\sqrt{m \mD v}$, which is nothing more than a director-zero circle that separates domains with opposite signs for $\scrd_{cv\bk}$, as illustrated in \fig{fig:directorchern}(a). Defining   $\be(\phi)$ as the directional vector of a vortex point at the director-zero coordinate $(k_x,k_y)=(k_r\cos \phi,k_r\sin \phi)$, $\be(\phi)$ rotates by $2\pi$ as $\phi$ is advanced by $2\pi$, as illustrated by the black arrows in \fig{fig:directorchern}(a).  If the $SO(2)$-symmetric mass-twisted Dirac Hamiltonian [\q{masstwistH}] is viewed as a quasi-2D Hamiltonian (over 3D $\bk$-space) with no $k_z$ dependence: $H_{\text{q2D}}(k_x,k_y,k_z)=H^{\sma{\text{D}}}_{k_x,k_y}$, then a generic perturbation to $H_{\text{q2D}}$ will transform the \dfield from having two polarized domains  to a skyrmionic vector field [\fig{fig:localskyrmion}(a)], in accordance with the skyrmion-vortex duality [\q{skyrmionvortexduality}]. \\

For a final and brief illustration in \fig{fig:directorchern}(c), the director-zero lines are three-fold symmetric and open for the massive Dirac Hamiltonian [\q{masstwistH}] with a $C_{3z}$-symmetric Twist [\tab{tab:twist}].

\section{Ballistic photovoltaic current in disordered Dirac-Weyl materials}\label{sec:current}

\subsection{Momentum-space vortices interplay with real-space crystalline disorder}\la{sec:interplay}

\newtcolorbox[auto counter]{mybox}[2][]{
    colback=white, colframe=black,
    title={1: #2}, #1,
    label={box:centraltheme}
}
\begin{mybox}{Central theme}
{Momentum-space vortices interplay with real-space crystalline disorder} to produce a {ballistic photovoltaic current}
that is {sensitive to the type of Dirac-Weyl fermion, disordering impurity and light polarization}.
\end{mybox}
Let us unpack this statement.\\

\noindent \textbf{Ballistic photovoltaic current} 

In a nonmagnetic medium, the bulk photovoltaic current  is a sum of two components: the shift current and the `ballistic photovoltaic current' (in short, \textit{ballistic photocurrent}).\cite{sturmanfridkin_book} 
While the shift current is an effect of the  band-off-diagonal elements of the velocity matrix and can be nonzero even if the photoexcited electron distribution is symmetric ($f_{\bk}=f_{-\bk}$),\cite{belinicher_kinetictheory} the 
ballistic photocurrent relies on the distribution being asymmetric ($f_{\bk}\neq f_{-\bk}$) and is proportional to the group velocity $v_{\bk}$ (the diagonal element of the velocity matrix):\cite{belinicher_ballistic}
\e{
&\text{Ballistic photocurrent:} \as \mathbf{j}_{\text{ballistic}}=-\frac{|e|}{2\mathcal{V}}\sum_{B}\bv_{B}(f_{B}-f_{-B}).\la{defineballistic}
}
Here $B=(b,\bk)$ is the composite index including the band index and wavevector, $-B=(b,-\bk)$ is its time-reversed partner, and $(f_{B}-f_{-B})$ is proportional to the photo-excitation light intensity. We will largely focus on two-band models in which $b=c$ (conduction) or $=v$ (valence). A necessary condition for $f_{B}\neq f_{-B}$ (in the steady state of a photoexcited medium) is that the \textit{transition rate matrix elements} $W_{B \lea  B'}$ [defined through
\e{
\text{(Collisional integral)}_{B}= \sum_{B'}\big(W_{B,B'}f_{B'}- W_{B',B}f_{B}\big)=\; \text{Incoming transition rate - Outgoing transition rate}\bigg]
\la{skew}
}
are asymmetric: $W_{B,B'} \neq W_{-B \lea -B'}$. As illustrated in \fig{fig:ballistic}(a), collisional mechanisms that result in such asymmetry will be called \textit{skew scattering}, and we will largely focus on skew scattering mediated by crystalline disorder/impurities.  \\

\noindent \textbf{Momentum-space vortices interplay with real-space crystalline disorder} 

2D $n$-Dirac and 3D $n$-Weyl fermions have topologically-enforced optical vortices [\s{sec:chernvortex}].
When a 1-Weyl semimetal is radiated by monochromatic light with electric-field amplitude $|\bE|$, complex polarization vector $\be$ and frequency $\omega$, photo-excitation occurs at all $\bk$ on the
\e{
\text{Excitation surface defined by}\as  E_{cv\bk}\equiv E_{c\bk}-E_{v\bk}=\hbar\omega.\la{defineexcsurface}
}
In this context, the excitation surface is spherical and encloses the Weyl point.  \fig{fig:shiftweyl} depicts how the $\vec{z}$-vortex line (for a 1-Weyl semimetal) emerges from the Weyl point and runs along the $k_z$ axis, thus intersecting the spherical excitation surface  at the north and south poles.  Because the optical affinity ($|\be\cdot \mathbf{A}_{cv,\bk}|^2=$ squared dipole matrix element ) vanishes along the vortex line, 
 the $\bk$-dependent photo-excitation rate (which is shown in \q{photo-excitationrate} to be proportional to the optical affinity)\footnote{In \q{photo-excitationrate}, we have  assumed $f_{v\bk}-f_{c\bk}\approx 1$ and therefore  the photo-excitation rate does not explicitly depend on the carrier distribution. This is justified if the thermal energy $k_BT$ is small compared to one-particle energies (measured relative to the chemical potential) for $\bk$ on the excitation surface, and if the laser intensity is not too high. More general expressions can be found in \app{app:ballisticphotocurrent}.}
is anisotropic over the excitation surface: minimal at the poles and maximal at the equator, as illustrated in \fig{fig:ballistic}(b). 
 \e{
\big(\text{Photoexcitation rate}\big)_{c\bk}= \frac{2\pi}{\hbar}e^2|\cale|^2|\be\cdot \mathbf{A}_{cv,\bk}|^2\delta(E_{cv\bk}-\hbar\omega)= -\big(\text{Photoexcitation rate}\big)_{v\bk}. \la{photo-excitationrate}
 }
 (The above formula holds for a driving monochromatic light wave whose electric field has the Fourier transform $\bcale_{\bq\omega}= |\bcale|\be,$ with $\be$ the polarization vector.) The vortex-induced anisotropy is quantified by the deviation of the photo-excitation rate from its iso-energy average: 
 \e{
 \big(\substack{\text{Deviating}\\ \text{photo-excitation rate}}\big)_{B}= \big(\text{Photoexcitation rate}\big)_{B}-\bigg[\big(\text{Photoexcitation rate}\big)_{B}\bigg];\la{deviatephotoexcite}\\
 \text{Iso-energy average}: \as [Fun_{B}]= \frac{\sum_{B'}Fun_{B'}\delta(E_{B'B})}{\sum_{B'}\delta(E_{B'B})}; \as E_{B'B}=E_{B'}-E_B,\la{isoenergyave}
  }
 because the excitation surface is also an iso-energy surface for $n$-Dirac/Weyl fermions with the electron-hole symmetry:  $E_{c\bk}=-E_{v\bk}$.\footnote{
$E_{c\bk}=-E_{v\bk}$ holds for any effective $\bk\cdot \bp$ Hamiltonian of  of the form: $H= \bd(\bk)\cdot \bsigma$. It follows that  the excitation surface is a surface of constant energy: $E_{c\bk}=\hbar\omega/2=-E_{v\bk}$.} \\

 Consider a hot electron that is photoexcited at a representative point ($\bk'$) on the equator (of the excitation surface) and subsequently collides with an impurity and scatters elastically to $\bk$ at the north pole. The quantum amplitude for the direct transition $(\bk\lea \bk')$ is first order in the impurity potential ($\Vim$); in Feynman-addition, there is also an indirect, second-order transition via an intermediate one-electron state: $(\bk\lea \bk''\lea \bk')$, as illustrated in \fig{fig:ballistic}(c). Quantum interference between the direct and indirect transitions is encoded in the gauge-invariant amplitude for an electron to make a loop:
 \e{
 \Vim_{\bk\bk''}\Vim_{\bk''\bk'}\Vim_{\bk'\bk}=\cbraopket{\psi_{\bk}}{\Vim}{\psi_{\bk''}}\cbraopket{\psi_{\bk''}}{\Vim}{\psi_{\bk'}}\cbraopket{\psi_{\bk'}}{\Vim}{\psi_{\bk}}\as (\text{band index suppressed}).\la{threepointBinvariant}
 }
Such interference is generically not symmetric in $(\bk \lea \bk') \ri (-\bk \lea -\bk')$ and results in third-order, elastic skew scattering which is proportional to the number ($\Nimp$) of impurities:
\e{
(\text{Asymmetric transition rate})_{\bk\lea\bk'}=  \tf{W_{\bk \lea \bk'} -W_{-\bk \lea -\bk'}}{2}= \tf{(2\pi)^2}{\hbar}\Nimp\delta(E_{\bk\bk'})   \imag \sum_{\bk''} \Vim_{\bk\bk''}\Vim_{\bk''\bk'}\Vim_{\bk'\bk}\delta(E_{\bk\bk''}).\la{skewscattering}
}\\

To recapitulate, the topologically-guaranteed momentum-space vortex  induces a deviating photo-excitation rate [\q{deviatephotoexcite}] which activates an impurity-mediated skew scattering [\q{skewscattering}]; 
the ballistic photocurrent is then a convolution of the deviating photo-excitation rate with the asymmetric transition rate:
\e{
 \mathbf{j}_{\text{ballistic}}=-\frac{|e|}{\calv}\sum_{B,B'}\big(\substack{\text{Group}\\ \text{velocity}}\big)_{B} \big(\substack{\text{Momentum}\\ \text{relaxation time}}\big)^2_{B} \big(\substack{\text{Asymmetric}\\ \text{transition rate}}\big)_{B,B'}\big(\substack{\text{Deviating}\\ \text{photo-excitation rate}}\big)_{B'},  \la{jballisticinwords}
}
with the $(\text{momentum-relaxation time})^2$ providing the saturation time for both rates.\footnote{The momentum-relaxation time is defined more precisely in \s{sec:nweylvortexintersection}. Going beyond the momentum-relaxation-time approximation, a more general expression for $\mathbf{j}_{\text{ballistic}}$ is available in \app{app:solvekinetic}.}
Assuming the  two photoexcited bands are electron-hole symmetric [see previous footnote], skew scattering is purely intraband with an equal contribution from both conduction and valence bands; then we may as well drop the band index and multiply the current by a factor of two:
\begin{equation}
\label{eq:jball}
\begin{aligned}
\mathbf{j}_{\text{ballistic}}&=-2_{\text{e-h}}\frac{4\pi^3|e|^3}{\hbar^2}(\tau^{mr}_{E_{ex}})^2 n_{\text{imp}}|\mathbf{E}|^2\sum_{\mathbf{k},\mathbf{k}^{\prime},\mathbf{k}^{\prime\prime}} \mathbf{v}_{\mathbf{k}}\;\operatorname{Im}(V^{\text{im}}_{\mathbf{k}^{\prime}\mathbf{k}}V^{\text{im}}_{\mathbf{k}\mathbf{k}^{\prime\prime}}V^{\text{im}}_{\mathbf{k}^{\prime\prime}\mathbf{k}^{\prime}})\delta(E_{\mathbf{k},\mathbf{k}''})\,\delta(E_{\mathbf{k},\mathbf{k}'})   \; \delta |\be\cdot \mathbf{A}_{cv,\mathbf{k}^{\prime}}|^2 \,\delta(E_{\mathbf{k}'}-E_{ex}).
\end{aligned}
\end{equation}
Here, $2_{\text{e-h}}=2$, with the subscript reminding us that  the factor originates from electron-hole symmetry;\footnote{A factor of half emerges from $\delta(E_{cv\bk}-\hbar\omega)= \delta(2E_{c\bk}-\hbar\omega)=\delta(E_{c\bk}-\hbar\omega)/2$.}  $\tau^{mr}_{E_{ex}}$ is the energy-dependent momentum relaxation time evaluated at the excitation energy $E_{ex}=\hbar\omega/2$;  $n_{\text{imp}}\equiv N_{\text{imp}}/\mathcal{V}$ is the impurity density; and
\e{
\text{Affinity deviant} \as  \delta |\be \cdot \mathbf{A}_{cv,\bk}|^2 \eq |\be \cdot \mathbf{A}_{cv,\bk}|^2-\big[\;|\be \cdot \mathbf{A}_{cv,\bk}|^2\;\big]\la{affinitydeviant}
}
is the deviation of the optical affinity from its iso-energy average.\\

In subsequent model calculations, it is convenient to  present not $\jball$ but the response tensor $\bB^{\text{de}}$ defined by
\e{
&j_{\text{ballistic},a} =\sum_{b,c}\big(\,B^{\text{de},l}_{abc}(\omega)+B^{\text{de},c}_{abc}(\omega)\,\big)e_b\overline{e_c}|\cale|^2; \as \bcale_{\bq\omega}= |\cale|\be; \la{Sigmaimdefine}\\
&\text{Linear defect-mediated ballistic photovoltaic tensor}\as B^{\text{de},l}_{abc}=B^{\text{de},l}_{acb} \in \R; \la{Sigmaimldefine}\\
&\text{Circular defect-mediated ballistic photovoltaic tensor}\as B^{\text{de},c}_{abc}=-B^{\text{de},c}_{acb}\in i \R, \la{Sigmaimcdefine}
}
The linear component is the $\jball$ response to linearly-polarized light; the circular component is the differential $\jball$ response to right-handed vs left-handed circularly-polarized light, and does not contribute in the response to unpolarized light.\cite{sturmanfridkin_book}\\

\noindent \textbf{Sensitivity to the type of Dirac-Weyl fermion, the type of disordering impurity and the light polarization}
 
 In conventional noncentric semiconductors,  the band gap sets the frequency threshold  for photo-excitation, i.e., a bulk photovoltaic current exists only for $\hbar\omega>$ band gap. In contrast, for topological semimetals, the vanishing of the band gap  is enforced by a combination of symmetry and topology, and the frequency threshold for photo-excitation is set not by the band gap  but by the chemical potential (which is measured relative to the Dirac-Weyl point, and is much smaller than semiconducting band gaps). Therefore, it is the low-frequency bulk photovoltaic response that distinguishes topological semimetals from conventional semiconductors. In this work, we focus on how the low-frequency ballistic photocurrent scales with frequency: 
 \e{
 \jball \propto \omega^{\text{exponent}} \la{exponent}
 }
 for impurity-mediated, intra-pocket/valley  scattering,
and elucidate how this \textit{frequency exponent} depends on  the type of Dirac-Weyl fermion, the type of disordering impurity and the light polarization. A representative result for $n$-Weyl fermions was given in \fig{fig:nweylexponent}  and will be greatly generalized in \s{sec:powerlaws}.  \\

Let us make several remarks to generalize the above story: 

\noi{i} The dictum of 
\e{
\substack{\text{Optical vortex in}\\ \text{Dirac-Weyl semimetal}} \as \ri\as  \substack{\text{Anisotropic photo-excitation \&} \\ \text{Asymmetric transition rate}} \as\ri\as \substack{\text{Ballistic}\\ \text{photocurrent}} \la{dictum}
}
generalizes to   3D $n$-Weyl semimetals [\s{sec:nweylvortexintersection}] and 2D $n$-Dirac [\s{sec:nDirac-all}], with the caveat that for massive $n$-Dirac semimetals, the vortex line is proximate to (but does not generically intersect) the excitation surface. \\

\noi{ii} The self-interference amplitude   in \q{threepointBinvariant} is one member of a broader class of  
\e{
\text{N-point self-interference amplitudes:}\as  V_{\bk_1\bk_2}V_{\bk_2\bk_3}\ldots V_{\bk_{N-1}\bk_{N}} V_{\bk_N\bk_1}; \as V_{\bk\bk'}=\cbraopket{\psi_{\bk}}{{V}}{\psi_{\bk'}}, \la{generalizedBargmann}
}
which are invariant under redefining $\cket{\psi_{\bk}}$ by a $\bk$-dependent phase. Our expressions for skew scattering [\q{skewscattering}] and the ballistic photocurrent [\q{eq:jball}]  apply in the lowest order of perturbation theory, which is third order in the impurity potential; in principle,  $(N\geq 4)$-point self-interference amplitudes make higher-order contributions to ballistic photocurrent,\cite{Konig-Levchenko:2021} which we neglect on the basis of weak impurity potentials and dilute impurity concentrations. 
For small-momentum scattering induced by a disorder potential that is smooth on the scale of the crystalline lattice period, the $N$-point amplitude simplifies  to a  product of the Fourier transform  ($\widetilde{V}_{\bk}$) of the impurity potential $V_{\br}$ and the N-point Bargmann\footnote{The impurity matrix element for small-momentum scattering is elaborated in \app{app:impuritymatrixelement}.  The Bargmann invariant was previously studied by Bargmann in the context of Wigner's theorem as a way to distinguish unitary transformations from antiunitary ones.\cite{bargmann_invariant}} invariant 
\e{
&\text{N-point self-interference amplitude} \approx \widetilde{V}_{\bk_1-\bk_2}\ldots \widetilde{V}_{\bk_N-\bk_1}\;\times\; \scrb_{\bk_1\ldots \bk_N}; \label{eq:smallangle}\\
&\text{N-point Bargmann invariant}: \as \scrb_{\bk_1\ldots \bk_N}=\braket{u_{\bk_1}}{u_{\bk_2}}\ldots \braket{u_{\bk_N}}{u_{\bk_1}}.\la{Bargmann}
}
The Bargmann invariant (i) is  generally non-reducible\footnote{If the quantum distance between every pair of $\{\psi_{\bk_1},\psi_{\bk_2},\psi_{\bk_3}\}$ is small,  then the Bargmann invariant is approximately a function of the intraband quantum metric tensor $g^{ij}$ and intraband Berry curvature $\bOmega$:
\begin{eqnarray}
\big|\braket{u_{\bk_1}}{u_{\bk_2}}\braket{u_{\bk_2}}{u_{\bk_3}}\braket{u_{\bk_3}}{u_{\bk_1}} \big|
&=& 1- \frac12 g_{12}^{ij}\delta k_{12}^i\delta k_{12}^j - \frac12  g_{23}^{ij}\delta k_{23}^i\delta k_{23}^j- \frac12 g_{31}^{ij}\delta k_{31}^i\delta k_{31}^j   +O(\delta k^3),\la{Bargmetric}\\
\arg \braket{u_{\bk_1}}{u_{\bk_2}}\braket{u_{\bk_2}}{u_{\bk_3}}\braket{u_{\bk_3}}{u_{\bk_1}} &=&  \bOmega_{123} \cdot d\bS_{123}/(2\pi)  + O(\delta k^3), \la{Bargcurv}
\end{eqnarray}
with repeated indices $i,j$ being summed over.  $g_{12}^{ij}$ is the quantum metric evaluated at the midpoint of $\bk_1$ and $\bk_2$; $\bOmega_{123}$ is the Berry curvature vector evaluated at the midpoint of the triangle with vertices $\bk_{1,2,3}$;  $dS_{123}=  \delta \bk_{12}\times (\delta \bk_{13}+ \delta \bk_{12})/4$ is the oriented triangular area element.
The above approximation of the Bargmann invariant is practically useless for skew scattering within an iso-energy surface enclosing a Dirac-Weyl point, which unavoidably involves large-angle scattering between one-electron states separated by a large quantum distance.} 
to a function of the quantum metric\cite{provost_riemannianstructure} and the Berry curvature\cite{berry_quantalphase}, and  (ii) is the fundamental quantum-geometric characteristic associated with skew scattering in the ballistic photocurrent, just as the excitation shift vector/connection\cite{belinicher_kinetictheory,ahn_riemanniangeometry} and associated intra/interband Berry phases\cite{morimoto_nonlinearoptic,aa_quantizationintrainter} (of which the optical vortex is a singularity thereof) are the quantum-geometric characteristics of the photo-excitation process in the bulk photovoltaic effect.

\subsection{Impurity-dependent power laws of the ballistic photocurrent}
\la{sec:powerlaws}

To recapitulate a few salient points in \s{sec:interplay}, the low-frequency ballistic photocurrent response offers a unique experimental diagnostic of topological semimetals, and we are interested in how the \textit{frequency exponent} in \q{exponent} depends on  the type of Dirac-Weyl fermion, the type of disordering impurity and the light polarization. Let us refer to this dependence as an \textit{impurity-dependent power law} for the ballistic photocurrent. A representative result for the 
\e{
\text{$n$-Weyl Hamiltonian}\as H^{w,n}_{\bk}= f_n \big[ k_{+}^n\sigma_- +\;h.c.\; \big]+ v k_z\sz; \as k_{\pm}=k_x\pm ik_y; \as \sigma_{\pm}=\sigma_x\pm i\sigma_y, \la{nweylham}
}
coupled to dipolar impurities was given in \fig{fig:nweylexponent}. This result will be generalized [in \tab{tab:powerlaw} below] to
\e{
\text{$n$-Dirac Hamiltonian}\as H^{d,n}_{\bk}=f_n \big[k_{+}^n\sigma_- +\;h.c. \big], \la{ndiracham}
}
and further generalized  to wider classes of impurities: 
\e{
\text{Monopolar} \as \widetilde{V}^{\text{m}}_{\bk}; \as \text{Dipolar}\as \widetilde{V}^{\text{d}}_{\bk};\as\text{Hybrid = monopolar + dipolar:} \as\widetilde{V}^{\text{h}}_{\bk}=\widetilde{V}^{\text{m}}_{\bk}+\widetilde{V}^{\text{d}}_{\bk}. \la{monodipolarhybrid} 
 }
For simplicity, we assume that all our impurity potentials are \textit{overscreened} by Dirac-Weyl fermions with a nonzero chemical potential, which means that the Thomas-Fermi screening wave vector greatly exceeds the $\bk$-radius of the excitation surface defined by \q{defineexcsurface} [see \app{app:overscreening} for details]. This implies that $\widetilde{V}^{\text{m}}$ is independent of $\bk$ while $\widetilde{V}^{\text{d}}$ is proportional to $\bd\cdot \bk$, with $\bd$ the oriented dipole moment of the impurity.\\


In the case of the monopolar impurity, $\jball$ vanishes due to the high symmetry of both the Dirac-Weyl Hamiltonians [\qq{nweylham}{ndiracham}] and the impurity potential; in this case, it is necessary to add symmetry-reducing corrections (e.g., warping, tilting) to the effective Hamiltonian so as to activate $\jball$, as elaborated in \s{sec:symmetryreduction}.  To ascertain which  response tensor elements of $\jball$ are symmetry-allowed given a certain Dirac-Weyl Hamiltonian and a certain impurity, we offer three symmetry theorems in \s{sec:symmetry}.\\

Why should the ballistic photocurrent scale with the frequency ($\omega$) of the light source? Impurity-mediated skew scattering [\q{skewscattering}] is elastic and occurs on the excitation surface of constant energy $\Eex=\hbar\omega/2$ (measured from the Dirac-Weyl node). This implies that $\jball$ depends on the dispersion and wave function of carriers (electrons and holes) with energy $\Eex$, and only with energy $\Eex$, as is evident from the energy-conserving delta functions in \q{eq:jball}. It follows that $\jball$ has a scaling relation with $\Eex$ and therefore with $\omega$.\\

To determine this scaling relation, we begin with the energy-momentum dispersion of the $n$-Dirac/Weyl fermions [\qq{nweylham}{ndiracham}]: 
\begin{align}
E=\begin{cases}
\sqrt{\left(f_n\sqrt{k_x^2+k_y^2}\right)^{2n}+(vk_z)^2}, &\text{for } n\text{-Weyl}, \\
\sqrt{\left(f_n\sqrt{k_x^2+k_y^2}\right)^{2n}}, &\text{for } n\text{-Dirac},
\end{cases}
\end{align}
With respect to energy scaling, $k_z$ (parallel to rotation axis) has dimensions of $\Eex$ (hence $\omega$) while $k_{x,y}$ (orthogonal to rotation axis) has dimensions of $\Eex^{1/n}$ (hence $\omega^{1/n}$). Both $k_z$ and $k_{x,y}$ enter the formula in \q{eq:jball} in a number of ways, including: (i) the monopole potential is $\bk$-independent while the dipolar potential scales as $k_z$ or $k_x$, depending on whether  the dipole moment $\bd$ is aligned parallel or orthogonal to the rotation axis, (ii) the optical affinity scales as $\partial_{k_j}^2 \sim (k_j)^{-2}$ for linear light polarized in the $j$ direction, and (iii) more ways besides, as detailed in \app{app:scaling}. Already with (i-ii), we appreciate that this scaling relation depends on the impurity type, the light polarization, and the order $n$ of the Dirac-Weyl fermion, which substantiates the main finding of \s{sec:interplay}. Reserving the full scaling analysis to \app{app:scaling}, we merely summarize our findings in \tab{tab:powerlaw}.\\

\begin{table}[ht]
	
\centering
		
\begin{tabular} {|c|c|c|c|} \hline
			
Fermion  & Impurity & $\bB^{\text{de},l/c}$ & Exponent \\  \hline \hline 

3D $n$-Weyl & Dipole ($\boldsymbol{\mathscr{d}}\parallel \vec{\bz}$) & $B^{\text{de},l}_{zzz}$ &  ${2}/{n}-3$ \\  \cline{3-4} 
 & & $B^{\text{de},l}_{zxx}=B^{\text{de},l}_{zyy}$ &  $-1$ \\  \cline{2-4}
 & Dipole ($\boldsymbol{\mathscr{d}}\parallel \vec{\bx}$) & $B^{\text{de},l}_{xxx/\{xyy\}}$ & $-2/n +1$ \\  \cline{3-4} 
 & & $B^{\text{de},l}_{xzz}$ &  $-1$ \\  \cline{2-4}
 & Hybrid ($\boldsymbol{\mathscr{d}}\parallel \vec{\bz}$) & $B^{\text{de},l}_{zzz}$ &  $3$ \\  \cline{3-4} 
 & & $B^{\text{de},l}_{zxx}=B^{\text{de},l}_{zyy}$ &  $2/n +1$ \\   \cline{2-4}
 & Hybrid ($\boldsymbol{\mathscr{d}}\parallel \vec{\bx}$) & $B^{\text{de},l}_{xxx/\{xyy\}}$ & $2/n +1$  \\  \cline{3-4} 
 & & $B^{\text{de},l}_{xzz}$ &  $4/n -1$ \\  \cline{3-4} 
 & & $B^{\text{de},l}_{zxy}$ &  $2/n$ for $n=1,2$ \\  \cline{3-4} \hline 
2D  $n$-Dirac 
& Dipole ($\boldsymbol{\mathscr{d}}\parallel \vec{\bx}$) & $B^{\text{de},l}_{xxx}$  
& $-2/n$ \\ \cline{3-4}  
&  & $B^{\text{de},l}_{\{xyy\}}$ & $-2/n$ for $n=1,3$ \\ \cline{2-4}  
& Hybrid ($\boldsymbol{\mathscr{d}}\parallel \vec{\bx}$) & $B^{\text{de},l}_{xxx/\{xyy\}}$ & $2/n$ \\ \cline{3-4} 
&  & $B^{\text{de},l}_{\{xyy\}}$ & $2/n$ for $n=1,3$ \\  \hline
\end{tabular}	
\caption{
The frequency exponent (fourth column) of the impurity-mediated ballistic photovoltaic tensor element [third column; cf. \q{Sigmaimdefine}] depends on the the type of Dirac-Weyl fermion (first column) and disordering impurity (second column).
} 
\label{tab:powerlaw}
\end{table}

For some cases in \tab{tab:powerlaw}, it may be seen that the frequency exponent is negative. This does not imply a divergence of $\jball$,\footnote{
Similarly, there is no divergence of the shift current in \ocite{ahn_lowfrequencydivergence}, despite their title.
} 
but it suggests $\jball$ can be large as $\omega$ approaches the photo-excitation threshold, this threshold being twice the absolute value of the chemical potential measured from the energy of the Weyl point.\footnote{For resonant photo-excitation, $\hbar\omega >2|\mu|$ by the Pauli exclusion principle. Sending $\mu \ri 0$ at fixed $\omega$ invalidates our assumption that the impurities are overscreened [\app{app:overscreening}], 
thus also invalidating the impurity-dependent power laws in \tab{tab:powerlaw}. However, sending $\mu \ri 0$ with fixed $\omega/|\mu|$ does not necessarily imply that the impurity loses its screening, because screening becomes more effective if the excitation surface is small [\app{app:overscreening}]. 
}\\

In comparison, previous works have calculated how the shift current (in Dirac-Weyl systems) scales with $\omega$.\cite{kim_shiftdiracsurface,ahn_lowfrequencydivergence,Raj-Fiete-PhotogalvanicResponseMultiWeyl-2024} We may say that $|\jball|\gg |\jshift|$ at low frequency if the frequency exponent ($\text{exponent}_{\text{ballistic}}$) for $\jball\propto \omega^{\text{exponent}_{\text{ballistic}}}$ is smaller than that for $\jshift\propto \omega^{\text{exponent}_{\text{shift}}}$. Whether $|\jball|\gg |\jshift|$ or $|\jball|\ll |\jshift|$ at low frequency depends on the order $n$ of the Dirac-Weyl fermion, the type of impurity, 
and  the material-dependent corrections to the effective Hamiltonian. 
To clarify the last point, some of the negative $\text{exponent}_{\text{shift}}$ reported in the literature\cite{ahn_lowfrequencydivergence} relies essentially on tilting corrections to the Weyl effective Hamiltonian; such tilting is symmetry-allowed for Weyl semimetals in which the Weyl points lie at generic wavevectors, but symmetry-forbidden if the Weyl points lie at time-reversal-invariant wavevectors (as in Kramers-Weyl semimetals). In contrast, the $\text{exponent}_{\text{ballistic}}$'s in  \tab{tab:powerlaw} apply generally to all materials in a given topological class ($n$-Dirac or $n$-Weyl), independent of material-dependent corrections to the effective Hamiltonian.

\subsubsection{Vortex-intersection effect in $n$-Weyl semimetals} 
\la{sec:nweylvortexintersection}

The central theme of \s{sec:interplay} [summarized in Box~\ref{box:centraltheme}  and \q{dictum}] is now expounded for $n$-Weyl fermions, generalizing the story in \s{sec:interplay} to $n>1$, to more impurity types and to arbitrary light polarizations.\\

\begin{figure*}[!ht]
\centering
\includegraphics[width=0.9\textwidth]{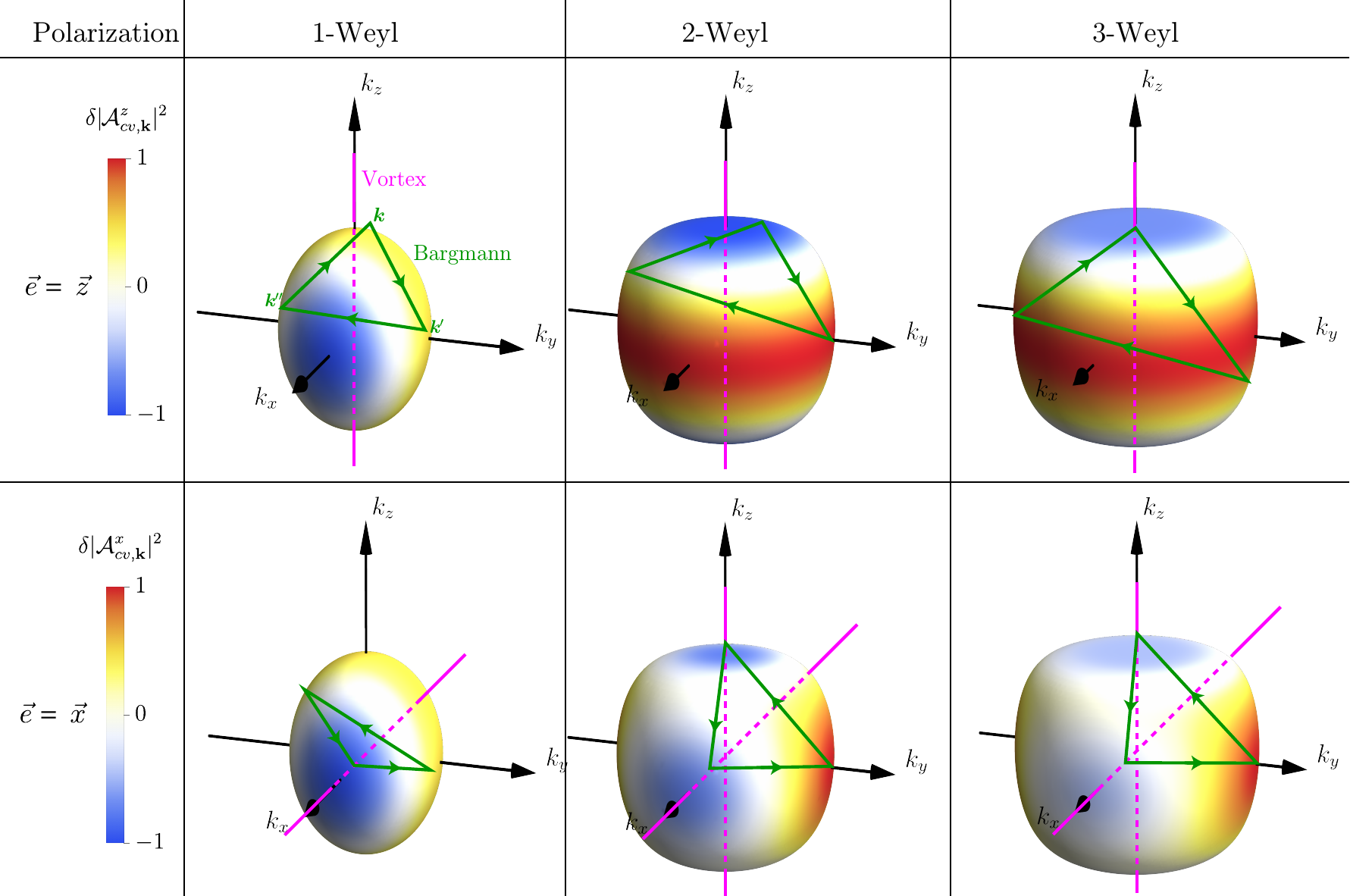}
\caption{
Excitation surface in an $n$-Weyl semimetal for three values of $n$, i.e., $n=1$ (left panel), $n=2$ (middle panel), and $n=3$ (right panel). The normalized by its maximal magnitude affinity deviant is shown in colors; see Eq.~\eqref{multiweyl-current-dev} for its definition. The light polarization vector is directed along the $z$-axis (top row) and the $x$-axis (bottom row).
In all panels, we use $v=2~\mbox{eV}\cdot \mbox{\AA}$, $f_n=2~\mbox{eV}\cdot \mbox{\AA}^n$, and $E_{\rm ex}=1~\mbox{eV}$.
}
\label{fig:multiweyl-aff-dev}
\end{figure*}

According to the Chern-vorticity theorem, each $n$-Weyl fermion has topologically-enforced optical vortex lines which radiate from the $n$-Weyl node [\s{sec:chernvortex}]. Such lines necessarily intersect the excitation surface, which we assume has the topology of a sphere and encloses the $n$-Weyl node. These intersections are illustrated in \fig{fig:multiweyl-aff-dev} for $n=1,2,3$ and for light polarization $\be=\bz$ (parallel to the rotation symmetry axis of the $n$-Weyl Hamiltonian [\q{nweylham}]) and $\bx$ (orthogonal to the rotation axis). \\

Because the optical affinity ($|\be\cdot \mathbf{A}_{cv,\bk}|^2=$ squared dipole matrix element) vanishes along each vortex line, the $\bk$-dependent photo-excitation rate  [\q{photo-excitationrate}] is anisotropic and vanishes at the aforementioned intersection points.
This anisotropy is quantified by the affinity deviant [\q{affinitydeviant}], which we plot over the excitation surfaces in \fig{fig:multiweyl-aff-dev}. \\
 
Consider a hot electron that is photo-excited at a representative point ($\bk'$ on the excitation surface) of higher optical affinity, within the conduction band. The electron subsequently collides with an impurity and scatters elastically to $\bk$ of lower affinity, i.e., closer to a vortex intersection point. Quantum interference between the direct ($\bk\lea \bk'$) and indirect transitions $(\bk\lea \bk''\lea \bk')$ is encoded in the three-point Bargmann invariant [defined in \q{Bargmann} and illustrated as green triangles in \fig{fig:multiweyl-aff-dev}]. Such interference is not symmetric in $(\bk \lea \bk') \ri (-\bk \lea -\bk')$ and results in skew scattering [\q{skewscattering}], which is the origin of the ballistic photocurrent in  \q{jballisticinwords}.  To recapitulate,
\e{
\text{Vortex-intersection effect:} \as \substack{\text{Vortex intersects excitation surface}\\ \text{in $n$-Weyl semimetal}} \as \ri\as  \substack{\text{Anisotropic photo-excitation \&} \\ \text{Asymmetric transition rate}} \as\ri\as \substack{\text{Ballistic}\\ \text{photocurrent}}, \la{vortexintersectioneffect}
}
which is one realization of the dictum in \q{dictum}.\\

To calculate $\jball$, we combine \q{eq:aff_2band}, \q{eq:jball} and \qq{generalizedBargmann}{Bargmann} with the explicit forms of the Bargmann invariant
\e{
\label{Bargmannrealimag}
\mbox{Re}{\scrb_{\bk\bk''\bk'}} \eq 
\frac{1}{8}\bigg( \;(\hat{\mathbf{d}}_{\bk}+ \hat{\mathbf{d}}_{\bk''} +  \hat{\mathbf{d}}_{\bk'})^2-1\; \bigg);\as \mbox{Im}{\scrb_{\bk\bk''\bk'}} = -\frac{1}{4}\hat{\mathbf{d}}_{\bk}\cdot \hat{\mathbf{d}}_{\bk''}\times \hat{\mathbf{d}}_{\bk'},
}
[with $\hat{\mathbf{d}}_{\bk}= \bd_{\bk}/|\bd_{\bk}|$ determined (in our context) by the $n$-Weyl Hamiltonian in \q{nweylham}: $H^{w,n}=\bd_{\bk}\cdot \bsigma$ and $\bsigma=(\sx,\sy,\sz)$],  and the matrix elements of the overscreened impurity potentials:
\e{
\substack{\text{Overscreened monopolar}\\ \text{impurity matrix element}} \as &V^{m,3D}_{\bk\bk'}=\widetilde{V}^{\text{m,3D}}_{\bk-\bk'}\langle u_{\bk} |u_{\bk'}\rangle_{\text{cell}}; \as \widetilde{V}^{\text{m,3D}}_{\bk-\bk'}=\frac{1}{\mathcal{V}}\frac{e^2}{\epsilon k_{\text{scr}}^2}; \la{overscreenedmonopole3D}
    \\
  \substack{\text{Overscreened dipolar}\\ \text{impurity matrix element}} \as  &V^{\text{d,3D}}_{\bk\bk'}= \widetilde{V}^{\text{d,3D}}_{\bk-\bk'}\langle u_{\bk} |u_{\bk'}\rangle_{\text{cell}}; \as \widetilde{V}^{\text{d,3D}}_{\bk}= -i\frac{1}{\mathcal{V}}\frac{e^2}{\epsilon k_{\text{scr}}^2}\boldsymbol{\mathscr{d}}\cdot\bk.\la{overscreeneddipole3D}
}
As a result,  the linear and circular components of the impurity-mediated ballistic photovoltaic tensor [\q{Sigmaimdefine}] are
\begin{equation}
\label{eq:conductivity_2band}
\begin{aligned}
B^{\text{de},l}_{abc}+B^{\text{de},c}_{abc}\eq -2_{\text{e-h}}\frac{4\pi^3|e|^3}{\hbar^2}(\tau^{mr}_{E_{ex}})^2 n_{\text{imp}}\sum_{\mathbf{k},\mathbf{k}^{\prime},\mathbf{k}^{\prime\prime}} \mathbf{v}_{\mathbf{k}}\;
\left\{\mbox{Re}{\scrb_{\bk\bk''\bk'}} \mbox{Im}{\widetilde{V}_{\mathbf{k}^{\prime}\mathbf{k}}\widetilde{V}_{\mathbf{k}\mathbf{k}^{\prime\prime}}\widetilde{V}_{\mathbf{k}^{\prime\prime}\mathbf{k}'} } 
+\mbox{Im}{\scrb_{\bk\bk''\bk'}} \mbox{Re}{\widetilde{V}_{\mathbf{k}^{\prime}\mathbf{k}}\widetilde{V}_{\mathbf{k}\mathbf{k}^{\prime\prime}}\widetilde{V}_{\mathbf{k}^{\prime\prime} \mathbf{k}'}} 
\right\}\\
&\times \left\{\left(g_{bc,\bk'}
-\left[g_{bc,\bk'}\right]
\right)+i\left(\frac{1}{2}\Omega_{bc,\bk'}
-\left[\frac{1}{2}\Omega_{bc,\bk'}\right]
\right)\right\}\delta(E_{\bk,\bk''})\,\delta(E_{\bk,\bk'})\delta(E_{\mathbf{k}'}-E_{ex}).
\end{aligned}
\end{equation}
Some remarks:\\

\noi{i} Bearing in mind that the quantum metric ($g_{bc,\bk}$) and curvature ($\Omega_{bc,\bk}$) are also expressible in terms of the Hamiltonian $\bd_{\bk}$-vector [Eq.~\eqref{eq:metric_Berrycuvature}], \q{eq:conductivity_2band} explicates how $\jball$ depends on the Hamiltonian eigenstates $\ket{u_{b\bk}}$ with manifest gauge invariance. \\

\noi{ii} \q{eq:conductivity_2band} also elegantly synthesizes three fundamental geometric characteristics: the $\bk$-local two-point distance and $\bk$-local curvature (associated to the $\bk\ri \bk$ photo-excitation) and $\bk$-nonlocal three-point distance (associated to the impurity-meditated large-momentum scattering).\\

Directly calculating \q{eq:conductivity_2band}, we find that $\jball$ mediated by monopolar impurities ($\widetilde{V}_{im}=\widetilde{V}^{m,3D}$) vanishes. This is ultimately because  the electron wave-function geometry of $n$-Weyl fermions lacks a polarity required for skew scattering by maximally-symmetric impurities, as elaborated in Sec.~\ref{sec:symmetryreduction}.\\

For both dipolar and hybrid [\q{monodipolarhybrid}]  impurities (whose dipole moments are uniformly oriented), the reduced symmetry of these impurity allows for a nontrivial current:
\begin{align}
\label{jballnweylhybrid-v1}
B^{\text{de}}_{abc} \eq -{2_{\text{p-h}}} |e|^3 \frac{n_{\rm imp} }{\hbar^2 (2\pi)^{6}} \left(\frac{e \mathscr{d}}{\varepsilon k_{\rm scr}^2}\right)^{2}\frac{E_{\rm ex}^{6/n}}{n^2v^2f_n^{6/n}} \times \begin{cases} \big(\tau^{mr,d}_{E_{\rm ex}}\big)^2\;\widetilde{B}^d_{abc}(\boldsymbol{\mathscr{d}}), \as \text{dipolar}; \\
\big(\tau^{mr,h}_{E_{\rm ex}}\big)^2\;\widetilde{B}^h_{abc}(\boldsymbol{\mathscr{d}}), \as \text{hybrid}.
\end{cases}
\end{align}
We list nontrivial components of the tensor $\widetilde{B}_{abc}(\bm{\mathscr{d}})$ in Tabs.~\ref{tab:j-ball-nWeyl-D} and \ref{tab:j-ball-nWeyl}. \\

\begin{table}[ht]
	
\centering
		
\begin{tabular} {|c|c|c|} \hline
			
Dipole direction  & $n$ & Response  \\  \hline \hline 

Dipole $\boldsymbol{\mathscr{d}}\parallel \vec{\bx}$ & $n=1$ &   $\widetilde{B}_{xxx} = -2\widetilde{B}_{xyy} = -2\dfrac{v^2}{f_1^2} \widetilde{B}_{xzz} = 4\widetilde{B}_{yxy} = 4\widetilde{B}_{yyx} 
= \left(\dfrac{e \mathscr{d}}{\varepsilon k_{\rm scr}^2}\right)  \dfrac{16\pi^3}{675} \dfrac{\Eex}{v}$  \\  \cline{2-3}
& $n=2$ &  $\widetilde{B}_{zxz}=\widetilde{B}_{zzx}= \left(\dfrac{e \mathscr{d}}{\varepsilon k_{\rm scr}^2}\right) \dfrac{\pi^5 v}{384 f_2} $  \\  
&  &  $\widetilde{B}_{xxx} = \left(\dfrac{e \mathscr{d}}{\varepsilon k_{\rm scr}^2}\right) \dfrac{\pi^4}{384} \dfrac{5152 -465\pi^2}{40} \dfrac{\Eex}{v}$, \,\,\,\, $\widetilde{B}_{xyy} = \left(\dfrac{e \mathscr{d}}{\varepsilon k_{\rm scr}^2}\right)  \dfrac{\pi^4}{384} \dfrac{2848 -441\pi^2}{24} \dfrac{\Eex}{v}$\\  
&  &  $\widetilde{B}_{xzz} = -\left(\dfrac{e \mathscr{d}}{\varepsilon k_{\rm scr}^2}\right) \dfrac{5\pi}{v^3}$, \,\,\,\,
$\widetilde{B}_{yxy} = \widetilde{B}_{yyx} = \left(\dfrac{e \mathscr{d}}{\varepsilon k_{\rm scr}^2}\right) \dfrac{76}{15} \dfrac{f_2 \Eex}{v^5}$\\ \cline{2-3}
& $n=3$ &  $\widetilde{B}_{xxx}= -\left(\dfrac{e \mathscr{d}}{\varepsilon k_{\rm scr}^2}\right) \dfrac{9\pi^{5/2}}{20\sqrt{3}}  \dfrac{3 \left[1092\, \Gamma^2{(1/3)} \Gamma^3{(7/6)} -1243 \pi^{3/2} \Gamma{(5/3)}\right]}{1001\, \Gamma{(5/3)}} \dfrac{\Eex}{v}$,  \\
&  &  $\widetilde{B}_{xyy} = \left(\dfrac{e \mathscr{d}}{\varepsilon k_{\rm scr}^2}\right) \dfrac{\left[29484\, \Gamma^2{(1/3)} \Gamma^3{(7/6)} -21395 \pi^{3/2} \Gamma{(5/3)}\right]}{5005\, \Gamma{(5/3)}} \dfrac{f_3^{2/3} \Eex}{v^5}$\\ 
&  &  $\widetilde{B}_{xzz} = \left(\dfrac{e \mathscr{d}}{\varepsilon k_{\rm scr}^2}\right) \dfrac{2 \pi^2 \Gamma{(1/3)}}{35\, \Gamma{(5/6)}} \dfrac{1}{v^3 E_{\rm ex}^{1/3}}$, \,\,\,\, $\widetilde{B}_{yxy} =\widetilde{B}_{yyx} = -\left(\dfrac{e \mathscr{d}}{\varepsilon k_{\rm scr}^2}\right) \dfrac{25\pi^{3/2}}{91} \dfrac{f_3^{2/3} \Eex}{v^5}$\\ 
&  &  $\widetilde{B}_{zxz} =\widetilde{B}_{zzx} = \left(\dfrac{e \mathscr{d}}{\varepsilon k_{\rm scr}^2}\right) \dfrac{\pi^2 \Gamma{(4/3)}}{14\, \Gamma{(17/6)}} \dfrac{1}{v^3 E_{\rm ex}^{1/3}}$\\  \cline{1-3}
Dipole $\boldsymbol{\mathscr{d}}\parallel \vec{\bz}$ & $n=1$ &  $\widetilde{B}_{xxz} =\widetilde{B}_{xzx} = \widetilde{B}_{yyz} =\widetilde{B}_{yzy} = \dfrac{f_1^2}{2 v^2} \widetilde{B}_{zzz} = -\widetilde{B}_{zxx} = -\widetilde{B}_{zyy} = \left(\dfrac{e \mathscr{d}}{\varepsilon k_{\rm scr}^2}\right) \dfrac{16\pi^3}{675} \dfrac{f_1^2 \Eex}{v^3}$ \\  \cline{2-3}
& $n=2$ &  $\widetilde{B}_{zzz}= \left(\dfrac{e \mathscr{d}}{\varepsilon k_{\rm scr}^2}\right) \dfrac{15\pi^6}{7680} \dfrac{\Eex}{v}$, \,\,\,\, 
$\widetilde{B}_{zxx} = \widetilde{B}_{zyy}= \left(\dfrac{e \mathscr{d}}{\varepsilon k_{\rm scr}^2}\right) \dfrac{64\pi^5}{7680} \dfrac{f_2 \Eex^{2}}{v^3}$  \\  \cline{2-3} 
& $n=3$ &  $\widetilde{B}_{zzz}= \left(\dfrac{e \mathscr{d}}{\varepsilon k_{\rm scr}^2}\right) \dfrac{18\pi^{9/2}}{15125\, \Gamma^3{(5/6)}} \Gamma^3{(1/3)} \dfrac{\Eex}{v}$, \,\,\,\, $\widetilde{B}_{zxx} = \widetilde{B}_{zyy}= \left(\dfrac{e \mathscr{d}}{\varepsilon k_{\rm scr}^2}\right) 110\, \sqrt{3}\, 2^{1/3}  \pi \dfrac{f_3^{2/3} \Eex^{7/3}}{v^3}$  \\    \hline
\end{tabular}	
\caption{
The explicit dependence of the nontrivial coefficients $\widetilde{B}_{abc}$ in the ballistic current given in Eq.~\eqref{jballnweylhybrid-v1} for the dipolar impurity potential $\Vim=V^{d,3D}$.
} 
\label{tab:j-ball-nWeyl-D}
\end{table}

\begin{table}[ht]
	
\centering
		
\begin{tabular} {|c|c|c|} \hline
			
Dipole direction  & $n$ & Response  \\  \hline \hline 

Dipole $\boldsymbol{\mathscr{d}}\parallel \vec{\bx}$ & $n=1$ &  $\widetilde{B}_{yxz}=\widetilde{B}_{yzx}= - \widetilde{B}_{zxy} = - \widetilde{B}_{zyx}= \left(\dfrac{e^2}{\varepsilon k_{\rm scr}^2}\right) \dfrac{4\pi^3}{135}$ \\  
&  &  $\widetilde{B}_{xxx} = -2\widetilde{B}_{xyy} = -2\dfrac{v^2}{f_1^2} \widetilde{B}_{xzz} = 4\widetilde{B}_{yxy} = 4\widetilde{B}_{yyx} 
= \left(\dfrac{e \mathscr{d}}{\varepsilon k_{\rm scr}^2}\right)  \dfrac{16\pi^3}{675} \dfrac{\Eex}{v}$  \\  \cline{2-3}
& $n=2$ &  $\widetilde{B}_{zxz}=\widetilde{B}_{zzx}= \left(\dfrac{e^2}{\varepsilon k_{\rm scr}^2}\right) \dfrac{3\pi^6}{1024}$  \\  
&  &  $\widetilde{B}_{xxx} = \left(\dfrac{e \mathscr{d}}{\varepsilon k_{\rm scr}^2}\right) \dfrac{\pi^4}{384} \dfrac{5152 -465\pi^2}{40} \dfrac{\Eex}{v}$, \,\,\,\, $\widetilde{B}_{xyy} = \left(\dfrac{e \mathscr{d}}{\varepsilon k_{\rm scr}^2}\right)  \dfrac{\pi^4}{384} \dfrac{2848 -441\pi^2}{24} \dfrac{\Eex}{v}$\\  
&  &  $\widetilde{B}_{xzz} = -\left(\dfrac{e \mathscr{d}}{\varepsilon k_{\rm scr}^2}\right) \dfrac{5\pi}{v^3}$, \,\,\,\,
$\widetilde{B}_{yxy} = \widetilde{B}_{yyx} = \left(\dfrac{e \mathscr{d}}{\varepsilon k_{\rm scr}^2}\right) \dfrac{76}{15} \dfrac{f_2 \Eex}{v^5}$\\ \cline{2-3}
& $n=3$ &  $\widetilde{B}_{xxx}= -\left(\dfrac{e \mathscr{d}}{\varepsilon k_{\rm scr}^2}\right) \dfrac{9\pi^{5/2}}{20\sqrt{3}}  \dfrac{3 \left[1092 \Gamma^2{(1/3)} \Gamma^3{(7/6)} -1243 \pi^{3/2} \Gamma{(5/3)}\right]}{1001 \Gamma{(5/3)}} \dfrac{\Eex}{v}$,  \\
&  &  $\widetilde{B}_{xyy} = \left(\dfrac{e \mathscr{d}}{\varepsilon k_{\rm scr}^2}\right) \dfrac{\left[29484 \Gamma^2{(1/3)} \Gamma^3{(7/6)} -21395 \pi^{3/2} \Gamma{(5/3)}\right]}{5005 \Gamma{(5/3)}} \dfrac{f_3^{2/3} \Eex}{v^5}$\\ 
&  &  $\widetilde{B}_{xzz} = \left(\dfrac{e \mathscr{d}}{\varepsilon k_{\rm scr}^2}\right) \dfrac{2 \pi^2 \Gamma{(1/3)}}{35 \Gamma{(5/6)}} \dfrac{1}{v^3 E_{\rm ex}^{1/3}}$, \,\,\,\, $\widetilde{B}_{yxy} =\widetilde{B}_{yyx} = -\left(\dfrac{e \mathscr{d}}{\varepsilon k_{\rm scr}^2}\right) \dfrac{25\pi^{3/2}}{91} \dfrac{f_3^{2/3} \Eex}{v^5}$\\ 
&  &  $\widetilde{B}_{zxz} =\widetilde{B}_{zzx} = \left(\dfrac{e \mathscr{d}}{\varepsilon k_{\rm scr}^2}\right) \dfrac{\pi^2 \Gamma{(4/3)}}{14 \Gamma{(17/6)}} \dfrac{1}{v^3 E_{\rm ex}^{1/3}}$\\  \cline{1-3}
Dipole $\boldsymbol{\mathscr{d}}\parallel \vec{\bz}$ & $n=1$ &  $\widetilde{B}_{xyz}= \widetilde{B}_{xzy} = -\widetilde{B}_{yxz} =-\widetilde{B}_{yzx} = \left(\dfrac{e^2}{\varepsilon k_{\rm scr}^2}\right) \dfrac{2\pi^3}{135}$ \\  
&  &  $\widetilde{B}_{xxz} =\widetilde{B}_{xzx} = \widetilde{B}_{yyz} =\widetilde{B}_{yzy} = \dfrac{f_1^2}{2 v^2} \widetilde{B}_{zzz} = -\widetilde{B}_{zxx} = -\widetilde{B}_{zyy} = \left(\dfrac{e \mathscr{d}}{\varepsilon k_{\rm scr}^2}\right) \dfrac{16\pi^3}{675} \dfrac{f_1^2 \Eex}{v^3}$ \\  \cline{2-3}
& $n=2$ &  $\widetilde{B}_{zzz}= \left(\dfrac{e \mathscr{d}}{\varepsilon k_{\rm scr}^2}\right) \dfrac{15\pi^6}{7680} \dfrac{\Eex}{v}$, \,\,\,\, 
$\widetilde{B}_{zxx} = \widetilde{B}_{zyy}= \left(\dfrac{e \mathscr{d}}{\varepsilon k_{\rm scr}^2}\right) \dfrac{64\pi^5}{7680} \dfrac{f_2 \Eex^{2}}{v^3}$  \\  \cline{2-3} 
& $n=3$ &  $\widetilde{B}_{zzz}= \left(\dfrac{e \mathscr{d}}{\varepsilon k_{\rm scr}^2}\right) \dfrac{18\pi^{9/2}}{15125 \Gamma^3{(5/6)}} \Gamma^3{(1/3)} \dfrac{\Eex}{v}$, \,\,\,\, $\widetilde{B}_{zxx} = \widetilde{B}_{zyy}= \left(\dfrac{e \mathscr{d}}{\varepsilon k_{\rm scr}^2}\right) 110\, \sqrt{3}\, 2^{1/3} \pi \dfrac{f_3^{2/3} \Eex^{7/3}}{v^3}$  \\    \hline
\end{tabular}	
\caption{
The explicit dependence of the nontrivial coefficients $\widetilde{B}_{abc}$ in the ballistic current given in Eq.~\eqref{jballnweylhybrid-v1} for the hybrid impurity potential $\Vim=V^{m,3D}+V^{d,3D}$. We list the dominant terms that can originate from the different combinations of the dipolar and monopolar potentials.
} 
\label{tab:j-ball-nWeyl}
\end{table}

The momentum-relaxation time {$\tau^{mr}_{E}$} [in Eq.~\eqref{jballnweylhybrid-v1}]
deserves further clarification. In thermal equilibrium, the  electron distribution $f_{\bk}$ (band index suppressed) is the Fermi-Dirac function and is \textit{iso-energy-symmetric}, meaning that the
\e{
\text{Distributional deviant}\as \delta f_{\bk}= f_{\bk}-[f_{\bk}] \la{distributionaldeviant}
}
vanishes; cf. \q{isoenergyave}. In the absence of a drive (e.g., a photoexciting light source), any non-equilibrium, non-iso-energy-symmetric distribution will relax to being iso-energy-symmetric on a time scale that we call the momentum-relaxation time:
\e{
\sum_{\bk'}\bigg( W^{im,s}_{\bk\bk'}-\delta_{\bk\bk'}\sum_{\bk''}W^{im,s}_{\bk\bk''}\bigg)\delta f_{\bk'} \approx -\tf{\delta f_{\bk}}{\tau^{mr}_{\Ek}}; \as  W^{im,s}_{\bk\bk'}= \tf{2\pi}{\hbar}N_{\text{imp}} |\Vim_{\bk\bk'}|^2\delta(E_{\bk\bk'}),\la{taumrdefine}
}
owing to collisions with impurities which act as a momentum bath.\footnote{
Phonons can also relax the electron's momentum, but we assume that impurities are the dominant momentum-relaxers. \q{taumrdefine} is an approximation; more precisely, the \textit{distributional deviant} $\delta f_{\bk}=\sum_h \delta f^h_{\bk}$ is decomposable into generalized harmonics (indexed by $h$) which are eigenfunctions of the impurity-mediated collisional integral, and each harmonic relaxes at a different rate $=-1/\tau^{mr}_{h,E}$:
\e{
\sum_{\bk'}\bigg( W^{im,s}_{\bk\bk'}-\delta_{\bk\bk'}\sum_{\bk''}W^{im,s}_{\bk\bk''}\bigg)\delta f^h_{\bk'}=  -\tf{\delta f^h_{\bk}}{\tau^{mr}_{h,\Ek}},\la{defineharmonic}
}
as detailed in \app{app:Iimp} and \app{app:rta}. In Hamiltonians with spherical symmetry, these harmonics are simply the real spherical harmonics [\app{app:realsphericalharmonic}]. The non-equilibrium distributional deviant is often dominated by one or a few harmonics, in which case $\tau^{mr}_{E}$ in \q{taumrdefine} is the relaxation time for that one harmonic, or an average over those few harmonics. As a case in point, z-polarized photo-excitation of the $n$-Weyl semimetal predominantly excites the $Y_{l=2,m=0}$ spherical harmonic, as illustrated in \fig{fig:multiweyl-aff-dev}. A more general expression for $\mathbf{j}_{\text{ballistic}}$ that encodes multiple harmonic-dependent relaxation times [\q{defineharmonic}] is available in \app{app:solvekinetic}.
}
For $n$-Weyl fermions coupled to dipolar impurities,  $\Vim$ in \q{taumrdefine} is the dipolar impurity potential [\q{overscreeneddipole3D}]; for $n$-Weyl fermions coupled to hybrid impurities, $\Vim$ in \q{taumrdefine} is approximated by the monopolar impurity potential [\q{overscreenedmonopole3D}].\footnote{ 
For small momentum scattering within a single valley of a Dirac-Weyl fermion, we assume that matrix elements ($V^d_{\bk\bk'}$) of the dipole potential are suppressed by a multiplicative factor $|\bk-\bk'|/$(Brillouin-zone period) compared to matrix elements of the monopolar potential. We therefore approximate $|V^m_{\bk\bk'}+V^d_{\bk\bk'}|^2\approx |V^m_{\bk\bk'}|^2$ within \q{taumrdefine}.
}\\

\subsubsection{Vortex-proximity effect in $n$-Dirac semimetals} 
\la{sec:nDirac-all}

The central theme of \s{sec:interplay} [summarized in Box~\ref{box:centraltheme}  and \q{dictum}] is now expounded for 2D $n$-Dirac fermions with the generalized Hamiltonian
\e{
\text{massive $n$-Dirac Hamiltonian}\as H^{d,n}_{\bk}=f_n \big[k_{+}^n\sigma_- + k_{-}^n\sigma_+\big]+ \mD v^2 \sz; \as \bk=(k_x,k_y), \la{ndirachammassive}
}
which includes \q{ndiracham} as a special case with Dirac mass $\mD=0$. This mass is tunable by different means: (a) magnetic field, strain, and dopants for the surface states of 3D topological insulators~\cite{Hasan-Rev, Bansil-Das-ColloquiumTopologicalBand-2016}, and (b) dual gating in multilayer graphene.\cite{McCann-Koshino:2012}.\\

Observe that replacing $\mD v \ri k_z$ in \q{ndirachammassive} gives us the $n$-Weyl Hamiltonian in \q{nweylham}. This structural similarity may be exploited: the $n$-Weyl vortex lines in  $(k_x,k_y,k_z)$-space [illustrated in \q{fig:multiweyl-aff-dev}] are exactly the $n$-Dirac vortex lines in $(k_x,k_y,\mD v)$-space. \\

\begin{figure}[H]
\centering
\includegraphics[height=0.3\textwidth]{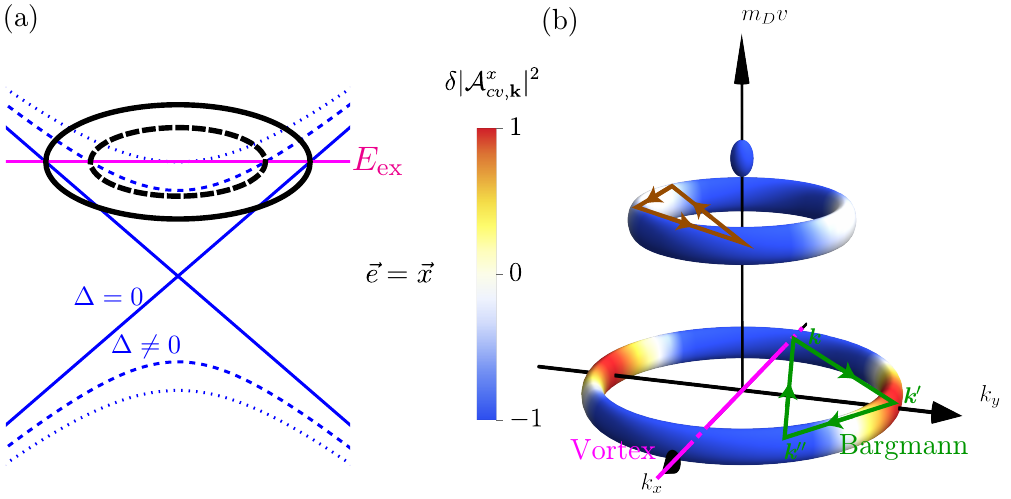}\\
\caption{Illustration of the vortex proximity effect for $1$-Dirac materials with the gap $\Delta = m_Dv^2$. We fixed $m_D v =0$, $m_D v =0.5\,\Eex/v$, and $m_D v \approx \Eex/v$ for the three excitation surfaces. For $(n{>}1)$-Dirac semimetals, there is an additional vortex line parallel to the $\mD v$-axis, which does not qualitatively change any of our conclusions.
}
\label{fig:vortexpromixity}
\end{figure}

There is one crucial difference between $n$-Weyl and $n$-Dirac fermions: the excitation surface is two-dimensional in the former case and one-dimensional in the latter case. A 2D excitation surface that encloses the $n$-Weyl point is necessarily intersected by the vortex lines that radiate from the $n$-Weyl point; in contrast, the intersection of a 1D excitation surface (henceforth called an \textit{excitation curve}) with a vortex line is non-generic, i.e., requires finetuning of Hamiltonian parameters. In particular, for light polarization $\be=\bx$, the excitation curve intersects the vortex line (parallel to $k_x$) if and only if $\mD=0$, as illustrated in \fig{fig:vortexpromixity}. If $\mD\neq 0$, the optical affinity is anisotropic over the excitation curve merely because the excitation curve is \textit{proximate} to the vortex line, as illustrated by the color plot of the affinity deviant in \fig{fig:vortexpromixity}.  \\

The same argument for skew scattering by quantum interference (for the electron loop $\bk'\lea \bk''\lea \bk \lea \bk'$) applies as in \s{sec:nweylvortexintersection}, with a new constraint that $\bk,\bk'$ and $\bk''$ all lie on an iso-energy excitation curve,  as illustrated in \fig{fig:vortexpromixity}. Altogether,
\e{
\text{Vortex-proximity effect:} \as \substack{\text{Vortex proximate to excitation curve}\\ \text{in $n$-Dirac semimetal}} \as \ri\as  \substack{\text{Anisotropic photo-excitation \&} \\ \text{Asymmetric transition rate}} \as\ri\as \substack{\text{Ballistic}\\ \text{photocurrent}}, \la{vortexproximityeffect}
}
which is a second realization of the dictum in \q{dictum}.\\

To calculate $\jball$, we plug into \q{eq:jball} the explicit forms of the Bargmann invariant
\e{
\real \scrb_{\bm k \bk'' \bk '} \eq  
\frac{f_{n}^2 k^{2n}}{4 E_{\bk}^2} \left\{1 +\frac{4\Delta^2}{f_{n}^2 k^{2n}} + \cos{[n(\phi - \phi'')]} +\cos{[n(\phi'' - \phi')]} +\cos{[n(\phi' - \phi)]}\right\} \label{nDiracrealBargmann}\\
\imag  \scrb_{\bm k \bk'' \bk '} \eq \frac{f_{n}^2 k^{2n} \Delta}{4 E_{\bk}^3} \left\{\sin{[n(\phi -\phi'')]}  +\sin{[n(\phi'' -\phi')]} +\sin{[n(\phi' -\phi)]} \right\},\label{nDiracimagBargmann}
}
[obtained from evaluating \q{Bargmannrealimag} with the Hamiltonian $\bd$-vector from \q{ndirachammassive}, in the polar angle representation: $(k_x,k_y)={(k\cos\phi,k\sin\phi)}$] and 
 the matrix elements of the overscreened 
 impurity potentials:
\e{
\substack{\text{Overscreened monopole}\\ \text{impurity matrix element}} \as &V^{m,2D}_{\bk\bk'}=\widetilde{V}^{\text{m,2D}}_{\bk-\bk'}\langle u_{\bk} |u_{\bk'}\rangle_{\text{cell}}; \as \widetilde{V}^{\text{m,2D}}_{\bk-\bk'}=\frac{1}{2\mathcal{V}}\frac{e^2}{\epsilon k_{\text{scr}}}; \la{overscreenedmonopole2D}
    \\
  \substack{\text{Overscreened dipolar}\\ \text{impurity matrix element}} \as  &V^{\text{d,2D}}_{\bk\bk'}= \widetilde{V}^{\text{d,2D}}_{\bk-\bk'}\langle u_{\bk} |u_{\bk'}\rangle_{\text{cell}}; \as \widetilde{V}^{\text{d,2D}}_{\bk}= -i\frac{1}{2\mathcal{V}}\frac{e^2}{\epsilon k_{\text{scr}}}\boldsymbol{\mathscr{d}}\cdot\bk,\la{overscreeneddipole2D}
}
[obtained from screening of a 3D potential by 2D crystalline electrons; cf. \app{app:screening}]. 
The final result for monopolar impurities is:
\e{
\Vim=V^{m,2D}: \as \mathbf{j}_{\rm ballistic} \eq 0, \la{jballndiracmonopole}
}
because the electron wave function geometry (of $n$-Dirac fermions) lacks a certain complexity that is needed for skew scattering mediated by maximally-symmetric monopole impurities, as elaborated in \s{sec:symmetryreduction}.\\

For dipolar and hybrid impurities (both with uniformly-oriented dipole moments),

\e{\label{jballnDirac}
B^{\text{de}}_{abc} \eq - {2_{\text{p-h}}} \frac{e^6 \mathscr{d}_{\perp}^3 (\tau^{mr}_{E_{\rm ex}})^2}{2^{12} \hbar^2 \varepsilon^3 k_{\rm scr}^3 E_{\rm ex}^4} \frac{\left(E_{\rm ex}^2 - \Delta^2\right)^{1+3/n}}{f_{n}^{6/n}} \widetilde{B}_{abc}.
}
We show the nonzero coefficients $\widetilde{B}_{abc}$ in Tab.~\ref{tab:j-ball-nDirac}.

For dipolar impurities, the momentum-relaxation time [$\tau^{mr}_{E}$ in \q{jballnDirac}] is determined by \q{taumrdefine} with
$\Vim$ equal to  the dipolar impurity potential [\q{overscreeneddipole2D}]; for  $n$-Dirac fermions coupled to hybrid impurities, $\tau^{mr}_{E}$ in  \q{jballnDirac} is determined by \q{taumrdefine} with
$\Vim$  approximated by the  monopolar impurity potential [\q{overscreenedmonopole2D}].\footnote{ 
This approximation is justified by the same reasoning that was given in a footnote in \s{sec:nweylvortexintersection}.  
}

\begin{table}[ht]
	
\centering
		
\begin{tabular} {|c|c|} \hline
			
$n$ & Response  \\  \hline \hline 
$n=1$ &  $\widetilde{B}_{xxx}= -\widetilde{B}_{xyy} = \widetilde{B}_{yxy} =\widetilde{B}_{yyx} = \dfrac{1}{4} \left(1 + \dfrac{16\Delta^2}{E_{\rm ex}^2 - \Delta^2}\right)$ \\  \cline{1-2}
$n=2$ &  $\widetilde{B}_{xxx}= -\widetilde{B}_{xyy} = \dfrac{1}{2} \left(1 + \dfrac{8\Delta^2}{E_{\rm ex}^2 - \Delta^2}\right)$  \\ \cline{1-2}
$n=3$ &  $\widetilde{B}_{xxx}= -\widetilde{B}_{xyy} = \dfrac{3}{4} \left(1 + \dfrac{16\Delta^2/3}{E_{\rm ex}^2 - \Delta^2}\right)$, \quad
$\widetilde{B}_{yxy} =\widetilde{B}_{yyx} = -\dfrac{9}{4}$  \\ \cline{1-2}
$n>3$ &  $\widetilde{B}_{xxx}= -\widetilde{B}_{xyy} = \left(1 + \dfrac{4\Delta^2}{E_{\rm ex}^2 - \Delta^2}\right)$  \\  \hline
\end{tabular}	
\caption{
The explicit dependence of the nontrivial coefficients $\widetilde{B}_{abc}$ in the ballistic current given in Eq.~\eqref{jballnDirac} for the hybrid impurity potential $\Vim=V^{m,2D}+V^{d,2D}$ in the $n$-Dirac case. We set $\bm{\mathscr{d}} \parallel \hat{\mathbf{x}}$.
} 
\label{tab:j-ball-nDirac}
\end{table}

\subsection{Symmetry-reduced Dirac-Weyl fermions and maximally-symmetric impurities}\la{sec:symmetryreduction}

The bulk photovoltaic current mediated by monopole impurities (with the maximal point-group symmetry $O(3)\times \Z_2^T$) is a case study that asks: if the symmetry-breaking (required for bulk photovoltaics)  does not originate from the impurity, whence does it come?  In part, it comes from the electronic wave function:

\begin{tcolorbox}[colback=white, sharp corners]
For nontrivial skew scattering mediated by nonpolar, isotropic monopole impurities, there must exist a preferred direction ( i.e., \textbf{polarity) for the Hamiltonian-vector} of the \textbf{complex-valued} electronic wave function.
\end{tcolorbox}

\noindent Therein we identify two geometric characteristics (of wave functions over $\bk$-space)  are measures of the required symmetry-breaking for the ballistic photocurrent, i.e.,\footnote{It was pointed out in [Fu,Sangyu,japanese] that nontrivial wave-function geometry can activate the low-frequency ballistic photocurrent via perturbing the Fermi surface in metals; here, we identify the precise geometric characteristics, and extend the theory to higher-frequency resonant photo-excitations.} these characteristics are necessary to activate monopole-impurity-mediated skew scattering [\q{skewscattering}].  Namely, we identify:\\

\noi{i} Complex (as opposed to real) values of the electronic energy eigenstates $\{\ket{u_{b\bk}}\}$.  For Schr\"odinger-type Hamiltonians, $\{\ket{u_{b\bk}}\}$ are  eigenstates of  $\emikr H^0 \eikr$; for an $m$-band effective Hamiltonian $H^0_{\bk}$, each $\ket{u_{b\bk}}$ is a vector in $\C^m$ vs $\R^m$, given an  $m$-dimensional Bloch-wave basis of L\"owdin-orthogonalized atomic orbitals.\cite{AA_wilsonloopinversion} For any two-band effective Hamiltonian [\q{twobandham}], 
$\ket{u_{b\bk}}\in \C^2$ if $H^0_{\bk}$ involves all three Pauli matrices, and otherwise $\ket{u_{b\bk}}\in \R^2$. In bundle-theoretic language,\cite{crystalsplit_AAJHWCLL,DeNittis_classifyAI}
skew scattering requires the vector bundle to be complex rather than real. \\

\noi{ii} The second required wave-function geometric characteristic is a nontrivial
\e{
\text{$\vec{H}$-vector polarization} \as [\hbd]_E = \tf{\sum_{\bk} \hbd_{\bk}\delta(E_{\bk}-E)}{ \sum_{\bk}\delta(E_{\bk}-E)}; \as \hbd_{\bk}= \tf{\bd_{\bk}}{|\bd_{\bk}|}. \la{Hpolarization}
}
This is essentially the iso-energy average of the normalized {Hamiltonian vector} (in short, $\vec{H}$-vector)  in  \q{twobandham}, which provides a notion of polarization in the Hilbert space spanned by the two L\"owdin-orthogonalized atomic orbitals.  \\

The necessity of both geometric characteristics (i-ii), as summarized colloquially in the above text box, is formally expressed by two theorems in  \s{sec:Hspecificsymmetry}. One implication of these theorems is that the monopole-impurity-mediated $\jball$ vanishes for both $n$-Weyl Hamiltonians [$H^{w,n}$ in \q{nweylham}] and $n$-Dirac Hamiltonians [$H^{d,n}$ in \q{ndiracham}]. For both classes of Hamiltonians, the required $\vec{H}$-vector polarization is symmetry-forbidden; for $n$-Dirac Hamiltonians, the required complexity (for $\ket{u_{b\bk}}$) is also symmetry-forbidden -- a double blow. These conclusions are summarized in the second and eighth rows of \tab{tab:nweyldiracsymmetryreducedmonopole}. \\

In the remainder of this section, we discuss various ways to activate the monopole-impurity-mediated $\jball$ by symmetry reduction of the $D_{\infty}\times \Z_2^T$-symmetric $n$-Weyl Hamiltonian and the $O(2)\times \Z_2^T$-symmetric $n$-Dirac Hamiltonian [cf. \app{app:symmetryndiracweyl}]. Let us add to $H^{w,n}$ (and $H^{d,n}$) various symmetry-reducing corrections:
\begin{align}
\text{Dirac mass}\as &H^{dm}=\mD v^2 \sz; \la{diracmass}\\
\text{x-tilt}\as &H^{xt}_{\bk}=v_{xt}k_x\iden; \la{xtilt}\\
\text{z-tilt}\as &H^{zt}_{\bk}=v_{zt}k_z\iden;\la{ztilt}\\
\text{3-warp}\as &H^{3}_{\bk}=\lambda(k_+^3+k_-^3)\sz; \la{threewarp}\\
\text{6'-warp}\as & H^{6',n}_{\bk}= \sum_{a,b,c}^{\N} h_{abc} k_{+}^{a}k_{-}^{b} \sigma_{+} +h.c. - H^{d,n}_{\bk};  \as b+2a+3c=n,
\la{sixpwarp}
\end{align}
with $h_{abc}\in \R$ being real Hamiltonian parameters indexed by natural numbers  $a,b,c\in \N=\{0,1,2,3,\ldots\}$;
in particular, $h_{0n0}=f_n$, which is the same `$f_n$'  in \q{ndiracham}. The exceptional complexity of the last Hamiltonian is motivated by its application to $n$-graphene, as elaborated in \s{sec:warp-nDirac}.\\

For these corrections in \qq{diracmass}{sixpwarp}, \tab{tab:nweyldiracsymmetryreducedmonopole}  shows the resultant symmetry-reduced point groups,  the complexity/reality of $\ket{u_{b\bk}}$, and the symmetry-allowed $\vec{H}$-vector polarizations and symmetry-allowed response tensor elements.\\

\begin{table}[H]
\centering
\renewcommand{\arraystretch}{1.4}
\begin{tabular}{c c c c c c c}
\hline\hline
 $H^{w,n}\;+\;$correction & $\as\as$ Group $\as\as$ & Elements & $\as\C^2$ vs $\R^2\as$ & $\as[\hbd]\as$ &$\bB^{\text{de},l}$ & $\bB^{\text{de},c}$  \\
\hline
0 &$\infty 2,1$ & $C_{2x},C_{2y},R_z^{\theta},T$ &  $\C$ & $0$   & 0 & 0  \\
z-tilt &$\infty2'$ & $C_{2x}T,C_{2y}T,R_z^{\theta}$ &  $\C$ & $\parallel 3$   & $B^{\text{de},l}_{xyz}=-B^{\text{de},l}_{yzx}$ & $B^{\text{de},c}_{xzx}=B^{\text{de},c}_{yzy}$  \\
x-tilt &$2'2'2$ & $C_{2x},C_{2y}T,C_{2z}T$ &  $\C$ & $\parallel 1$   & $B^{\text{de},l}_{\{xyz\}}$ & $B^{\text{de},c}_{yxy/zxz}$  \\
3-warp &  $321'$ & $C_{3z},C_{2y},T$ &  $\C$ & $0$ & $0$ & 0\\
6'-warp & $6'22'$ & $C_{3z}, C_{2x}, C_{2y}T, C_{2z}T$ &  $\C$ & $0$&  $0$ & 0 \\ [5pt]
\hline\hline 
 $H^{d,n}\;+\;$correction & $\as\as$ Group $\as\as$ & Elements &$\C^2$ vs $\R^2$ & $[\hbd]$ & $\bB^{\text{de},l}$ & $\bB^{\text{de},c}$  \\
\hline
0 & $\infty m,1$ &  $M_x,M_y,R_z^{\theta},T$ &  $\R$ & $0$ & 0& 0\\ 
Dirac mass & $\infty m'm'$ &  $M_xT,M_yT,R_z^{\theta}$ &  $\C$ & $\parallel 3$ & 0& 0\\ 
x-tilt &$m'm2'$ & $M_y,C_{2z}T,M_xT$ &  $\R$ & $\parallel 1$&0& $0$  \\
Dirac mass + x-tilt & $m'$ &  $M_xT$ &  $\C$ & $\parallel 13$ & $B^{\text{de},l}_{yyy/\{xxy\}}$  & $B^{\text{de},c}_{yxy}$ \\
3-warp &  $3m1'$ & $C_{3z},M_x,T$ &  $\C$ & $0$ & $0$ & 0\\
6'-warp & $6'mm'$ & $C_{3z}, M_y, M_xT, C_{2z}T$&  $\R$ & $0$ &0& 0 \\
Dirac mass + 3/6'-warp &   $3m'$ &  $C_{3z},M_xT$ &  $\C$ & $\parallel 3$& $B^{\text{de},l}_{yyy}=-B^{\text{de},l}_{xxy} = -B^{\text{de},l}_{xyx}=-B^{\text{de},l}_{yxx}\as$  &  $ 0$\\ [5pt] \hline \hline
\end{tabular}
\caption{Top sub-table: $n$-Weyl Hamiltonian [\q{nweylham}] with various symmetry-reducing corrections (left-most=first  column and cf. \qq{diracmass}{sixpwarp}) and the corresponding point groups (second column), symmetry elements (non-exhaustively listed in the third column, and with symmetry representations that can be read off from \tab{tab:nWeylsymmetryreps}). The fourth column describes if $\ket{u_{b\bk}}$ is symmetry-allowed to be complex or symmetry-constrained to be real. Fifth column describes the $\vec{H}$-vector polarization ($[\hbd]$ defined in \q{Hpolarization}); if nonzero, $[\hbd]\para 3$ means that the 3-component of $[\hbd]=([\hd_1],[\hd_2],[\hd_3])$ is symmetry-allowed to be nonzero, with $d_3$ multiplying $\sz$ in \q{twobandham}; $[\hbd]\para 13$ means that $[\hbd]$ is symmetry-allowed to lie in the 13 plane.
For the \textbf{monopole-impurity}-mediated ballistic photocurrent, we list the symmetry-allowed linear tensor elements (in the sixth column) and symmetry-allowed circular tensor elements (seventh column). Bottom sub-table: $n$-Dirac Hamiltonian with various symmetry-reducing corrections, and with symmetry representations that can be read off from \tab{tab:nDiracsymmetryreps}.
The names of groups have been given in the Hermann-Mauguin notation; in particular, $\infty 2,1=D_{\infty}\times \Z_2^T$, $\infty m,1=O(2)\times \Z_2^T$,  $321'\equiv D_{3}\times \Z_2^T$ and  $3m1'\equiv C_{3v}\times \Z_2^T$. All the symmetry-allowed tensor elements here  are derivable from combining \qq{SO2allowedlinear}{SO2allowedcircular}, \qq{C3allowedlinear}{C3allowedcircular}  and \tab{tab:compindex}. $B^{\text{de}}_{abc/def}$ means both $B^{\text{de}}_{abc}$ and $B^{\text{de}}_{def}$ are symmetry allowed; $B^{\text{de}}_{\{abc\}}$ means all permutations of the indices `$abc$' are symmetry-allowed. }
\label{tab:nweyldiracsymmetryreducedmonopole}
\end{table}

\subsubsection{Tilt-activated $\jball$ for symmetry-unpinned Weyl fermions}\la{sec:tiltactivated}

For Kramers-Weyl semimetals (e.g., $\beta$-Ag$_2$Se~\cite{zhang_betaAg2Se} and (TaSe$_4$)$_2$I~\cite{kim2025signatures}, Ag$_3$BO$_3$ and AgBi(Cr$_2$O$_7$)$_{2}$~\cite{chang2018topological}, and TaGe$_{2}$~\cite{Li_TaGe2}  ), the Weyl band-touching point is symmetry-pinned to a time-reversal-invariant $\bk$-point, which forbids tilting corrections like those in \qq{xtilt}{ztilt}.  Such symmetry-pinned  Weyl fermions can sustain a $\jball$ mediated by dipole/hybrid impurities [\q{jballnweylhybrid-v1}] but not by overscreened monopole impurities [\tab{tab:nweyldiracsymmetryreducedmonopole}].\\

In other Weyl semimetals like TaAs~\cite{yang2015weyl,xu2015discovery,lv2015experimental,lv2015observation} and WTe$_2$~\cite{Wang:WTe2,wu:WTe2}, the Weyl band-touching points are not symmetry-pinned and lie at generic $\bk$-points. The lower symmetry (of the group of a generic $\bk$-point) allows for tilting corrections, and can sustain a $\jball$ mediated by overscreened monopole impurities [\q{overscreenedmonopole3D}], as we now demonstrate for a model of z-tilted Weyl fermions:
 \begin{equation}
\label{eq: Weyl_tilted}
H^{1,w}_{\bk}\big|_{f_1=v}+ H^{zt}_{\bk}=  v \bk \cdot \boldsymbol{\sigma}+ v_{zt} k_z\sigma_{0} .
\end{equation}
The only nonzero elements of the $\text{photovoltaic}^{\sma{\text{im}}}_{\sma{\text{ballistic}}}$ tensor are derived in Appendix~\ref{app:tilted Weyl} to be
\begin{equation}
\label{eq:conduc_tiltedWeyl}
\begin{aligned}
&B_{xyz}^{\text{de},l}=-B_{yzx}^{\text{de},l}= \frac{1}{2} f_0(\omega,v)\,f^l\big(\tf{v_{zt}}{v}\big);\as B_{xzx}^{\text{ed},c}=B_{yzy}^{\text{de},c}= \frac{i}{2} f_0(\omega,v)\, f^c\big(\tf{v_{zt}}{v}\big);\\
&f_0(\omega,v)=\frac{|e|^3}{(2\pi)^3}\tau_{\text{eff}}^2 \left(\frac{e^2}{\epsilon}\right)^3\frac{n_{\text{imp}}}{k_{scr}^6}\left(\frac{\omega}{2v}\right)^{2}\tf1{v^2};\\ 
&f^l(\alpha) = \frac{1}{105}\frac{(\text{arctanh}(\alpha)-\alpha/(1-\alpha^2))(7+3\alpha^2)}{\alpha(1-\alpha^2)^2}= -\frac{2}{45}  \alpha^2 + O(\alpha^{4});
\\
&f^c(\alpha) = \frac{35+3\alpha^2(14+\alpha^2)}{105}\frac{(\alpha/(1-\alpha^2)-\text{arctanh}(\alpha))}{4\alpha^2(1-\alpha^2)^2}= \frac{1}{18}\alpha + O(\alpha^{3}),
\end{aligned}
\end{equation}
as is consistent with the symmetry analysis in  Table~\ref{tab:nweyldiracsymmetryreducedmonopole}.  The dimensionless tilt-dependent functions vanish as the tilt velocity $v_{zt}$ goes to zero, with $f^l \propto v_{zt}^2$ and $f^c \propto v_{zt}$ in the lowest-order Taylor truncation; apparently, the circular $\jball$ is more sensitive to tilting than the linear component. \\

\subsubsection{Warp-activated $\jball$ in $n$-graphene}
\la{sec:warp-nDirac}

6'-warped $n$-Dirac fermions [\q{ndiracham}, \q{sixpwarp}] are materialized by $n$-graphene,\cite{Min-MacDonald-ChiralDecompositionElectronic-2008,Koshino-McCann-TrigonalWarpingBerrys-2009, Zhang-Zaliznyak-ExperimentalObservationQuantum-2011, Lui-Heinz-ObservationElectricallyTunable-2011, Han-Ju-CorrelatedInsulatorChern-2024a} as is evident from two observations:\\

\noi{i} $n$-graphene is described by four flavors of $n$-Dirac fermions (distinguished by a  valley index $\xi=\pm $ and a spin index $s=\pm$) subject to   the (valley+spin)-dependent Hamiltonian:
\e{
H^{\xi,s}_{\bk}= \sum_{a,b,c}^{\N} h_{abc} (k^{\xi}_{+})^{a}(k^{\xi}_{-})^{b} \sigma_{+} +h.c. + \Delta_{\xi,s} \sz; \as k^{\xi}_{\pm}= \xi k_x\pm ik_y; \as b+2a+3c=n.\la{Hvalley}
}
This is essentially\footnote{
We emphasize `essentially', because \q{Hvalley} corrects an error in Eq. (14) of \ocite{Koshino-McCann-TrigonalWarpingBerrys-2009}. Our \q{Hvalley} is consistent with spatial inversion being represented as $\widehat{P}=\tau_1\sx s_0$, with $s_0$ being the identity matrix in the spin subspace, $\tau_j$ being a Pauli matrix in the valley subspace; the eigenvalues of $\tau_3$ correspond to the valley index $\xi$. The proportionality of $\widehat{P}$ to $\sx$ is because spatial inversion interchanges the orbitals $\psi_{A,1}$ and $\psi_{B,n}$ with a trivial phase factor.\cite{Koshino-McCann-TrigonalWarpingBerrys-2009} Each orbital here is labelled by a sublattice index (to the left) and a layer index (to the right); the sublattice index $A,B$ corresponds to the $\pm 1$ eigenvalues of $\sz$.
} 
the Koshino-McCann effective Hamiltonian for a single valley of $n$-graphene,\cite{Koshino-McCann-TrigonalWarpingBerrys-2009} with the inter/intralayer coupling parameters ($\gamma_j$) absorbed into the real coefficients $h_{abc}$. It should be understood that $H^{+,s}_{\bk}$  is the small-$|\bk|$ effective Hamiltonian centered at $C_{3z}$-invariant wavevector $K$, while $H^{-,s}_{\bk}$ is the small-$|\bk|$ effective Hamiltonian centered at the other inequivalent $C_{3z}$-invariant wavevector $K'$. The term $\Delta_{\xi,s}\sz$, if nonzero, opens a band gap; in the absence of this [potentially (valley+spin)-dependent] gapping term, $H^{+,s}$ reduces to $H^{d,n}+H^{6',n}$ [\q{ndiracham} and \q{sixpwarp}]; \\

\noi{ii} The layer group of rhombohedral $n$-graphene is $p\bar{3}m1 \times \Z_2^T$,\cite{Mu-Zhou-ValleydependentGiantOrbital-2025} with point group $D_{3d}\times \Z_2^T$. This point group contains  elements:  $C_2'$ (two-fold rotation with an in-plane axis), $P$ (spatial inversion), $T$ and $S_6=M_xC_{6z}$ (six-fold roto-rotation); the aforementioned four elements map $K$ to $K'$, i.e., they exchange valleys. Therefore, only pairwise compositions (like $TP$) belong to the 
\e{
\text{Little group of  K} \;= {\bar{3}'m}= C_{3v}\cup \{ TP, 3TC_{2}', 2TS_6\},
}
which are the set of point group symmetries that map $K$ to $K$.
If we ignore the $z$ coordinate in these three-dimensional spatial transformations, the little group of K projects to 
\e{
\bar{3}'m \as \substack{\text{projects}\\ \longrightarrow}\as  6'mm' = C_{3v}\cup \{ TC_{2z}, 3TM_x, 2TC_{6z}\}. 
}
Thus the two groups are isomorphic, with $TP$ corresponding to $TC_{2z}$ and $TC_2'$ corresponding to $TM_x$, etc. The distinction between the two groups is irrelevant to symmetry constraints of photovoltaic tensors for in-plane currents and electric fields.  This is why $6'mm'$-symmetric $H^{d,n}+H^{6',n}$ [cf. \tab{tab:nweyldiracsymmetryreducedmonopole}] can also be viewed as being ${\bar{3}'m}$-symmetric, which is the little-group symmetry of $n$-graphene.\\

In estimating $\jball$ for $n$-graphene, we assume for longer-wavelength disorder that intra-valley electron-defect scattering dominates over inter-valley scattering, and for non-magnetic defects that electron-defect scattering is spin-preserving,  such that the ballistic photovoltaic response tensor decomposes into a sum of intra-valley and intra-spin contributions: $\Bde\approx \sum_{\xi,s}\Bde_{\xi,s}$. Each contribution [derived in Appendix~\ref{app:current-ndirac-warped}] reads as
\begin{eqnarray}
\la{13-multiWeyl-nDirac-w-jball-fin}
B^{\text{de}}_{abc, \xi,s} = 2_{\rm p-h}|e|^3\frac{(2\pi)^3}{\hbar^2} n_{\rm imp} (\widetilde{V}^{m}_{\mathbf{0}})^3 \frac{(\tau^{mr}_{E_{\rm ex}})^2}{2}
\widetilde{B}_{abc, \xi,s}^{(n)}.
\end{eqnarray}
We explicate $\widetilde{B}_{abc, \xi}^{(n)}$ for  a  few representative values of $n$ (number of graphene layers) and $h_{abc}$ in Tab.~\ref{tab:j-ball-nDirac-warp}.

\begin{table}[ht]
	
\centering
		
\begin{tabular} {|c|c|c|} \hline
			
$n$ & Warping $H^{6',n}_{\bk}$ & Response  \\  \hline \hline 
$n=1$ & $h_{200} k_{+}^{2} \sigma_{+} +h.c.$ &
$\widetilde{B}_{xxy, \xi}^{(1)} = \widetilde{B}_{xyx, \xi}^{(1)} =\widetilde{B}_{yxx, \xi}^{(1)} = -\widetilde{B}_{yyy, \xi}^{(1)} =\dfrac{h_{200} \Delta_{\xi,s} \left(E_{\rm ex}^2 -\Delta_{\xi,s}^2\right)^2 \left(\Delta_{\xi,s}^2 +E_{\rm ex}^2\right) }{2^8\, \pi^3 f_1^5 E_{\rm ex}^3}$ \\  \hline
$n=2$ & $h_{100} k_{+} \sigma_{+} +h.c.$ &
$\widetilde{B}_{xxy, \xi}^{(2)} = \widetilde{B}_{xyx, \xi}^{(2)} =\widetilde{B}_{yxx, \xi}^{(2)} = -\widetilde{B}_{yyy, \xi}^{(2)} =\dfrac{h_{100} \Delta_{\xi,s} \left(E_{\rm ex}^2 -\Delta_{\xi,s}^2\right)^{3/2}}{2^{10}\, \pi^3 f_2^2 E_{\rm ex}^3}$ \\  \hline
$n=5$ & $h_{130} k_{+}k_{-}^3 \sigma_{+} +h.c.$ &
$\widetilde{B}_{xxy, \xi}^{(5)} = \widetilde{B}_{xyx, \xi}^{(5)} =\widetilde{B}_{yxx, \xi}^{(5)} = -\widetilde{B}_{yyy, \xi}^{(5)} 
=\dfrac{h_{130}^3 \Delta_{\xi,s} \left(E_{\rm ex}^2 -\Delta_{\xi,s}^2\right)}{100\,(2)^{8}\, \pi^3 f_5^2 E_{\rm ex}^3}$ \\  \hline
\end{tabular}	
\caption{
The explicit dependence of the nontrivial coefficients $\widetilde{B}_{abc, \xi}^{(n)}$ in the ballistic current response tensor given in Eq.~\eqref{13-multiWeyl-nDirac-w-jball-fin} for the warped $n$-Dirac case. 
} 
\label{tab:j-ball-nDirac-warp}
\end{table}

 By inspection, $\jball$
is proportional to $(h_{abc})^{\sma{\text{positive integer}}}\Delta_{\xi, s}$, which reflects that combining both the gap $\Delta_{\xi, s}$ and the 6'-warp is crucial to activate the intravalley $\jball^{\xi}$, and that neither gap nor warp alone is sufficiently activating, in accordance with the symmetry analysis in \tab{tab:nweyldiracsymmetryreducedmonopole}. In the presence of an out-of-plane displacement field, the gap $\Delta_{\xi,s}$ is valley- and spin-independent, in which case $\sum_{\xi,s}\jball^{\xi,s}$ adds up rather than canceling out, in accordance with the displacement field breaking centrosymmetry. More interesting possibilities for the gap $\Delta_{\xi,s}$ are discussed in \s{sec:correlated}.

\section{Symmetry theorems for bulk photovoltaic response tensors}\la{sec:symmetry}

This is a self-contained, read-alone or read-without interlude on the symmetry constraints of the photovoltaic response tensor.\\

Suppose a homogeneous medium is subject to a dynamic electric field [$\cale(\br t)=\cale_{\bq\omega}e^{i\bq\cdot\br-i\omega t}+c.c.$] associated to a light wave.
The  bulk photovoltaic response tensor $\bSigma$ is defined by
\e{
j_{a} =\sum_{b,c}\big(\Sigma^l_{abc}+\Sigma^c_{abc}\big)\cale_{\bq\omega,b}\overline{\cale_{\bq\omega,c}}; \as \Sigma^l_{abc}=\Sigma^l_{acb}; \as \Sigma^c_{abc}=-\Sigma^c_{acb}, \la{bPVtensordefine}
}
with $\bj$ being the static homogeneous current (`direct current') density. The linear and circular components of $\bSigma$ are respectively defined to be symmetric and antisymmetric under interchange of the last two spatial indices. $\bSigma$ implicitly depends on $\omega$ but does not depend at all on $\bq$, hence ruling out photon drag effects\cite{danishevskii_dragging}.\\

We are interested in how $\bSigma$ is constrained by the Hamiltonian being symmetric with respect to a point-group transformation $g$. We distinguish between spatial point-group transformations ($g=g^s$) and magnetic point-group transformations ($g=Tg^s=g^sT$, meaning  the composition of time-reversal with a spatial point-group transformation, which could be the trivial spatial transformation). Many of the point-group transformations we will encounter are \textit{order-two}, meaning that $g^2$ has a trivial action on spacetime; we list them  in \tab{tab:pointgroup}.  For any $g$ which acts on spacetime, order-two or not, we define $g^s_{ab}$ as an $O(3)$ matrix for the reduced  action of $g$ on 3D space alone, as exemplified by the last column in  \tab{tab:pointgroup}.\\

\begin{table}[H]	
\centering
\begin{tabular} {|c|l|l|l|} \hline	  
Type & Name & Spacetime action & $g^s_{ab}$ \\ \hline \hline
Spatial    $g=g^s$    & Mirror reflection $M_x$ & $(x,y,z,t)\ri (-x,y,z,t)$ &  $\text{diag}(-1,1,1)$ \\ \cline{2-4}
   & Two-fold rotation $C_{2z}$ & $(x,y,z,t)\ri (-x,-y,z,t)$  &  $\text{diag}(-1,-1,1)$ \\ \cline{2-4}
     & Parity $P$ & $(x,y,z,t)\ri (-x,-y,-z,t)$   &  $\text{diag}(-1,-1,-1)$\\ \hline
 Magnetic  $g=Tg^s$  & Time reversal   $T$ &  $(x,y,z,t)\ri (x,y,z,-t)$  &  $\text{diag}(1,1,1)$ \\ \cline{2-4}
  & Magnetic mirror reflection $TM_x$ & $(x,y,z,t)\ri (-x,y,z,-t)$  &  $\text{diag}(-1,1,1)$ \\ \cline{2-4}
     & Magnetic two-fold rotation $TC_{2z}$ & $(x,y,z,t)\ri (-x,-y,z,-t)$   &  $\text{diag}(-1,-1,1)$\\ \cline{2-4}
     & Magnetic parity $TP$ & $(x,y,z,t)\ri (-x,-y,-z,-t)$   &  $\text{diag}(-1,-1,-1)$\\ \hline
\end{tabular}		
\caption{   A dictionary of commonly-used, order-two point-group transformations. In the last column, `diag$(i,j,k)$' means a diagonal $O(3)$ matrix with diagonal elements equal to $(i,j,k)$.  \label{tab:pointgroup}}
\end{table}

If a Hamiltonian $H$ acts in the Hilbert space of a certain set of wave functions, we say the Hamiltonian is $g$-symmetric if $\hg$ (the representation of $g$ in said Hilbert space) commutes with the Hamiltonian, or equivalently
\e{
\text{$g$-symmetric Hamiltonian:} \as H^g=H; \as H^g\equiv \hg H\hgmo.
}
$\hg$ is a unitary operator if $g=g_s$, and an antiunitary operator if $g=Tg_s$. In the photovoltaic context, $H=H^0+V$ is the sum of a zeroth-order Hamiltonian $H^0$ (typically a mean-field Hamiltonian of electron quasiparticles in a perfect crystalline medium) and a perturbation $V$ that encodes the coupling to a photo-exciting light wave (in the dipole approximation) and the coupling to an environment (e.g., impurities and a thermal background of photons and phonons, which play the role of a source/sink of energy or  momentum  or both).\\

 In the specific application to the impurity-mediated $\jball$,   $H^0$  reduces  (in $\bk\cdot \bp$ theory) to the effective Hamiltonians for $n$-Weyl fermions [$H^{w,n}$ in \q{nweylham}] and $n$-Dirac fermions [$H^{d,n}$ in \q{ndiracham}], which  respectively have the point groups  $D_{\infty}\times \Z_2^T$ and
$O(2)\times \Z_2^T$ respectively, with the specific symmetry representations  detailed in \app{app:symmetryndiracweyl}. $\Z_2^T$ is the $\Z_2$ group generated by time reversal, $O(2)$ is the orthogonal group in two spatial dimensions, and $D_{\infty}$ is the dihedral group $D_2$ (generated by two-fold rotations $C_{2x}$ and $C_{2y}$ about the x- and y-axes respectively) with an additional continuous rotation symmetry about the z axis.

\subsection{General symmetry constraints}\la{sec:generalsymmetryconstraints}

As derived in \app{app:symmetrytheorems}, a symmetric Hamiltonian has the following implications for the photovoltaic tensor:
\e{
\text{Spatial-symmetric photovoltaic theorem}: \as H^{g^s}\eq H \imp  \bSigma_{abc}= g^s_{ad} g^s_{be} g^s_{cf} \bSigma_{def}; \la{gsconstrainsSigma2}\\
\text{Magnet-symmetric photovoltaic theorem}: \as H^{Tg^s}\eq H \imp  \bSigma_{abc}= \text{Ind}^T_{\Sigma} g^s_{ad} g^s_{be} g^s_{cf}\bSigma_{def}. \la{TgsconstrainsSigma2}
}
The second theorem\footnote{A precursor of this theorem  can be found in \ocite{ahn_lowfrequencydivergence}, which is the source of our inspiration. Their formulation lacked a clarity and generality that our theorem and derivation aim to provide.} 
applies: (i) whether or not $H^T=H$, (ii) whether or not $H^{g^s}=H$,  and (iii) to  components of the bulk photovoltaic tensor with a well-defined\footnote{Any response tensor $\Sigma$ can separated into components $\Sigma_{\pm}$ with compatibility index $\pm 1$:
\e{
\Sigma(H) = \Sigma_+ + \Sigma_-; \as \Sigma_{\pm}= \tf{\Sigma(H)\pm \Sigma(H^T)}{2}.
}
For most of the manuscript, we do not explicitly manifest the subscript $\pm$.} \textit{compatibility index under time reversal}:
\e{
\Sigma\big(H^T\big)= \text{Ind}^T_{\Sigma}\cdot \Sigma\big(H\big); \as \text{Ind}^T_{\Sigma}=\pm 1.\la{compatibilityindexT2}
}
The meaning of $\Sigma(H)$ is that we view $\Sigma$ (the response tensor) as the output of a calculational algorithm (also denoted $\Sigma$) which inputs the Hamiltonian. For instance, the calculational algorithm for \q{eq:jball} manifestly inputs the eigen-energies and eigenstates of $H^0$, as well as inputs  the impurity potential and light polarization vector of $V$. $\text{Ind}^T_{\Sigma}$, the compatibility index of $\Sigma$ under time reversal, is defined in \q{compatibilityindexT2} as the sign factor difference between $\Sigma(H)$ and $\Sigma(H^T)$. The advantage of knowing this index is that if $\text{Ind}^T_{\Sigma}=+1$, $\Sigma$ can be nonzero if $H^T=H$, while if $\text{Ind}^T_{\Sigma}=-1$, $\Sigma$ must be zero if $H^T=H$; we say that $\Sigma$ is compatible with a time-reversal-symmetric Hamiltonian in the former case, and incompatible in the latter case.  Postponing an explicit calculation of the compatibility indices to \app{app:symmetrytheorems}, we merely list the compatibility indices for various components of the bulk photovoltaic tensor in \tab{tab:compindex}.\\

\begin{table}[ht]
\centering
\begin{tabular} {|c|c|c|c|c|} \hline			
  & SInjection$^l$ & SInjection$^c$ &  Shift$^l$/Ballistic$^l$ & Shift$^c$/Ballistic$^c$ \\  \hline \hline 
  $\text{Ind}^T_{\Sigma}$ &$-$&$+$&$+$&$-$ \\ \hline
$\text{Ind}^P_{\Sigma}$ &$-$&$-$&$-$&$-$\\ \hline
$\text{Ind}^{PT}_{\Sigma}$ &$+$&$-$&$-$&$+$ \\ \hline\hline
$M_j$ & $N_{j}$ even & $N_{j}$ even & $N_{j}$ even & $N_{j}$ even \\ \hline
$TM_{j}$  & $N_{j}$ odd & $N_{j}$ even & $N_{j}$ even & $N_{j}$ odd \\ \hline
 $C_{2j}$  & $N_{j}$ odd & $N_{j}$ odd & $N_{j}$ odd & $N_{j}$ odd \\ \hline
  $TC_{2j}$  & $N_{j}$ even & $N_{j}$ odd & $N_{j}$ odd & $N_{j}$ even \\ \hline
   $P$  & 0 & 0 & 0 & 0 \\ \hline
    $PT$  & No constraint & 0 & 0 &  No constraint \\ \hline
\end{tabular}
\caption{In a possibly-magnetic medium, the bulk photovoltaic response tensor can be split into shift, ballistic and saturated injection (SInjection) components;\cite{sipe_secondorderoptical}[Junting Yu, AA] each of these three components can be further split into linear and circular components (indicated by $l$ and $c$ superscripts) based on the symmetry of the last two spatial indices. $\text{Ind}^{g}_{\Sigma}$ is the compatibility index of response tensor $\Sigma$ under $g$  symmetry; this index is defined by \q{compatibilityindexT2} with $T$ replaced by $g$.  In the lower portion of the table, we list the symmetry constraints  arising from the order-two point-group symmetries in \tab{tab:pointgroup}, with  $M_j$   having the generalized meaning of inverting the $j$ spatial coordinate, and  $C_{2j}$  meaning a two-fold rotation about the $j$ axis. `0' means that no response is allowed for that component. $N_j$ is the number of `$abc$' indices (in a particular bulk photovoltaic tensor element $\Sigma_{abc}$) that equals $j$.  	\label{tab:compindex}}
\end{table}

As our first application of the theorems in \qq{gsconstrainsSigma2}{TgsconstrainsSigma2}, the symmetry constraints arising from the order-two symmetries in \tab{tab:pointgroup} are listed in the bottom sub-table in \tab{tab:compindex}. These constraints are of three kinds: (a) trivial, (b) complete, meaning all tensor elements (in a specific class of tensors) vanish, as indicated by `0' in   \tab{tab:pointgroup}, or (c) a parity constraint on $N_j$, which is defined as the number of `$abc$' indices (in a particular bulk photovoltaic tensor element $\Sigma_{abc}$) that equals $j$. For illustration, let us consider how the magnetic reflection $TM_x$ constrains the circular component of the shift current; we look up the $T$-compatibility index from the top sub-table of \tab{tab:compindex}, namely $-1$, and plug this index into our second theorem in \q{TgsconstrainsSigma2}:
\e{
 H^{TM_x}\eq H \imp  \bSigma^{\text{shift},c}_{abc}= - (M_x)_{ad} (M_x)_{be} (M_x)_{cf}\bSigma^{\text{shift},c}_{def}; \as (M_x)=\text{diag}(-1,1,1).
}
It follows that any nonzero $\bSigma^{\text{shift},c}_{abc}$ must have  
 an even number of $x$ indices, e.g., $(abc)=(xxy)$ and $(xyx)$ and $(yzz)$ are allowed but $(xxx)$ is not.\\

In the specific application of Dirac-Weyl fermions coupled to impurities, the symmetry of $H$ is the intersection of the symmetry of $H^0$ and the symmetry of the impurity potential $\Vim$. This means that we only apply the theorems in \qq{gsconstrainsSigma2}{TgsconstrainsSigma2} if  $g$ is simultaneously a symmetry of $H^0=H^{0,g}$ and $\Vim=V^{\text{im,g}}$. The set of all $g$ forms a group specified in the third column of \tab{tab:nweyldiracfermions}. For instance, the monopolar impurity potential $V^m$ has the highest symmetry: $O(d) \times \Z_2^T$ in $d$ spatial dimensions, so the symmetry of $H$ is simply the symmetry of $H^0$, which is sufficiently large to extinguish all ballistic photocurrents for $H^0=H^{d,n}$, and sufficiently large to leave only one independent tensor element ($\Sigma^l_{xyz}$) for $H^0=H^{w,n}$; this is the statement of the second and fifth rows in \tab{tab:nweyldiracfermions}. For completeness, we also provide [in \tab{tab:nweyldiracsymmetryreducedgeneral}] the point groups and symmetry-allowed tensor elements for symmetry-reduced $n$-Dirac/Weyl fermions [\qq{diracmass}{sixpwarp}]; the majority of these  point groups are  magnetic point groups which do not contain time reversal $T$ as an element, but contain various magnetic point-group elements ($Tg^s$), thereby necessitating the use of the second theorem in \q{TgsconstrainsSigma2}. Both tables [\tab{tab:nweyldiracfermions} and \tab{tab:nweyldiracsymmetryreducedgeneral}] are more broadly applicable outside the context of Dirac-Weyl fermions; that is to say, one can erase the leftmost columns and view what remains as \textit{general symmetry constraints} on the photovoltaic tensor given a symmetry group of $H$.

\begin{table}[H]
\centering
\begin{tabular} {|c|c|c|c|c|} \hline
 Fermion & Impurity &  Group & Shift$^l$/Ballistic$^l$ &   Shift$^c$/Ballistic$^c$\\  \hline 
 $H^{w,n}$ & $V^m$  & $D_{\infty}\times \Z_2^T$  & $\Sigma^l_{xyz}=-\Sigma^l_{yzx}$ &0\\ \cline{2-5}				
 &  $V^{d\para z}$ or $V^m+V^{d\para z}$&   $SO(2)\times \Z_2^T$ &$\Sigma^l_{zzz}; \as\Sigma^l_{zxx}=\Sigma^l_{zyy};\as\Sigma^l_{xxz}=\Sigma^l_{yyz};\as \Sigma^l_{xyz}=-\Sigma^l_{yzx}$  & 0\\  \cline{2-5}				
 & $V^{d\para x}$ or $V^m+V^{d\para x}$&   $\Z_2^{C_{2x}}\times \Z_2^T$ & $\Sigma^l_{xxx/\{xyy\}/ \{xzz\} /\{xyz\}}$  & 0\\ \hline		
 $H^{d,n}$ & $V^m$ &  $O(2)\times \Z_2^T$ & 0&0\\ \cline{2-5}
 & $V^{d\para x}$ or  $V^m+V^{d\para x}$& $\Z_2^{M_y}\times \Z_2^{T}$  & $\Sigma^l_{xxx/\{xyy\}}$ & 0\\ \hline
\end{tabular}
\caption{Symmetry-allowed shift/ballistic photovoltaic tensor elements for both linear ($\Sigma^l$) and circular ($\Sigma^c$) components, and for  $n$-Weyl ($H^{w,n}$) and $n$-Dirac ($H^{d,n}$) fermions, coupled to monopolar ($V^m$) or j-oriented dipolar ($V^{d\para j}$) impurities or their combination. $\Sigma_{\{abc\}}$ means all permutations of $(abc)$ are symmetry-allowed;  `0' means all tensor elements vanish.	
\label{tab:nweyldiracfermions}}
\end{table}

\begin{table}[H]
\centering
\renewcommand{\arraystretch}{1.4}
\begin{tabular}{c c c c c c}
\hline\hline
 $H^{w,n}\;+\;$correction & $\as\as$ Group $\as\as$ & Shift$^l$/Ballistic$^l$ & Shift$^c$/Ballistic$^c$ & $\as\as$ Materials $\as\as$ \\
\hline
z-tilt &$\infty2'$ & $\Sigma^l_{xyz}=-\Sigma^l_{yzx}$ & $\Sigma^c_{xxz}=\Sigma^c_{yyz}$ & ?\\
x-tilt &$2'2'2$ & $\Sigma^l_{\{xyz\}}$ & $\Sigma^c_{yxy/zxz}$ &  TaAs~\cite{lv2015observation}\\
3-warp &  $321'$ & $\as\Sigma^l_{yyy}=-\Sigma^l_{xxy}=-\Sigma^l_{yxx};\as \Sigma^l_{xyz}=-\Sigma^l_{yzx}\as$ & 0 &  Ag$_3$BO$_3$~\cite{chang2018topological}\\
6'-warp & $6'22'$ & $\Sigma^l_{xyz}=-\Sigma^l_{yzx}$ & 0 & ?\\ [5pt]
\hline\hline 
 $H^{d,n}\;+\;$correction & $\as\as$ Group $\as\as$ & Shift$^l$/Ballistic$^l$ & Shift$^c$/Ballistic$^c$ & Materials \\
\hline
Dirac mass & $\infty m'm'$ & 0& 0 &  -\\ 
x-tilt &$m'm2'$ & 0& $\Sigma^c_{yxy}$ & ? \\
Dirac mass + x-tilt & $m'$ &   $\Sigma^l_{yyy/\{xxy\}}$  & $\Sigma^c_{yxy}$ & Bilayer WTe$_2$ \cite{LMAA}\\
3-warp &  $3m1'$ &  $\Sigma^l_{yyy}=-\Sigma^l_{xxy}=-\Sigma^l_{yxx}$ & 0 & ?\\
6'-warp & $6'mm'$ & 0& 0 & $n$-graphene \cite{Koshino-McCann-TrigonalWarpingBerrys-2009}\\
Dirac mass + 3/6'-warp &   $3m'$ & $\Sigma^l_{yyy}=-\Sigma^l_{xxy}=-\Sigma^l_{yxx}\as$  &  $ 0$ &  Displaced  $n$-graphene\\ [5pt] \hline \hline
\end{tabular}
\caption{Top sub-table: $n$-Weyl Hamiltonian [\q{nweylham}] with various symmetry-reducing corrections (left-most=first  column and cf. \qq{diracmass}{sixpwarp}) and the corresponding point groups (second  column). We list the symmetry-allowed linear shift/ballistic tensor elements (in the third column) and  symmetry-allowed circular shift/ballistic  tensor elements (fourth column). The last column lists some candidate materials that we are aware of. Bottom sub-table: $n$-Dirac Hamiltonian with various symmetry-reducing corrections, and with symmetry representations that can be read off from \tab{tab:nDiracsymmetryreps}.
The  names of groups have been given in the Hermann-Mauguin notation; in particular, $\infty 2,1=D_{\infty}\times \Z_2^T$, $\infty m,1=O(2)\times \Z_2^T$,  $321'\equiv D_{3}\times \Z_2^T$ and  $3m1'\equiv C_{3v}\times \Z_2^T$. All the symmetry-allowed tensor elements here  are derivable from combining \qq{SO2allowedlinear}{SO2allowedcircular}, \qq{C3allowedlinear}{C3allowedcircular}  and \tab{tab:compindex}. $\Sigma_{abc/def}$ means both $\Sigma_{abc}$ and $\Sigma_{def}$ are symmetry allowed; $\Sigma_{\{abc\}}$ means all permutations of the indices `$abc$' are symmetry-allowed. $n$-graphene means rhombohedral $n$-layer graphene; `displaced $n$-graphene' means $n$-graphene with an out-of-plane displacement field. }
\label{tab:nweyldiracsymmetryreducedgeneral}
\end{table}

\subsection{Hamiltonian-specific symmetry constraints}\la{sec:Hspecificsymmetry}

In contrast with \textit{general symmetry constraints} that apply to all photovoltaic tensors independent of specific details of the Hamiltonian (beyond the assumption that the Hamiltonian is symmetric: $H^g=H^e$), we introduce here the notion of \textit{$H$-specific symmetry constraints} (short for Hamiltonian-specific), which apply to specific components of the photovoltaic tensor and depend on specific structural details of the Hamiltonian. Here are two examples:

\begin{tcolorbox}[colback=white, sharp corners]
\textbf{$TC_{2z}$ theorem:} For 2D systems with $TC_{2z}$-symmetric Hamiltonians, the  ballistic photocurrent (mediated by centrosymmetric impurities) vanishes.    \\

\textbf{$\vec{H}$-polarity theorem:} For any two-band $H^0$ with a point group $G$, if there exists no invariant vector in the Hamiltonian-vector representation of $G$, then the  ballistic photocurrent (mediated by overscreened monopole impurities) vanishes.    
\end{tcolorbox}

\noindent What's `specific' here is that (i) both theorems apply to the impurity-mediated $\jball$, which is only one component of the photovoltaic response; (ii) both theorems assume a coupling to a specific class of impurities, namely  centrosymmetric impurities ($\Vim_{\br}=\Vim_{-\br}$) in the first theorem, and overscreened monopole impurities [\q{overscreenedmonopole3D} and \q{overscreenedmonopole2D}] in the second;  (iii) the second theorem applies to two-band Hamiltonians [\q{twobandham}] which are essentially dot products of a Hamiltonian vector $\bd$ with the vector of Pauli matrices.\\

The formal notion of an `invariant vector in the Hamiltonian-vector representation' is described in
\app{app:dvectorep};  the detailed derivation of the theorems are relegated to \app{app:symconstraintsfirstkind}. Here, we will informally convey the essence of the derivation: for long-wavelength skew scattering mediated by either class of impurities,   the $(\text{Asymmetric transition rate})_{\bk\lea\bk'}$ [\q{skewscattering}] is proportional to 
\e{
 \sum_{\bk''}   \imag\braket{u_{\bk}}{u_{\bk''}}   \braket{u_{\bk''}}{u_{\bk'}} \braket{u_{\bk'}}{u_{\bk}}  \delta(E_{\bk''}-\Eex) \refeq{Bargmannrealimag} \frac{1}{4} \hat{\mathbf{d}}_{\bk}\times \hat{\mathbf{d}}_{\bk'}  \cdot [{\hbd}]_{\Eex},
}
which is proportional to the $\vec{H}$-vector polarization [\q{Hpolarization}] at the excitation energy $\Eex$ (the energy of the excitation surface).
Being proportional to the asymmetric transition rate [\q{eq:jball}], the ballistic photocurrent evidently vanishes if either the wave function is real ($\ket{u_{\bk}}\in \R^m$) or if the $\vec{H}$-vector polarization is trivial ($[{\hbd}]_{\Eex}=0$).   \\

Certain symmetries render the wave function to be real for all $\bk$ in the Brillouin zone. For illustration, the {massless} $n$-Dirac Hamiltonian in \q{ndiracham} is symmetric under $C_{2z}T$ (the composition of two-fold rotation and time-reversal):
\e{
[H^{d,n}_{k_x,k_y},\widehat{C_{2z}T}]=0; \as \widehat{C_{2z}T}= (\text{unimportant phase factor})\, \sx K; \as K \;\text{implements complex conjugation},
}
which ensures that $H^0$ involves only two of three Pauli matrices; a unitary transformation can always be chosen to make these two Pauli matrices real. In contrast, the $n$-Weyl Hamiltonian involves all three Pauli matrices, which reflects the absence of
$PT$ symmetry (composition of parity and time-reversal).\footnote{In 3D systems with negligible spin-orbit coupling,  $\widehat{PT}$ squares to the identity operator in its action on $\ket{u_{b\bk}}$, which ensures $\ket{u_{b\bk}}$ can be made real.} These observations underlie the above $TC_{2z}$ theorem.   \\

 Any symmetry $g$ of $H^0$ [\q{twobandham}] has three representations encapsulated in:
\e{
 H^0_{g\cdot \bk}=\hg H^0_{\bk} \hgmo= (\overg \cdot \bd_{\bk})\cdot  \bsigma +\delta E_{\bk}\iden_2,\la{twobandH0symmetry}
}
namely (i) $g\circ$ is an orthogonal matrix acting in $\bk$-space, (ii) $\hg$ is a (possibly antiunitary) matrix acting on complex wave functions $\in \C^2$, and (iii) $\overg$ is an orthogonal, three-by-three matrix acting on the Hamiltonian $\bd$-vector. As a case in point, for $g=T$ (time-reversal), $g\cdot \bk=-\bk$, $\hat{T}=i\sy K$ acts on (pseudo)spin-half wave functions, and $\overT\cdot \bd_{\bk}=-\bd_{\bk}$ because time-reversal flips all (pseudo)spin-half components. More generally, for any symmetry $g$ of $H^0$ (time-reversing or not), \q{twobandH0symmetry} implies that the $\vec{H}$-vector polarization [\q{Hpolarization}] transforms as a scalar under $g$: $[\hbd]_E= \overg \cdot [\hbd]_E$, meaning it is invariant under the action of $g$ in the Hamiltonian-vector representation. In the case of $\hg=\hat{T}=i\sy K$, this evidently implies there can be no polarization: $[\hbd]_E=- [\hbd]_E=0$, which extinguishes the ballistic photocurrent. This observation, when generalized to all magnetic point groups and taken to its logical conclusion, leads to the above $\vec{H}$-polarity theorem.  \\

As applications of the two aforementioned theorems,  \tab{tab:nweyldiracsymmetryreducedmonopole}  shows which of our case studies have $TC_{2z}$ symmetry and which have an invariant vector in the Hamiltonian-vector representation; a `0' in the column headed by $[\hbd]$ means that  no invariant vector exists; `$\para 3$' means that it exists and lies parallel to the 3 direction. Accordingly, $\jball=0$ in \tab{tab:nweyldiracsymmetryreducedmonopole} wherever the  assumptions of the two theorems hold true.\\

Our final example of a $\vec{H}$-specific symmetry constraint, which we will call the `impurity-parity theorem', again applies to the impurity-mediated ballistic photovoltaic tensor: $B^{\text{de}}$. Unlike previous examples wherein $H^g=H$, the impurity-parity theorem   relies on  $(H=H^0+V)$ being  $g$-\textbf{a}symmetric in a particular way, namely, with $H^0$ being $g$-symmetric but the impurity potential $\Vim$ being $g$-\textbf{anti}symmetric: $(\Vim)^g= \hg \Vim \hgmo=-\Vim$. (We may say the impurity has odd $g$-parity.) Under these conditions, a third theorem states, for $g=g^s$ being a spatial symmetry, 
\e{
\text{Impurity-parity theorem}: \as
H^{0,g^s}=H^{0}, \as V^{\text{im},g^s}=- V^{\text{im}}:\as  \Sigma^{\text{im}}_{\text{ballistic},abc}=  - g^s_{ad} g^s_{be} g^s_{cf} \Sigma^{\text{im}}_{\text{ballistic},def}. \la{secondtypespatial}
}
As detailed in \app{app:symconstraintssecondkind},
\q{secondtypespatial} derives from  the impurity-mediated $\jball$ [\q{eq:jball}] being third order in $\Vim$ (at lowest-order perturbation theory), and hence is odd under $\Vim \ri -\Vim$. To describe one application, the impurity potential (of a dipole moment oriented in the $x$ direction) is antisymmetric under a two-fold rotation ($C_{2z}$) about the $z$-axis; $g^s=C_{2z}$ is a symmetry of both $n$-Weyl and $n$-Dirac Hamiltonians, either of which can be $H^0$ in \q{secondtypespatial}; the impurity-parity theorem then imposes $H$-specific constraints (in addition to the general constraints in \s{sec:generalsymmetryconstraints}) which are summarized in \tab{tab:nweyldiracfermions}; cf. \app{app:symconstraintsnweyl}-\ref{app:symconstraintsndirac}. This concludes the interlude on symmetry.

\begin{table}[H]
\centering
\begin{tabular} {|c|c|c|c|c|} \hline
   Fermion & Impurity &  Group & $\bB^{\text{de},l}$  & $\bB^{\text{de},c}$ \\  \hline 
 $H^{w,n}$ & $V^m$ & $D_{\infty}\times \Z_2^T$  & 0 & 0 \\ \cline{2-5}						
 &  $V^{d\para z}$ or $V^m+V^{d\para z}$&   $SO(2)\times \Z_2^T$ &$B^{\text{de},l}_{zzz}; \as B^{\text{de},l}_{zxx}=B^{\text{de},l}_{zyy};\as B^{\text{de},l}_{xxz}=B^{\text{de},l}_{yyz};\as B^{\text{de},l}_{xyz}=-B^{\text{de},l}_{yzx}$   & 0 \\ \cline{2-5}	
 & $V^{d\para x}$&   $\Z_2^{C_{2x}}\times \Z_2^T$ & $B^{\text{de},l}_{xxx/\{xyy\}/\{xzz\}}$ &0 \\ \cline{2-5}	
 & $V^m+V^{d\para x}$&  $\Z_2^{C_{2x}}\times \Z_2^{T}$  & $B^{\text{de},l}_{xxx/\{xyy\}/\{xzz\}/\{xyz\}}$ &0  \\ \hline			 
  $H^{d,n}$ & $V^m$ &  $O(2)\times \Z_2^T$ & 0&0\\ \cline{2-5}
 & $V^{d\para x}$ or $V^m+V^{d\para x}$&  $\Z_2^{M_y}\times \Z_2^T$  &  $B^{\text{de},l}_{xxx/\{xyy\}}$ & 0\\ \hline
\end{tabular}
\caption{Symmetry constraints specific to the impurity-mediated ballistic photovoltaic tensor,  for both linear ($\Sigma^l$) and circular ($\Sigma^c$) components, and  for  $n$-Weyl ($H^{w,n}$) and $n$-Dirac ($H^{d,n}$) fermions,  coupled to monopolar ($V^m$) or j-oriented dipolar ($V^{d\para j}$) impurities or their combination.  $B^{\text{de}}_{\{abc\}}$ means all permutations of $(abc)$ are symmetry-allowed;  `0' means all tensor elements vanish.\label{tab:nweyldiracconstraintspecific}}
\end{table}

\section{Discussion, generalization, outlook}\la{sec:discussion}

Herein we present a partial summary of our results from a broader perspective, discuss their generalizations, and present an outlook for future theoretical and experimental progress.

\subsection{Optical vortices in materials both topological and non-topological}

Topological band touchings are known to be sources/sinks of the intraband Berry curvature;\cite{TKNN,Haldane1988,wan_weylsemimetal} here we uncover their dual property as sources/sinks of vorticity in the interband Berry phase, meaning that the topological band touching $\bk$-point is a wave-function-singular point from which optical vortex lines radiate, see Fig.~\ref{fig:ballistic}(a). This follows from a generalized Chern-vorticity theorem  [\q{chernvortex}] which guarantees optical vorticity in every topological material ($\equiv$ insulator + metal) associated with nontrivial Chern numbers, including $n$-Weyl semimetals, higher-angular-momentum multifold fermions, and Chern insulators. \\

This is not to say that optical vortices are absent in non-topological materials; on the contrary, vortices likely exist \textit{generically} in materials -- topological or not. What the theorem [\q{chernvortex}] implies (for materials with trivial Chern numbers) is that the net vorticity of generically-occurring vortex points vanish. Moreover, it is not possible to infer the $\bk$-locations of generically-occurring vortex points by topological arguments.\\

It is also worth emphasizing having a nontrivial Chern number is not the defining characteristic of all topological materials. After all, the Chern number is a quantized Abelian intraband Berry phase associated to energy-nondegenerate bands, and some topological materials are characterized by a quantized non-Abelian intraband Berry phase associated to bands with symmetry-enforced energy degeneracy. Examples of the latter category include 3D Dirac semimetals~\cite{young_3Ddiracsemimetal} and 2D and 3D $\Z_2$ topological insulators.~\cite{kane2005B} It may be interesting to  explore the implications of a postulated, non-Abelian generalization of the Chern-vortex theorem to the bulk photovoltaic effect in broader classes of topological materials.

\subsection{Vortex-enhanced photovoltaic current in both bulk and surface}

Optical vorticity makes the photoexcitation rate anisotropic over the excitation surface, see Fig.~\ref{fig:ballistic}(c). For $n$-Weyl semimetals, such anisotropy follows from a topologically-guaranteed intersection of the excitation surface by vortex lines [\q{vortexintersectioneffect}]; for $n$-Dirac materials, such anisotropy follows from a topologically-guaranteed proximity of the excitation curve to vortex points [\q{vortexproximityeffect}]. An anisotropic photoexcitation rate can also result from an accidental intersection/proximity to generically-occurring vortex lines/points in non-topological materials; whether such accidents happen or not depends on material details, and can only be determined on a case-by-case basis, in contrast with the general certitude
gained with topological materials. \\

The aforementioned \textit{vortex-induced photoexcitation anisotrop}y activates  several photovoltaic effects:\\

\noi{i} The first effect is that when electrons collide with phonons/impurities and scatter from $\bk$-regions of higher photoexcitation rate to $\bk$-regions of lower photoexcitation rate, there is an accompanying shift associated to the electron scattering transitions,\cite{belinicher_kinetictheory} resulting in a shift photovoltaic current.\\

\noi{ii} Secondly, the same scattering (from high- to low-photoexcitation regions) has a skew component that makes the electron distribution asymmetric: $f_{\bk}\neq f_{-\bk}$. Thus vortices enhance the ballistic photovoltaic current.  We have focused on  skew scattering  mediated by crystalline defects, which results in the following: 
\begin{equation}
\label{eq:ballisticcurrent2}
\text{Defect-mediated}\; \bj_{\text{ballistic}}=-\frac{|e|}{\mathcal{V}}\sum_{BB'B''B'''} 
\bv_{B}(W^{mr})^{-1}_{BB'} W^{de,a}_{B'B''} 
(W^{mr})^{-1}_{B''B'''} \, \delta I^{ex,s}_{B'''}.
\end{equation}
Here, $B=(b\bk)$ and $B'=(b'\bk')$ are labels for electronic Bloch states, $\bv_B$ is the group velocity, $W^x_{BB'}$ is the transition rate matrix elements associated to momentum relaxation [for $x=mr$] and defect-mediated skew scattering [for $x=(de,a)$], and $\delta I^{ex,s}_B$ is the anisotropic component of the photoexcitation rate, with explicit expressions in App.~\ref{app:solvekinetic}.  \q{eq:ballisticcurrent2} is a more general expression for $\jball$ which generalizes our previous expressions [\qq{jballisticinwords}{eq:jball}] beyond the assumption of electron-hole symmetry and the momentum-relaxation-time approximation; in the last regard,  \q{eq:ballisticcurrent2} accounts for the distinct relaxation times of the different harmonics of a momentum-excited electron distribution, as explained in \ref{app:rta}. In addition to defect-mediated processes, other mechanisms of skew scattering  (e.g., phonon-mediated\footnote{See Eq. (20) in \cite{belinicher_phononmechanism}. }) can also be activated by the vortex-induced photoexcitation anisotropy. \\

\noi{iii} Beyond bulk photovoltaic effects, the vortex-induced photoexcitation anisotropy should combine with diffuse surface scattering to produce a surface photovoltaic effect.~\cite{Alperovich-Terekhov:1981} In 2D systems, the above sentence should be modified by replacing `bulk' with `2D-bulk' and `surface' with `edge'.\\

It would be interesting to investigate the above vortex-enhanced photovoltaic effects more generally in topological materials, beyond the present study of the defect-mediated $\jball$.

\subsection{Bulk-photovoltaic probe of topological semimetallic and correlated insulating phases}\la{sec:correlated}

As a representative application of our theory of the defect-mediated $\jball$, we 
consider $n$-graphene (rhombohedral $n$-layer graphene), which hosts both topological semimetallic and correlated insulating phases depending on the value of $n$, the choice of substrate, and the externally applied displacement field.\cite{Lui-Heinz-ObservationElectricallyTunable-2011,Han-Ju-CorrelatedInsulatorChern-2024a}.\\

We begin with the uncorrelated semimetallic phases in free-standing $n$-graphene, which is centrosymmetric ($P$-symmetric) and does not admit a bulk photovoltaic current in its pristine form. If impurities are implanted (by defect engineering) with a uniformly-oriented electric dipole moment, then these impurities can mediate $\jball \propto \omega^{\text{exponent}}$ in the low-frequency response.  This frequency exponent is $-2/n$ if the impurities are pure dipoles with negligible net charge, and equals $+2/n$ if the impurities are charged dipoles [\tab{tab:powerlaw}], according to the impurity-dependent power laws more generally described in \s{sec:powerlaws}.\\

If we restrict ourselves to plain-vanilla, monopolar-charged impurities with no dipolarity, then free-standing, uncorrelated $n$-graphene will not admit a defect-mediated $\jball$, according to the symmetry theorems in \s{sec:Hspecificsymmetry}. However, an out-of-plane displacement field will generate a $P$-breaking Dirac mass and induce a band gap, and concomitantly activate $\jball$. Theorists be forewarned! A naive modeling based on lowest-order effective Hamiltonians can give the illusion of a null $\jball$; higher-order-in-$\bk$ warping terms are key to dispelling this illusion, as explicated in \s{sec:warp-nDirac}.  \\

Let us move on to the correlated insulating phases of $n$-graphene, some of which have manifested experimentally for $n\geq 4$,\cite{Han-Ju-CorrelatedInsulatorChern-2024a} according to \tab{tab:correlated}. To the extent that intra-valley and intra-spin scattering dominate for longer-wavelength, non-magnetic disorder, the total $\jball$ is a sum of contributions from each valley and spin flavor [\q{13-multiWeyl-nDirac-w-jball-fin}], and each of these contributions depends on the flavor-dependent Dirac mass listed in \tab{tab:correlated}. Thus do we determine which phases admit $\jball$ (mediated by monopolar impurities), and which phases admit a spin-differentiated $\jball^s$ (defined by spin-up $\jball$ minus spin-down $\jball$). This illustrates the utility of bulk photovoltaics in diagnosing correlated phases, as will be elaborated in the following experimental program.

\begin{table}[H]
\centering
\begin{tabular}{lccccc c c}
\hline
Correlated insulator & $\as$ Observed for $n$ $\as$ & Dirac mass $\as$ & $T$ & $P$ & $PT$ & $\as\jball\as$ & $\as\jball^s\as$\\
\hline
Layer antiferromagnet      & $4,5$   & $\Delta\,\sigma_3 s_3$        & broken    & broken    & preserved & 0  & nonzero\\
Quantum anomalous Hall     & $4,5,6$ & $\Delta\,\sigma_3 \tau_3$     & broken    & preserved & broken  &  0  & 0\\
Quantum valley Hall        & ---     & $\Delta\,\sigma_3$            & preserved & broken    & broken   &  nonzero & 0\\
Quantum spin Hall          & ---     & $\Delta\,\sigma_3 \tau_3 s_3$ & preserved & preserved & preserved & 0 & 0\\
\hline
\end{tabular}
\caption{  
Each correlated insulating phase (first column) is described by valley- and spin-dependent Dirac mass (third column). `$\Delta\,\sigma_3 s_3$' means that the gapping parameter $\Delta_{\xi,+}=-\Delta_{\xi,-}=\Delta$ [\q{Hvalley}] is valley-independent but spin-dependent;  `$\Delta\,\sigma_3 \tau_3$' means that $\Delta_{+,s}=-\Delta_{-,s}=\Delta$ is spin-independent but valley-dependent. The second column indicates the number of layers $(n)$ for which the insulating phase has been experimentally observed in charge-neutrality $n$-graphene. The symmetries of the insulating phase are indicated in the next three columns. The seventh column specifies if $\jball$ (mediated by monopole impurities, summing over all valleys and spin) is symmetry-allowed; the last column specifies if  $\jball^s$ (defined by spin-up  $\jball$  minus spin-down $\jball$) is symmetry-allowed. 
\la{tab:correlated}
}
\end{table}

\subsection{Tri-pronged experimental program}

Consider a tri-pronged experimental program that  combines:

\noi{i} Photoconductivity measurements: polarization-variability of the optical setup  is necessary to test the symmetry constraints of the photovoltaic tensor [\s{sec:symmetry}] so as to experimentally confirm that the photocurrent is a bulk photovoltaic current; probing the low-energy properties of topological semimetals requires a  mid- to far-infrared laser; to test the impurity-dependent power laws [\s{sec:powerlaws}], frequency tunability of the laser is desirable;

\noi{ii} Crystalline defect characterization, by transmission electron and/or scanning tunneling microscopy;  

\noi{iii} Defect engineering by variation of material synthesis parameters, e.g., dopant type and concentration, stochiometric ratios in flux growths.\\

\noindent Such a tri-pronged program serves three purposes:

\noi{a} A phenomenological framework to rigorously diagnose dirty topological materials. The sharpest experimental signatures of topological materials rely on the materials being clean.\cite{AALG_fermiologyreview} In contrast, the defect-mediated bulk photovoltaic effect dominates when materials are dirty.

\noi{b} Pin down the dominant mechanism of the bulk photovoltaic current. A bulk photovoltaic current that grows in proportionality to the defect concentration is certainly defect-mediated. It is desirable to distinguish between the two components of the  defect-mediated bulk photovoltaic current: shift vs ballistic.\cite{belinicher_kinetictheory,belinicher_ballistic} The impurity-dependent power laws [\s{sec:powerlaws}] and the Hamiltonian-specific symmetry constraints of the photovoltaic tensor [\s{sec:Hspecificsymmetry}] are unique to the defect-mediated ballistic photovoltaic effect.

\noi{c} Defect engineering allows to move in the material phase space; such motion can be theory-guided toward optimizing optoelectronic functionalities (e.g., polarization-sensitive photodetection)  based on the bulk photovoltaic effect.\\

\begin{acknowledgments}
{\it Acknowledgments.}
{P.S. acknowledges useful communications with Dmitry Chichinadze who brought his recent work~\cite{Chichinadze-Nussinov-GiantNonlinearConductivity-2026} to our attention.
P.S. is grateful to the QuSpin Center of Excellence at NTNU for warm hospitality, where part of this work was completed.  P.Z. was primarily supported by the Center for Emergent Materials, an NSF MRSEC, under award number DMR-2011876.
} Elio K\"onig kindly taught us about the  `diffractive' and `Gaussian' skew scatterings, and Ryan Baumbach shared his immense experience in defect engineering. Conversations with Wahaj Ayub have been helpful in generalizing the Chern-vorticity theorem.
\end{acknowledgments}
\vspace{2cm}
\appendix

\noindent Here is an outline of the Appendix.\\

\centerline{\underline{\app{app:preliminaries}: Preliminaries}}
\vspace{3mm}
\begin{itemize}
    \item \ref{app:crystallineHam}: A review of the basic vocabulary underlying Hamiltonians with crystallographic symmetry.

\item \ref{app:effectiveHam}: We transition from exact Hamiltonians to effective few-band Hamiltonians, explaining their strengths, limitations and point-group symmetries. Emphasis is placed on the effective Hamiltonians of $n$-Dirac/Weyl fermions.

\item \ref{app:impuritymatrixelement}: A review of basic properties of matrix elements of screened impurity potentials. Extra care is taken for 3D impurity potentials screened by 2D electrons. 

\item \ref{sec:kineticmodel}: This appendix elucidates and solves the kinetic model which applies  to disordered bulk-photovoltaic materials. 

\item \ref{app:ballisticphotocurrent}: Expressions for the impurity-mediated ballistic photocurrent are derived, with and without the momentum relaxation time approximation.
\end{itemize}

\centerline{\underline{\app{app:vortices}: Appendix to `Optical vortices in topological semimetals and Dirac-Weyl materials'}}

\begin{itemize}
    \item \ref{app:chernvortex2sphere}: The Chern-vorticity theorem is proven for a spherical, enclosing surface. 

\item \ref{app:affinity2band}: The optical affinity of two-band effective Hamiltonians is expressed in terms of the quantum metric and Berry curvature, with an application to the isotropic Weyl fermion.

    \item \ref{app:symmetryshift}:  The point-group-symmetric constraint on the photonic shift vector field is derived. 

    \item \ref{app:models-multi}: Based on the effective Hamiltonians of multifold fermions, their Chern numbers and optical vorticity are calculated and demonstrated to satisfy the Chern-vorticity theorem.

 \item \ref{app:director}: Equivalent expressions of the director field are discussed. A duality is proven between optical vortices and skyrmions of the director field, with an application to the isotropic Weyl fermion.
    
\end{itemize}

\centerline{ \underline{ \app{app:jballdiracweyl}: Appendix to  `Ballistic photovoltaic current in disordered Dirac-Weyl materials'}}

\begin{itemize}
    \item \ref{app:scaling}: The impurity-dependent power laws of the ballistic photocurrent are derived by scaling/dimensional analysis.

    \item \ref{app:argument}: A more general argument is presented for the vortex-enhanced, defect-mediated ballistic photocurrent. The argument extends the vortex-intersection/proximity arguments specific to Dirac-Weyl systems.

    \item \ref{app:vortexintersectionnweyl}-\ref{app:vortexpromityndirac}: Calculational details are provided for the `Vortex-intersection effect in $n$-Weyl semimetals' [\s{sec:nweylvortexintersection}] and    `Vortex-proximity effect in $n$-Dirac semimetals' [\s{sec:nDirac-all}].

    \item \ref{app:tilted Weyl}-\ref{app:current-ndirac-warped}: Calculational details are provided for the `{Tilt-activated $\jball$ for symmetry-unpinned Weyl fermions}' [\s{sec:tiltactivated}] and `{Warp-activated $\jball$ for $n$-graphene}' [\s{sec:warp-nDirac}].
    
\end{itemize}

\centerline{ \underline{ \app{app:symmetrytheorems}:  Appendix to `Symmetry theorems for bulk photovoltaic response tensors'} }

\begin{itemize}

\item \ref{app:responsetensorsymmetric}: What it means for a response tensor to be point-group-symmetric, with emphasis on the bulk photovoltaic tensor.

\item \ref{app:algorithm}: How to view a response tensor as a calculational algorithm that inputs the one-electron distribution, Hamiltonian and collisional integral.

\item \ref{app:gtransformcollisionalintegral}: How the aforementioned collisional integral transforms under point-group symmetry.

\item \ref{app:compatibility}: What is a symmetry compatibility index of a response tensor.

\item \ref{app:generalconstraints}:  Derivation of the general symmetry  theorems on photovoltaic tensors [\s{sec:generalsymmetryconstraints}].

\item \ref{sec:equivariance}: Determination of the symmetry-allowed photovoltaic response tensor elements, by viewing a $G$-symmetric response tensor as a $G$-equivariant linear map.

\item \ref{app:specificsymconstraints}: Derivation of the Hamiltonian-specific symmetry theorems on photovoltaic tensors [\s{sec:Hspecificsymmetry}].

\item \ref{app:symconstraintsnweyl}: Application of all the preceding theorems to determine the symmetry-allowed photovoltaic tensor elements of $n$-Weyl fermions,

\item \ref{app:symconstraintsndirac}: Application of all the preceding theorems to determine the symmetry-allowed photovoltaic tensor elements  of $n$-Dirac fermions.

\end{itemize}

\section{Preliminaries}\la{app:preliminaries}

\subsection{Crystalline Hamiltonians and their point-group symmetries}\la{app:crystallineHam}

Let $\cheH^0$ be the mean-field Hamiltonian of electronic quasiparticles in a 3D crystalline medium of volume $\calv=L_xL_yL_z$ (which will be taken to infinity). Because this medium has discrete translational symmetry in three independent directions, the eigenstates of  $\cheH^0$ can be chosen to have a Bloch-wave form\cite{ashcroft_mermin} in the Schr\"odinger representation, 
\e{
(\cheH^0-E_B)\cket{\psi_B}=0; \as \psi_{B}(\br)=\cbraket{\br}{\psi_B} = \tf{\eikr}{\rootv}u_{B}(\br); \as u_{B}(\br)=u_{B}(\br+\bR). \la{Blochwave}
}
meaning that each wave function is  product of a plane-wave phase factor and a \textit{cell-periodic function} that is periodic under translation by any Bravais-lattice vector ($\bR$). Here and throughout the text, we have employed the composite index
\e{
B=(b,\bk); \as B'=(b',\bk'); \as B''=(b'',\bk'')
}
to combine the band index and wavevector.\\

In addition to having discrete translational symmetry, a crystalline medium may also have point-group symmetries. A general  point-group symmetry ($g$) acts on spacetime
\e{
g \circ (\br,t)= (g^{s,\mo}\cdot \br,i_gt); \as g^{s,\mo}\cdot \br = g^{s,\mo}_{ab}r_b; \as i_g \in \{-1,1\}; \la{gspacetime}
}
with $g^s_{ab}$ being an $O(3)$ matrix. $i_g$ is a $\Z_2$ index that  distinguishes between a
\e{
\text{Spatial point-group transformation:} \as g=g^s,\as i_{g^s}=1;\\
\text{Magnetic point-group transformation:} \as g=Tg^s=g^sT,\as i_{Tg^s}=-1, \la{spatialvsmagnetic}
}
with $Tg_s$ meaning  the composition of time-reversal with a spatial point-group transformation.\\

When the same symmetry $g$ acts on Schr\"odinger-type wave functions [$\psi(\br)=(\br|\psi)$] and the first-quantized  position operator $\cher$, $g$  is represented by an operator $\check{g}$:
\e{
  \cheg \cher_a \chegmo = g^{s,\mo}_{ab}\cher_b; \as (\br|\cheg= K^g(g^{s,\mo}\cdot \br| \imp (\br|\cheg|\psi)= \overline{(g^{s,\mo}\cdot \br|\psi)}^g.\la{checkg}
}
$\check{g}$ is unitary if $i_g=1$ and otherwise is antiunitary; this manifests in \q{checkg} with
 $K^g=K$ (operator of complex conjugation) if $g$ inverts time,  and otherwise $K^g=1$. In particular, for the exponentiated position operator,
\e{
\cheg e^{i\bk\cdot\chebr}\chegmo= e^{i(g\cdot \bk)\cdot\chebr}; \as  g\cdot \bk = i_g g^s\cdot \bk.\la{expposition}
}
Our crystalline Hamiltonian has a point-group symmetry $g$ if
\e{
[\cheg, \cheH^0]=0 \imp \cheg \cheH^0_{\bk}\chegmo \refeq{expposition} \cheH^0_{g\cdot \bk}; \as \cheH^0_{\bk}\equiv e^{-i\bk\cdot\chebr}\cheH^0 e^{i\bk\cdot\chebr}.    
}

It  is convenient to  separate the positional coordinate  into an intracellular coordinate (parametrizing a primitive unit cell) and a Bravais-lattice vector:
\e{
\text{Position}\as \br= \;(\text{Intracellular} \;\btau)+ \;(\text{Bravais-lattice vector}\;  \bR) \in \R^d.
} 
Bloch's theorem allows to view $\cheH^0_{\bk}$ as a Hamiltonian acting on wave functions of the intracellular coordinate defined within a primitive unit cell.\cite{ashcroft_mermin} In the transition from the Hilbert space of wave functions over $\R^3$ to intracellular wave functions, we concomitantly modify our notations for operators and bra-kets:
\e{
(\hH^0_{\bk}-E_{b\bk})\ket{u_{b\bk}}=0; \as \hg \hH^0_{\bk}\hgmo = \hH^0_{g\cdot \bk}.\la{symmetricbloch}
}
Our normalization is
\e{
\cbraket{\psi_B}{\psi_{B'}}=\int_{\bR^d}|\psi_B(\br)|^2 d\br   = \delta_{BB'} \iff  \braket{u_B}{u_B'}= \int_{\vcell}  \overline{u_{B}(\btau)}u_{B'}(\btau)\tf{d\btau}{\vcell}=\delta_{BB'}, \la{normalizeBloch}
}
with $\delta_{BB'}=\delta_{bb'}\delta_{\bk\bk'}$ being a product of dimensionless Kronecker delta functions. If spin is an important degree of freedom, then both inner products in \q{normalizeBloch} should include a summation over spin indices.

\subsection{Effective Hamiltonians and their point-group symmetries}\la{app:effectiveHam}

\subsubsection{The topological value of effective Hamiltonians}

A minimal model for a geometrically nontrivial wave function involves a two-band, Luttinger-Kohn-type, effective Hamiltonian:\cite{Luttinger_Kohn_function}
\e{
H^0_{\bk}=\bd_{\bk}\cdot \bsigma +\delta E_{\bk}\iden_2; \as \bd \in \R^3; \as\bsigma=(\sx,\sy,\sz);\as  \iden_2=\diagmatrixtwo{1}{1}.  \la{twobandham2}
}
Such a Hamiltonian acts in a two-dimensional Hilbert space with basis vectors expressible as $\ket{u_{1,\bk_0}}$ and $\ket{u_{2,\bk_0}}$, with $\ket{u_{b,\bk_0}}$ being an energy eigenstate of the exact Hamiltonian $\hat{H}^0_{\bk}$ [\q{symmetricbloch}] at a (typically high-symmetry) wavevector $\bk_0$, which in our context is the wavevector of the Dirac-Weyl node. \\

By diagonalizing the effective Hamiltonian as 
\e{
U_{\bk}^{\mone} H_{\bk} U_{\bk}= \delta \Ek\iden_{2} + |\bd_{\bk}|\sigma_3,
}
the columns of the unitary matrix $U_{\bk}$ can be used to construct:
\e{
\text{Approximate energy eigenstates} \as \ket{u_{\pm,\bk}}=\sum_{\alpha=1}^2 \ket{u_{\alpha\bk_0}}(U_{\bk})_{\alpha \pm}. \la{LKapprox}
}
In contrast, the exact energy eigenstates (at $\bk\neq \bk_0$) are not generally expressible  as linear combinations of the exact energy eigenstates (at $\bk_0$). The error in the Luttinger-Kohn approximation [\q{LKapprox}] becomes more severe as $\bk$ deviates further from $\bk_0$. It follows that eigenstate-dependent quantities (like the Berry connection) calculated in the Luttinger-Kohn approximation will deviate quantitatively from their exact values.\cite{AA_wilsonloopinversion}   \\

Nevertheless, not all is doom and gloom. The value of effective Hamiltonians like \q{twobandham2} is that they accurately encode topological quantities like the Chern number (for a $\bk$-sphere enclosing $\bk_0$), which implies [by the Chern-vorticity theorem in \q{chernvortex}] that effective Hamiltonians also accurately encode the optical vorticity (of said $\bk$-sphere). Because of quantitative errors in the interband Berry connection,\cite{AA_wilsonloopinversion} the $\bk$-position of a vortex line (approximated by an effective Hamiltonian) may deviate from the exact $\bk$-position. But this  will not change the topological fact that vortex lines must intersect the $\bk$-sphere, with a net vorticity determined by the Chern numbers of conduction and valence bands [\q{chernvortex}]. To further reduce the gloom,  certain point-group-symmetric effective Hamiltonians will exactly encode the $\bk$-position of the vortex line, e.g., in cases when the vortex line is  symmetry-pinned to a rotation-invariant $\bk$-line [cf. \q{symmetryshift2}]. This motivates identifying the point groups of the effective Hamiltonians we use to model $n$-Dirac/Weyl fermions.

\subsubsection{Point groups of $n$-order Dirac-Weyl Hamiltonians}\la{app:symmetryndiracweyl}

The $n$-Weyl Hamiltonian\cite{chen_multiweyl} is
\e{
H^{w,n}_{\bk}= f_n \big[ k_{+}^n\sigma_- +\;h.c.\; \big]+ v k_z\sz; \as k_{\pm}=k_x\pm ik_y; \as \sigma_{\pm}=\tf{\sigma_x\pm i\sigma_y}{2}.\la{Hwnapp}
}
The point group is $D_{\infty}\times \Z_2^T$, with $\Z_2^T$ being the $\Z_2$ group generated by time reversal, and $D_{\infty}$ being the dihedral group $D_2$ (generated by two-fold rotations $C_{2x}$ and $C_{2y}$ about the x- and y-axes respectively) with an additional continuous rotation symmetry about the z axis denoted $R_z^{\theta}$: 
\e{
\widehat{R_z^{\theta}} H^{w,n}_{k_{\pm},k_z} \widehat{R_z(-\theta)} = H^{w,n}_{e^{\pm i\theta}k_{\pm},k_z}; \as   e^{-i\tf{\sz}{2}\phi}\sigma_{\pm}e^{i\tf{\sz}{2}\phi}=e^{\mp i \phi} \sigma_{\pm}; \as R_z\big(\tf{2\pi}{m}\big)=C_{mz}; \as m=2,3,4,6.\la{ctsrotsymmetry}
}
The direct product structure in $D_{\infty}\times \Z_2^T$ encodes the fact that time-reversal commutes with all spatial transformations. If $v=f_1$, then the point group of $H^{w,1}$ is upgraded to $SO(3)\times \Z_2^T$.  Spatial inversion $P$ is not a symmetry; if it were, $PT$ would be, ensuring either a Kramers degeneracy  or zero Abelian Berry curvature at each $\bk$, depending on whether the representation of $PT$ is projective or linear; either case is incompatible with n-Weyl fermions. Mirror reflection is not a symmetry; if it were, mirror composed with two-fold rotation gives $P$ symmetry, which has been demonstrated to be absent. \\

The specific Hamiltonian and symmetry representations for each $n$-Weyl fermions are given in \tab{tab:nWeylsymmetryreps}. For odd $n$, the symmetry representations of $D_{\infty}\times \Z_2^T$ are realized projectively  (e.g., $T^2=$identity is projectively represented as $T^2=-\iden$), meaning that the Weyl fermion is naturally interpreted as having  half-odd-integer (pseudo)spin.\\

\begin{table}[h]
\centering
\renewcommand{\arraystretch}{1.4}
\begin{tabular}{c c c c c c c c c}
\hline\hline
$n$ & $H^{w,n}_{\bk}$ & $\widehat{C_{2x}}$ & $\widehat{C_{2y}}$ & $\widehat{C_{2z}}$ & $\widehat{R_z^{\theta}}$ & $\widehat{T}$ & $\widehat{C_{2z}T}$ & $\widehat{C_{2y}T}$\\
\hline
1 & $f_1(k_x\sx + k_y\sy) + v k_z\sz$ & $-i\sx$ & $-i\sy$ & $-i\sz$  & $e^{-i\sz\theta/2}$ & $-i\sy K$ & $i\sx K$ & $-K$ \\
2 & $f_2[(k_x^2+k_y^2)\sx + 2k_xk_y\sy] + v k_z\sz$ & $\sx$ & $\sx$ &  $-\iden$ & $e^{-i\sz\theta}$ & $\sx K$ & $-\sx K$ & $K$ \\
3 & $f_3[(k_x^3-3k_xk_y^2)\sx + (3k_x^2k_y-k_y^3)\sy] + v k_z\sz$ & $i\sx$ & $i\sy$ & $i\sz$   & $e^{-i3\sz\theta/2}$ & $i\sy K$ & $-i\sx K$ & $-K$ \\
\hline\hline
\end{tabular}
\caption{For $n$-Weyl fermions with $D_{\infty}\times \Z_2^T$ symmetry, we list the Hamiltonians and symmetry representations in the complex vector space ($\C^2$) of wave functions.   The representation of any time-inverting symmetry is antiunitary and proportional to $K$, which implements complex conjugation.}
\label{tab:nWeylsymmetryreps}
\end{table}

The massless $n$-Dirac Hamiltonian\cite{Koshino-McCann-TrigonalWarpingBerrys-2009, Zhang-MacDonald-BandStructureABCStacked-2010} is
\e{
 H^{d,n}_{\bk}=f_n \big[k_{+}^n\sigma_- +\;h.c.\big].\la{Hdnapp}
}
The point group of $H^{d,n}$ is not just the two-dimensional projection (namely, $SO(2)\times \Z_2^T$) of the symmetry group of $H^{w,n}$; the absence of any term proportional to $\sz$ in $H^{d,n}_{\bk}$ implies that $H^{d,n}$ has mirror reflection symmetries (denoted $M_x\cdot (x,y)=(-x,y)$ and  $M_y\cdot (x,y)=(x,-y)$) which are absent in $H^{w,n}$. The point group of $H^{d,n}$ is therefore $O(2)\times \Z_2^T$, with the specific symmetry representations in \tab{tab:nDiracsymmetryreps}. 

\begin{table}[h]
\centering
\renewcommand{\arraystretch}{1.4}
\begin{tabular}{c c c c c c c c c c}
\hline\hline
$n$ & $H^{d,n}_{\bk}$ & $\widehat{M_{y}}$ & $\widehat{M_{x}}$ & $\widehat{R_z^{\theta}}$ & $\widehat{T}$ & $\widehat{M_xT}$ & $\widehat{C_{2z}T}$ & $\widehat{C_{3z}}$ & $\widehat{C_{2z}}$ \\
\hline
1 & $f_1(k_x\sx + k_y\sy)$ & $-i\sx$ & $-i\sy$ & $e^{-i\sz\theta/2}$ & $-i\sy K$  & $K$ & $i\sx K$ &  $e^{-i\pi \sz/3}$ & $-i\sz$ \\
2 & $f_2[(k_x^2 -k_y^2)\sx + 2k_xk_y\sy]$ & $\sx$ & $\sx$ & $e^{-i\sz\theta}$ & $\sx K$ & $K$ & $-\sx K$  & $e^{-i2\pi \sz/3}$  &   $-\iden$\\
3 & $f_3[(k_x^3-3k_xk_y^2)\sx + (3k_x^2k_y-k_y^3)\sy]$ & $i\sx$ & $i\sy$ & $e^{-i3\sz\theta/2}$ & $-i\sy K$ & $K$ & $-i\sx K$ & $-\iden$ & $i\sz$\\
4 & $f_4[(k_x^4-6k_x^2k_y^2+k_y^4)\sx
    +4k_xk_y(k_x^2-k_y^2)\sy]$ & $-\sx$ & $\sx$ & $e^{-i2\sz\theta}$ & $\sx K$ & $K$ & $\sx K$ & $e^{-i4\pi \sz/3}$ & $\iden$ \\
5 & $f_5[k_x(k_x^4-10k_x^2k_y^2+5k_y^4)\sx
    +k_y(5k_x^4-10k_x^2k_y^2+k_y^4)\sy]$ & $-i\sx$ & $i\sy$ & $e^{-i5\sz\theta/2}$ & $-i\sy K$ & $K$ & $i\sx K$ & $e^{-i5\pi \sz/3}$ & $-i\sz$ \\
\hline\hline
\end{tabular}
\caption{For massless $n$-Dirac fermions with $O(2)\times \Z_2^T$ symmetry,  we list the Hamiltonians and symmetry representations in the complex vector space ($\C^2$) of wave functions. 
}
\label{tab:nDiracsymmetryreps}
\end{table}

\tab{tab:nweyldiracsymmetryreducedmonopole} shows the resultant point groups when we add  to $H^{w,n}$ (and $H^{d,n}$) various symmetry-reducing corrections [\qq{diracmass}{sixpwarp}].
In verifying if a Hamiltonian is $C_{mz}$-symmetric with the specific representation ($R_z^n$) of $C_{mz}$ of $n$-Dirac/Weyl fermions:
\e{
R^n_z\big(\tf{2\pi}{m}\big) H_{k_{\pm},k_z} R^n_z\big(-\tf{2\pi}{m}\big)  = H_{e^{\pm i{2\pi}/{m}}k_{\pm},k_z} \as\text{with}\as  R^n_z(\theta)=\exp\big({-i\tf{n\sz}{2}\theta}\big),
\la{Rz-symmetry}
}
it is useful to recognize that the  general Hamiltonian term proportional to 
\e{
&(k_+^a k_-^b \sigma_+ +h.c.) \; \text{is $C_{mz}$-symmetric}\; \imp a-b+n \in m \Z; \lin
&k_+^c k_-^d \sz \; \text{is $C_{mz}$-symmetric}\; \imp c-d \in m \Z, 
\la{cmz-symmetry}
}
which follows from the rotation-induced transformation of $\sigma_{\pm}$ in \q{ctsrotsymmetry}. A useful corollary is that
\e{
a-b+n \;\text{odd,}\as \text{or}\as c-d \;\text{odd}\as \imp C_{2z}\text{-symmetry broken,}
}
which is a necessary condition for any bulk photovoltaic current in a 2D system, according to the symmetry theorem in \q{gsconstrainsSigma2}. \\

For pedagogy and demonstrated rigor, we carry out the symmetry-group identification for our most complex effective Hamiltonian, namely, the $n$-Dirac Hamiltonian plus a $6'$-warp correction, which sums to the Koshino-McCann effective Hamiltonian for a single valley of $n$-graphene:\cite{Koshino-McCann-TrigonalWarpingBerrys-2009}
\e{
H^{d,n}_{\bk}+ H^{6',n}_{\bk} \equiv  H^{KM,n}_{\bk}=\sum_{a,b,c}^{\N} h_{abc} k_{+}^{a}k_{-}^{b} \sigma_{+} +h.c.; \as b+2a+3c=n; \as h_{abc}\in \R;\as a,b,c\in \N=\{0,1,2,3,\ldots\}.
 \la{Hdn-warp}
}

\begin{itemize}
    \item The condition on the natural numbers in \q{Hdn-warp} is equivalent to
\e{
a-b+n= 3(a+c) \in 3\Z,  \la{naturalcondition}
}
which certainly satisfies the condition for $C_{3z}$-symmetry [\q{cmz-symmetry} with $m=3$].

\item Given that the summation in \q{Hdn-warp} contains terms with $(a+c)$ being odd, \q{naturalcondition} implies that $C_{2z}$-symmetry is broken, according  to \q{cmz-symmetry} with $m=2$. 

\item Let us verify that the Koshino-McCann Hamiltonian has $M_xT$ symmetry:
\e{
\widehat{TM_x}H^{KM,n}_{\bk}\widehat{TM_x}^{\mo}= H^{KM,n}_{TM_x \cdot \bk}; \as \widehat{M_xT}=K.
}
This holds because conjugation by $K$ maps $k_{\pm} \ri k_{\mp}$, and so does $TM_x\cdot \bk =(k_x,-k_y)$ map $k_{\pm} \ri k_{\mp}$. 

\item The Koshino-McCann Hamiltonian also has $M_y$ symmetry represented by $\sx$ (modulo multiplicative phase factor):
\e{
 \widehat{M_y}  H^{KM,n}_{\bk}\widehat{M_y}^{\mo}= \sum_{a,b,c}^{\N} h_{abc} k_{+}^{a}k_{-}^{b} \sigma_{-} +h.c. =H^{KM,n}_{k_x,-k_y}.
}

\item  $T$-transforming the Hamiltonian, 
\e{
\text{Even} \; n: \as \widehat{T}=\sx K; \as \widehat{T} H^{KM,n}_{\bk} \widehat{T}^{\mo}= \sum_{a,b,c}^{\N} h_{abc} k_{-}^{a}k_{+}^{b} \sigma_{-} +h.c.
}
which only equals $H^{KM,n}_{-\bk}$ is $(a+b)\in 2\Z$. Given that the  summation in \q{Hdn-warp} contains terms with $(a+b)$ being odd, $T$ symmetry is broken. One arrives at the same condition for $n$ odd, albeit with a different representation for time reversal: $\widehat{T}=-i\sy K$. 

\item Since $TM_x$ is preserved but $T$ broken, $M_x$ must also be broken. 

\end{itemize}

Altogether, the preserved symmetries are $C_{3z},M_y$ and $M_xT$, which generate the magnetic point group $6'mm'$. \\

Let us check that this result is consistent with the layer group of $n$-graphene being $p\bar{3}m1 \times \Z_2^T$,\cite{Mu-Zhou-ValleydependentGiantOrbital-2025} with point group $D_{3d}\times \Z_2^T$. This point group contains  elements:  $C_2'$ (two-fold rotation with an in-plane axis), $P$ (spatial inversion), $T$ and $S_6=M_xC_{6z}$ (six-fold roto-rotation); the aforementioned four elements   map $K$ to $K'$, i.e., they exchange valleys. Therefore, only pairwise compositions (like $TP$) belong to the 
\e{
\text{Little group of  K} \;= {\bar{3}'m}= C_{3v}\cup \{ TP, 3TC_{2}', 2TS_6\}
}
If we ignore the $z$ coordinate in these three-dimensional spatial transformations, the little group of K projects to 
\e{
\bar{3}'m= C_{3v}\cup \{ TP, 3TC_{2}', 2TS_6\} \as \substack{\text{projects}\\ \longrightarrow}\as  6'mm' = C_{3v}\cup \{ TC_{2z}, 3TM_x, 2TC_{6z}\}. 
}
Thus the two group are isomorphic, with $TP$ corresponding to $TC_{2z}$ and $TC_2'$ corresponding to $TM_x$, etc. The  distinction between the two groups is irrelevant to symmetry constraints of photovoltaic tensors; this crucially justifies our application of $6'mm'$ to $n$-graphene. \\

\subsubsection{The Hamiltonian-vector representation}\la{app:dvectorep}

For any two-band Hamiltonian $H^0$ [\q{twobandham2}] with the symmetry of the point group $G$, we will introduce a \textit{Hamiltonian-vector representation} of $G$, which represents each element of $G$ as
an $O(3)$-matrix, which should not be confused with  the defining $O(3+1)$-matrix representation of $G$ acting on spacetime.\\

 What it means for $g\in G$ to be a symmetry of $H^0$ is
\e{
\text{Two-band Hamiltonian symmetry}:\as H^0_{g\cdot \bk}=\hg H^0_{\bk} \hgmo=\bd_{\bk}\cdot   \hg  \bsigma\hgmo+\delta E_{\bk}\iden_2= (\overg \cdot \bd_{\bk})\cdot  \bsigma +\delta E_{\bk}\iden_2, \la{twobandsymmetry}
}
with $\overg$ an $O(3)$ matrix acting on vectors in $\R^3$. In fact, the set of all $\overg$ forms an  $O(3)$-matrix representation of $G$ with the property that $\overg$ is improper if and only if $g$ is time-inverting ($i_g=-1)$:
\e{
\det \overg =i_g  =-1 \as \text{if $g$ inverts time, and otherwise equals}\as +1.\la{deterdvec}
}
(Even the parity  transformation   which is an improper rotation in the defining representation [$P\circ (x,y,z)= (-x,-y,-z)$] is a proper rotation in the Hamiltonian-vector representation.) 
The reason for \q{deterdvec} is that if $g$ inverts time, the representation $\hg$ acting on complex wave functions (in $\C^2$) is antiunitary, meaning $\hg=U_gK$, with $U_g$ a unitary two-by-two matrix, and $K$ implementing complex conjugation. On one hand, because of the group homomorphism between $U(2)$ and $SO(3)$,\cite{tinkhambook} 
\e{
U_g\bd\cdot\bsigma U_g^{-1}= [(\text{Proper rotation matrix})\cdot  \bd]\cdot \bsigma.  \la{properrot}
}
On the other hand, because $K\sigma_2K=-\sigma_2$,
\e{
K\bd\cdot\bsigma K= (d_1,-d_2,d_3)\cdot \bsigma= [(\text{Improper rotation matrix})\cdot  \bd]\cdot \bsigma. \la{Kimproper}
}
Bringing both hands together by multiplying the proper rotation matrix [\q{properrot}] with the improper rotation matrix [\q{Kimproper}], we deduce that $\overg$ must be improper. We present the mapping between $\hg$ and $\overg$ in \tab{tab:hgoverg}.

\begin{table}[H]
	
\centering
		
\begin{tabular} {c|c c c c c c c} \hline 
			
$\hg$  & $\sx$ & $\sy$ &  $\sz$ &  $\sx K$ & $\sy K$ &  $\sz K$ &  $e^{-i\sz \phi/2}$  \\ [5 pt]  \hline  $\overg$  & $\diagmatrixthree{1}{-1}{-1}$ & $\diagmatrixthree{-1}{1}{-1}$ &  $\diagmatrixthree{-1}{-1}{1}$ & $\diagmatrixthree{1}{1}{-1}$ & $\diagmatrixthree{-1}{-1}{-1}$ &  $\diagmatrixthree{-1}{1}{1}$ &  $ \matrixthree{\cos (\phi)}{-\sin (\phi)}{0}{\sin(\phi)}{\cos(\phi)}{0}{0}{0}{1}$  \\  [5pt] \hline

\end{tabular}
		
\caption{Homomorphism between two-by-two (anti)unitary matrices and three-by-three orthogonal matrices. This mapping is preserved even if we multiply $\hg$ by an arbitrary complex phase factor.	\label{tab:hgoverg}}
\end{table}

For illustration, for the 1-Weyl Hamiltonian in \q{nweylham} with $\hg$ tabulated in \tab{tab:nWeylsymmetryreps}, 
\e{
\overlr{C_{2x}}=\diagmatrixthree{1}{-1}{-1}; \as \overlr{C_{2y}}=\diagmatrixthree{-1}{1}{-1}; \as\overlr{C_{2z}}=\diagmatrixthree{-1}{-1}{1}; \as\overlr{T}=\diagmatrixthree{-1}{-1}{-1}. \la{dvectorrep1}
}
This Hamiltonian-vector representation $\overg$ is linear (e.g. $\overT$ squares to the identity matrix), even though the representation $\hg$ is projective. \\

In some cases, the Hamiltonian-vector representation contains the trivial representation as a subrepresentation. This means that there exists an \textit{invariant vector} $\bv$ which is invariant under all $O(3)$ transformations:
\e{
\text{Invariant vector:}\as\overg \cdot \bv =\bv \in \R^3 \as \text{for all}\as g\in G, \la{invariantvectordrep}
}
and is a basis vector in the trivial representation. \\

For all $n$-order Dirac-Weyl Hamiltonians, there are no trivial $\bd$-vector subrepresentations, i.e., no invariant vectors. To prove this for odd $n$, we observe from \tab{tab:nWeylsymmetryreps}-\ref{tab:nDiracsymmetryreps} that time reversal is represented as $\hat{T} \propto \sy K$, which implies $\overlr{T}$ has the form given by the right-most equality in \q{dvectorrep1}, and that $\overlr{T}\circ \bv=\bv \imp \bv=\bze.$ For even $n$, 
\e{
\overlr{T}=\diagmatrixthree{1}{1}{-1} \iand \overlr{R_z^{\theta}} = \matrixthree{\cos (n\theta)}{-\sin (n\theta)}{0}{\sin(n\theta)}{\cos(n\theta)}{0}{0}{0}{1};
}
the invariance of $\bv=(v_1,v_2,v_3)$ under $\overlr{T}$ implies  $v_3=0$; the invariance under $\overlr{R_z^{\theta}}$ implies $v_1=v_2=0$; altogether, $\bv=\bze$ for all $n$. \\

As a final application, the two- and three-point Bargmann invariants [defined in \q{Bargmann}, with $\ket{u_{\bk}}$ being the higher-energy eigenstate] 
 are expressible in terms of the Hamiltonian vector as in \q{Bargmannrealimag} and 
\e{
\scrb_{\bk\bk'}\eq  \tf{1}{2}\bigg(1+\hat{\mathbf{d}}_{\bk}\cdot \hat{\mathbf{d}}_{\bk'}\bigg); \as \hat{\mathbf{d}}_{\bk}= \bd_{\bk}/|\bd_{\bk}|\la{2Barg}
}
If $g$ is a Hamiltonian symmetry, then   $\bd_{g\cdot \bk}=\overg \cdot \bd_{\bk}$ [from \q{twobandsymmetry}] implies the following constraints on the Bargmann invariants: 
\e{
\scrb_{g\cdot \bk,g\cdot \bk'}\eq \scrb_{\bk\bk'}; \as \mbox{Im}{\scrb_{g\cdot \bk,g\cdot \bk'',g\cdot \bk'}} =i_g \mbox{Im}{\scrb_{\bk \bk'' \bk '}}; \as \mbox{Re}{\scrb_{g\cdot \bk,g\cdot \bk'',g\cdot \bk'}}=\mbox{Re}{\scrb_{\bk \bk'' \bk '}},\la{magsymBargmann}
}
bearing in mind that time-inverting symmetries ($i_g=-1$) are represented as an improper rotation on the $\bd$ vector [\q{deterdvec}], which leaves dot products invariant, but inverts the orientation of the triple product.

\subsection{Impurity potentials}\la{app:impuritymatrixelement}

\subsubsection{Matrix elements of general scalar potentials}\la{app:scalarmatrixelement}

In general, matrix elements of a scalar potential $V$ (in the Bloch basis) are written as
\e{
\text{Scalar potential matrix element}\as V_{B,B'}= \cbra{\psi_{B}}V\cket{\psi_{B'}}. \la{VimBBp}
}  
For small-momentum scattering induced by a scalar potential that is smooth on the scale of the crystalline lattice period,
\e{
\text{Small-momentum scattering:} \as V_{BB'}\approx \widetilde{V}_{\bk-\bk'}\braket{u_{B}}{u_{B'}}; \as \text{Fourier transform}\as \tilde{V}_{\bk}\equiv \frac1{\calv} \int_{\R^d} d\br  e^{-i\bk\cdot \br} V_{\br}.\la{VimBBpsmallk}
}
How come? Decomposing the total crystalline volume $\calv=\ncell\vcell$ as the product of the number of primitive unit cells times the volume of each cell, and separating the spatial coordinate $\br=\btau+\bR$ into an intracellular coordinate and a Bravais-lattice vector,
\begin{equation}
V_{BB'}
= \int_{\text{cell}} \frac{d\bm{\tau}}{\vcell} e^{i(\bk' -\bk)\bm{\tau}} \overline{u_{B}(\bm{\tau})}u_{B^{\prime}}(\bm{\tau})
\frac{1}{\ncell} \sum_{\mathbf{R}} e^{i(\bk' -\bk)\cdot \mathbf{R}} V_{\bm{\tau}+\mathbf{R}},
\end{equation}
We approximate $V_{\bm{\tau}+\mathbf{R}}\approx V_{\mathbf{R}}$ because it is smooth on the scale of the crystalline lattice period.
\begin{equation}
V_{BB'} \approx \int \frac{d\bm{\tau}}{\vcell} e^{i(\bk' -\bk)\bm{\tau}}  \overline{u_{B}(\bm{\tau})}u_{B^{\prime}}(\bm{\tau})\times 
\frac{1}{\ncell} \sum_{\bR} e^{i(\bk' -\bk)\cdot \mathbf{R}} V_{\mathbf{R}}.\la{VBBproof}
\end{equation}
Within the left integral, $e^{i(\bk' -\bk)\bm{\tau}}\approx 1$ with a correction that is bounded by $|\bk-\bk'|/$(Reciprocal period); by `small-momentum', we mean precisely that $|\bk-\bk'|/$(Reciprocal period) is small. The discrete Fourier sum  in \q{VBBproof} can be approximated by a Fourier integral for `small-momentum':
\e{
\frac{1}{\ncell} \sum_{\bR} e^{i(\bk' -\bk)\cdot \mathbf{R}} V_{\mathbf{R}} \condapprox{\text{small momentum}}  \frac1{\calv} \int_{\calv} d\br  e^{i(\bk' -\bk)\cdot \br} V_{\br} \equiv \tilde{V}_{\bk-\bk'},
}
bearing in mind $\calv$ is eventually taken to infinity. Altogether, these imply that \q{VBBproof} reduces to \q{VimBBpsmallk}.

\subsubsection{Symmetry classes and overscreening of impurity potentials}\la{app:overscreening}

Let us distinguish between three symmetry classes of impurity potentials:
\e{
&\text{Centrosymmetric}\as \Vim = V^s_{\br} = V^{s}_{-\br}; \la{class1}\\
&\text{Noncentric}\as \Vim = V^a_{\br} = -V^{a}_{-\br};\\
&\text{Hybrid}\as \Vim = V^s_{\br}+V_{\br}^a = V^{s}_{-\br}-V^{a}_{-\br}, \la{class3}
}
whose respective Fourier transforms ($\widetilde{V}^s_{\bk}, \widetilde{V}^a_{\bk}$ and $\widetilde{V}^s_{\bk}+\widetilde{V}^a_{\bk}$) are  real, imaginary and complex.\\

In this work, we focus on the  two leading contributions in the multipole expansion of a three-dimensional (3D) impurity potential. Depending on whether the monopole/dipole impurity  is screened by 3D $n$-Weyl fermions or  by 2D $n$-Dirac fermions, its matrix elements and Fourier transforms are given as:
\e{
\label{eq:monopoleanddipole3D}
\text{3D screening:}\as &V^{\text{m,3D}}_{\bk\bk'}=\widetilde{V}^{\text{m,3D}}_{\bk-\bk'}\langle u_{B} |u_{B'} \rangle; \as \widetilde{V}^{\text{m,3D}}_{\bk}\equiv \frac{1}{\mathcal{V}}\frac{e^2}{\epsilon}\frac{1}{|\bk|^2 +k_{\text{scr}}^2};    \\
    &V^{\text{d,3D}}_{\bk\bk'}=\widetilde{V}^{\text{d,3D}}_{\bk-\bk'}\langle u_{B} |u_{B'} \rangle; \as \widetilde{V}^{\text{d,3D}}_{\bk}\equiv -i\frac{1}{\mathcal{V}}\frac{e^2}{\epsilon}\frac{\boldsymbol{\mathscr{d}}\cdot\bk}{|\bk|^2 +k_{\text{scr}}^2};\label{eq:monopoleanddipole3D2}\\
\text{2D screening:}\as &V^{\text{m,2D}}_{\bk\bk'}=\widetilde{V}^{\text{m,2D}}_{\bk-\bk'}\langle u_{B} |u_{B'} \rangle; \as \widetilde{V}^{\text{m,2D}}_{\bk}\equiv \frac{1}{\mathcal{V}}\frac{e^2}{2\epsilon}\frac{1}{|\bk| +k_{\text{scr}}};  \label{eq:monopoleanddipole2D}  \\
    &V^{\text{d,2D}}_{\bk\bk'}=\widetilde{V}^{\text{d,2D}}_{\bk-\bk'}\langle u_{B} |u_{B'} \rangle; \as \widetilde{V}^{\text{d,2D}}_{\bk}\equiv -i\frac{1}{\mathcal{V}}\frac{e^2}{2\epsilon}\frac{\boldsymbol{\mathscr{d}}\cdot\bk}{|\bk| +k_{\text{scr}}},  \la{eq:monopoleanddipole2D2}
}
where $|e|\boldsymbol{\mathscr{d}}$ is the dipole moment of the dipole impurity,  $k_{\text{scr}}$ is the generalized Thomas-Fermi  wavevector due to screening by fermions [explicated for 2D screening in \q{dielectric2D}], and $\epsilon$ is a dielectric constant that heuristically accounts for screening by lattice deformations. \\

For intra-pocket/valley impurity-mediated scattering in a photoexcited Dirac-Weyl material, $|\bk-\bk'|$ has an upper bound comparable to the $\bk$-radius of the excitation surface [\q{defineexcsurface}], which is much smaller than the Brillouin zone period. For a sufficiently low frequency of the driving light source, the $\bk$-radius of the excitation surface will be much smaller than generalized Thomas-Fermi wavevector, which implies that we can replace $|\bk|^2+k_{\text{scr}}^2 \ri k_{\text{scr}}^2$ in \qq{eq:monopoleanddipole3D}{eq:monopoleanddipole3D2} and $|\bk|+k_{\text{scr}} \ri k_{\text{scr}}$ in \qq{eq:monopoleanddipole2D}{eq:monopoleanddipole2D2}. This is what we mean for an impurity to be \textit{overscreened}.

\subsubsection{Screening of 3D potential by 2D crystalline electrons}\la{app:screening}

While \qq{eq:monopoleanddipole3D}{eq:monopoleanddipole3D2} are attainable from standard texts,\cite{ashcroft_mermin} \qq{eq:monopoleanddipole2D}{eq:monopoleanddipole2D2} deserve a careful derivation which we provide here.\\

We begin by setting up the basic problem of electronic screening in an arbitrary medium. Let $H^0$ be the mean-field Hamiltonian of electron quasiparticles, and let $V$ be a perturbing scalar potential:
\e{
H=H^0+V; \as V_{\br}=-|e|\phi_{\br}.
}
$\phi_{\br}$ here is the screened electric potential; its Fourier transform $\widetilde{\phi}_{\bp}$ 
relates to the externally-applied potential $\widetilde{\phi}_{ext\bp}$ by the
\e{
\text{Inverse dielectric matrix:}\as \widetilde{\phi}_{\bp}=\sum_{\bp'}\eps^{\mone}_{\bp\bp'} \widetilde{\phi}_{ext\bp'}.\la{dielectricinhomogeneous3}
}
This is a matrix whose elements are indexed by the canonical momentum $\bp \in \R^3$, so $\sum_{\bp}=\calv\int_{\R^3}d\bp/(2\pi)^3$. The dielectric matrix relates\cite{shamziman_electronphonon} to the
\e{
\text{Susceptibility matrix}\as \chi_{\bp\bp'}\eq \tf{|e|^2}{\calv}  \sum_{BB'}\;\braopket{\psi_{B'}}{e^{-i\bp\cdot\hbr}}{\psi_B}\braopket{\psi_{B}}{e^{i\bp'\cdot\hbr}}{\psi_{B'}}\f{f^{\text{FD}}_{BB'} }{E_{BB'}}; \as f^{\text{FD}}_{BB'}=f^{\text{FD}}_B-f^{\text{FD}}_{B'}, \la{suscepmatrix}\\
\text{by}\as \eps_{\bp\bp'}= \delta_{\bp,\bp'}- \tf{1}{\bp^2}\chi_{\bp\bp'}.
}
($\eps_{\bp\bp'}$ is unrelated to the $\eps$ in \qq{eq:monopoleanddipole3D}{eq:monopoleanddipole2D2}.)
Here, $\ket{\psi_B}$ is an eigenstate of $H^0$, $\hbr$ is the first-quantized position operator and  $f^{\text{FD}}$ is the Fermi-Dirac function [\q{fermidirac}]. The susceptibility matrix relates the induced charge density (screened charge density minus externally-applied charge density) to the screened potential:
\e{
\text{Induced charge density}\as \widetilde{\delta \rho}_{\bp} \eq  \sum_{\bp'} \chi_{\bp\bp'}\widetilde{\phi}_{\bp'}. \la{deltarhoq2}
}

To the extent that $B$ is not necessarily a label for a Bloch wave in \q{suscepmatrix}, the above formulation generally applies no matter the form or symmetry of $H^0$. In the specific context of `Screening of 3D potential by 2D crystalline electrons', 
we assume that:\\

\noi{i} Within an energy window $[\mu-k_BT,\mu+k_BT]$, the eigenstates of $H^0$ (labelled by $\ket{\psi_B}$) are two-dimensional (2D), meaning that  $B=(b,\bkper)$ with $\bkper=(k_x,k_y)\in rBZ$ in a reduced Brillouin zone, and the wave function $\psi_B(x,y,z)$ has negligible width in the $z$ direction. This implies that screening is dominated by   virtual excitations of 2D $H^0$-eigenstates, because  the susceptibility matrix [\q{suscepmatrix}] is proportional to $f^{\text{FD}}_{BB'}/E_{BB'}$.\\

\noi{ii} Since the perturbing scalar potential $V_{\br}=-|e|\phi_{\br}$ is slowly varying with respect to $z$ (relative to the localization length of the 2D $H^0$-eigenstates), in evaluating the matrix element 
\e{
\cbraopket{\psi_B}{V_{\hbr}}{\psi_{B'}}\approx  -|e|\cbraopket{\psi_B}{{\phi}_{\hbx,\hby, \hbz}}{\psi_{B'}}\big|_{\hbz \ri 0}
}
between two 2D eigenstates, we may as well set the scalar potential to its value at $z=0$:
\e{
\phi_{\brper,z=0} \eq \sum_{\bp} \widephi_{\bp}e^{i\bpper\cdot\brper}= \sum_{\bpper} e^{i\bpper\cdot\brper}\widephi^{\per}_{\bpper}; \as \widephi^{\per}_{\bpper}= \sum_{p_z} \widephi_{\bp}; \as \bpper=(p_x,p_y); \as \brper=(x,y);\lin
\sum_{\bp}\eq \calv\int_{\bR^3} \tf{d\bp}{(2\pi)^3};\as \sum_{\bpper}=L_xL_y\int_{\bR^2} \tf{d\bpper}{(2\pi)^2}; \as \sum_{p_z}=L_z\int_{\bR} \tf{dp_z}{2\pi}; \as \calv=L_xL_yL_z.
}
Under these assumptions, long-wavelength, intraband screening of the potential leads to an effective dielectric function
\e{
\widephi^{\per}_{ext\bqper}\equiv  \sum_{p_z} \widephi_{ext\bqper p_z}= \eps^{\per}_{\bqper}\widephi^{\per}_{\bqper}; \as  \eps^{\per}_{\bqper}= 1+\tf{k_{scr}}{|\bqper|};\as k_{scr}= \tf{e^2}{2} \partial_{\mu}n^{\per}; \as n^{\per}= \int_{rBZ} \tf{d\bkper}{(2\pi)^2}\sum_b f^{\text{FD}}_{b\kper}, \la{dielectric2D}
}
with $n^{\per}$ being the number density (per unit area) of 2D eigenstates, and $\partial_{\mu}n^{\per}$ being the derivative with respect to the chemical potential (or `quasi-Fermi level' in the non-equilibrium steady state).\\

To prove \q{dielectric2D}, we begin from the general susceptibility matrix in \q{suscepmatrix} and split the canonical momentum as
\e{
\text{canonical momentum} \;\bp= (\bqper+\bGper,p_z); \as \bqper \in rBZ; \as \bGper\in rRL,
}
with $\bqper$ (or sometimes $\bkper$) a 2D crystal momentum in the reduced BZ, and $\bGper$ a 2D reciprocal lattice vector. Letting $B=(b\bkper)$ and $B'=(b'\bkper')$ label our 2D eigenstates, 
\e{
&\cbraopket{\psi_{B'}}{e^{-i\bp\cdot\hbr}}{\psi_B}\cbraopket{\psi_{B}}{e^{i\bp'\cdot\hbr}}{\psi_{B'}}= 
\cbraopket{\psi_{b'\bkper'}}{e^{-i(\bqper+\bGper)\cdot\hbrper}e^{-ip_z\hz}}{\psi_{b\bkper}}\cbraopket{\psi_{b\bkper}}{e^{i(\bqper'+\bGper')\cdot\hbrper}e^{ip_z'\hz}}{\psi_{b'\bkper'}}.
}
Because of assumption (ii), we may set $e^{-ip_z\hz}\ri 1$ within a matrix element. By 2D crystalline momentum conservation,
\e{
\cbraopket{\psi_{B'}}{e^{-i\bp\cdot\hbr}}{\psi_B}\cbraopket{\psi_{B}}{e^{i\bp'\cdot\hbr}}{\psi_{B'}}= 
\cbraopket{\psi_{b'\bkper'}}{e^{-i(\bqper+\bGper)\cdot\hbrper}}{\psi_{b\bkper'+\bqper}}\cbraopket{\psi_{b\bkper'+\bqper}}{e^{i(\bqper'+\bGper')\cdot\hbrper}}{\psi_{b'\bkper'}}\delta_{\bqper,\bqper'}\delta_{\bkper,\bkper'+\bqper}.
} 
Substituting this into \q{suscepmatrix}, we find that the susceptibility matrix 
\e{
\chi_{\bqper+\bGper,p_z; \bqper'+\bGper',p_z'} \eq  \chi^{\bqper}_{\bGper,\bGper'}\delta_{\bqper,\bqper'} \;\text{is independent of}\;p_z\;\text{and}\;p_z'; \\
\chi^{\bqper}_{\bGper,\bGper'} \equiv &\; \tf{|e|^2}{\calv}  \sum_{bb'}\sum_{\bkper}^{rBZ}\braopket{u_{b'\bkper}}{e^{-i\bGper\cdot \boldsymbol{\hat{\tau}}_{\per}}}{u_{b\bkper+\bqper}}\braopket{u_{b\bkper+\bqper}}{e^{i\bGper'\cdot \boldsymbol{\hat{\tau}}_{\per}}}{u_{b'\bkper}}\f{f^{\text{FD}}_{(b\bkper+\bqper)(b'\bkper)} }{E_{(b\bkper+\bqper)(b'\bkper)}},\la{suscepmatrix33}
}
with $\hat{\btau}_{\per}$ the intracell component of $\hbrper$, and $\sum_{\bkper}^{rBZ}= L_xL_y\int_{rBZ}d\bkper/(2\pi)^2$. Because the susceptibility matrix is diagonal in $\bqper$ and independent of $p_z$ and $p_z'$, the induced charge density [\q{deltarhoq2}]
\e{
\widetilde{\delta \rho}_{\bqper+\bGper,p_z} \eq  \sum_{\bqper'\bGper'p_z'} \chi_{\bqper+\bGper p_z,\bqper'+\bGper'p_z'}\widetilde{\phi}_{\bqper'+\bGper'p_z'} = \sum_{\bGper'}\chi^{\bqper}_{\bGper,\bGper'} \sum_{p_z'} \widetilde{\phi}_{\bqper'+\bGper'p_z'} \; \text{is independent of}\; p_z, \la{deltarhoq3}
}
which implies $\delta \rho_{\br}\propto \delta(z),$  in accordance with the screening electrons being in the $z=0$ layer. \\

If we focus on the intraband contribution ($b=b'$) and ignore Umklapp scattering, the susceptibility matrix reduces to a scalar function:
\e{
\text{Intraband, long-wavelength susceptibility} \as \chi^{\text{intra}\bqper}_{\bze,\bze} \eq  -\tf{e^2}{L_z}\partial_{\mu}n^{\per} + \tf{|e|^2}{\calv}\bqper\cdot\sum_{b}\sum_{\bkper}^{rBZ}f_{b\bkper}  f^{\text{FD}}_{b\bkper}\;  \bg_{b\bkper} \cdot \bqper +O(\bqper^3). \la{thomasfermichi}
}
The first term  is analogous to the Thomas-Fermi susceptibility; the $O(\bqper^2)$ correction involves the quantum metric $\bg_B$ defined by
\e{
\big|\braket{u_{b\bkper}}{u_{b\bkper+\bqper}}\big|^2 = 1 - \bqper\cdot \bg_{b\bkper}\cdot \bqper+O(\bqper^3). \la{metricdefine}
}
To derive \q{thomasfermichi}, we begin from \q{suscepmatrix33} and approximate the right-most fraction by the first-order Taylor approximation of both numerator and denominator:
\e{
\f{f^{\text{FD}}_{(b\bkper+\bqper)(b\bkper)} }{E_{(b\bkper+\bqper)(b\bkper)}} \approx \f{ \bqper\cdot \nab_{\bkper} f^{\text{FD}}_b  }{ \bqper\cdot \nab_{\bkper} E_b}=  -\partial_{\mu} f^{\text{FD}}_{b\bkper}, \la{fFDE}
}
which is valid for $b=b'$ and not otherwise. Then 
\e{
 \chi^{\text{intra}\bqper}_{\bze\bze} \eq - \tf{|e|^2}{\calv}  \sum_{b}\sum_{\bkper}^{rBZ}\cbraopket{\psi_{b\bkper}}{e^{-i\bqper\cdot\hbr}}{\psi_{b\bk+\bqper}}\cbraopket{\psi_{b\bk+\bqper}}{e^{i\bqper\cdot\hbr}}{\psi_{b\bkper}}\partial_{\mu} f_{b\bkper}=  -\tf{|e|^2}{\calv}\partial_{\mu} \sum_{b}\sum_{\bkper}^{rBZ}f_{b\bkper} \big|\braket{u_{b\bkper}}{u_{b\bk+\bqper}}\big|^2. \notag
} 
By expanding the last inner product in powers of $\bqper$ [\q{metricdefine}] and applying $\calv=L_xL_yL_z$, we arrive at \q{thomasfermichi}.\\

The general dielectric matrix [\q{dielectricinhomogeneous3}] reduces in our context to
\e{
\eps_{\bqper+\bGper p_z,\bqper'+\bGper'p_z'}= \delta_{\bqper+\bGper p_z,\bqper'+\bGper'p_z'}- \tf{1}{(\bqper+\bGper)^2+p_z^2}\chi^{\bqper}_{\bGper,\bGper'}\delta_{\bqper,\bqper'}.
}
Keeping only the leading, Thomas-Fermi-like term  in \q{thomasfermichi},
\e{
\eps_{\bqper p_z,\bqper'p_z'}\appr  \delta_{\bqper,\bqper'}\bigg(\delta_{p_z,p_z'}+ \tf{A}{\bqper^2+p_z^2}\bigg); \as A =  \tf{ e^2 \partial_{\mu}n^{\per}}{L_z} \lin
&\imp \widephi_{ext\bqper p_z} \refeq{dielectricinhomogeneous3} \sum_{\bqper'p_z'}\eps_{\bqper p_z,\bqper'p_z'}\widephi_{\bqper'p_z'}= \widephi_{\bqper p_z} +  \tf{A}{\bqper^2+p_z^2}\sum_{p_z'} \widephi_{\bqper p_z'}.
}
Integrating the above equation over $p_z$,
\e{
\int \tf{dp_z}{2\pi}\tf{1}{\bqper^2+p_z^2}= \tf{1}{2|\bqper|} \imp
\widephi^{\per}_{ext\bqper}= \sum_{p_z} \widephi_{ext\bqper p_z} =  \widephi^{\per}_{\bqper} \bigg(1+\tf{  e^2 \partial_{\mu}n^{\per}}{2|\bqper|} \bigg),
}
which completes the proof of \q{dielectric2D}.\\

As an application, we consider the 3D potential of an electric dipole with an in-plane dipole moment ($|e|\mathscr{d}$): 
\e{
\phi_{ext\br}\eq  \tf{|e|^2\boldsymbol{\mathscr{d}}\cdot \br}{r^3} = -\boldsymbol{\mathscr{d}} \cdot \nabr \tf{|e|^2}{r} \imp \widephi_{ext\bp} =-i\tf{\boldsymbol{\mathscr{d}} \cdot \bpper}{\calv} \tf{ |e|^2}{\bp^2}\imp 
\widephi^{\per}_{ext\bpper}= \sum_{p_z}\widephi_{ext\bp} = -i\tf{|e|^2}{2L_xL_y}\tf{\boldsymbol{\mathscr{d}} \cdot \bpper}{|\bpper|}; \lin
&\imp \widephi^{\per}_{\bqper}= \tf{\widephi^{\per}_{ext\bqper}}{1+k_{scr}/|\bqper|} = -i\tf{|e|^2}{2\calv}\tf{\boldsymbol{\mathscr{d}} \cdot \bqper}{|\bqper|+k_{scr}},
}
which forms the basis for \qq{eq:monopoleanddipole2D}{eq:monopoleanddipole2D2}.

\subsection{Kinetic model for the impurity-mediated ballistic photocurrent}\la{sec:kineticmodel}

This appendix elucidates and solves the kinetic model which underlies the impurity-mediated ballistic photocurrent. After introducing the electron distribution in \app{app:electrondistribution}, the basic kinetic equation is set up in \app{app:decomposecollisionalintegral}. The ingredients of the kinetic equation are the collisional integrals due to photo-excitation [\app{app:excitationcollision}], electron-impurity scattering [\app{app:Iimp}], and electron-phonon scattering. The latter two scatterings relax the energy and momentum of the hot photo-excited electrons, as elaborated in \app{app:rta}; therein, we also introduce a generalized harmonic approach to solving the eigenmodes of the impurity-mediated collisional integral. The generalized harmonics reduce to spherical harmonics in isotropic cases [\app{app:realsphericalharmonic}]. A linear algebra of generalized harmonics is formalized in \app{app:braketharmonics}. Finally, the solution of the kinetic equation  in \app{app:solvekinetic} gives us the electron distribution on the excitation surface, which is an essential ingredient to calculate the ballistic photocurrent. We present two solutions: an exact one which accounts for the harmonic-dependent momentum relaxation of the generalized harmonics, as well as a simpler solution which employs the momentum-relaxation time approximation. The   meaning and limitations of this approximation will be discussed.

\subsubsection{Electron distributions}\la{app:electrondistribution}

A key ingredient in the kinetic model, the electron distribution $f_B$  can be viewed  as a diagonal  element of the reduced one-electron density matrix, in the band basis. It will prove useful to decompose the distribution into $B$-symmetric and $B$-asymmetric components:
\e{
&\text{Symmetry-decomposed electron distribution:}\as f_B=f^s_B+\delta f^a_B; \lin
&\text{$B$-symmetric}\as f^s_B= f^s_{-B};\lin
&\text{$B$-asymmetric}\as \delta f^a_B =-\delta f^a_{-B}.\la{symmf2}
}
The meaning of $-B$ is $(b,-\bk)$, which 
can be thought of as the time-reversed partner of $B=(b,\bk)$; we are interested in bulk photovoltaic materials with broken centrosymmetry, so energy-eigenstates are energy-nondegenerate at generic $\bk$, allowing us to choose our band indices such that time reversal just inverts $\bk$ while preserving $b$.\\

An important $B$-symmetric distribution is the
\e{
\text{Fermi-Dirac distribution}\as f^{\text{FD}}_B=\f{1}{e^{\beta (E_B-\mu)}+1}= f^{\text{FD}}_{-B}, \la{fermidirac}
}
with $E_B=E_{-B}$ guaranteed by time-reversal symmetry. The Fermi-Dirac distribution applies not just to materials in thermodynamic equilibrium, but also (in an approximate sense) to photo-excited materials in quasi-equilibrium.\cite{wurfel_solarcells} In the latter case, the temperature $T$ and `quasi-Fermi level' $\mu$ depends on the source-driven light intensity.\\

The Fermi-Dirac distribution exemplifies an \textit{iso-energy-symmetric} distribution, meaning it depends on $B$ only through its dependence on $E_B$. Equivalently, an iso-energy-symmetric distribution equals its iso-energy average; more generally, for any function $Fun_B$ of $B$
\e{
\text{Iso-energy-symmetric}\as Fun_B = [Fun_B] \as \text{iso-energy average}.\la{isoenergysymmetric}
}
The iso-energy average at $B$ is defined as the average of $Fun_{B'=(b'\bk')}$ over the hypersurface (in $\bk'$-space) defined by $E_{B'}=E_B$:
\begin{equation}
\text{Iso-energy average}: \as [Fun_{B}]=[Fun]_{E_B}=\sum_{B'}Fun_{B'} \frac{\delta(E_{B'}-E_{B})}{D_{E_B}},\label{eq:isoEave}
\end{equation}
with $D_E=\sum_{B}\delta(E_B-E)$ the volume-extensive density of states per unit energy. For any function $Fun_{\bk}$, we will refer to 
\e{
\delta Fun_{\bk}=Fun_{\bk}-[Fun_{\bk}] \as \text{as an iso-energy deviant (noun)},\la{deviant}
}
 or as being  iso-energy-deviating (adjective).

\subsubsection{Symmetric decomposition of the collisional integral} \la{app:decomposecollisionalintegral}

In the kinetic theory approach to bulk photovoltaics,\cite{belinicher_kinetictheory,sturmanfridkin_book} all electron quasiparticle transitions are modeled by the collision integral $I_B(f)$ which is a functional of the non-equilibrium steady-state electron distribution $f_B$. Because $f_B$ is time-independent, we set the collisional integral to zero:
\e{
\text{Kinetic equation}: \as I_{B}(f)=I_{B}^{ex}(f)+I_{B}^{pn}(f)+I_{B}^{im}(f)=0.\la{eq:kineticequation}
}
This integral is decomposed into three terms with superscripts $ex,pn,im$ which respectively account for  photo-\textit{ex}citation (due to the electron's interaction with a monochromatic  light wave), electron-\textit{p}hono\text{n} scattering, and electron-\text{im}purity scattering.\footnote{In the general case, we may consider a fourth term ($I_B^{re}$) that accounts for \textbf{radiative} electron-hole recombination.\cite{sturmanfridkin_book} $I^{re}$ is ignorable for topological semimetals, because electron-hole recombination is dominated by much-faster electron-phonon scattering and accounted for by $I^{pn}$. For (massive-Dirac) semiconductors, $|I_B^{re}|$ is typically significant for $B$ near a semiconducting band edge, where photo-excitation is absent and intraband collisions are substantially reduced owing to the formation of a quasi-thermal Boltzmann/Fermi-Dirac gas;\cite{zhuAA_anomalousshift} conversely, $|I_B^{re}|$ is typically insignificant for $B$ on the excitation surface, in comparison with the much faster electron-phonon/impurity scatterings. The application to the impurity-mediated ballistic photocurrent only requires that we solve the kinetic equation for $B$ on the excitation surface.}\\

The collisional integral and its  components can be further split into \textit{$B$-symmetric and $B$-asymmetric} components\cite{belinicher_ballistic}
 \e{
&I^x_B=I_B^{x,a}+I_B^{x,s};\as I^{x,a}_B(f^s) = -I^{x,a}_{-B}(f^s); \as I^{x,s}_B(f^s) = I^{x,s}_{-B}(f^s); \as 
x= ex, pn, im,
\la{IsIa}\\
& I_B=I_B^a+I_B^s; \as I_B^{a}=\sum_x I_B^{x,a}; \as I_B^{s}=\sum_x I_B^{x,s}, \la{tnop2}
}
$I^s_B(f)$ exemplifies a $B$-symmetric functional of $f_B$, a notion that is distinct from (but overlaps with) a $B$-symmetric function of $B$ [\q{symmf2}]: $I^s_B(f)$ can be viewed as a $B$-symmetric function only if the input $f$ is $B$-symmetric [\q{tnop2}], and otherwise,
\e{
I^a_B(\delta f^a) = I^a_{-B}(\delta f^a); \as I^s_B(\delta f^a) = -I^s_{-B}(\delta f^a).
}
The significance of symmetrically decomposing the collisional integral is that without a $B$-asymmetric component $I^a_B$, a $B$-symmetric distribution $f^s_B$ will remain $B$-symmetric for all times. Conversely, a $B$-asymmetric $\delta f^a_B$ [\q{symmf2}] can only be generated from an initially $B$-symmetric distribution if $I^a_B$ is nontrivial. For instance, assuming that a non-magnetic material is in thermal equilibrium before photo-excitation, it is guaranteed that the pre-excitation distribution, being the Fermi-Dirac distribution $f^{\text{FD}}_B$ [\q{fermidirac}], is $B$-symmetric and iso-energy-symmetric [\q{isoenergysymmetric}]. We will speak synonymously of the $B$-symmetric (resp.\ $B$-asymmetric) collisional integral as resulting in \textit{symmetric (resp.\ skew) scattering}.\\

The precise forms of $I^{ex}$ and $I^{im}$ are respectively elaborated in \app{app:excitationcollision} and \app{app:Iimp}; the effect of the $B$-symmetric component of $I^{pn}+I^{im}$ is to relax energy and momentum [\app{app:rta}].

\subsubsection{The photo-excitation collisional integral}\la{app:excitationcollision}

\e{
\text{The dipole coupling to a classical light wave:}\as \Vex=|e|\bcale_{\bq\omega}\cdot \chebr  \la{dipolecoupling}
}
is responsible for photo-excitations. Ignoring spontaneous emission, the $\bB$-symmetric component of the photo-excitation collisional integral  is, at lowest nontrivial order in $\Vex$,
\e{
I^{ex,s}_B
\eq \sum_{B'} W^{ex}_{BB'}f_{B'B}; \as W^{ex}_{BB'}=W^{ex}_{B'B}=\tf{2\pi}{\hbar} \sum_{B'}\dekkp |\Vex_{BB'}|^2\bigg[ \delta(E_{BB'}-\hbar\omega) +\delta(E_{B'B}-\hbar\omega) \bigg];\as f_{BB'}=f_B-f_{B'}. \la{Iexgeneral}
}
In a two-band model with band labels $c$ (for conduction) and $v$ (for valence), \q{Iexgeneral} simplifies to
\begin{equation}
I^{\text{ex},s}_{c\bk}(f)=\frac{2\pi}{\hbar}e^2(f_{v\bk}-f_{c\bk})|\mathbf{E}|^2|\be\cdot \mathbf{A}_{cv,\bk}|^2\delta(E_{cv\bk}-\hbar\omega)= -I^{\text{ex},s}_{v\bk}(f).\la{Iexcs}
\end{equation}
The electric field ($\bE$) induced by the light source is assumed to be sufficiently strong that spontaneous emission is negligible compared to induced emission.\cite{zhuAA_anomalousshift} The energy-conserving delta function in \q{Iexcs} confines collisions (with source-produced photons) to the excitation surface [\q{defineexcsurface}]. With the weaker laser intensities typical of continuous wave experiments,  the non-equilibrium steady $f_B$ is typically  nondegenerate (meaning $f_{v\bk}-f_{c\bk}\approx 1$) on the excitation surface,
\footnote{
For semiconductors, this is justified experimentally\cite{zakharchenya_photoluminescence} and theoretically.\cite{zhuAA_anomalousshift} For semimetals, we assume that $f_B$ quasi-equilibrates with $k_BT$ that is small compared to the one-particle excitation energies measured relative to the chemical potential, i.e., $k_BT \ll |E_{c\bk}-\mu|$ and $\ll |E_{v\bk}-\mu|$, with $E_{c/v,\bk}$ satisfying  \q{defineexcsurface}.
}
in which case \q{Iexcs} simplifies to:
\e{
\text{Nondegenerate distribution}: \as I^{\text{ex},s}_{c\bk}(f)\approx \frac{2\pi}{\hbar}e^2|\mathbf{E}|^2|\be\cdot \mathbf{A}_{cv,\bk}|^2\delta(E_{cv\bk}-\hbar\omega).\la{Iexcsnondeg}
}
We will ignore the  $B$-asymmetric $I^{ex,a}$ which makes a different contribution to the ballistic photocurrent than the one we focus on.\cite{belinicher_ballistic,belinicher_phononmechanism,zhenbang_phononballistic,zhenbang_electronholeballistic}

\subsubsection{The impurity-mediated collisional integral} \la{app:Iimp}

Elastic collisions with impurities are generally described by a collisional integral which is linear in $f$, no matter the order of the Born approximation:\cite{sturman_collisionintegral}
\e{
\text{Impurity-mediated, elastic:}\as I^{im}_B(f)=\;\text{Incoming rate - Outgoing rate}\;=\sum_{B'}\bigg(W^{im}_{BB'}f_{B'}-W^{im}_{B'B}f_B\bigg). \la{Iimp}
}
We refer to $W^{x}_{B,B'}$ as the \textit{transition rate matrix element} for the transition $B\lea B'$ mediated by $x$. $W^{im}$  is derivable from Lippmann-Schwinger scattering theory; assuming incoherent scattering with $N_{\rm imp}$ number of impurities, the transition rates add up as
\begin{equation}
\label{diffusion-skew-W-def}
W^{im}_{B,B'} = \frac{2\pi}{\hbar} N_{\rm imp} |T_{B,B'}|^2 \delta{\left(E_{BB'}\right)}; \as E_{BB'}=E_{B} -E_{B'},
\end{equation}
with $T_{B,B'}$ being the $T$ matrix element. Assuming that the impurity is non-magnetic, time-reversal symmetry gives the 
\e{
\text{Reciprocity theorem:}\as |T_{B,B'}|= |T_{- B',-B}| \imp W^{im}_{B,B'} =W^{im}_{- B',-B}.\la{reciprocity}
}

Let us split the transition rate matrix element into components which are even and odd under the simultaneous  inversion of both $B$ and $B'$:
\e{
& W^{\text{im}}_{B,B'}=W^{\text{im},s}_{B,B'}+W^{\text{im},a}_{B,B'}; \la{WsWa} \\
& (BB')\text{-inversion-symmetric} \as W^{\text{im},s}_{B,B'}=W^{\text{im},s}_{-B\lea -B'}; \la{Ws}\\ 
& (BB')\text{-inversion-asymmetric} \as W^{\text{im},a}_{B,B'}=-W^{\text{im},a}_{-B\lea -B'}. \la{Wa}
}
Because $I^{\text{im}}$ is linear in $W^{\text{im}}$ [\q{Iimp}], it splits into $B$-symmetric and $B$-asymmetric components [\q{IsIa}] according to the decomposition of $W^{\text{im}}$ [\q{WsWa}]. An explicit expression for  $W^{\text{im},s}$ is attained expanding the T matrix in powers of the impurity potential energy $V$ and retaining the lowest nontrivial order, this being second order:  
\begin{eqnarray}
\label{diffusion-skew-W-S-def}
W_{B,B'}^{im,s} = \frac{2\pi}{\hbar} N_{\rm imp} \delta{\left(E_{BB'}\right)} \left|V^{\text{im}}_{B,B'}\right|^2,
\end{eqnarray}
with $V^{\text{im}}_{B,B'}$  being the the impurity potential matrix element [\q{VimBBp}].
The lowest-order approximation of $W^{im,a}$ is third order:\cite{Sturman:1984}
\begin{eqnarray}
\label{WimaBBp}
W_{B,B'}^{im,a} = \frac{(2\pi)^2}{\hbar} N_{\rm imp} \delta{\left(E_{BB'}\right)} \sum_{B''}  \imag{V^{\text{im}}_{B,B''} V^{\text{im}}_{B'',B'} V^{\text{im}}_{B',B} } \delta{\left(E_{BB''}\right)}.
\end{eqnarray}

To motivate the form of \q{WimaBBp}, 
the net amplitude ($T_{\bk\bk'}$) for a transition from $\bk'\ri \bk$ (band index suppressed) is a Feynman sum of individual amplitudes $a_j$ for the possible histories of the transition, including $a^1_{\bk'\ri\bk}$ for the direct one-step scattering from $\mathbf{k}'$ to $\mathbf{k}$,  $a^2_{\bk'\ri \bk''\ri \bk}$ for a two-step scattering from $\mathbf{k}'$ to $\mathbf{k}''$ and then to $\mathbf{k}$ [illustrated in Fig.~\ref{fig:exc&iso}(a)], and $a^{j\geq 3}$ for higher-step scattering processes.  The total transition rate $W^{im}_{\bk\bk'}\propto |T_{\bk\bk'}|^2$ (via the standard T matrix theory) is proportional to $|a^1|^2$ at lowest perturbative order, and $\sum_{\bk''}(a^2\overline{a^1}+\overline{a^2}a^1)$ to the next leading order.  The third-order $a^2\overline{a^1}$ can be interpreted as the quantum-interfered amplitude for two Feynman paths connecting $\bk'$ to $\bk$, or it can be interpreted as self-interference of an electron making a loop $\bk' \ri \bk'' \ri \bk \ri \bk'$ if we view $\overline{a^1}_{\bk'\ri \bk}=a^1_{\bk\ri\bk'}$. Interchanging $\bk\leftrightarrow\bk'$ in $\sum_{\bk''}a^1_{\bk\ri\bk'}a^2_{\bk'\ri \bk''\ri \bk}$ is equivalent to taking the complex conjugate, hence only the imaginary component of this sum contributes to skew scattering, which is the essence of \q{WimaBBp}. This is the lowest-order contribution to skew scattering because $|a^1_{\bk'\leftarrow \bk}|^2$ is manifestly symmetric under interchanging $\bk\leftrightarrow\bk'$. The truncation of \q{WimaBBp} to third order (in the impurity potential term) applies to dilute/sparse concentrations of defects with weak impurity potentials.\footnote{With denser concentrations of strong potentials, so-called `diffractive' and `Gaussian' skew scatterings become relevant at fourth order in the potential.\cite{Konig-Levchenko:2021}}\\

Combining the reciprocity theorem [\q{reciprocity}] with the ($BB'$)-inversion (a)symmetry [\qq{Ws}{Wa}], one deduces that $W^{im,s}$ (resp. $W^{im,a}$) is also symmetric (resp. antisymmetric) under interchange of $B$ and $B'$
\e{
 &  (BB')\text{-interchange symmetry:}\as W^{\text{im},s}_{B,B'}= W^{\text{im},s}_{B',B} = \half\big(W^{\text{im}}_{B,B'}+W^{\text{im}}_{B',B}\big); \la{Wimps}\\
&(BB')\text{-interchange asymmetry:}\as W^{\text{im},a}_{B,B'}=-W^{\text{im},a}_{B',B} =\half\big({W^{\text{im}}_{B,B'}-W^{\text{im}}_{B',B}  }\big).\la{Wimpa}
}
The interchange (a)symmetry allows to simplify our expressions for the impurity collisional integral:
\begin{eqnarray}
I_B^{im,s}(f) &=&  \sum_{B'} W^{im,s}_{BB'}(f_{B'}-f_B),\la{Iims}\\
\label{diffusion-skew-Icoll-asymm}
I_B^{im,a}(f)&=&  \sum_{B'} W^{im,a}_{BB'}(f_{B'}+f_B) =  \sum_{B'} W^{im,a}_{BB'}f_{B'}.\la{Iima}
\end{eqnarray}
The last step of \q{Iima} follows from a relation arising from unitarity of the scattering matrix:\footnote{This closely relates to  the optical theorem of elastic scattering~\cite{Davydov-QuantumMechanics-1991}
.}
\begin{equation}
\label{diffusion-skew-optical}
\sum_{B'}W^{im}_{B'B}= \sum_{B'}W^{im}_{BB'},
\end{equation}
and the trivial identity $\sum_{B'}W^{im,s}_{B'B}= \sum_{B'}W^{im,s}_{BB'}$, 
which together imply
\e{ \sum_{B'}W^{im,a}_{B'B}= \sum_{B'}W^{im,a}_{BB'}=0.\la{asper} }
The impurity collisional integral has a trivial action on iso-energy-symmetric distributions:\footnote{
The first vanishing may be seen from $f_{B'}-f_B=[f_{B'}]-[f_B]=0$ in \q{Iims} for elastic scattering; the second vanishing follows from \q{Iima} being proportional to $2[f_{B}]\sum_{B'}W^a_{BB'}=0$, as per \q{asper}. 
}
\e{I_{B}^{im,s}([f])=I_{B}^{im,a}([f])=0.\la{impvanish}}

\subsubsection{Energy and momentum relaxation}\la{app:rta}

Momentum relaxation (or `isotropization'\cite{belinicher_polaraxis}) is the smoothening of any iso-energy-asymmetries in the distribution; {e}nergy {r}elaxation is  the relaxation of one-electron energies toward a quasi-thermal distribution, or more precisely, it relaxes a generic, non-equilibrium,  iso-energy-symmetric distribution to a specific, quasi-equilbrium iso-energy-symmetric distribution ($f^{\text{FD}}$), with a temperature and chemical potential (or `quasi-Fermi level') that depends on the light wave intensity. Static impurities scatter electrons elastically and contribute to momentum relaxation but not energy relaxation; electron-phonon scattering contributes to both momentum and energy relaxation. As will be explained, only the $B$-symmetric component of the collisional integral leads to relaxation of energy/momentum. Altogether,
\e{
        &I^{pn,s}_{B}(f)+I^{im,s}_{B}(f)\approx I^{er}_B(f)+I^{mr}_{B}(f);\\
        \text{\textit{E}nergy \textit{r}elaxation}: \as & I^{er}_B(f)=\sum_{E'}dN_{E'} W_{EE'}^{er} [f]_{E'}; \as W_{EE'}^{er}=W_{E\lea E'}- \delta_{E,E'}\sum_{E''}W_{E''\lea E};\la{Ier}\\
        \text{\textit{M}omentum \textit{r}elaxation}: \as & I^{mr}(f)=\sum_{B'}W^{mr}_{BB'} \delta f_{B'}; \as \delta f_{B'}=f_{B'}-[f_{B'}]\la{Imr}
}
In \q{Ier}, we have represented an energy integral as a discrete sum, with a newly introduced factor $dN_E$ that encodes the density of states per unit energy, as elaborated in \q{dNE} below. Thus $\sum_{E'}dN_{E'}W_{E\lea E'} [f]_{E'}$ can be understood as the net transition rate of one-electron states from the energy interval $E'+dE$ to the energy interval $E+dE$. If the phonon energy is small compared to $k_BT$ and any other energy scale for the variation of $[f]_E$, then the action of $W_{EE'}^{er}$ can be expressed as a differential operator on $[f]_E$ in the Fokker-Planck approximation.\cite{esipov_temperatureenergy} If the phonon energy is not small, as may be for optical phonons, then other models for $W_{EE'}^{er}$ are appropriate.\cite{esipov_temperatureenergy} If we are interested only in the long-term distribution, at times $t \gg \tau^{E_{exc}}_{er}$ (the \textit{energy relaxation time} for electrons at the excitation energy), we may sacrifice accuracy in the transient dynamics by crudely adopting the 
\e{
\text{Energy relaxation time approximation (RTA):}\as I^{er}_{\bk}([f])=-\f{[f_{\bk}]-f^{\text{FD}}_{\bk}}{\tau^{E_{\bk}}_{er}},\la{energyrta}
}
with $f^{\text{FD}}$ the quasi-equilibrium Fermi-Dirac function [\q{fermidirac}]. \\

For disordered materials at low temperature, we make the simplifying assumption  that momentum relaxation is dominated by impurities alone:
\e{
I^{mr}_{\bk}=I^{im,s}_{\bk} \imp {W}^{mr}_{\bk\bk'}\refeq{Iims} W^{im,s}_{\bk\bk'}-\sum_{\bk''}W^{im,s}_{\bk''\bk}\dekkp. \la{impalone}
}
Because $I^{im,s}$ has a trivial action on iso-energy-symmetric distributions [\q{impvanish}], $I^{im,s}(f)=I^{im,s}(\delta f)$, in accordance with \q{Imr}.\\

Now $W^{mr}_{\bk\bk'}$ is symmetric under interchanging $\bk$ with $\bk'$, because both $W^{im,s}_{\bk\bk'}$ and $\dekkp$ are interchange symmetric [\q{Wimps}]. If we view $\bk$ as a discrete wavevector, then  $W^{mr}_{\bk\bk'}$ is a real, symmetric, negative-trace matrix. Being `real, symmetric' allows to express $W^{mr}_{\bk\bk'}$ as a sum over real eigenmodes with real eigenvalues.  These eigenvalues are guaranteed to be non-positive by the Gershgorin circle theorem.\footnote{
By the Gershgorin theorem, every eigenvalue of $W^{mr}$ lies within at least one Gershgorin disc associated with a certain $\bk$, which is centered at $W^{mr}_{\bk\bk}$ and has radius $R_{\bk} = \sum_{\bk'\neq \bk}W^{mr}_{\bk\bk'}$ in the complex plane. Since $W^{im,s}_{\bk'\bk}\geqslant 0$ and $W^{im,s}_{\bk'\bk}=W^{im,s}_{\bk\bk'}$, we have
\[
W^{mr}_{\bk\bk}
=
-\sum_{\bk'\neq\bk}W^{im,s}_{\bk'\bk}\equiv -R_{\bk},
\]
Therefore, for any eigenvalue $\lambda$ of $W^{mr}$, there exists a momentum $\bk$ such that
\[
\left|
\lambda+R_{\bk}
\right|
\leqslant
R_{\bk} \Rightarrow -2R_{\bk}
\leqslant
\lambda
\leqslant
0,
\]
where $\Rightarrow$ used the fact that $\lambda$ is a real number because $W^{mr}$ is a normal matrix.
} 
To reformulate a previous observation, any iso-energy-symmetric distribution is an eigenmode in the kernel of $W^{mr}$; it follows that any eigenmode with nonzero (and necessarily negative) eigenvalue must be an iso-energy-deviant [\q{deviant}], leading to 
\e{
\text{Multi-harmonic momentum relaxation:}
\as W^{mr}_{\bk,\bk'} =-\sum_{hE} \f{h^E_{\bk}h^E_{\bk'}}{\tau_{hE}^{mr}}.\la{harmonicmomentumrta} 
}
As indexed by $(hE)$, each eigenmode is a  real function of $\bk$ (denoted $h_{\bk}^{E}$) which  (i) is nonzero only if $E=E_{\bk}$, and (ii) has zero iso-energy average: $[h_{\bk}^E]=0$, just as $[\delta f_{\bk}]=0$. 
We further presume that the set of all $(hE)$-eigenmodes form a  
\e{
\text{Complete basis for iso-energy deviant functions:}\as \delta f_{\bk}= \sum_{\bk'}\sum_{hE} {h^E_{\bk}h^E_{\bk'}}\delta f_{\bk'}.
}
We will refer to $\{h_{\bk}^E\}$ as \textit{generalized harmonics}, because they reduce to real spherical harmonics in isotropic cases, as elaborated in \app{app:realsphericalharmonic}. The meaning of \q{harmonicmomentumrta} is that impurity-mediated collisions diminish each harmonic by a factor $1/e$ in time $\tau^{mr}_{hE}$, which is harmonic-dependent. Because $\tau^{mr}_{hE}\geq 0$, the harmonics do not grow in time, i.e., there is no momentum excitation, in accordance with the H theorem (of non-decreasing entropy) proven for elastic scattering in \ocite{sturman_collisionintegral}.\\

In a given application, it may be that only a few harmonics are generated by a driving force (e.g., light source), and for these few harmonics, their relaxation times may not be very different, in which case 
 \q{harmonicmomentumrta} reduces to
\e{
\text{Momentum relaxation time approximation (RTA):}\as I^{mr}_{\bk}(f)\approx -\f{\delta f_{\bk}}{\tau_{E_{\bk}}^{mr}},\la{momentumrta}
}
with  the \textit{momentum relaxation time} ($\tau_{E}^{mr}$) being some weighted average of $\tau_{hE}^{mr}$ over the relevant harmonics at fixed energy $E$.  Warning: `momentum relaxation time' may be defined differently in other literature,  e.g., for transport calculations in \ocite{lundstrom_book}.

\subsubsection{Interlude: Braket formulation of generalized harmonics}\la{app:braketharmonics}

To clarify the linear algebraic structure of the space of generalized harmonics, we introduce a braket notation with a symmetric inner product: $\braket{a}{b}=\braket{b}{a}$,  and the resolution of identity for real, smooth functions of $\bk$:
\e{
\iden= \sum_{\bk}\ketbra{\bk}{\bk}=\sum_{E}\bigg\{\ketbra{oE}{oE}+\sum_{h}\ketbra{hE}{hE}\bigg\}, \la{resolution}
}
with each ket an orthonormal basis vector:
\e{
\braket{\bk}{\bk'}=\dekkp; \as \braket{oE}{oE'}=\delta_{EE'}; \as \braket{hE}{h'E'}=\delta_{h,h'}\delta_{E,E'}; \as \braket{hE}{oE}=0.\la{orthonormal}
}
$\sum_E$ can be viewed as a discrete sum if we split $\bk$-space into infinitesimally-separated iso-energy contours: $\ldots, E-dE/2, E+dE/2,E+3dE/2,\ldots$; our `smooth' functions are assumed to vary negligibly over $dE/|\nabk E|$.   The number of $\bk$-points within one \textit{iso-energy band} $[E-dE/2,E+dE/2)$ sandwiched by two neighboring contours is 
\e{dN_E=D_E dE =\sum_{\bk}\delta_{E_{\bk},E}, \as D_E=\sum_{\bk}\delta(E-E_{\bk}), \la{dNE}}
with $D_E$ the volume-extensive density of states per unit energy. 
The \textit{energy Kronecker delta function} in \q{orthonormal} should be understood as a projector to $\bk$-points within an iso-energy band:
\e{ 
\delta_{E_{\bk},E}=1 \as \text{if and only if}\as E_{\bk} \in [E-dE/2,E+dE/2).\la{kroneckerE}
}
This allows us  to view discrete sums and continuous integrals interchangeably:
\e{
\la{dEE1sumE}
\f{\delta_{EE'}}{dN_E}=\f{\delta(E-E')}{D_E}; \as \sum_{E} dN_E = \int dE \,D_E.
}

The quantity in the curly brackets of \q{resolution} is the resolution of identity for real functions that have support (i.e., are nonzero) only in the iso-energy band $[E-dE/2,E+dE/2)$. In particular, $\ket{oE}$ is the \textit{breathing mode} which is constant within $[E-dE/2,E+dE/2)$ but is zero everywhere else:
\e{
o_{\bk}^E=\braket{\bk}{oE}=\f{\delta_{\Ek E}}{\sqrt{dN_E}}; \as 1=\braket{oE}{oE}=\sum_{\bk}(o_{\bk}^E)^2, \la{oE}
}
As motivated in \q{harmonicmomentumrta}, $\ket{hE}$ is both an eigenmode of the operator $\widehat{W^{mr}}$ and a  basis vector for the space of real-valued, smooth, iso-energy-deviating functions [\q{deviant}];
$\ket{hE}$ relates to the previously defined amplitudes $h_{\bk}^E$ as
\e{
\braket{\bk}{hE}=h_{\bk}^E; \as 1=\braket{hE}{hE}=\sum_{\bk}(h_{\bk}^E)^2. \la{hE}
}

Just as we split $f_{\bk}=[f_{\bk}]+\delta \fbk$, we may view $Fun_{\bk}=\braket{\bk}{Fun}$ and split $\ket{Fun}= \ket{[Fun]}+\ket{\delta Fun}$ into iso-energy-symmetric and iso-energy deviating components with corresponding projectors $\hat{P}_{[]}$ and $\hat{P}_{\delta}$:
\e{
\ket{[Fun]}\eq \hat{P}_{[]}\ket{Fun};\as \hat{P}_{[]}=\sum_{E}\ket{oE}\bra{oE}; \as  \braket{oE}{Fun}=\sqrt{dN_E}[Fun]_E; \as \braket{hE}{[Fun]}=0;  \la{oEproduct}\\
\ket{\delta Fun}\eq  \hat{P}_{\delta}\ket{Fun}; \as \hat{P}_{\delta}= \sum_{hE}{\ketbra{hE}{hE}};\as \braket{oE}{\delta Fun}=0. \la{projdev}
}
In particular, any odd function is an {iso-energy deviant}, because of the time-reversal constraint $(E_{\bk}=E_{-\bk})$ on iso-energy contours:
\e{
\braket{\bk}{Fun}=-\braket{-\bk}{Fun} \imp \braket{oE}{Fun}=0.\la{odddeviant}
}

Iso-energy averages [\q{eq:isoEave}] can be equivalently expressed in terms of the energy Kronecker function:
\begin{equation}
\text{Iso-energy average}: \as [Fun_{\bk}]=[Fun]_{E_{\bk}}=\sum_{\bk'}Fun_{\bk'} \frac{\delta_{E_{\bk}E_{\bk'}}}{dN_{E_{\bk}}},\label{eq:isoEave2}
\end{equation}
which implies for the functional derivatives:
\e{
\p{\fbk}{\fbkp}=\dekkp; \as \p{[\fbk]}{\fbkp}=\frac{\delta_{E_{\bk}E_{\bk'}}}{dN_{E_{\bk}}}=\sum_E\frac{\delta_{E_{\bk}E}}{\sqrt{dN_{E}}}\frac{\delta_{E_{\bkp}E}}{\sqrt{dN_{E}}} = \sum_E \braket{\bk}{oE}\braket{oE}{\bkp}.\la{dfdf}
}

Integration over $\bk$ can be equivalently expressed in terms of inner products with breathing modes:
\e{
\sum_{\bk}\braket{\bk}{Fun}= \sum_E dN_E [Fun]_E=\sum_{E}\sqrt{dN_E}\braket{oE}{Fun}.
}
Let us apply this equivalence to \q{asper}, which we can rewrite as
\e{ 
\sum_{\bk'}W^{im,a}_{\bk'\bk}\delta_{E_{\bk}E_{\bk'}}=0= \sum_{\bk'}W^{im,a}_{\bk\bk'}\delta_{E_{\bk}E_{\bk'}} \imp \bra{oE}\widehat{W^{im,a}}=0=\widehat{W^{im,a}}\ket{oE},\la{asper2} 
}
because impurity-mediated scattering is elastic.\footnote{Because of this elasticity, it may be seen that (i) $\bra{oE}\widehat{W^{im,a}}\ket{\bk} =0$ if $E\neq \Ek$. Conversely, integration over $\bk'$ in \q{asper} gives  
\e{
0=\sum_{E}\sqrt{dN_E}\bra{oE}\widehat{W^{im,a}}\ket{\bk}\delta_{E\Ek} = \sqrt{dN_{\Ek}} \bra{o\Ek}\widehat{W^{im,a}}\ket{\bk}.\la{further}
}
Combining (i) and (ii) gives
\e{
\bra{oE}\widehat{W^{im,a}}\ket{\bk} =0, \as \text{for all}\; E,\bk.\la{Wimadev}
}
}
This further implies that the impurity-mediated skew-scattering only acts nontrivially on iso-energy deviant functions:
\e{
\hat{P}_{[]}\widehat{W^{im,a}}=0=\widehat{W^{im,a}}\hat{P}_{[]}\imp \widehat{W^{im,a}}=\hat{P}_{\delta}\widehat{W^{im,a}}\hat{P}_{\delta}.\la{WimaPdev}
}

The functional derivatives of the collisional integrals for energy [\q{energyrta}] and momentum relaxation [\q{harmonicmomentumrta}] can be viewed as linear operators:
\e{
 \p{I^{er}_{\bk}}{\fbkp} \eq \bra{\bk}\widehat{\partial_f I^{er}}\ket{\bk'}; \as \widehat{\partial_f I^{er}}=-\sum_{E}\f{\ketbra{oE}{oE}}{\tau_{er}^{E}}; \la{partialfIer} \\
 \p{I^{mr}_{\bk}}{\fbkp} \eq   \bra{\bk}\widehat{\partial_f I^{mr}}\ket{\bk'}; \as \widehat{\partial_f I^{mr}}=\widehat{W^{mr}}=-\sum_{E,h}\f{\ketbra{hE}{hE}}{\tau^{mr}_{hE}};\la{Wmrsop}
}
with help from \q{hE} and \q{dfdf}. Assuming none of $\{\tau^{mr}_{hE}\}$ is zero or infinite, we can  define an inverse operator in the restricted space of iso-energy deviants: 
\e{
\widehat{W^{mr}}^{\mo}=\widehat{\partial_f I^{mr}}^{-1}= -\sum_{hE}\tau^{mr}_{hE}{\ketbra{hE}{hE}}; \as  \widehat{\partial_f I^{mr}}^{-1}\widehat{\partial_f I^{mr}}=\hat{P}_{\delta}, \la{inverseWmr}
}
with $\hat{P}_{\delta}$ being the  deviant projector [\q{projdev}]. (Warning: $\widehat{\partial_f I^{mr}}^{-1}$ is not the inverse operator in the full space of real-valued functions of $\bk$.) Thus if the inverse is applied to the momentum-relaxing collisional integral [\q{harmonicmomentumrta}],
\e{
\widehat{W^{mr}}^{-1}\ket{I^{mr}(f^s)}= \widehat{W^{mr}}^{-1}\widehat{W^{mr}}\ket{f^s}= \ket{\delta f^s}.\la{iden23}
}

\subsubsection{Isotropic eigensolution of the monopolar-impurity-mediated collisional integral}\la{app:realsphericalharmonic}

To develop intuition for the generalized harmonics, we consider an isotropic model in which the eigenmodes of the momentum-relaxation collision integral ($I^{mr}$) can be obtained analytically and are given by the real spherical harmonics.

Specifically, we consider the Kramers--Weyl fermion described by
\begin{equation}
    H_{KW} = \hbar v\mathbf{k}\cdot \boldsymbol{\sigma},
\end{equation}
subject to an overscreened monopole impurity potential, of which the matrix element in the Bloch basis is given by
\begin{equation}
    V_{\bk\bk'}=\frac{1}{\mathcal{V}}\frac{e^2}{\epsilon}\frac{\braket{u_{\bk}}{u_{\bk'}}}{k_{\text{scr}}^2}.
\end{equation}
In the following, we show that the momentum-relaxation transition-rate matrix
\begin{equation}
\label{eq:start}
W_{\bk\bk'}^{mr}  = \frac{2\pi}{\hbar}N_{\text{imp}}\left(\frac{1}{\mathcal{V}}\frac{e^2}{\epsilon k_{\text{scr}}^2}\right)^2\left(|\braket{u_{\bk}}{u_{\bk'}}|^2\delta(E_{\bk\bk'})-\delta_{\bk\bk'}\sum_{\bk''}|\braket{u_{\bk}}{u_{\bk''}}|^2\delta(E_{\bk\bk''})\right)
\end{equation}
can be decomposed into the harmonic form of Eq.~\eqref{harmonicmomentumrta},
\begin{equation}
\label{eq:goal}
\begin{aligned}
\boxed{W_{\bk\bk'}^{mr} =-\sum_{E,h} \frac{h^{E}_{\bk} h^{E}_{\bk'}}{\tau^{mr}_{hE}}, \quad  \text{with}\ h^{E}_{\bk}=\sqrt{4\pi}\frac{\delta_{E,E_{\bk}}}{\sqrt{dN_E}}Y_{lm}(\Omega_{\bk}), \ \frac{1}{\tau^{mr}_{hE}}= a_l\frac{\pi}{\hbar}\frac{n_{\text{imp}}}{k_{\text{scr}}^4}\left(\frac{e^2}{\epsilon}\right)^2 \rho_{E_{\bk}}.}
\end{aligned}
\end{equation}
Here, $Y_{lm}(\Omega_{\bk})$ denotes the real spherical harmonics and $h=(l,m)$. Furthermore, $n_{\text{imp}}=N_{\text{imp}}/\mathcal{V}$ is the impurity density, $\rho_{E_{\bk}} = E^2_{\bk}/(2\pi^2(\hbar v_{F})^3)$ is the density of states, and the coefficients are given by 
\begin{equation}
a_{0}=0, \quad a_1=2/3,\quad a_{l\geqslant 2}=1. 
\end{equation}
The vanishing of $a_0$ reflects particle-number conservation within each energy shell under elastic scattering. Finally, $\delta_{E,E_{\bk}}$ and $dN_{E}$ are defined through Eqs.~\eqref{dNE} and \eqref{dEE1sumE}.\\ 

To rewrite Eq.~\eqref{eq:start} in the form of Eq.~\eqref{eq:goal}, we first note that for the Kramers--Weyl fermion described by $H_{KW}$, the Bloch-state overlap factor $|\braket{u_{\bk}}{u_{\bk'}}|^2$ depends only on the angle between $\bk$ and $\bk'$, denoted by $\gamma_{\bk\bk'}$. Explicitly,
\begin{equation}
    |\braket{u_{\bk}}{u_{\bk'}}|^2=\frac{1+\cos\gamma_{\bk\bk'}}{2}, \quad \cos \gamma_{\bk\bk'}=\cos\theta_{\bk}\cos\theta_{\bk'}+\sin\theta_{\bk}\sin\theta_{\bk'}\cos(\varphi_{\bk}-\varphi_{\bk'}),
\end{equation}
where $(\theta_{\bk},\varphi_{\bk})$ is the spherical angle of $\bk$.
Using this result, the second term of $W_{\bk\bk'}^{mr}$ can be evaluated as
\begin{equation}
\begin{aligned}
  \sum_{\bk''}|\braket{u_{\bk}}{u_{\bk''}}|^2\delta(E_{\bk\bk''}) &= \mathcal{V} \int\frac{k''^2dk''d\Omega''}{(2\pi)^3} \frac{1+\cos\theta}{2} \frac{\delta(k-k'')}{\hbar v} =\frac{\mathcal{V}k^2}{(2\pi)^2\hbar v}.
\end{aligned}
\end{equation}
Furthermore, the Kronecker delta can be written as
\begin{equation}
\delta_{\bk\bk'}=\frac{(2\pi)^3}{\mathcal{V}}\delta^{(3)}(\bk-\bk'),
\end{equation}
which implies
\begin{equation}
\frac{\mathcal{V}}{(2\pi)^2}\frac{k^2\delta_{\bk\bk'}}{\hbar v}=2\pi\delta(E_{\bk\bk'})\delta^{(2)}(\Omega_{\bk}-\Omega_{\bk'}).
\end{equation}
Substituting these results into the expression for $W_{\bk\bk'}^{mr}$ yields
\begin{equation}
\label{eq:wharmonic}
W_{\bk\bk'}^{mr}  = \frac{2\pi}{\hbar}\frac{N_{\text{imp}}}{k_{\text{scr}}^4}\left(\frac{1}{\mathcal{V}}\frac{e^2}{\epsilon}\right)^2\delta(E_{\bk\bk'})\left(\frac{1+\cos\gamma_{\bk\bk'}}{2}-2\pi\delta^{(2)}(\Omega_{\bk}-\Omega'_{\bk'})\right).
\end{equation}

Since Eq.~\eqref{eq:wharmonic} depends on the spherical solid angles of $\bk$ and $\bk'$, a harmonic expansion of it should naturally involve the spherical harmonics [c.f. Eq.~\eqref{eq:goal}]. According to the  identity equation and the orthonormality of spherical harmonics:
\begin{equation}
\begin{aligned}
&\sum_{m=-l}^{m=l}Y_{lm}(\Omega)Y_{lm}(\Omega')=\frac{2l+1}{4\pi}P_{l}(\cos\gamma),\ \sum_{l}\frac{2l+1}{4\pi}P_{l}(\cos\gamma)=\delta^{(2)}(\Omega-\Omega')=\delta(\phi-\phi')\delta(\cos\theta-\cos\theta'),
\\
&\int d\Omega Y_{lm}(\Omega)Y_{l'm'}(\Omega)=\delta_{ll'}\delta_{mm'},
\end{aligned}
\end{equation}
where $P_{l}(\cos \gamma
)$ is the Legendre polynomials,
it is straightforward to show that $h^{E}_{\bk}$ defined in Eq.~\eqref{eq:goal}  satisfy:
\begin{equation}
 \sum_{\bk'}h^{E}_{\bk'}h'^{E'}_{\bk'}=\delta_{h,h'}\delta_{E,E'}, \quad \sum_{hE} h^{E}_{\bk}h^{E}_{\bk'} =\delta_{\bk\bk'}
\end{equation}

Note that
\begin{equation}
    1+\cos\gamma_{\bk\bk'} = P_{0}(\cos \gamma)+P_{1}(\cos\gamma)=\sum_{l=0}^{1}\frac{4\pi}{2l+1}\sum_{m=-l}^{l}Y_{lm}(\Omega_{\bk})Y_{lm}(\Omega_{\bk'}).
\end{equation}
It is straightforward to rewrite $W_{\bk\bk'}^{mr}$ in Eq.~\eqref{eq:wharmonic} as
\begin{equation}
\begin{aligned}
W_{\bk\bk'}^{mr} &= \frac{\pi}{\hbar}\frac{N_{\text{imp}}}{k_{\text{scr}}^4}\left(\frac{1}{\mathcal{V}}\frac{e^2}{\epsilon}\right)^2\delta(E_{\bk\bk'})\left(\sum_{l=0}^{1}\frac{4\pi}{2l+1}\sum_{m=-l}^{l}Y_{lm}(\Omega_{\bk})Y_{lm}(\Omega_{\bk'})-4\pi\sum_{l=0}^{+\infty}\sum_{m=-l}^{l}Y_{lm}(\Omega_{\bk})Y_{lm}(\Omega_{\bk'})\right)
\\
&=-\frac{\pi}{\hbar}\frac{N_{\text{imp}}}{k_{\text{scr}}^4}\left(\frac{1}{\mathcal{V}}\frac{e^2}{\epsilon}\right)^2D_{E_{\bk}}\sum_{E}\frac{\delta_{E,E_{\bk}}\delta_{E,\bk'}}{dN_{E}}\sum_{l=0}^{+\infty}a_{l}\sum_{m=-l}^{l}4\pi Y_{lm}(\Omega_{\bk})Y_{lm}(\Omega_{\bk'})
\\
&=-\sum_{E,h}\frac{\pi}{\hbar}\frac{n_{\text{imp}}}{k_{\text{scr}}^4}\left(\frac{e^2}{\epsilon}\right)^2 \rho_{E_{\bk}} a_l  h^{E}_{\bk} h^{E}_{\bk'},
\end{aligned}
\end{equation}
where 
\begin{equation}
a_l = \left\{
\begin{array}{c}
2l/(2l+1) \quad \text{for}\ l=0,1
\\
1 \quad \text{otherwise}
\end{array}\right .
\end{equation}

\subsubsection{Solution of kinetic model} \la{app:solvekinetic}

Let us derive the $B$-asymmetric electron distribution that solves the kinetic equation in \q{eq:kineticequation}.\\

As a rule of thumb,\footnote{The first inequality can be justified by perturbation theory: for the linear photogalvanic effect, $I^{x,a}$ is always higher-order in interaction couplings than  $I^{x,s}$,\cite{belinicher_ballistic} as elaborated for the impurity-mediated collisions in \app{app:Iimp}.} 
the effect of skew scattering  is weaker than of symmetric scattering for each scattering mechanism:
\e{
\text{Rule of thumb:}\as [|I^{x,a}_B|]\ll [|I^{x,s}_B|];\as \text{steady}\;  f_B=f_B^s+\delta f_B^a; \as [|\delta f^a|]\ll [|f^s|].\la{rulethumb} 
}
Because the initial distribution before photo-excitation is a thermal and symmetric [\q{fermidirac}], any $B$-asymmetric distribution $\delta f^a$ must be dynamically generated by skew scattering [cf. second equality in \q{IsIa}]; the smallness of $I^{a}$ relative to $I^{s}$ then guarantees that the steady distribution is dominated by its symmetric component [cf. last inequality in \q{rulethumb}].\\

The smallness of $I^{a}$ relative to $I^{s}$ [\q{rulethumb}] then guarantees that the steady distribution is dominated by its symmetric component, i.e., $[|\delta f^{a}|]\ll [f^{s}]$, because $\delta f^a$ is dynamically generated by skew scattering ($I^{im,a}_B$), as discussed in \app{app:decomposecollisionalintegral}. This leads to an ansatz: we first solve \q{eq:kineticequation} for the symmetric $f^s$, which then determines the $B$-asymmetric $\delta f^a$ by:
\e{
\text{Ansatz}: \as I^s_B(f^s)=0; \as \sum_{B'}\f{\partial I_B^s}{\partial f_{B'}}\bigg|_{f^s} \delta f_{B'}^a+I_B^a(f^s)=0. \la{ansatz}
}
The term proportional to $\delta f_{B'}^a$ in the second equation should be understood as the first-order term in a functional Taylor expansion, i.e.,  $\partial I_B^s/\partial f_{B'}|_{f^s}$ is a functional derivative\cite{feynman_thesis} evaluated at $f^s$ that solves $I_B^s(f^s)=0$. \q{ansatz} corrects an error in \ocite{sturmanfridkin_book}.\\

Combining our results in \qq{Ier}{harmonicmomentumrta}, our ansatz equations in \q{ansatz} can be expressed as
\e{
0\eq I^s_{\bk}(f^s) = I^{ex,s}_{\bk}(f^s) + I_{\bk}^{er}([f])+ I_{\bk}^{mr}( f^s);\la{ansatz1}\\
0 \eq I_{\bk}^a(f^s)+\sum_{\bk'}\big(\partial_f I^s\big)_{\bk\bk'} \delta f_{\bk'}^a,\la{ansatz2}
}
with the band index $c$ (for conduction) implicit in the above equations.
The functional derivative $\partial_f I^s$  is evaluated at $f^s$ which solves \q{ansatz1}. The first ansatz equation [\q{ansatz1}] can be iso-energy averaged to give
\e{
0\eq [I^{ex,s}_{\bk}(f^s)] + I_{\bk}^{er}([f])+ [I_{\bk}^{mr}( f^s)],\la{isoansatz1}
}
bearing in mind the averaging procedure does nothing to the energy-relaxing collisional integral, which is already iso-energy-symmetric [\q{Ier}]. The last term in \q{isoansatz1} is now demonstrated to vanish, according to \q{oEproduct} and \q{Wmrsop}:
\e{
\sqrt{dN_{\Ek}}[I_{\bk}^{mr}( f^s)]= \braket{o\Ek}{I^{mr}(f^s)}  = -\sum_{hE}\f{\braket{o\Ek}{hE}\bra{hE}}{\tau^{mr}_{hE}}=0 \imp [I^{ex,s}_{\bk}(f^s)] =- I_{\bk}^{er}([f]).\la{Iexsiden}
}
Plugging this into the first ansatz equation [\q{ansatz1}] gives
\e{
0= \delta I^{ex,s}_{\bk}(f^s) + \sum_{\bk'}W^{mr}_{\bk\bk'} f^s_{\bk'}; \as \as\delta I^{ex,s}=I^{ex,s}-[I^{ex,s}].\la{ansatz1simple}
}

The functional derivative $\partial_f I^s$ in the second ansatz equation [\q{ansatz2}] is contributed by  $I^{ex,s}, I^{er}$, and $I^{mr}$, according to \q{ansatz1}.\footnote{Don't be misled by \q{Iexcsnondeg}; one should directly differentiate \q{Iexcs}.} Except for extremely intense laser pulses (with electric field strengths comparable to or greater than  $10^9 \,V/m$), the time scale for collisions with the source-produced photons is much larger than the  momentum-relaxation time ($\tau_{mr}$), which is typically 100fs or less.  Conversely, to describe continuous-wave laser or solar radiation experiments, it is typically safe to neglect the excitation contribution to  
\e{
\widehat{\partial_f I^s}\ket{\delta f^a} \appr \big(\widehat{\partial_f I^{mr}} +\widehat{\partial_f I^{er}}\big)\ket{\delta f^a} =  \widehat{\partial_f I^{mr}}\ket{\delta f^a}.
} 
In the last step, we applied that $\ket{\delta f^a}$ is in the kernel of $\widehat{\partial_f I^{er}}$, in accordance with \q{partialfIer}, so that our second ansatz equation [\q{ansatz2}] simplifies to
\e{
0 \approx I_{\bk}^a(f^s)+\sum_{\bk'}\big(\partial_f I^{mr}\big)_{\bk\bk'} \delta f_{\bk'}^a; \as \partial_f I^{mr}=W^{mr}.\la{ansatz2simple}
}
 
 By applying $(W^{mr})^{-1}$ [\q{inverseWmr}] to both ansatz equations [\q{ansatz1simple} and \q{ansatz2simple}], we derive
\e{
&B\text{-symmetric, iso-energy-deviating}\as\delta f^s_{\bk} \equiv\; f^s_{\bk}-[f_{\bk}]=  -\sum_{{\bk}'}  \big(W^{mr}\big)^{-1}_{{\bk}{\bk}'}\delta I^{ex,s}_{{\bk}'}(f^s);   \la{deltafs}\\
&B\text{-asymmetric}\as\delta f^a_{\bk} = -\sum_{{\bk}'}  \big(W^{mr}\big)^{-1}_{{\bk}{\bk}'}  I^{a}_{{\bk}'}\big(\,[f]+\delta f^s\,\big).\la{deltafa33}
}
In principle,  $I^a=\sum_x I^{x,a}$ is contributed by each scattering mechanism ($x=ex,pn,im,re$); in practice, we focus in this work on electron-impurity scattering ($x=im$). \q{deltafa33} is then recast as
\e{
 \delta f^{a}_{\bk} &\refeq{impvanish} -\sum_{{\bk}'}  \big(W^{mr}\big)^{-1}_{{\bk}{\bk}'}  I^{im,a}_{{\bk}'}(\delta f^s)\refeq{Iima} -\sum_{{\bk}',\bk''}  \big(W^{mr}\big)^{-1}_{{\bk}{\bk}'}\big(W^{im,a}\big)_{{\bk}'{\bk}''} \delta f^s_{\bk''} \lin
&\refeq{deltafs} \sum_{{\bk}'{\bk}''{\bk}'''}  \big(W^{mr}\big)^{-1}_{{\bk}{\bk}'} \big(W^{im,a}\big)_{{\bk}'{\bk}''}\big(W^{mr}\big)^{-1}_{{\bk}''{\bk}'''}\delta I^{ex,s}_{{\bk}'''}(f^s).\la{deltafima1}
}
Assuming the nonequilibrium electron distribution is nondegenerate [\q{Iexcsnondeg}], $\delta I^{ex,s}_{{\bk}'''}$ does not depend on $f^s$.   \\

On one hand, to motivate the form of \q{deltafs},  an optical vortex intersecting (or proximate) to the excitation surface makes the  photo-excitation rate $I^{ex,s}_{\bk}$ anisotropic and results in a nonzero iso-energy deviant: $\delta I_{\bk}^{ex,s} \equiv  I_{\bk}^{ex,s}-[ I_{\bk}^{ex,s}] \neq 0$. A deviating photo-excitation rate tends to create a deviating electron distribution $\delta f_{\bk}^s$, but this is counterbalanced by momentum-relaxing transitions; namely, symmetric, impurity-mediated scattering (encoded in $W^{mr}$) tends to relax the distribution to being iso-energy-symmetric. \q{deltafs} describes how this counterbalancing determines the steady-state  deviant $\delta f^{s}_{\bk}$.\\

On the other hand, to motivate the form of \q{deltafima1}, a deviating  $\delta f^{s}_{\bk}$ activates the impurity-mediated skew scattering (encoded in  $W^{im,a}$). Such skew scattering tends to create  an inversion-asymmetric deviant $\delta f^a_{\bk}$, but this is counterbalanced  by the same symmetric, momentum-relaxing transitions encoded in $W^{mr}$. \q{deltafima1} describes how this  counterbalancing determines the steady-state  $\delta f^{a}_{\bk}$. \\

Bringing both hands together, \qq{deltafs}{deltafima1} show how $\delta f^a$ is  enhanced  by the intersection/proximity of optical vortices. This implies that the ballistic photocurrent, which is a linear functional of $\delta f^a$ [\q{defineballistic}], is also vortex-enhanced. \\

In the momentum RTA [\q{momentumrta}], \qq{deltafs}{deltafima1} reduce to\footnote{
To see this reduction, first note that \q{inverseWmr} reduces to $(W^{mr})^{-1}\approx - \tau^{mr}_E \hat{P}_{\delta}$, which is simply proportional to the deviant projector. The deviant projector acts trivially on $\delta I^{ex,s}$, which is manifestly iso-energy deviant. The deviant projector also acts trivially on $W^{im,a}_{\bk'\bk''}$, according to \q{WimaPdev}.
}
\e{
\text{Momentum RTA:}\as \delta f^s_{\bk} \appr \tau^{mr}_{\Ek}\delta I^{ex,s}_{{\bk}}(f^s); \as \delta f^{a}_{\bk}  \approx (\tau^{mr}_{\Eex})^2\sum_{{\bk}'}  W^{im,a}_{{\bk},{\bk}'}\delta I^{ex,s}_{{\bk}'}(f^s),\la{momentumRTAdeltaf}
}
bearing in mind that the energy-conserving delta functions [implicitly within $W^{im,a}$ and $\delta I^{ex,s}$] restrict $E_{\bk}=E_{\bk'}=\Eex$ (the excitation energy).
The first $\approx$ states that  $\delta f^s$ is just proportional to the iso-energy asymmetry of the excitation process; in contrast, the more general expression for $\delta f^s$ in \q{deltafs} shows that an iso-energy asymmetry of $f$ can also arise from an iso-energy asymmetry of the momentum-relaxation process; this arising is artificially precluded in the momentum RTA. Thus, \q{momentumRTAdeltaf} should be understood as an over-simplified estimate; while it offers a useful physical insight, the insight is gained at the cost of narrowing our viewport. If greater quantitative accuracy is desired, one can just calculate \q{deltafima1} with the harmonic-dependent  momentum-relaxation times.

\subsection{Impurity-mediated ballistic photocurrent}
\la{app:ballisticphotocurrent}

\q{eq:jball} for the impurity-mediated ballistic photocurrent forms the basis for most of our model calculations in \s{sec:current}. The goal is to derive not just \q{eq:jball} but also a more general formula [\q{eq:ballisticcurrent}] that applies beyond the momentum relaxation time approximation (RTA) [\q{momentumRTAdeltaf}] implicitly used  in \q{eq:jball}.\\

Unlike the shift current which originates from band-off-diagonal elements of the velocity matrix, the ballistic photocurrent is essentially the band-diagonal velocity matrix (meaning the group velocity $\bv_B$) times the $B$-asymmetric component $\delta f^a_B$ [\q{symmf2}] of the electron distribution:
\e{
&\text{Ballistic photocurrent:} \as \mathbf{j}_{\text{ballistic}}=-\frac{|e|}{\mathcal{V}}\sum_{B}\bv_{B}\delta f_{B}^{a},\la{app-defineballistic}
}
with $\delta f_{B}^{a}$ being proportional to the light intensity, cf. Eq.~\eqref{defineballistic}. As explained in \app{sec:kineticmodel},
 $\delta f^a_B$ arises generally from \textit{skew scattering} in a noncentric medium, and more specifically (in our context) from impurity-mediated skew scattering encoded in the transition rate matrix elements $W^{im,a}_{BB'}$ [\q{WimaBBp}]. 
 Inputting our solution for $\delta f^a$ [\q{deltafima1}] into \q{app-defineballistic},
\begin{equation}
\label{eq:ballisticcurrent}
\bj_{\text{ballistic}}=-\frac{|e|}{\mathcal{V}}\sum_{BB'B''B'''} 
\bv_{B}(W^{mr})^{-1}_{BB'} W^{im,a}_{B'B''} 
(W^{mr})^{-1}_{B''B'''} \, \delta I^{ex,s}_{B'''}.
\end{equation}
This simplifies in the momentum RTA [\q{momentumRTAdeltaf}] to 
\begin{equation}
\label{eq:app-ballisticcurrent2}
\bj_{\text{ballistic}}=-\frac{|e|}{\mathcal{V}}
(\tau^{mr}_{\Eex})^2\sum_{B,B'}  \bv_{B}W^{im,a}_{B,B' }\delta I^{ex,s}_{B'},
\end{equation}
which is equivalent to \q{eq:jball}. \\

For the three symmetry classes of impurity potentials
in \qq{class1}{class3} and for small-momentum scattering [\q{VimBBpsmallk}], the general transition rate matrix element in \q{WimaBBp} reduces to
\e{
&\text{Centrosymmetric}:\as W^{im,a}_{B,B' }=\frac{(2\pi)^2}{\hbar}N_{\text{im}}\sum_{B''}\widetilde{V}^s_{\bk'-
\bk} \widetilde{V}^s_{\bk- 
\bk''}\widetilde{V}^s_{\bk''-
\bk'}\text{Im} \scrb{BB'B''}\delta(E_{BB ''})\delta(E_{BB'});\la{Wimasymmimp}\\
&\text{Noncentric}:\as W^{im,a}_{B,B' }=\frac{(2\pi)^2}{\hbar}N_{\text{im}}\sum_{B''}\widetilde{V}^a_{\bk'-
\bk} \widetilde{V}^a_{\bk- 
\bk''}\widetilde{V}^a_{\bk''-
\bk'}\text{Re} \scrb{BB'B''}\delta(E_{BB ''})\delta(E_{BB'});\la{Wimanonsymmimp}\\
&\text{Hybrid}:\as W^{im,a}_{B,B' }= \frac{(2\pi)^2}{\hbar}N_{\text{im}}\sum_{B''}\bigg(\widetilde{V}^s\widetilde{V}^s\widetilde{V}^s+\widetilde{V}^s\widetilde{V}^a\widetilde{V}^a+\widetilde{V}^a\widetilde{V}^s\widetilde{V}^a+\widetilde{V}^a\widetilde{V}^a\widetilde{V}^s\bigg)\text{Im} \scrb{BB'B''}\delta(E_{BB ''})\delta(E_{BB'})\lin
&\as\as\as\as\as+ \frac{(2\pi)^2}{\hbar}N_{\text{im}}\sum_{B''}\bigg(\widetilde{V}^a\widetilde{V}^a\widetilde{V}^a+\widetilde{V}^a\widetilde{V}^s\widetilde{V}^s+\widetilde{V}^s\widetilde{V}^a\widetilde{V}^s+\widetilde{V}^s\widetilde{V}^s\widetilde{V}^a\bigg)\text{Re} \scrb{BB'B''}\delta(E_{BB ''})\delta(E_{BB'}), \la{Wimageneric}
}
with the implicit momentum arguments: $\widetilde{V}_{\bk'-\bk}\widetilde{V}_{\\bk-\bk''}\widetilde{V}_{\bk''-\bk'}$ for each term in \q{Wimageneric}. If the electron wave function is geometrically trivial, as was assumed by Belinicher et al,\cite{belinicher_polaraxis} then Im$\scrb{BBB''}=0$ and skew scattering requires the impurity potential to have an asymmetric $V^a$ component. The case of a symmetric impurity and nontrivial  electron wave function [\q{Wimasymmimp}] was studied in \ocite{isobe_rectification} for 2D metals. However, any realistic impurity in a noncentric medium  has an asymmetric $V^a$ component, requiring either \q{Wimanonsymmimp} or \q{Wimageneric} in the final analysis.

\section{Appendix to `Optical vortices in topological semimetals and Dirac-Weyl materials'}\la{app:vortices}

\subsection{Chern-vorticity theorem}\la{app:chernvortex2sphere}

Let us prove the
\e{\text{Chern-vorticity theorem:}\as C_v-C_c= \text{Vor}^{\be}_{cv} =\sum_{vortex}\oint_{\partial \text{vortex}} \bS^{\be}_{cv\bk}\cdot \tf{d\bk}{2\pi}, \label{chernvortexapp}}
with $\bS^{\be}_{cv\bk}$ being the  \textit{shift vector} defined in \q{photonicshift}; $\text{Vor}^{\be}_{cv}$ is the net circulation of the interband Berry phase over all $\be$-vortex points in a closed 2D $\bk$-manifold $\bSigma$, which can be a 2D Brillouin zone, or a two-toroidal/spherical cut of the 3D Brillouin zone. This theorem was previously proven in \cite{zhuAA_anomalousshift} for a $\bk$-surface that is a two-torus; we provide a proof for a two-sphere here.\\

\begin{figure}[H]
\centering
\includegraphics[width=9 cm]{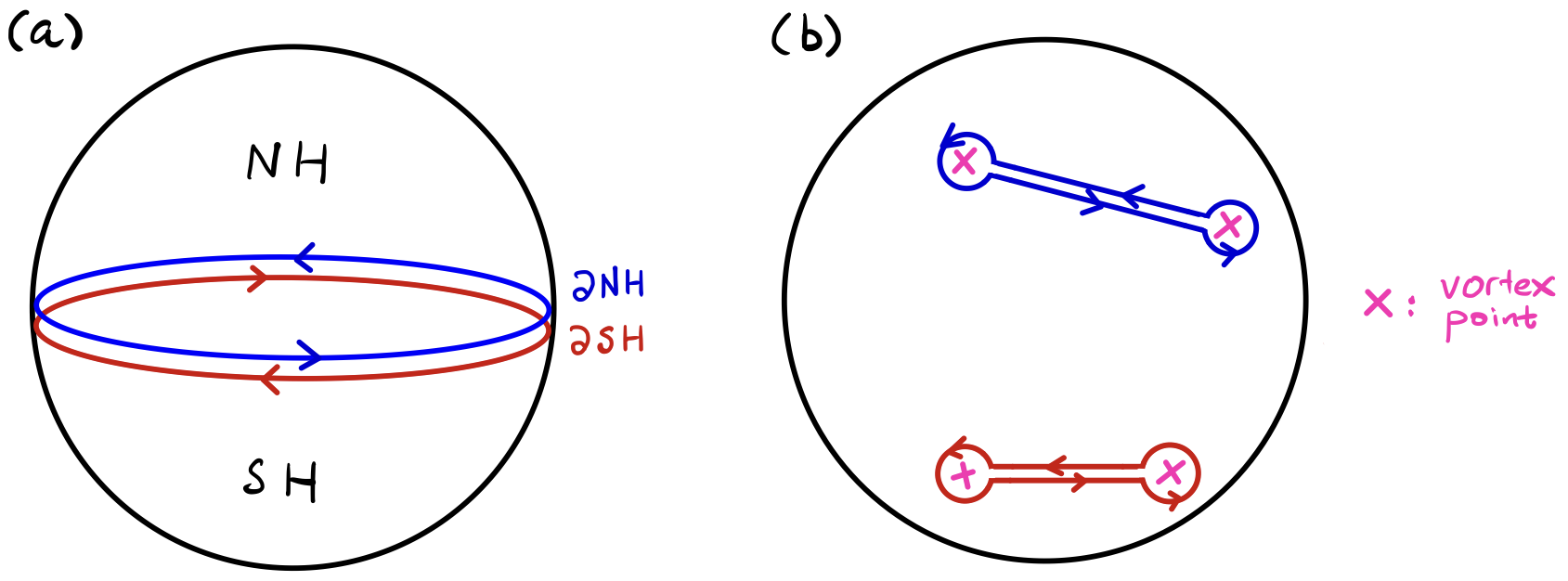}
\caption{ (a) Partitioning into hemispheres. (b) Deformed loops around vortex centers.}
\label{fig:hemisphere}
\end{figure}

Over a two-sphere, vorticity generically occurs at points, hence one may find a great circle which avoids all these vortex points, implying  the shift vector is well-defined everywhere along the great circle. We may also choose this great circle to separate the north and south hemispheres (NH and SH), with the wave function $\ket{u_{c/v\bk}}$ being analytically defined in each hemisphere, and being necessarily discontinuous across the great circle if the Chern number of the two-sphere is nontrivial.  By choice of momentum coordinates, we may as well have this great circle lie on the equator, as    in \fig{fig:hemisphere}(a). \\

One then defines  two line integrals of the shift vector, one along the boundary  [$\partial$NH; cf. blue loop in \fig{fig:hemisphere}(a)] of NH and another along the boundary  [$\partial$SH; cf. red loop in \fig{fig:hemisphere}(a)] of SH.
\e{
\oint_{\partial \text{NH}/\partial\text{SH}} \bS^{\be}_{cv\bk}\cdot \tf{d\bk}{2\pi}\refeq{photonicshift}
\eq -\oint_{\partial \text{NH}/\partial \text{SH}}\nabk \arg \{\be \cdot\bA_{cv}\}+ \int_{\text{NH}/\text{SH}} \bigg(\curl \bA_{cc\bk}-\curl\bA_{vv\bk}\bigg)\cdot \tf{d^2\bk}{2\pi}.\la{equator} 
}
In the last step, we applied that the wave function is analytic over each hemisphere, allowing to convert the line integral (of the intraband Berry connections) to an area integral (of the Berry curvature over each hemisphere) by Stokes theorem. The two line integrals in \q{equator} sum to zero because $\partial$NH and $\partial$SH are identical loops except that the orientations are opposite, as is evident from \fig{fig:hemisphere}(a). The area integrals [on the right-hand side of \q{equator}] sum to an area integral of the Berry curvature over the whole sphere, which is just the Chern number $C$, hence 
\e{
\oint_{\partial \text{NH}} \nab \arg \{\be \cdot \bA_{cv}\} \cdot \f{d\bl}{2\pi}+ \oint_{\partial \text{SH}} \nab \arg \{\be \cdot \bA_{cv}\}\cdot \f{d\bl}{2\pi} =C_c-C_v. \la{aboveline}
}
Though the wave functions $\ket{u_{c\bk}}$ and $\ket{u_{v\bk}}$ are analytic functions of $\bk$ within each hemisphere, $\arg \{\be \cdot \bA_{cv}\}$ is singular at isolated vortex points (if they exist) wherever $\be \cdot \bA_{cv}=0$.\\ 

Each of the  line integral in \q{aboveline} is a winding number for the phase of a complex number, so the line integral is invariant to continuous deformations of the red or blue loop, so long as the deformed loop never crosses a vortex point where $\arg \{\be \cdot \bA_{cv}\}$ becomes singular. Taking care to avoid such crossings, we deform the red loop (resp. blue loop) to encircle all vortex points in NH (resp. SH), as illustrated in \fig{fig:hemisphere}(b); \q{aboveline} then reduces to
\e{
\sum_{vortex \,\in\,\text{NH} }\oint_{\partial \text{vortex}}  \nab \arg \{\be \cdot \bA_{cv}\} \cdot \f{d\bl}{2\pi} + \sum_{vortex \,\in\,\text{SH} }\oint_{\partial \text{vortex}}  \nab \arg \{\be \cdot \bA_{cv}\} \cdot \f{d\bl}{2\pi}=C_c-C_v. \la{northsouth}
}
Making the circle around each vortex point infinitesimally small, we can convert each phase winding number to a line integral of the shift vector:
\e{
-\oint_{\partial \text{vortex}}  \nab \arg \{\be \cdot \bA_{cv}\} \cdot \f{d\bl}{2\pi} = \oint_{\partial \text{vortex}}  \bS^{\be}_{cv} \cdot \f{d\bl}{2\pi},\la{convertwinding}
}
the difference in the two quantities being an infinitesimal line integral of $\bA_{cc\bk}$ and $\bA_{vv\bk}$, which are both analytic functions within each hemisphere. Substituting \q{convertwinding} into \q{northsouth}, we derive \q{chernvortex}, as desired.

\subsection{Optical affinity of two-band effective Hamiltonians}
\la{app:affinity2band}

For two-band effective Hamiltonians [\q{twobandham2}], the optical affinity admits a compact expression in terms of the quantum metric and Berry curvature [Eq.~\eqref{eq:aff_2band}], as we now derive.\\

By definition, $\mathbf{A}_{cv,\bk}$ is given by 
\begin{equation}
A_{cv,\bk}^{j} = i\braket{u_{c\bk}}{\partial_j u_{v\bk}},
\end{equation}
where $\partial_{j}$ represents for $\partial_{k_j}$.
Note that

\begin{align*}
\bra{u_{c\bk}}H\ket{\partial_j u_{v\bk}}=E_{c\bk}\braket{u_{c\bk}}{\partial_j u_{v\bk}}&=\partial_{j}\bra{u_{c\bk}}H\ket{u_{v\bk}}-E_{v\bk}\braket{\partial_j u_{c\bk}}{u_{v\bk}}-\bra{u_{c\bk}}\partial_j H\ket{u_{v\bk}} 
\\
&\Rightarrow \braket {u_{c\bk}}{\partial_j u_{v\bk}}=\frac{\bra{u_{c\bk}}\partial_j H\ket{u_{v\bk}}}{E_{c\bk}-E_{v\bk}},
\end{align*}
where in $\Rightarrow$ we used the fact that $\bra{c}H\ket{v}=0$ and $\braket{\partial_j c}{v}=-\braket{c}{\partial_jv}$. For two-band models, $E_{c,\bk}-E_{v,\bk}=2d_{\bk}$ where $d_{\bk}=|\bd_{\bk}|$. Thus, $A_{cv,\bk}^j$ can be rewritten as
\begin{equation}
A_{cv,\bk}^{j} =\frac{i\bra{u_{c\bk}}\partial_j H\ket{u_{v\bk}}}{2d_{\bk}}= \frac{i \partial_j \bd_{\bk}}{2d_{\bk}}\cdot \bra{u_{c\bk}}\boldsymbol{\sigma}\ket{u_{v\bk}}.
\end{equation}
Since  $|\hat{e}\cdot \mathbf{A}_{cv,\bk}|^2=\sum_{a,b}e_a \overline{e}_b A^{a}_{cv,\bk}\bar{A}^{b}_{cv,\bk}$, we first look at $A^{a}_{cv,\bk}\bar{A}^{b}_{cv,\bk}$:
\begin{equation}
\begin{aligned}
A^{a}_{cv,\bk}\bar{A}^{b}_{cv,\bk}&= \frac{1}{4d_{\bk}^2} \partial_a \bd_{\bk}\cdot\bra{u_{c\bk}}\boldsymbol{\sigma}\ket{u_{v\bk}}\partial_b \bd_{\bk}\cdot\bra{u_{v\bk}}\boldsymbol{\sigma}\ket{u_{c\bk}}
\\
&=\frac{1}{4d_{\bk}^2} \partial_a d^m_{\bk} \partial_b d^n_{\bk}\bra{u_{c\bk}}\sigma_m\ket{u_{v\bk}}\bra{u_{v\bk}}\sigma_n\ket{u_{c\bk}}
\\
&=\frac{1}{4d_{\bk}^2} \partial_a d^m_{\bk} \partial_b d^n_{\bk}\bra{u_{c\bk}}\sigma_m(I-\ket{u_{c\bk}}\bra{u_{c\bk}})\sigma_n\ket{u_{c\bk}}
\\
&=\frac{1}{4d_{\bk}^2} \partial_a d^m_{\bk} \partial_b d^n_{\bk}\left\{ \delta_{mn}+i\epsilon_{mnp}\bra{u_{c\bk}}\sigma_p\ket{u_{c\bk}}-\bra{u_{c\bk}}\sigma_m\ket{u_{c\bk}}\bra{u_{c\bk}}\sigma_n\ket{u_{c\bk}}\right\}
\end{aligned}
\end{equation}
Given that 
\begin{equation}
\bra{u_{c\bk}}\sigma_p\ket{u_{c\bk}}=\hat{d}^{p}_{\bk},
\end{equation}
we can derive 
\begin{equation}
\label{eq:Across}
\begin{aligned}
A^{a}_{cv,\bk}\bar{A}^{b}_{cv,\bk}&=\frac{1}{4d_{\bk}^2} \left\{ \partial_a \bd_{\bk}\cdot \partial_b \bd_{\bk}+i \hat{\mathbf{d}}_{\bk}\cdot ( \partial_a \bd_{\bk}\times \partial_b \bd_{\bk})- \hat{\mathbf{d}}_{\bk}\cdot\partial_a \bd_{\bk} \hat{\mathbf{d}}_{\bk}\cdot \partial_b \bd_{\bk}\right\}
\\
&= \frac{1}{4}\left(\partial_a\hat{\mathbf{d}}_{\bk}\cdot\partial_b\hat{\mathbf{d}}_{\bk}+i\hat{\mathbf{d}}_{\bk}\cdot(\partial_a\hat{\mathbf{d}}_{\bk}\times\partial_b\hat{\mathbf{d}}_{\bk})\right)
\end{aligned}
\end{equation}

Note that the real and imaginary parts of Eq.~\eqref{eq:Across} correspond respectively to the quantum metric tensor and the Berry curvature tensor, as defined in \q{eq:metric_Berrycuvature}. Consequently, $|\hat{e}\cdot \mathbf{A}_{cv,\bk}|^2$ can be expressed in terms of these two tensors
\begin{equation}
     |\hat{e}\cdot \mathbf{A}_{cv,\bk}|^2  = \sum_{ab} \left(g_{ab,v\bk}+\frac{1}{2}\Omega_{ab,v\bk}\right)e_a\overline{e}_b=\sum_{ab} 
\left(g_{ab,v\bk}\real(e_a\overline{e}_b)-\frac{1}{2}\Omega_{ab,v\bk}\imag(e_a\overline{e}_b)\right),
\end{equation}
where in the second step we used that $g_{ab,\bk}$ is symmetric under exchange of $a$ and $b$, whereas $\Omega_{ab,\bk}$ is antisymmetric. 

For purely real $\be$, $|\hat{e}\cdot \mathbf{A}_{cv,\bk}|^2 $ reduces to $\sum_{ab}g_{ab,\bk}e_a e_b$, which is equivalent to the formula given in Ref.~\cite{aa_topologicalprinciple}:
\begin{equation}
\label{eq:oldaffform}
|\hat{e}\cdot \mathbf{A}_{cv,\bk}|^2 ={\frac{1}{4}|\hat{\mathbf{d}}_{\bk} \times \pe \hat{\mathbf{d}}_{\bk} |^2},
\end{equation}
where $\pe \equiv \be \cdot \partial_{\bk}$.
This can be seen directly from 
\begin{equation}
\begin{aligned}
{|\hat{\mathbf{d}}_{\bk} \times \pe \hat{\mathbf{d}}_{\bk} |^2}&=\sum_{ab}\sum_{i}\sum_{jl,j'l'}\epsilon_{ijl}\hat{d}^{j}_{\bk}e_{a}\partial_{a}\hat{d}^{l}_{\bk}\epsilon_{ij'l'}\hat{d}^{j'}_{\bk}e_{b}\partial_{b}\hat{d}^{l'}_{\bk}
\\
&=\sum_{ab}(\partial_a\hat{\mathbf{d}}_{\bk}\cdot\partial_b\hat{\mathbf{d}}_{\bk}-\hat{\mathbf{d}}_{\bk}\cdot \partial_a\hat{\mathbf{d}}_{\bk}\hat{\mathbf{d}}_{\bk}\cdot\partial_b\hat{\mathbf{d}}_{\bk}) e_a e_b
\\
&=\sum_{ab}\partial_a\hat{\mathbf{d}}_{\bk}\cdot\partial_b\hat{\mathbf{d}}_{\bk} e_a e_b
\end{aligned}
\end{equation}
The last step is due to the fact
\begin{equation}
\hat{\mathbf{d}}_{\bk}\cdot \partial_a\hat{\mathbf{d}}_{\bk} =\frac{1}{2}\partial_a \left(\hat{\mathbf{d}}_{\bk}\cdot \hat{\mathbf{d}}_{\bk}\right) =\frac{1}{2}\partial_a 1 =0
\end{equation}

\subsubsection{Optical affinity of isotropic 1-Weyl fermion}\la{app:1Weylaffinity}

We want to characterize the optical vorticity of the isotropic 1-Weyl fermion  with the effective Hamiltonian [reproduced from \q{hamiltonianisotropicweyl}]:
\e{
H^{i,w}_{\bk}= \bd_{\bk}\cdot \bsigma; \as \bd_{\bk}=v\bk; \as \bsigma=(\sigma_1,\sigma_2,\sigma_3). \la{hamiltonianisotropicweyl2}
}
A necessary condition for an $\be$-vortex to lie at $\bk$ is that the optical $\be$-affinity [defined as $\text{Aff}_{cv\bk}^{\be}= |\be\cdot \bA_{cv\bk}|^2$ in \q{affinityisotropicweyl}] vanishes at $\bk$. The optical affinity has the form
\e{
\text{Aff}_{cv\bk}^{\be}=\tf{1-(\hbd_{\bk}\cdot \be)^2}{4k^2}; \as \hbd_{\bk}=\tf{\bd_{\bk}}{d_{\bk}}=\tf{\bk}{k}; \as k=|\bk|; \as d=|\bd|,\la{affinityweyl}
}
for real $\be$ and for the isotropic Weyl Hamiltonian [\q{hamiltonianisotropicweyl2}], as we now prove.\\

Applying Eq.~\eqref{eq:oldaffform} for real polarization vector to the isotropic Weyl Hamiltonian [\q{hamiltonianisotropicweyl2}], 
\e{ 4|\be\cdot\bA_{cv,\bk}|^2\eq {|\hat{\mathbf{d}}_{\bk}\times \pe \hat{\mathbf{d}}_{\bk} |^2}\condeq{\pe \hat{\mathbf{d}}_{\bk}=[\be -\hbd(\hat{\mathbf{d}}_{\bk}\cdot \be)]/{k}}\f{|\hat{\mathbf{d}}_{\bk}\times \be|^2}{k^2}=\f{1-(\hat{\mathbf{d}}_{\bk}\cdot \be)^2}{k^2}, \la{affinityisotropicweyl2}}
which is the desired expression in \q{affinityweyl}.

\subsection{Symmetry of the shift vector field}\la{app:symmetryshift}

\begin{tcolorbox}[colback=white, sharp corners]
If $g$ is a point-group symmetry of the Hamiltonian [\q{gspacetime} and \q{symmetricbloch}], then  the photonic shift vector [\q{photonicshift}] is constrained as:
\e{
 \bS^{\be}_{cv,\bk} 
\eq g^{s,\mo}\cdot \bS^{g^s\cdot\overline{\be}^g}_{cv,\gk}; \as \gk= i_g g^s\cdot \bk,\la{symmetryshift2}
}
with $\overline{A}^g=\overline{A}$ = (complex conjugate of $A$) if $g$ inverts time, and otherwise $\overline{A}^g=A$. \q{symmetryshift2} is reproduced from \q{symmetryshift}.
\end{tcolorbox}

To prove \q{symmetryshift2}, we begin from the fundamental equation for a point-group-symmetric Hamiltonian [\q{symmetricbloch}] and decompose  $\hg$  as $\hg=\hU K^g$, with   $\hU$ a unitary operator, multiplied with an additional $K$ if $g$ inverts time.  Then the Hamiltonian symmetry in \q{symmetricbloch} implies that the eigenstates at $\bk$ and $g\cdot \bk$ are related as
\e{
\ket{u_{b,g\cdot \bk}}= \lambda_{b\bk}\hU \overline{\ket{u_{b\bk}}}^g; \as \lambda_{b\bk}= e^{i\phi_{b\bk}},\la{ugk}
}
with $\lambda$ a unimodular phase factor encoding the gauge freedom, i.e., arbitrariness in the phase of eigenstates.\\

Interlude on clear notation: for any function $f$ of $\bk$, we write    $\nabla_{()}f_{()}|_{\bk}$ as the gradient of $f$ evaluated at $\bk$. The derivative evaluated at $g\cdot \bk$ satisfies
\e{
\nabla_{()}f_{()}\big|_{g\cdot \bk} = g\cdot \nabla_{()}f_{g\cdot ()}\big|_{\bk},\la{difff}
}
which should be understood in component form as
\e{
\vec{x}\cdot \nabla_{()}f_{()}\big|_{g\cdot \bk} = \limit{\Delta \ri 0} \tf{f_{g\cdot \bk+\Delta \vec{x}}-f_{g\cdot \bk}}{\Delta}=  \limit{\Delta \ri 0} \tf{f_{g(\cdot \bk+\Delta \gmo \cdot \vec{x})}-f_{g\cdot \bk}}{\Delta}=\vec{x} \cdot g \cdot  \nabla_{()}f_{g\cdot ()}\big|_{\bk},
}
bearing in mind $g_{ab}=i_g g^s_{ab}$ is an orthogonal matrix because $g^s_{ab}$ is an orthogonal matrix. \\

Applying \q{difff} to the gradient of an energy eigenstate [\q{ugk}], and suppressing the band index for now,
\e{
\ket{  \nabla_{()}u_{()}  }\big|_{g\cdot \bk} \refeq{difff} g\cdot \nabla_{()}\ket{   u_{g\cdot ()}}\big|_{\bk}\refeq{ugk} =  g\cdot \nabla_{()} \big\{  \lambda_{()}\hU \overline{\ket{u_{()}}}^g \big\}\big|_{\bk}= \hU g\cdot \bigg( \nabk \lambda \overline{\ket{u_{\bk}}}^g + \lambda_{\bk} \overline{\ket{\nabk u}}^g \bigg),\la{nabugk}
}
with $\ket{\nabk u}$ a shorthand for $\ket{\nab_{()}u_{()}}\big|_{\bk}$.
The  Berry connection evaluated at $\gk$ is
\e{
\bA_{bb'g\cdot \bk}= \braket{u_{b()}}{i\nab_{()}u_{b'()}}\big|_{g\cdot\bk}\reftwoeq{ugk}{nabugk} g\cdot \bigg(  \overline{\braket{u_{b\bk}}{u_{b'\bk}}}^g \overline{\lambda_{b\bk}}i\nabk \lambda_{b'} + i_g \overline{\lambda_{b\bk}} \lambda_{b'\bk}  \overline{\braket{u_{b\bk}}{i\nabk u_{b'}}}^g \bigg).
}
The intraband Berry connection is real while the interband Berry connection is complex, hence
\e{
\bA_{bb,g\cdot \bk}\eq  -g\cdot\nabk \phi_b +g^s\cdot \bA_{bb\bk}; \as \bA_{cv,g\cdot \bk}= \overline{\lambda_{c\bk}} \lambda_{v\bk}g^s\cdot \overline{\bA_{cv\bk}}^g.\la{Berrygk}
}
Further manipulations of the interband component gives
\e{
&\arg\big\{ \be\cdot \bA_{cv,g\cdot \bk}\big\} = -\phi_{c\bk}+\phi_{v\bk}+i_g\arg\big\{ \overline{\be}^g\cdot g^s\cdot \bA_{cv\bk} \big\};   \la{argAcv}\\
&\nab_{()}\arg\big\{ \be\cdot \bA_{cv,()}\big\}\big|_{\gk} \refeq{difff} g\cdot  \nab_{()}\arg\big\{ \be\cdot \bA_{cv,g\cdot()}\big\}\big|_{\bk} \refeq{argAcv} g\cdot \bigg(   -\nabk\phi_{c}+\nabk\phi_{v}+i_g\nabk\arg\big\{ \overline{\be}^g\cdot g^s\cdot \bA_{cv}\big\} \bigg),\la{nabargAcvgk}
}
bearing in mind $\arg \overline{A}^g=i_g \arg A$.\\

We utilize \qq{Berrygk}{nabargAcvgk} to evaluate the shift vector at $\gk$:
\e{
\bS^{\be}_{cv\gk}= -\nab_{()}\arg\{ \be\cdot \bA_{cv,()}\}\big|_{\gk} +\bA_{cc,g\cdot \bk}-\bA_{vv,g\cdot \bk} = g^s\cdot \bigg( -\nabk\arg\big\{ \overline{\be}^g\cdot g^s\cdot \bA_{cv}\big\}   +\bA_{cc,\bk}-\bA_{vv,\bk} \bigg)= g^s\cdot \bS^{g^{s,\mo}\cdot \overline{\be}^g}_{cv\bk},\notag
}
which finally leads to \q{symmetryshift2}. The cancellation of arbitrary phases $\phi_{c/v\bk}$ reflects the gauge invariance of the shift vector.

\subsection{Effective Hamiltonians and Chern-vorticity of multifold fermions}
\la{app:models-multi}

\subsubsection{Pseudospin-1 Weyl fermions}

Let us choose the following representation of the spin-1 matrices in Eq.~\eqref{multifold-3D-H}:
\begin{equation}
\label{multifold-3D-S}
    \hat{L}_{x}=\begin{pmatrix}
        0 & i & 0\\
        -i& 0 & 0\\
        0 & 0 & 0\\
    \end{pmatrix},\quad  \hat{L}_{y}=\begin{pmatrix}
        0 & 0 & -i\\
        0& 0 & 0\\
        i & 0 & 0\\
    \end{pmatrix},\quad \hat{L}_{z}=\begin{pmatrix}
        0 & 0 & 0\\
        0& 0 & i\\
        0 & -i & 0\\
    \end{pmatrix}.
\end{equation}

The energy spectrum of the pseudospin-1 Hamiltonian $\mathcal{H}^{s1}$ given in Eq.~\eqref{multifold-3D-H} contains three isotropic bands:
\begin{equation}
\label{multifold-spin1-3D-eps}
\epsilon_{\pm 1} = \pm v k, \quad \quad \epsilon_{0} = 0.
\end{equation}

The Berry curvature for the model \eqref{multifold-3D-H} is
\begin{equation}
\label{multifold-3D-Omega}
\bm{\Omega}_0 = \mathbf{0}, \quad \quad \bm{\Omega}_{-1} = -\bm{\Omega}_{1}= - \frac{\mathbf{k}}{k^3}.
\end{equation}
Therefore, the dispersive conduction $n=1$ (valence $n=-1$) band has the Chern number $-2$ ($2$) and the flat band has the Chern number $0$.

The inter-band Berry connection is nontrivial for the transitions involving the flat band, i.e., $\mathbf{A}_{-1,0}=\mathbf{A}_{0,1} \neq \mathbf{0}$ while $\mathbf{A}_{-1,1}=\mathbf{0}$, leading to the following optical affinity 
\e{
\text{Aff}_{-1,0;\bk}^{\be \in \R}=\text{Aff}_{0,1;\bk}^{\be \in \R} =  \frac{1-(\hbd\cdot \be)^2}{2k^2}, \quad \text{Aff}_{-1,1;\bk}^{\be \in \R} =0,
\la{multifold-aff}
}
cf. Eq.~\eqref{affinityisotropicweyl} for Weyl fermions. Note that the vanishing optical affinity $\mathbf{A}_{-1,1}=\mathbf{0}$ is an artifact of the linearized model \eqref{multifold-3D-H}, Introduction of the quadratic corrections to the Hamiltonian can allow for a nontrivial  $\mathbf{A}_{-1,1}$.

The shift vector for the transitions allowed in the linearized model reads
\begin{eqnarray}
\label{multifold-3D-Sv0}
\mathbf{S}_{-1,0}^{(\hat{\mathbf{e}})} = \mathbf{S}_{0,1}^{(\hat{\mathbf{e}})} = \frac{\left(\mathbf{k}\cdot \hat{\mathbf{e}}\right) \left[\mathbf{k}\times \hat{\mathbf{e}}\right]}{k_{\perp}^2 k}.
\end{eqnarray}
The interband vorticity of the shift vectors $\mathbf{S}_{0,1}^{(\hat{\mathbf{e}})}$ and $\mathbf{S}_{0,1}^{(\hat{\mathbf{e}})}$ is $-2$. Therefore, if one considers the transitions involving the dispersive and the flat bands, the structure of the vortex lines for pseudospin-1 systems is similar to that in a Weyl semimetal. 


The director field for pseudospin-1 fermions is
\begin{equation}
\label{multifold-sv-director}
\mathbf{D}_{-1,0} = \mathbf{D}_{0,1} = \frac{1}{2} \frac{\mathbf{k}}{k^3}, \quad \mathbf{D}_{-1,1} = 0.
\end{equation}

Let us exemplify the second-order corrections to the pseudospin-1 Hamiltonian in the case of the No.~198 or P$2_1 3$ symmetry group, which is relevant to CoSi. One derives the following Hamiltonian in the vicinity of the $\Gamma$ point of CoSi~\cite{Ni-Wu-GiantTopologicalLongitudinal-2021}
\begin{eqnarray}
\label{multifold-3D-2-H}
    \mathcal{H}_2 &=& v \left(-\lambda_2 k_x +\lambda_5 k_y -\lambda_7 k_z\right) + ak^2 \lambda_0 +b\left[\lambda_1 k_x k_y +\lambda_4 k_x k_z +\lambda_6 k_y k_z\right] \nonumber\\
    &+& c \left[ \left(-\frac{1}{2} \lambda_3 +\frac{\sqrt{3}}{2} \lambda_8\right) \left(k_x^2 -k_y^2\right) +\left(-\frac{\sqrt{3}}{2} \lambda_3 -\frac{1}{2} \lambda_8\right) \frac{2k_z^2 -k_x^2 -k_y^2}{\sqrt{3}}\right] \nonumber\\
    &+& d \left[ -\left(-\frac{1}{2} \lambda_3 +\frac{\sqrt{3}}{2} \lambda_8\right) \frac{2k_z^2 -k_x^2 -k_y^2}{\sqrt{3}} +\left(-\frac{\sqrt{3}}{2} \lambda_3 -\frac{1}{2} \lambda_8\right) \left(k_x^2 -k_y^2\right) \right],
\end{eqnarray}
where $\lambda_i$ are standard Gell-Mann matrices.

Assuming $d=0$~\cite{Ni-Wu-GiantTopologicalLongitudinal-2021}, the approximate energy spectrum reads
\begin{eqnarray}
\label{multifold-3D-2-eps-app-1}
\epsilon_{-1} &=& -v k +ak^2 -\left(b +\frac{2c}{3}\right) \frac{k_x^2 k_y^2 +k_y^2 k_z^2 +k_z^2 k_x^2}{k^2} +\frac{2c}{3} \frac{k_x^4 +k_y^4 +k_z^4}{k^2},\\
\label{multifold-3D-2-eps-app-2}
\epsilon_{0} &=& ak^2 +2\left(b +\frac{2c}{3}\right) \frac{k_x^2 k_y^2 +k_y^2 k_z^2 +k_z^2 k_x^2}{k^2} -\frac{4c}{3} \frac{k_x^4 +k_y^4 +k_z^4}{k^2},\\
\label{multifold-3D-2-eps-app-3}
\epsilon_{1} &=& v k +ak^2 -\left(b +\frac{2c}{3}\right) \frac{k_x^2 k_y^2 +k_y^2 k_z^2 +k_z^2 k_x^2}{k^2} +\frac{2c}{3} \frac{k_x^4 +k_y^4 +k_z^4}{k^2}.
\end{eqnarray}

To make an analytical advance, let us assume $a=b=0$ and expand the optical affinity up to the leading order in $c$. We obtain the following optical affinity for the $-1,1$ bands:
\begin{eqnarray}
\label{multifold-3D-2-Affcv}
\left| \mathbf e\cdot \mathbf{A}_{-1,1}\right|^2
&=& \frac{c^2}{9v^2} \frac{ \left[ k_x^4+k_y^4-k_y^2k_z^2+k_z^4-k_x^2(k_y^2+k_z^2) \right]^2} {k_{\perp}^2k^{10}} \nonumber\\
&\times&
\left\{ k_{\perp}^2 \Big( \mathbf{e}_{\perp}\cdot\bk_{\perp}-e_zk_z \Big)^2 + 4k_z^2 \Big[ e_{\perp}^2k^2 -e_zk_z (\mathbf{e}_{\perp}\cdot\bk_{\perp}) \Big] \right\}.
\end{eqnarray}
Therefore, quadratic corrections reduce the artificially high symmetry of the Hamiltonian $\mathcal{H}^{s1}$ given in Eq.~\eqref{multifold-3D-H}, and allow for the optical affinity for all inter-band transitions. We checked numerically that the corresponding shift vector $\mathbf{S}_{-1,1}^{(\hat{\mathbf{e}})}$ acquires the vorticity $-4$, which agrees with the Chern-vorticity theorem. The results for the optical vorticity are summarized in Tab.~\ref{fig:shift-multi-weyl}.

\subsubsection{Pseudospin-3/2  Rarita-Schwinger-Weyl fermions}

Let us choose the following representation of the spin-3/2 matrices:
\begin{equation}
\label{multifold-rsw-3D-J}
    \hat{J}_{x} = \frac{1}{2} \begin{pmatrix}
        0 & \sqrt{3} & 0 & 0\\
        \sqrt{3} & 0 & 2 & 0\\
        0 & 2 & 0 & \sqrt{3}\\
        0 & 0 & \sqrt{3} & 0\\
    \end{pmatrix},\quad  \hat{J}_{y} = \frac{i}{2} \begin{pmatrix}
        0 & -\sqrt{3} & 0 & 0\\
        \sqrt{3} & 0 & -2 & 0\\
        0 & 2 & 0 & -\sqrt{3}\\
        0 & 0 & \sqrt{3} & 0\\
    \end{pmatrix},\quad \hat{J}_{z} = \frac{1}{2} \begin{pmatrix}
        3 & 0 & 0 & 0\\
        0 & 1 & 0 & 0\\
        0 & 0 & -1 & 0\\
        0 & 0 & 0 & -3\\
    \end{pmatrix}.
\end{equation}

The energy spectrum of the pseudospin-3/2 Hamiltonian $\mathcal{H}^{s3/2}$ given in Eq.~\eqref{multifold-3D-H} contains four isotropic bands
\begin{equation}
\label{multifold-rsw-3D-eps}
\epsilon_{\pm 1/2} = \pm \frac{1}{2} v k, \quad \quad \epsilon_{\pm 3/2} = \pm \frac{3}{2} v k.
\end{equation}

The corresponding Berry curvature is
\begin{equation}
\label{vortex-rsw-3D-Omega}
\bm{\Omega}_{\pm 1/2} = \pm \frac{\mathbf{k}}{2k^3}, \quad \quad \bm{\Omega}_{\pm 3/2} =\pm 3 \frac{\mathbf{k}}{2k^3}.
\end{equation}
Therefore, bands $\pm 1/2$ and $\pm 3/2$ have the Chern numbers $\pm1$ and $\pm3$, respectively.

The nontrivial components of the optical affinity are dictated by the angular momentum of the states, i.e., only the transitions between the states whose angular momentum differs by $\pm1$ are allowed in the linearized model \eqref{multifold-3D-H}. We have the following optical affinity:
\e{
\text{Aff}_{-3/2, -1/2; \bk}^{\be \in \R}= \text{Aff}_{-1/2, -3/2; \bk}^{\be \in \R} = \text{Aff}_{1/2, 3/2; \bk}^{\be \in \R}= \text{Aff}_{3/2, 1/2; \bk}^{\be \in \R} = \frac{3}{4} \frac{1-(\hbd\cdot \be)^2}{k^2}, 
\quad \text{Aff}_{-1/2, 1/2; \bk}^{\be \in \R}= \text{Aff}_{1/2, -1/2; \bk}^{\be \in \R} =\frac{1-(\hbd\cdot \be)^2}{k^2}, 
\la{multifold-rsw-aff}
}
cf. Eq.~\eqref{multifold-aff} for pseudospin-1 fermions. Therefore, the structure of the vortex lines in pseudospin-3/2 systems is similar to that in Weyl semimetals and in pseudospin-1 systems, i.e., there is a vortex line along the light polarization.

The director field reads 
\begin{equation}
\label{vortex-rsw-3D-director}
\mathbf{D}_{-3/2,-1/2} = \mathbf{D}_{-3/2, 1/2} =  \frac{3}{4} \frac{\mathbf{k}}{k^3}, \quad \quad \mathbf{D}_{-1/2,3/2} = \mathbf{D}_{1/2,3/2} = -\frac{3}{4} \frac{\mathbf{k}}{k^3}.
\end{equation}

As for the pseudospin-1 fermions, the optical affinity between the states whose angular momentum differs by more than 1 can be nonzero if quadratic corrections are included in the effective Hamiltonian. Similar to the pseudospin-1 case, we derive the following Hamiltonian for pseudospin-3/2 fermions with the No.~198 or P$2_1 3$ symmetry group
\begin{eqnarray}
\label{vortex-rsw-quad-1}
\mathcal{H}^{s3/2}(\mathbf{k}) &=& v \mathbf{k}\cdot\hat{\mathbf{J}}+ a k^2 1_4 + b
\left[\frac{\left(2k_z^2-k_x^2-k_y^2\right)\left(2\hat{J}_z^2-\hat{J}_x^2-\hat{J}_y^2\right)}{3}+ \left(k_x^2-k_y^2\right) \left(\hat{J}_x^2-\hat{J}_y^2\right) \right] \nonumber\\
&+& d\left[\frac{2k_z^2-k_x^2-k_y^2}{\sqrt{3}} \left(\hat{J}_x^2-\hat{J}_y^2\right) - \left(k_x^2-k_y^2\right) \frac{2\hat{J}_z^2-\hat{J}_x^2-\hat{J}_y^2}{\sqrt{3}}\right] \nonumber\\
&+& c\left[k_x k_y\left\{\hat{J}_x,\hat{J}_y\right\} + k_x k_z\left\{\hat{J}_x,\hat{J}_z\right\} +k_y k_z \left\{\hat{J}_y,\hat{J}_z \right\} \right],
\end{eqnarray}
where $\{\hat{A},\hat{B}\}=\hat{A}\hat{B}+\hat{B}\hat{A}$ is an anticommutator. We numerically checked that, including even weak second-order terms, one can obtain a nonzero optical affinity for all types of inter-band transitions. Therefore, in addition vortex lines with vorticity $-1$ for $-3/2 \to -1/2$ and $-1/2 \to 1/2$ transitions, quadratic corrections allow for the transitions $-3/2 \to 1/2$ and $-3/2 \to 3/2$ with vorticities $-2$ and $-3$, respectively, see Tab.~\ref{fig:shift-multi-weyl} for the summary of all transitions.

\subsection{Appendix to `Interlude: Optical vortices from the director field'}\la{app:director}

\subsubsection{Equivalent formulations of  the director field}

The $i$'th component of the director field $D^i_{cv\bk}$ [\q{directorfield}] is essentially $\sum_{jk}\eps_{ijk} F_{jk}^{cv}$, with $\eps$ being the Levi-Cevita tensor and $F$ being the imaginary component of Ahn's Hermitian metric.\cite{ahn_riemanniangeometry}\\

For (and only for) two-band effective Hamiltonians, the {director field} $\bD_{cv\bk} =-\half \curl \bA_{vv\bk}$ reduces to half the intraband Berry curvature taken with the sign minus.\cite{berry_quantalphase} The proof follows directly once we write  $\bR_{cv\bk}=\frac{1}{2}\left(i\braket{u_c}{\partial_{\bk}u_v}-i\braket{\partial_{\bk}u_v}{u_c}\right)$ and $\bI_{cv\bk}=\frac{1}{2}\left(\braket{u_c}{\partial_{\bk}u_v}+\braket{\partial_{\bk}u_v}{u_c}\right)$, and recall that $\nabk \times \mathbf{A}_{vv\bk}=\sum_{b\neq v}i\braket{\nabk u_v}{u_{b}}\times \braket{u_{b}}{\nabk u_v}$, which reduces to $i\braket{\nabk u_v}{u_c}\times \braket{u_c}{\nabk u_v}$ for two-band Hamiltonians. Then, 
\e{
\bD_{cv\bk}=\bR_{cv\bk}\times\bI_{cv\bk}=\frac{i}{4}(\braket{u_c}{\nabk u_v}\times\braket{\nabk u_v}{u_c}-\braket{\nabk u_v}{u_c}\times \braket{u_c}{\nabk u_v})=-\frac{1}{2}\nabk \times \mathbf{A}_{vv\bk}.
}

\subsubsection{Proof of skyrmion-vortex duality}
\la{app:skyrmionvortex}

Collecting and reproducing some results [\qq{nomisdirection}{skyrmionvortexduality}] from the main text for convenience,
\e{
&\big(\;|\bD_{cv\bk}|\notin \{0,\infty\} \iff \bscrd_{cv\bk}=\tf{\bD_{cv\bk}}{|{\bD_{cv\bk}}|}\equiv \be \as \text{has unit-norm}\;\big) \imp  \bigg(\;\be\cdot \bA_{cv\bk}=0 \iff \text{$\be$-zero at $\bk$}\;\bigg);  \la{Dcvnonzero}\\
&\text{Skyrmion density}\as \bscrf_{cv\bk} = \eps_{abc}\scrd^a_{cv} \nabk \scrd^b_{cv}\times \nabk \scrd^c_{cv}\bigg|_{{\bk}}; \la{skyrmiondensity2}\\
&\text{Skyrmion-vortex duality:}  \as \bscrd_{cv\bk_0}=\be \as\text{and}\as  \be\cdot \bscrf_{cv\bk_0} \neq  0 \iff \;\be\text{-zero at $\bk_0$}\;= \;\text{order-1}\;\be\text{-vortex at }\bk_0.  \la{skyrmionvortexduality2}
}
(In the remainder of this appendix,  the subscript $cv$ [on $\bscrd_{cv\bk_0}$ and $\bscrf_{cv\bk_0}$] is omitted to simplify notation.) We say there is an order-one $\be$-vortex at $\bk_0$ if (i) $\be\cdot \bA_{cv\bk_0}=0$ along a $\bk$-line intersecting $\bk_0$, and (ii) the shift vector has unit winding  over an infinitesimal ellipse [denoted $\text{Ell}_{\bk_0}$] centered at $\bk_0$ and linked with said $\bk$-line:
\e{
\text{Vor}^{\be}_{\text{Ell},\bk_0} =\oint_{\text{Ell}_{\bk_0}} \bS^{\be}_{cv\bk}\cdot \tf{d\bk}{2\pi} \refeq{convertwinding} -\oint_{\text{Ell}_{\bk_0}}  \nabk \arg \{\be \cdot \bA_{cv}\} \cdot \tf{d\bk}{2\pi} =  \pm 1.\la{vorticityellipse}
}\\

The duality in \q{skyrmionvortexduality2} follows directly from the following skyrmion-vortex identity:
\e{
\text{If}\;\;\bscrd_{\bk_0}=\be \;\text{has unit-norm},\as \be \cdot \bscrf_{\bk_0}  \eq -\sqrt{J^{\perp \be}_{\bk_0}}\,\text{Vor}^{\be}_{\text{Ell},\bk_0}= \mp \sqrt{J^{\perp \be}_{\bk_0}},\la{skyrmionvortex}
}
with $\text{Ell}_{\bk_0}$ centered at $\bk_0$ and in the plane normal to $\be$;  $J^{\perp \be}$ is the determinant  of the 2-by-2 Hessian matrix:
\e{
\text{Hessian determinant}\as  J^{\perp \be}_{\bk} = \det \bigg[\tf{\partial^2}{\partial k_a\partial k_b}\big\{\tf{|\be\cdot \bA_{cv}|^2}{|\bD|}\big\}\bigg]_{\bk},\la{hessiandet}
}
with the derivatives $\partial/\partial k_a$ and $\partial/\partial k_b$ taken in the plane orthogonal to $\be$ [cf.\ \q{jacobian}]. In interpreting \q{skyrmionvortex}, it should be recognized that
\e{
\bscrd_{\bk_0}=\be \;\text{has unit-norm} \refimp{Dcvnonzero} \tf{|\be\cdot \bA_{cv}|^2}{|\bD|}\big|_{\bk_0}=0.
}
Moreover, because $|\be\cdot \bA_{cv}|^2/|\bD|$ is a non-negative function, its second-order derivatives (evaluated at $\bk_0$) cannot be negative, hence the determinant of the Hessian matrix cannot be negative, which makes $\sqrt{J^{\perp \be}_{\bk_0}}$ well-defined. Then it may be seen from \q{skyrmionvortex} that  ($\bscrd_{\bk_0} =\be$ and  $\be\cdot \bscrf_{\bk_0}\neq 0$) imply that  ($\be\cdot \bA_{cv\bk_0}=0$ and $\text{Vor}^{\be}_{\text{Ell},\bk_0}=\pm 1$), which are  defining properties of an order-1  $\be$-vortex. \\

Let us prove \q{skyrmionvortex} for $\be=\bz$. (The coordinates can always be transformed such that $\be=\bz$.)  In a preliminary step, it is convenient to define
\e{
\text{Skyrmion density two-form}\as F^{jk}\eq \scrf^{jk}_{\bk}dk_j\wedge dk_k\as \text{(no summation)},\la{skyrmiontwoform}\\
\text{Skyrmion density vector} \as \scrf_i \eq  \eps_{ijk}\scrf^{jk} \as \text{(repeated indices summed)}; \as    \scrf^{jk}_{\bk}= \bscrd \cdot \nabla_j\bscrd \times \nabla_k \bscrd. 
}
If the skyrmionic density 
is nonzero at $\bk_0$, we would like to infer a nontrivial phase winding of $\bz\cdot \bA_{cv}$ around an ellipse (of infinitesimal radius $\sim \eps$) centered at $\bk_0$ and contained in the xy plane. On any point on this ellipse, we estimate the order of magnitude for the director components as
\e{
\sum_j\scrd_j^2=1 \imp \scrd_{1,2}=O(\eps), \as\scrd_3=1-O(\eps^2) \imp  \nabla_{1,2} \scrd_{1,2}=O(1),\as \nabla_{1,2}\scrd_3=O(\eps), \la{estimatescrd}
}
hence the skyrmionic density simplifies to a leading term of order one:
\e{
\scrf^{12} \eq  \scrd_3 \big(\; \nabla_1\scrd_1 \nabla_2 \scrd_2 - \nabla_1\scrd_2 \nabla_2 \scrd_1\;\big) +O(\eps^2) 
= \eps_{3ab}\nabla_1\scrd_a \nabla_2 \scrd_b  +O(\eps^2),\lin
F^{12}\eq \scrf^{12} dk_1\times dk_2 \approx \half \eps_{3ab}\nabla_i\scrd_a \nabla_j \scrd_b  dk_i\times dk_j
}
with repeated indices $i,j$ summed over $\{1,2\}$. The advantage of normalizing the director field ($\scrd_3 \approx 1$) is so that the asymptotic form of $F^{12}$ can be expressed as the exterior derivative of a one-form:\footnote{The exterior derivative acting on a one-form $A=A_i dk_i$ gives $dA= \nabla_iA_j dk_i\wedge dk_j$ with $dk_i\wedge dk_j=-dk_j\wedge dk_i$.}
\e{
F^{12} \eq dA; \as A =\half \eps_{3ab}  \scrd_a  \nabla_j \scrd_b\, dk_j.
}
By Stokes theorem,
\e{
\int F^{12} = \oint A = \half \oint \eps_{3ab}  \scrd_a  \nabk \scrd_b\cdot d\bk = \half \oint \big(  \scrr_z \nabk \scri_z - \scri_z \nabk \scrr_z\big)\cdot d\bk + O_r(\eps).\la{applystokes}
}
In the last step, we applied
\e{
\eps_{3ab}  \scrd_a  \nabk \scrd_b \eq \big(\scrr_y\scri_z-\scrr_z \scri_y\big)\nabk  \big(\scrr_z\scri_x-\scrr_x \scri_z\big) -  \big(\scrr_z\scri_x-\scrr_x \scri_z\big) \nabk \big(\scrr_y\scri_z-\scrr_z \scri_y\big) \lin
\eq \nabk \scrr_x\big( -\scri_z\scrd_x\big) +\nabk \scrr_y\big( -\scri_z\scrd_y\big)+\nabk \scrr_z\big( -\scri_z\scrd_z\big) +\nabk \scri_x\big( \scrr_z\scrd_x\big) +\nabk \scri_y\big( \scrr_z\scrd_y\big)+\nabk \scri_z\big( \scrr_z\scrd_z\big).\la{toss}
}
Use was made of Leibniz's product rule, which assumed that both $\scrr_z$ and $\scri_x$ are differentiable functions; this is true of first-order vortices, but for higher-order vortices these functions are discontinuous.\footnote{This originates from $\bD$ vanishing at the vortex center.}
Because $\scri_j$ and $\scrr_j$ are not constrained to be semipositive definite,
\e{
|\bz\cdot \bA_{cv}|^2_{\bk_0} = I_z^2+R_z^2= 0 \imp \scri_z =O(\eps), \as \scrr_z=O(\eps),\lin \nabla_{1,2}\scri_{x,y,z}=O(1), \as \nabla_{1,2}\scrr_{x,y,z}=O(1).
}
Combining these order-of-magnitude estimates with our previous estimates for $\scrd_j$ [\q{estimatescrd}], it follows that the leading terms in \q{toss} are of order $\eps$ and proportional to $\nabk \scrr_z$ and to $\nabk\scri_z$: 
\e{
\eps_{3ab}  \scrd_a  \nabk \scrd_b  \eq \scrd_z\big( \scrr_z\nabk \scri_z -\scri_z\nabk \scrr_z  \big) +O(\eps^2).
}
Further replacing $\scrd_z=1-O(\eps^2),$ one obtains \q{applystokes}. The gradient of the interband Berry phase is expressible in terms of the real and imaginary components of $\bz\cdot \bA_{cv}$: 
\e{
\nabk \arg \bz\cdot \bA_{cv} = \nabk \tan^{-1}\f{I_z}{R_z} = \nabk \tan^{-1}\f{\scri_z}{\scrr_z} = \f{\scrr_z\nabk \scri_z-\scri_z \nabk \scrr_z}{\scri_z^2+\scrr_z^2} = \f{\scrr_z\nabk \scri_z-\scri_z \nabk \scrr_z}{|\bz\cdot \bA_{cv}|^2/|\bD|}; \la{nabkarg} 
}
use of the quotient rule assumed that both $\scri_z$ and $\scrr_z$ are differentiable, which rules out higher-order vortices. We insert \q{nabkarg} in \q{applystokes} to obtain
\e{
 \int F^{12} = \half \oint_{\text{Ell}_{\bk_0}} \f{|\bz\cdot \bA_{cv}|^2}{|\bD|} \nabk \arg (\bz\cdot \bA_{cv}) \cdot d\bk + O_r(\eps)= \f{|\bz\cdot \bA_{cv}|^2}{2|\bD|}\bigg|_{ellipse} \oint_{\text{Ell}_{\bk_0}}  \nabk \arg \bz\cdot \bA_{cv} \cdot d\bk + O_r(\sqrt{A_{ellipse}}).\la{lasteq1}
}
We have applied that $|A^z_{cv}|^2/|D|$ vanishes at the optical zero,\footnote{For an $n$'th order vortex, $|A^z_{cv}|^2$ vanishes as $(A_{ellipse})^n$ while $|D|$ vanishes as $(A_{ellipse})^{n-1}$.} and that iso-contours of $|A^z_{cv}|^2/|D|$ are elliptical with area $A_{ellipse}$, and that $\oint$  integrates over into an infinitesimal ellipse (${\text{Ell}_{\bk_0}}$) matching the iso-affinity contour. This means that $|A^z_{cv}|^2/|D|$ (evaluated on this contour) is an integration constant. \\

The area integral $\int F^{12}$ gives the skyrmionic density [defined in \q{skyrmiondensity2} and evaluated at the center of the ellipse] multiplied by the (assumed-infinitesimal) area of the ellipse:
\e{
\int F^{12}= \half \big(\bscrf_{\bk_0} \cdot \bz\big)\, A_{ellipse}, \la{lasteq2}
}
with the factor of half emerging from $\scrf^{12}= \scrf_3/2.$
Combining \qq{lasteq1}{lasteq2} in the limit $A_{ellipse}\ri 0^+$, and assuming that
\e{
\f{|\be\cdot \bA_{cv}|^2}{|\bD|}\bigg|_{ellipse}\ri \f1{2\pi} \sqrt{J^{\perp \be}_{\bk_0}} A_{ellipse} \as\text{as}\as A_{ellipse}\ri 0^+, \la{jacobian}
} 
we arrive at 
\e{
\half \big(\bscrf_{\bk_0} \cdot \bz\big)\, A_{ellipse}=\sqrt{J^{\perp \be}_{\bk_0}} A_{ellipse}\oint_{\text{Ell}_{\bk_0}}  \nabk \arg \bz\cdot \bA_{cv} \cdot \tf{d\bk}{2\pi},\la{assumethat}
}
with $J^{\perp \be}_{\bk_0}$ being the Hessian determinant [\q{hessiandet}]. Identifying the line integral in \q{assumethat}  with the vorticity [\q{vorticityellipse}], we obtain the skyrmion-vortex identity [\q{skyrmionvortex}].\\

The rest of this section will be spent proving \q{jacobian}.
Indeed, $|\be\cdot \bA_{cv}|^2/|\bD|$ is both semipositive-definite and real-analytic with respect to $\bk$; since $|\be\cdot \bA_{cv}|^2/|\bD|$ vanishes at $\bk_0$ (a global mininum), it admits the convergent Taylor expansion:
\e{
f=\f{|\be\cdot \bA_{cv}|^2}{|\bD|}\bigg|_{ellipse}= \half \bk \cdot f''\cdot \bk +\ldots  ; \as (f'')_{ab}= \f{\partial^2}{\partial k_a\partial k_b}\f{|\be\cdot \bA_{cv}|^2}{|\bD|}. \la{convergent}
}
Because $f''$ is a real, symmetric, 2-by-2 matrix, it may be diagonalized with real eigenvalues (denoted $e_{1,2}$) and real right-eigenvectors which equal (the transpose of) the left-eigenvectors (denoted $\bv_{1,2}$):
\e{
f=\half \big( e_1k_1^2 + e_2k_2^2\big); \as k_j=\bk \cdot \bv_j.\la{convergent2}
}
This expresses an ellipse:
\e{
1= \f{k_1^2}{a_f^2}+\f{k_2^2}{b_f^2}; \as a_f=\sqrt{2f/e_1}; \as b_f=\sqrt{2f/e_2},
}
with elliptical radii $a_f$ and $b_f$. The ratio of the maximal values of $k_1$ and $k_2$ (on an ellipse) are independent of $f$:
\e{
\f{k_1^{max}}{k_2^{max}}=\f{a_f}{b_f}=\f{a_1}{b_1}=\sqrt{e_2/e_1} \imp 
f=\half e_2(k^{max}_2)^2=\half e_1(k^{max}_1)^2=\half \sqrt{e_1e_2}k^{max}_1 k^{max}_2.
}
Given that the area of the ellipse is $\pi k^{max}_1 k^{max}_2$ and $e_1e_2$ is nothing more than the Hessian determinant [compare \qq{convergent}{convergent2} with \q{hessiandet}], we arrive at \q{jacobian}.

\subsubsection{Skyrmion-vortex duality of isotropic 1-Weyl fermion}\la{app:1Weylduality}

 Let us verify the skyrmion-vortex identity [\q{skyrmionvortex}] for the isotropic Weyl Hamiltonian [\q{hamiltonianisotropicweyl2}]. That $\text{Vor}^{\be}_{\text{Ell},\bk_0} =1$ has been determined  by independent calculations, e.g., by application of the Chern-vorticity theorem in \s{sec:isotropicweyl}.  Thus verifying that  $ \be \cdot \bscrf/\sqrt{J^{\perp \be}}=-1$  is a nontrivial check of the skyrmion-vortex identity. We will calculate (i) $\bD$, then (ii) $\bscrf$, then (iii) $J^{\perp \be}$, with the final result:
\e{
|\bD|=\tf1{4k^2};\as \be \cdot \bscrf =-\tf{2}{k^2}=-\sqrt{J^{\perp \be}}.\la{toproveisotropicweyl}
}

\noi{i} We calculate $\bD$ by applying the relation between the director field and the Berry curvature field of the valence band: $\bD=-\half \bOmega_v$,  which applies for two-band Hamiltonians [cf. footnote in \s{sec:isotropicweyl}]. The Berry curvature vector can be expressed in terms of an antisymmetric tensor: $\Omega_{v,c} =\half \eps_{cab}\Omega_v^{ab}$, which is the skyrmion density of the $\hbd$ field:
 \e{
 \Omega_v^{ab}\eq \half \hbd \cdot \nabla_a \hbd \times\nabla_b\hbd=\f1{2k^2}{\hbd\cdot \ba \times \bb} \imp \bOmega_v = \f{\hbd}{2k^2}.
 }
 Here, $\nabla_a$ and $\nabla_b$ are shorthand for $\partial/\partial k_a=\ba\cdot \nabk$ and  $\partial/\partial k_b=\bb\cdot \nabk$, with $\ba$ and $\bb$ being unit-norm, directional vectors along a Cartesian (x,y,z) direction.
 Our expression for $\bOmega_v$ integrates (over a sphere of area $4\pi k^2$) to $2\pi$, corresponding to the Chern number  $C_v=+1$. This implies for the director field that
\e{
\bD=-\thalf \bOmega_v= -\tf{\hbd}{4k^2} \imp |\bD|= \tf1{4k^2}.\la{Disotropicweyl}
}

\noi{ii} Applying $\bscrd=-\hbd$ from \q{Disotropicweyl}, the skyrmion density of $\bscrd$ is
\e{
\ba\cdot \bscrf  =\eps_{abc}\bscrd \cdot \nabla_b \bscrd \times \nabla_c \bscrd= -\eps_{abc}\hbd \cdot \nabla_b \hbd \times \nabla_c \hbd =  -\eps_{abc}\tf{\hbd\cdot \bb \times \bc}{k^2}= -2 \tf{\hbd \cdot \ba}{k^2}\imp \bscrf= -2\tf{\hbd}{k^2}. 
}
In particular, $\be \cdot \bscrf$ evaluated at $\bk_0$ (where $\be=\hbd$) is just $-\tf{2}{k^2}$. \\

\noi{iii} Combining our expressions for the affinity [\q{affinityisotropicweyl2}] and director field [\q{Disotropicweyl}], we begin to evaluate the Hessian matrix:
\e{
\text{Hess}_{ab}=\nabla_a\nabla_b\f{|\be\cdot \bA_{cv\bk}|^2}{|\bD|} = -\nabla_a\nabla_b(\hbd\cdot \be)^2=2\nabla_a
 \f{(\be \cdot\hbd)^2(\hbd\cdot \bb)-(\hbd\cdot \be)(\be \cdot\bb)}{k}.
}
This is only nonvanishing  if $\nabla_a$ acts on $\hbd \cdot \bb$, because the Hessian matrix is to be evaluated at $\bk=\bk_0$, where $\be=\hbd$ and $\ba\cdot \hbd=\bb\cdot \hbd=0$.
\e{
\text{Hess}_{ab}\bigg|_{\bk_0}\eq \f{2}{k}\nabla_a
 (\hbd\cdot \bb)=\f{2}{k}\bb \cdot \f{\ba-\hbd(\hbd\cdot \ba)}{k} = 2 \f{\ba\cdot \bb}{k^2}.
}
This means the Hessian is a 2-by-2 diagonal matrix with each diagonal element equal to $2/k^2$, which must then also be equal to the square root of the determinant. This completes the proof of \q{toproveisotropicweyl}.

\section{Appendix to  `Ballistic photovoltaic current in disordered Dirac-Weyl materials'}\la{app:jballdiracweyl}

\subsection{Scaling analysis for `Impurity-dependent power laws of the ballistic photocurrent'}\la{app:scaling}

The `impurity-dependent power law' refers to how the low-frequency ballistic photocurrent ($\jball$) scales with the source-driven light frequency: 
 \e{
 \jball \propto \omega^{\text{exponent}} \la{exponent2}
 }
 for impurity-mediated, intra-pocket/valley  scattering. By scaling/dimensional analysis of our explicit expression for $\jball$ [Eq.~\eqref{eq:jball}], we will demonstrate that this \textit{frequency exponent} depends on  the type of Dirac-Weyl fermion (as classified by the $n$-order [\qq{Hwnapp}{Hdnapp}] and dimensionality $d$ of $\bk$-space), the type and orientation of disordering impurity, and the light polarization vector $\be$. This dependence is summarized in \tab{tab:powerlaw}, which is meant to be justified by this appendix.\\

Since impurity-mediated skew scattering occurs on the excitation surface, the relevant energy scale is the excitation energy $\Eex=\hbar\omega/2$. Diagonalizing  \qq{Hwnapp}{Hdnapp} gives us the energy-momentum dispersion of the $n$-Dirac/Weyl fermions as
\begin{align}
E=\begin{cases}
\sqrt{(f_n\kper)^{2n}+(vk_z)^2}, &\text{for } n\text{-Weyl}, \\
\sqrt{(f_n\kper)^{2n}}, &\text{for } n\text{-Dirac},
\end{cases}
\end{align}
where $\kper=\sqrt{k_x^2+k_y^2}$ and  all energies and momenta are measured from the Dirac–Weyl node.
With respect to energy scaling, $k_z$ has dimensions of $E$ while $k_{x,y}$ has dimensions of $E^{1/n}$:  
 \e{
 k_z=\tf{E}{v}\widetilde{k}_z \iand k_{x,y}=\tf{E^{1/n}}{f_n}\widetilde{k}_{x,y}, \la{kscaling}
 }
 with $\widetilde{k}_i$ being dimensionless. The corresponding momentum derivatives become
 \e{
 \nabla_{k_z} =  \tf{v}{E} \nabla_{\widetilde{k}_z}\iand \nabla_{k_{x,y}}=\tf{f_n}{E^{1/n}}\nabla_{\widetilde{k}_{x,y}}.\la{kderiv}
 }

The integrand in \q{eq:jball} (for $\jball$) consists of multiple factors (listed below), each of which contributes to the overall dimensional scaling with respect to the excitation energy $\Eex$ and thus the light-frequency $\omega$:
\begin{enumerate}
  
    \item \textbf{Group velocity}: $v_z=\nabla_{k_z}E_{\bk}$ does not scale with $\Eex$, and $v_x=\nabla_{k_x}E_{\bk}$ scales as $\Eex^{1-1/n}$, according to \q{kderiv}. Whether we apply $v_z$ or $v_x$ depends on whether we are interested in the photocurrent in the direction parallel ($v_z$) or perpendicular ($v_x$) to the rotation axis.\\

    \item \textbf{Energy-conserving delta function}: Each delta function $\delta(\Eex-E_{\bk})=\delta(1-E_{\bk}/\Eex)/\Eex$ scales as $1/\Eex$. There are three such delta functions (reflecting that third-order scattering is the minimal order for skew scattering), hence altogether $\Eex^{-3}$.\\
    
    \item \textbf{Momentum integration}: For 3D n-Weyl semimetals, $\sum_{\bk}\propto \int dk_z\,d\kper \,\kper d\theta$ scales as $\Eex^{1+2/n}$, while for 2D n-Dirac semimetals $\sum_{\bk}\propto \int d\kper \,\kper d\theta$ scales as $\Eex^{2/n}$, according to \q{kscaling}. We may combine both statements as $\Eex^{d-2+2/n}$, with $d=2$ (resp. 3) for Dirac (resp. Weyl). There are three such explicit momentum integrations for third-order scattering, giving a total scaling of $\Eex^{3d-6+6/n}$.\\
    
    \item The \textbf{N-order Bargmann invariant} is dimensionless for any $N$, because the eigenvector $\ket{u_{\bk}}$ is normalized as $\braket{u_{\bk}}{u_{\bk}}=1$ and is therefore dimensionless. The integrand of $\jball$ includes a factor of the third-order Bargmann invariant. \\

    \item The scaling of the \textbf{impurity scalar potential} $\tilde{V}_{ \bk\bk'}  \propto \Eex^{D_{\sma{V}}}$ depends on the impurity type/orientation and the strength of the screening.  Assuming that impurities are overscreened by carriers of the topological semimetal at nonzero chemical potential [\app{app:overscreening}], 
    \begin{align}
        &\text{Overscreened monopolar impurity:}\as \tilde{V}_{\bk}\reftwoeq{overscreenedmonopole3D}{overscreenedmonopole2D}\tilde{V}^{\text{m,2D/3D}}_{\bk} \propto \Eex^0 \imp \DV=0; \la{nv1}\\
        &\text{Overscreened dipolar impurity:} \as \tilde{V}_{\bk} \reftwoeq{overscreeneddipole3D}{overscreeneddipole2D} \tilde{V}^{\text{d,2D/3D}}_{\bk} \propto\begin{cases}
             k_z \imp \DV=1 &(\text{dipole}\;\parallel\; \text{to rotation axis}); \\
             \kper \imp \DV=\tf{1}{n} &(\text{dipole}\;\perp\; \text{to rotation axis}).
        \end{cases} \la{nv2}
    \end{align}
    There being three explicit powers of $\tilde{V}_{\bk}$ in the integral for the photocurrent, the combined scaling is $\Eex^{3D_{\sma{V}}}$. For the hybrid impurity potential $\tilde{V}_{\bk\bk'}=\tilde{V}_{\bk\bk'}^{\text{m}}+\tilde{V}_{\bk\bk'}^{\text{d}}$ that is the sum of a monopole and dipole potential, then the photocurrent is generally expressible as a sum of terms proportional to $(\tilde{V}_{\bk\bk'}^{\text{d}})^3$ or $\tilde{V}_{\bk\bk'}^{\text{d}}(\tilde{V}_{\bk\bk'}^{\text{m}})^2$, etc. [as in Eq.~\eqref{Wimageneric}], but only one of these terms (proportional to $\tilde{V}^{a}\tilde{V}^b\tilde{V}^c$) is dominant (at low frequency) and contributes a scaling of $\Eex^{3D_{\sma{V}}}$ with $3D_{\sma{V}}=D_{\sma{V^a}}+D_{\sma{V^b}}+D_{\sma{V^c}}$. \\

    \item The scaling of the \textbf{affinity deviant} depends on the orientation of the light polarization vector: 
    \begin{align}
         \delta|\be\cdot \bA_{cv\bk}|^2 \propto\begin{cases}
            \nabla_{k_z}^2 \propto \Eex^{-2} &(\be\;\parallel\; \text{to rotation axis}); \\
            \nabla_{k_{x,y}}^2 \propto \Eex^{-2/n} &(\be\;\perp\; \text{to rotation axis}),
        \end{cases}
    \end{align}
   bearing in mind $\ket{u_c}$ and $\ket{u_v}$ (in the definition of $\bA_{cv}$) are both dimensionless.\\

\item The scaling of the inverse \textbf{momentum-relaxation time} $1/\tau^{mr}_{E}$ depends on dimensionality of $\bk$-space and the nature of the impurity:
    \e{
    \tf1{\tau^{mr}_{\Eex}} \propto \Eex^{d-3+2/n+2\DVmr},
    }
which is derived by combining items 2-5 with the observation that the scaling of $1/\tau^{mr}_{E}$ is identical to the scaling of
\e{
\sum_{\bk'}\big| \tilde{V}_{\bk'\bk}\braket{u_{\bk'}}{u_{\bk}}\big|^2\delta(E_{\bk'}-\Eex)
}
according to \q{diffusion-skew-W-S-def}, \q{impalone} and \q{momentumrta}. $\DVmr$ is the energy dimension of the impurity potential predominantly responsible for momentum relaxation; if the impurity potential has a multipole expansion, we assume that it is the lowest-order monopole potential that predominates momentum relaxation.\footnote{This is a phenomenological assumption, i.e., a plausible hypothesis to be experimentally confirmed.} In particular, it is not necessary for $\DVmr$ to equal $D_{\sma{V}}$ from item 5. In particular, for the hybrid impurity potential which sums both monopolar and dipolar components, we assume momentum relaxation is governed by the monopolar component. 
The photocurrent is proportional to two powers of $\tau^{mr}_{\Eex}$ and hence also proportional to  $\Eex^{6-2d-4/n-4\DVmr}$.  \\
\end{enumerate}

Collecting all seven items and remembering that $\Eex=\hbar\omega/2$, 
\begin{align}
\label{j-ball-scaling-z}
&\jball^z \propto \omega^{Z}; \as Z=d-3+\tf{2}{n}+3\DV -4\DVmr +\begin{cases} -2, &\be\parallel \bz \\ -\tf{2}{n};& \be\perp \bz; \end{cases}\\
\label{j-ball-scaling-x}
&\jball^x \propto \omega^{X}; \as X=Z+1-\tf{1}{n},
\end{align}
with $3\DV$ and $4\DVmr$ being impurity-dependent rational numbers specified in items 5 and 7. In particular, $j^{z}_{\text{ballistic}}$ exists only for 3D n-Weyl semimetals, hence  we can simply  set $d=3$  in \q{j-ball-scaling-z}. By applying \qq{j-ball-scaling-z}{j-ball-scaling-x} to the various possibilities for $n$, $\be$ and impurity type, and bearing in mind only certain response tensor elements are symmetry-allowed [\q{sec:symmetry}], we derive Table~\ref{tab:powerlaw}.

\subsection{Generalized argument for the vortex-enhanced, defect-mediated ballistic photocurrent}\la{app:argument}

Let us elaborate and generalize our arguments for the vortex-intersection effect in $n$-Weyl semimetals [\q{vortexintersectioneffect}] and the vortex-proximity effect in $n$-Dirac materials [\q{vortexproximityeffect}]. Optical vortices may also exist (and likely exists generically) in non-Dirac/Weyl materials, but the $\bk$-locations of these vortex lines are no longer inferrable by topological arguments such as the Chern-vorticity theorem [\q{chernvortex}]. No matter the origin of the vortex, we argue that a vortex line intersecting (or in proximity) to the excitation surface positively correlates with the defect-mediated $\jball$.\\

Such $\jball$ arises from  skew electron-impurity scattering, which is trivial for any iso-energy-symmetric [\q{isoenergysymmetric}] electron distribution:
\e{
I^{im,a}_B([f])\refeq{impvanish} 0. \la{triviality}
}
The triviality condition holds to any order in the impurity potential,\cite{sturman_collisionintegral} follows from the optical theorem and the elasticity of impurity-mediated collisions, as we reviewed in \app{app:Iimp}. Crucially, all iso-energy-symmetric distributions  are $B$-symmetric  but not all $B$-symmetric distributions [\q{symmf2}] are iso-energy-symmetric. In particular, $I^{im,a}_B(f^s_B)$ is generically nonzero if $f^s_B$ is not iso-energy-symmetric.   To recapitulate,\footnote{\q{isoenergyasymmetry} does not imply that the phonon-mediated ballistic photocurrent vanishes if $f^s_B$ is iso-energy-symmetric. Because electron-phonon scattering is inelastic, a condition like \q{triviality} does not hold for $I^{pn,a}$.} 
\e{
\text{Necessary condition of iso-energy asymmetry:}\as \text{impurity-mediated}\;\; \mathbf{j}_{\text{ballistic}} \neq \bze  \imp f^s_B-[f_B]\neq 0.\la{isoenergyasymmetry}
}\\

\begin{figure}[h]
\centering
\includegraphics[width=11 cm]{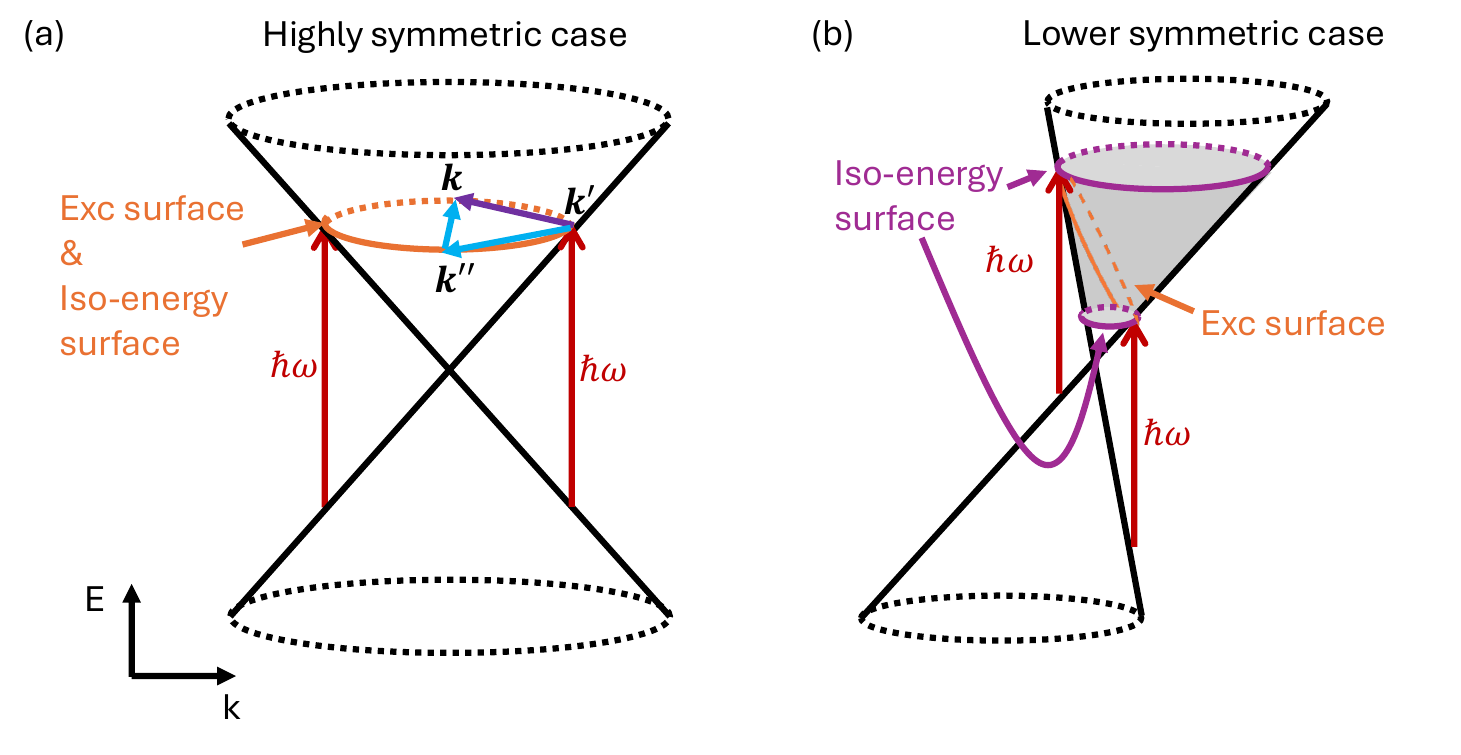}
\caption{Excitation and iso-energy surfaces in (a) highly symmetric and (b) lower symmetric cases. In (a), the purple arrow and blue arrows indicate the one-step and two-step Feynman path, respectively. In (b), the shaded cone segment represents the excitation energy interval for a specific photon energy $\hbar\omega$.}
\label{fig:exc&iso}
\end{figure}

The optical vortex is one way to generate a large iso-energy asymmetry for $f^s_B$. When an optical vortex line intersects the excitation surface [e.g., \fig{fig:multiweyl-aff-dev}], the optical affinity is anisotropic over the excitation surface; such anisotropy is also generated when a vortex line merely lies proximate to the excitation surface [e.g., \q{fig:vortexpromixity}].  In highly symmetric cases, the excitation surface coincides with an iso-energy surface, as illustrated in Fig.~\ref{fig:exc&iso}(a). With lower symmetry, the excitation surface covers a nonzero energy interval (henceforth referred to as the \textit{excitation energy interval}) which is typically small compared to energy band widths [Fig.~\ref{fig:exc&iso}(b)]. Whatever the symmetry, we reason that an affinity deviant [\q{affinitydeviant}] over the excitation surface correlates with an affinity deviant over all iso-energy surfaces with $E$ in the excitation energy interval. The latter anisotropy implies that the $B$-symmetric electron distribution is iso-energy-asymmetric ($f^s_B-[f_B]\neq 0$) for energies in the excitation energy interval, thus fulfilling a requirement [\q{isoenergyasymmetry}] for the \textit{steady}, impurity-mediated $\jball$. In the dynamical evolution of the \textit{transient} $\jball(t)$, the pre-photo-excitation Fermi-Dirac distribution evolves (at early times $t<100 fs$) into a $B$-symmetric (but iso-energy-asymmetric) distribution, owing to photo-excitation in the presence of vortices. At later times ($t > 100fs$),  the distribution further evolves to develop an $B$-asymmetric component $\delta f^a_B$, owing to impurity-mediated skew scattering. This `$\delta f^a_B$' is proportional to the source-driven light intensity and activates $\jball$ via \q{app-defineballistic}. 

\subsection{Appendix to `Vortex-intersection effect in $n$-Weyl semimetals'}\la{app:vortexintersectionnweyl}

In this section, we detail the calculations of the ballistic photocurrent in $n$-Weyl semimetals whose Hamiltonian is given in Eq.~\eqref{multiWeyl-H}. We focus on the case of an unperturbed $n$-Weyl fermions, i.e., we ignore the effects of tilt, warping, inter-valley scattering, etc. \\

The ballistic photocurrent is defined in Eq.~\eqref{eq:jball} and is determined by three nontrivial parts: the Bargmann invariant $\scrb{\bk \bk'' \bk'}$, the affinity deviant $\delta|\be\cdot\cA_{cv,\bk}|^2$, and the combination of three impurity potentials $\tilde{V}_{a,\mathbf{k}^{\prime}\mathbf{k}} \tilde{V}_{b,\mathbf{k}\mathbf{k}^{\prime\prime}} \tilde{V}_{c,\mathbf{k}^{\prime\prime}\mathbf{k}^{\prime}}$. In what follows, we consider each of these components separately.\\

Using the polar angle representation [$\bk=(k_x,k_y,k_z)=(k\cos\phi,k\sin\phi, k_z)$], the Bargmann invariant \eqref{Bargmann} reads as
\begin{eqnarray}
\label{nweyl-all-B}
\scrb_{\bm k' \bk \bk''} &\equiv& \braket{u_{\bk'}}{u_{\bk}} \braket{u_{\bk}}{u_{\bk''}} \braket{u_{\bk''}}{u_{\bk'}} 
=\frac{1}{8E_{\bk}^3 (E_{\bk}+vk_z')(E_{\bk}+vk_z)(E_{\bk}+vk_z'')}
\Bigg\{ (E_{\bk}+vk_z')^2(E_{\bk}+vk_z'')^2(E_{\bk}+vk_z)^2 \nonumber\\
&+& f_n^6(k_{\perp}'k_{\perp}''k_{\perp})^2 +f_n^2 \Big[
(E_{\bk}+vk_z')(E_{\bk}+vk_z'') \left[f_n^2 (k_{\perp}'k_{\perp}''(k_{\perp})^2)^n +(k_{\perp}'k_{\perp}'')^n(E_{\bk}+vk_z)^2 \right] \cos{[n(\phi' -\phi'')]} \nonumber\\
&+&(E_{\bk}+vk_z')(E_{\bk}+vk_z) \left[f_n^2 (k_{\perp}'k_{\perp}(k_{\perp}'')^2)^n +(k_{\perp}'k_{\perp})^n(E_{\bk}+vk_z'')^2 \right] \cos{[n(\phi' -\phi)]} \nonumber\\
&+&(E_{\bk}+vk_z'')(E_{\bk}+vk_z) \left[f_n^2 (k_{\perp}''k_{\perp}(k_{\perp}')^2)^n +(k_{\perp}''k_{\perp})^n(E_{\bk}+vk_z')^2 \right] \cos{[n(\phi'' -\phi)]} 
\Big]
\Bigg\} \nonumber\\
&+&i \frac{f_n^2}{8E_{\bk}^3} \Bigg\{
\frac{(k_{\perp}'k_{\perp}'')^2(E_{\bk} +vk_z)^2 -f_n^2\left(k_{\perp}'k_{\perp}''(k_{\perp})^2\right)^n}{E_{\bk} +v k_z} \sin{[n(\phi' -\phi'')]} \nonumber\\
&+&\frac{(k_{\perp}''k_{\perp})^2(E_{\bk} +vk_z')^2 -f_n^2\left(k_{\perp}''k_{\perp}(k_{\perp}')^2\right)^n}{E_{\bk} +v k_z'} \sin{[n(\phi'' -\phi)]}
+\frac{(k_{\perp}'k_{\perp})^2(E_{\bk} +vk_z'')^2 -f_n^2\left(k_{\perp}'k_{\perp}(k_{\perp}'')^2\right)^n}{E_{\bk} +v k_z''} \sin{[n(\phi -\phi')]}
\Bigg\}, \nonumber\\
\end{eqnarray}
bearing in mind that all energies are equal ($E_{\bk}=E_{\bk'}=E_{\bk''}$) as enforced by the $\delta$-functions in Eq.~\eqref{eq:jball}.\\

The affinity deviant given in Eq.~\eqref{affinitydeviant} is,  
\begin{equation}
\label{multiweyl-current-dev}
\delta|\be\cdot\cA_{cv,\bk}|^2 = -\frac{v^2}{2(n+2)} \frac{|e_z|^2}{E_{\bk}^2} -\frac{n f_n^{2/{n}} \Gamma{\left(\frac{1}{2} +\frac{1}{n}\right)}}{3\sqrt{\pi} \Gamma{\left(1 +\frac{1}{n}\right)}} \frac{|e_x|^2 +|e_y|^2}{E_{\bk}^{2/n}}
+\frac{f_n^2 k_{\perp}^{2({n}-1)}}{4E_{\bk}^4} 
\left|e_z v k_{\perp} -nvk_z (\bm{e}_{\perp}\cdot\hat{\bk}) +iE_{\bk} [\bm{e}\times\hat{\bk}]_z\right|^2.
\end{equation}
It is plotted over the excitation surface in Fig.~\ref{fig:multiweyl-aff-dev} for $n=1,2,3$ and two characteristic directions of the light polarization vector, $\be \parallel \hat{\bm{z}}$ and $\be \perp \hat{\bm{z}}$. The largest values of the affinity deviant agree with the positions of optical vortices, see Sec.~\ref{sec:isotropicweyl}.\\

The DOS for an $n$-Weyl semimetal needed for the isoenergy average \eqref{eq:isoEave} in the affinity deviant reads
\begin{eqnarray}
\label{multiweyl-current-nu}
\nu(\epsilon) &=& \int \frac{d^3k}{(2\pi)^3} \df{\epsilon - E_{\bk}} 
= \frac{1}{2\pi^2} \int k_{\perp} dk_{\perp} \frac{|\epsilon|}{v \sqrt{\epsilon^2 - f_n^2 k_{\perp}^{2n}}} \nonumber\\
&\stackrel{u=f_n^2 k_{\perp}^{2n}/\epsilon^2}{=}& \frac{1}{(2\pi)^2} \frac{1}{n v} \left(\frac{|\epsilon|}{f_n}\right)^{2/{n}} \int_0^{1} du\, u^{1/n -1} \left(1 -u\right)^{-1/2}
=\frac{1}{(2\pi)^2 n v} \left(\frac{|\epsilon|}{f_n}\right)^{2/{n}} B\left(\frac{1}{n}, \frac{1}{2}\right),
\end{eqnarray}
where $B\left(x,y\right) = \Gamma(x) \Gamma(y)/\Gamma(x+y) = \int_0^{1}du\,u^{x-1} (1-u)^{y-1}$ is the beta function.\\

Before moving to the discussion of the ballistic photocurrent, let us consider the impurity potential part of the current. In our calculations, we account for monopolar and dipolar components of the impurity potential and use the overscreened impurity potentials defined in Eq.~\eqref{overscreenedmonopole3D}. 
As we discuss in detail in Sec.~\ref{app:symconstraintssecondkind}, are only two combinations of the impurity potential components that could potentially contribute to the ballistic photocurrent: 
$\sim \tilde{V}_{m} \tilde{V}_{d}^2$ and $\sim \tilde{V}_{d}^3$. 
In the case of the purely monopolar part, $\sim \tilde{V}_{m}^3$, according to Eq.~\eqref{eq:jball}, the current is determined by the imaginary part of the Bargmann invariant $\sim \hat{\mathbf{d}}_{\bk}\cdot \hat{\mathbf{d}}_{\bk''}\times \hat{\mathbf{d}}_{\bk'}$ and vanishes after integrating over the momentum $\int d\bk''$. A combination of monopolar and dipolar terms $\sim \tilde{V}_{m}^2\tilde{V}_{d}$ vanishes after summing over all possible permutations of indices $a$, $b$, and $c$, i.e., $\sum_{a,b,c=\left\{m,m,d\right\}}\tilde{V}_{a,\mathbf{k}^{\prime}\mathbf{k}} \tilde{V}_{b,\mathbf{k}\mathbf{k}^{\prime\prime}} \tilde{V}_{c,\mathbf{k}^{\prime\prime}\mathbf{k}^{\prime}} =0$. This result is a property of an overscreened dipolar potential, see Eq.~\eqref{dSigmal}.

By using the obtained expressions for the Bargmann invariant, affinity deviant, and impurity potential, let us now consider the ballistic photocurrent. The photocurrent depends on the relative orientation of the impurity dipole $\bm{\mathscr{d}}$, light polarization $\bm{e}$, and the $C_{nz}$ rotation axis $//z$. The natural orientation of the impurity dipole $\bm{\mathscr{d}}$ is along the $z$-axis. Different orientations of the dipoles may be possible, but require a special engineering of the impurity potential.

As we discussed above, the ballistic photocurrent contains two contributions determined by $\sim V_mV_d^2$ and $\sim V_d^3$ terms. The ballistic photocurrent due to the scattering off the hybrid potential $\sim V_mV_d^2$ is nonzero for $n=1$ and $n=2$ Weyl semimetals,
\begin{equation}
\label{multiweyl-mon-dip-current-j-all}
B_{abc; md^2}^{\rm de}= -2_{\text{p-h}} |e|^3 \frac{n_{\rm imp} (\tau^{mr}_{E_{\rm ex}})^2}{\hbar^2 (2\pi)^{6}} \left(\frac{e \mathscr{d}}{\varepsilon k_{\rm scr}^2}\right)^{2}\left(\frac{e^2}{\varepsilon k_{\rm scr}^2}\right) \frac{E_{\rm ex}^{6/n}}{n^2v^2f_n^{6/n}} \widetilde{B}_{abc; md^2}^{(n)}(\bm{\mathscr{d}}).
\end{equation}

Nonzero components of the tensor $\widetilde{B}^{(n)}_{abc; md^2}(\bm{\mathscr{d}})$ are lisetd in Tab.~\ref{tab:multiweyl-mon-dip-current-j-all}.
%

\begin{table}[ht]
	
\centering
		
\begin{tabular} {|c|c|c|} \hline
			
Dipole direction  & $n$ & Response  \\  \hline \hline 

Dipole $\boldsymbol{\mathscr{d}}\parallel \vec{\bx}$ & $n=1$ &  $\widetilde{B}_{yxz; md^2}^{(1)} = \widetilde{B}_{yzx; md^2}^{(1)} = -\widetilde{B}_{zxy; md^2}^{(1)} = -\widetilde{B}_{zyx; md^2}^{(1)} =\dfrac{4\pi^3}{135}$ \\   \cline{2-3}
& $n=2$ &  $\widetilde{B}_{zxy; md^2}^{(1)} = \widetilde{B}_{zyx; md^2}^{(2)} =\dfrac{3\pi^6}{1024}$  \\  \cline{1-3}
Dipole $\boldsymbol{\mathscr{d}}\parallel \vec{\bz}$ & $n=1$ &  $\widetilde{B}_{xyz; md^2}^{(1)} = \widetilde{B}_{xzy; md^2}^{(1)} = -\widetilde{B}_{yxz; md^2}^{(1)} = -\widetilde{B}_{yzx; md^2}^{(1)} =\dfrac{2\pi^3}{135}$  \\    \hline
\end{tabular}	
\caption{
The explicit dependence of the nontrivial coefficients $\widetilde{B}_{abc}$ in the ballistic current given in Eq.~\eqref{multiweyl-mon-dip-current-j-all} for the hybrid potential $\sim V^{m,3D}(V^{,3D})^2$.
} 
\label{tab:multiweyl-mon-dip-current-j-all}
\end{table}

\subsection{Appendix to `Vortex-proximity effect in $n$-Dirac semimetals'}\la{app:vortexpromityndirac}

As in the case of $n$-Weyl semimetals, we focus on the case of untilted and unwarped $n$-Dirac fermions. The Hamiltonian of $n$-Dirac semimetals is similar to that of $n$-Weyl semimetals, see Eq.~\eqref{multiWeyl-H}, albeit with $vk_z\to \Delta$, where $\Delta$ is a gap. 

Let us present the necessary nontrivial ingredients in the ballistic photocurrent \eqref{eq:jball}. Using the polar angle representation [$\bk=(k_x,k_y)={k(\cos\phi,\sin\phi)}$], the Bargmann invariant [\q{Bargmannrealimag}] reads as
\begin{eqnarray}
\label{nDirac-B}
\scrb_{\bk' \bk \bk''} &=& \frac{f_{n}^2 k^{2n}}{4 E_{\bk}^2} \left\{1 +\frac{4\Delta^2}{f_{n}^2 k^{2n}} + \cos{[n(\phi' - \phi)]} +\cos{[n(\phi - \phi'')]} +\cos{[n(\phi'' - \phi')]}\right\} \nonumber\\
&+&i\frac{f_{n}^2 k^{2n} \Delta}{4 E_{\bk}^3} \left\{\sin{[n(\phi' -\phi)]}  +\sin{[n(\phi -\phi'')]} +\sin{[n(\phi'' -\phi')]} \right\},
\end{eqnarray}
where all energies are equal due to the $\delta$-functions in Eq.~\eqref{eq:jball}.

The optical affinity deviant \eqref{affinitydeviant} is
\begin{equation}
\label{nDirac-affinity-deviant}
\delta|\be\cdot\bA_{cv,\bk}|^2 =  n^2 \frac{f_{n}^4 k^{2(2n-1)}}{4 E_{\bk}^4} \left\{ \frac{|\left[\be\times\bk\right]_z|^2}{k^2} -\frac{1}{2} \right\}
\end{equation}
with $|E_{\bk}| \geq |\Delta|$, and $\be$ being the light source's polarization vector. Assuming $\Delta>0$, the affinity deviant at the excitation surface reads $\delta|\be\cdot\bA_{cv,\bk}|^2 \sim k_{\rm ex}^{2(2n-1)} \sim \left(E_{\rm ex}^2 - \Delta^2\right)^{2 -1/n}$ with $k_{\rm ex} = \left[(E_{\rm ex}^2 - \Delta^2)/f_{n}^2\right]^{1/(2n)}$. Note that affinity deviant $\delta|\be\cdot\bA_{cv,\bk}|^2$ growth with $n$ as $\propto n^2$. As one can see from Eq.~\eqref{nDirac-affinity-deviant}, regardless of the value of $n$, the optical affinity deviant is smaller away from the vortex line, i.e., at $\Delta>0$. 

The ballistic photocurrent in a gapped $n$-Dirac semimetal strongly depends on the composition of the impurity potential. The only nontrivial contribution in the case of an overscreened potential comes from $\sim V_d^3$ terms. The reason for the vanishing contributions from the $\sim V_m^2V_d$ and $\sim V_m^3$ terms is the same as for $n$-Weyl semimetals, see Sec.~\ref{sec:nweylvortexintersection}. The vanishing contribution from the $\sim V_mV_d^2$ terms has the symmetry origin, see the detailed discussion in Sec.~\ref{app:symconstraintssecondkind}.

\subsection{Appendix to `Tilt-activated $\jball$ for symmetry-unpinned Weyl fermions'}
\label{app:tilted Weyl}

\noindent\textit{Isotropic Weyl fermion with tilting. --}
Let us first derive the conductivity shown in Eq.~\eqref{eq:conduc_tiltedWeyl}. The tilting term breaks the SO(3) but does not change the wavefunction and thus the optical affinity.  Using Eq.~\eqref{eq:aff_2band}, the optical affinity of the tilted Weyl can be explicitly calculated as 
\begin{equation}
|\hat{e}\cdot \mathbf{A}_{cv,\mathbf{q}}|^2= \frac{1}{4q^2}(1-|\hat{e}\cdot \hat{q}|^2+i(\hat{e}\times\hat{\overline{e}})\cdot \hat{q}),
\end{equation}
where $\hat{q}\equiv\bq /|\bq|$ and $
\hat{e}$ is the unit modular light polarization.
With the tilting, the excitation surface is an excitation surface, thus the ballistic photocurrent is no longer proportional to $\delta|\hat{e}\cdot\mathbf{A}_{cv,\bk}|^2$. We need to start from Eq.~\eqref{eq:ballisticcurrent2} to derive the following formula for our calculations:

\begin{equation}
\begin{aligned}
\mathbf{j}_{\text{ballistic}}&=-\frac{8\pi^3|e|^3}{\hbar^2}\tau_{\text{eff}}^2 n_{\text{imp}}|\mathbf{E}|^2\sum_{\mathbf{q},\mathbf{q}^{\prime},\mathbf{q}^{\prime\prime}} \mathbf{v}_{c\mathbf{q}}(|\hat{e}\cdot \mathbf{A}_{cv,\mathbf{q}^{\prime}}|^2\delta(
2\hbar vq'-\hbar\omega)-[|\hat{e}\cdot \mathbf{A}_{cv,\mathbf{q}^{\prime}}|^2\delta(
2\hbar vq'-\hbar\omega)])
\\
& \ \ \ \ \times\operatorname{Im}(V_{\mathbf{q}^{\prime}\mathbf{q}}V_{\mathbf{q}\mathbf{q}^{\prime\prime}}V_{\mathbf{q}^{\prime\prime}\mathbf{q}^{\prime}})\delta(E_{c\mathbf{q}}-E_{c\mathbf{q}^{\prime\prime}})\delta(E_{c\mathbf{q}}-E_{c\mathbf{q}^{\prime}})+c\rightarrow v,
\end{aligned}
\end{equation}
where the $\delta(2\hbar v q'-\hbar\omega)$ can no longer be taken out of the iso-energy average as in other cases.

Let us focus only on the contribution of the electrons in the conduction band, and the calculations of the contribution of the holes in the valence band are similar. The term containing the optical affinity (instead of its isoenergy average) in the ballistic current can be rewritten as
\begin{equation}
\begin{aligned}
&\frac{1}{(2\pi)^6}\frac{|e|^3}{\hbar^2}\tau_{\text{eff}}^2 \left(\frac{e^2}{\epsilon}\right)^3n_{\text{imp}}\frac{1}{k_{scr}^6}|\mathbf{E}|^2\int d\mathbf{q}d\mathbf{q}'d\mathbf{q}''\big(v\sin\theta\cos\phi \hat{x}+v\sin\theta\sin\phi \hat{y}+(v\cos\theta+v_{zt})\hat{z}\big)
\\
&\times\frac{1}{4q'^2}(1-|\hat{e}\cdot \hat{q}'|^2+i(\hat{e}\times\hat{\overline{e}})\cdot \hat{q}')\operatorname{Im}(\braket{u_{c\mathbf{q}'}}{u_{c\mathbf{q}}}\braket{u_{c\mathbf{q}}}{u_{c\mathbf{q}''}}\braket{u_{c\mathbf{q}''}}{u_{c\mathbf{q}'}})\delta(2\hbar v q'-\hbar\omega)\delta(E_{c\mathbf{q}}-E_{c\mathbf{q}''})\delta(E_{c\mathbf{q}}-E_{c\mathbf{q}'}).
\end{aligned}
\end{equation}
Note that if we define $\alpha =v_{zt}/v$, the Dirac-delta funciton can be expressed as
\begin{equation}
 \delta(E_{c\mathbf{q}}-E_{c\mathbf{q}'})=\frac{1}{\hbar v}\delta(q+\alpha q\cos\theta-q'-\alpha q'\cos\theta')=\frac{1}{\hbar v}\frac{1}{1+\alpha\cos\theta}\delta(q-q'\frac{1+\alpha\cos\theta'}{1+\alpha\cos\theta}).
\end{equation}
Given that 
\begin{equation}
\operatorname{Im}(\braket{u_{\mathbf{q}'}}{u_{\mathbf{q}}}\braket{u_{\mathbf{q}}}{u_{\mathbf{q}''}}\braket{u_{\mathbf{q}''}}{u_{\mathbf{q}'}})=-\frac{1}{4}\hat{\mathbf{d}}_{\mathbf{q}''}\cdot \hat{\mathbf{d}}_{\mathbf{q}'}\times \hat{\mathbf{d}}_{\mathbf{q}} = -\frac{1}{4}\hat{q}''\cdot \hat{q}'\times \hat{q}
\end{equation}
we can then integrate over $q$, $q'$, and $q''$ to integrate out the Dirac delta functions: 
\begin{equation}
\label{eq:angleintegral}
\begin{aligned}
&\int  d\Omega d\Omega' d\Omega''  v\big(\sin\theta\cos\phi \hat{x}+\sin\theta\sin\phi \hat{y}+(\cos\theta+\alpha)\hat{z}\big)\frac{1}{4q'^2}(1-|\hat{e}\cdot \hat{q}|^2+i(\hat{e}\times\hat{\overline{e}})\cdot \hat{q})
\\
&\times\left(-\frac{1}{4}\cos\theta ''\sin\theta'\sin\theta(\cos\phi'\sin\phi-\sin\phi'\cos\phi)\right)
\\
&\times\int q^2 q''^2dq dq' dq''\delta(2\hbar v q'-\hbar\omega)\delta(E_{c\mathbf{q}}-E_{c\mathbf{q}''})\delta(E_{c\mathbf{q}}-E_{c\mathbf{q}'})
\\
&=\int  d\Omega d\Omega' d\Omega'' v\big( \sin\theta\cos\phi \hat{x}+\sin\theta\sin\phi \hat{y}+(\cos\theta+\alpha)\hat{z}\big)\frac{1}{4q'^2}(1-|\hat{e}\cdot \hat{q}|^2+i(\hat{e}\times\hat{\overline{e}})\cdot \hat{q})
\\
&\times\left(-\frac{1}{4}\cos\theta ''\sin\theta'\sin\theta(\cos\phi'\sin\phi-\sin\phi'\cos\phi)\right)
\\
&\times\frac{1}{2(\hbar v)^3}\frac{1}{1+\alpha\cos\theta}\frac{1}{1+\alpha\cos\theta''}\left(\frac{\omega}{2v}\frac{1+\alpha\cos\theta'}{1+\alpha\cos\theta}\right)^2\left(\frac{\omega}{2v}\frac{1+\alpha\cos\theta'}{1+\alpha\cos\theta''}\right)^2 
\\
&=-\frac{1}{32}\left(\frac{\omega}{2v}\right)^{2}\frac{1}{\hbar(\hbar v)^2}\int  d\Omega d\Omega' d\Omega'' \big(\sin\theta\cos\phi \hat{x}+\sin\theta\sin\phi \hat{y}+(\cos\theta+\alpha)\hat{z}\big)(1-|\hat{e}\cdot \hat{q}|^2+i(\hat{e}\times\hat{\overline{e}})\cdot \hat{q})
\\
&\times\cos\theta ''\sin\theta'\sin\theta(\cos\phi'\sin\phi-\sin\phi'\cos\phi)\frac{(1+\alpha\cos\theta')^4}{(1+\alpha\cos\theta)^3(1+\alpha\cos\theta'')^3}.
\end{aligned}
\end{equation}
Note that in the first step of the above equation, we only keep $\hat{q}''_z(\hat{q}'\times\hat{q})_z$ for the imaginary part of the Bargamann invariant, because the Bargamann invariant is the only term in the integrand that depends on $\phi''$ and $\int d\phi '' \hat{q}''_{x/y} =0 $ \footnote{This is also true for n-Weyl for $n>1$, where we should replace $\hat{q}''_{x/y}$ by $(\hat{d}_{\mathbf{q}''})_{x/y}$. }. Moreover, $(1-|\hat{e}\cdot\hat{q}'|^2)$ can be explicitly evaluated as 
\begin{equation}
\begin{aligned}
1-|\hat{e}\cdot\hat{q}'|^2
={}&
1
-|e_x|^2\sin^2\theta'\cos^2\phi'
-|e_y|^2\sin^2\theta'\sin^2\phi'
-|e_z|^2\cos^2\theta'
\\
&-2\,\real\!\left(e_x \overline{e}_y\right)\sin^2\theta'\cos\phi'\sin\phi'
\\
&-2\,\real\!\left(e_x \overline{e}_z\right)\sin\theta'\cos\theta'\cos\phi'
\\
&-2\,\real\!\left(e_y \overline{e}_z\right)\sin\theta'\cos\theta'\sin\phi'. 
\end{aligned}
\end{equation}
Note that besides $1-\left|\mathbf e\cdot \hat{\mathbf q}'\right|^2$, $\phi$ and $\phi'$ only appears in $(\cos\phi'\sin\phi - \sin\phi'\cos\phi)$ and in the velocity. If we look at $j_z$, then $\int d\phi d\phi' (1-\left|\mathbf e\cdot \hat{\mathbf q}'\right|^2)(\cos\phi'\sin\phi - \sin\phi'\cos\phi) =0 $. If we look at $j_x$, only the term $-2\,\real\!\left(e_y \overline{e}_z\right)\sin\theta'\cos\theta'\sin\phi'$ in $(1-\left|\mathbf e\cdot \hat{\mathbf q}'\right|^2)$  and $-2\imag(e_z\overline{e}_x)\sin\theta'\sin\phi'$ in $i(\hat{e}\times \hat{\overline{e}})\cdot \hat{q}'$ will contribute, and thus

\begin{equation}
\begin{aligned}
    j_x= & \frac{1}{(2\pi)^6}\frac{|e|^3}{\hbar^2}\tau_{\text{eff}}^2 \left(\frac{e^2}{\epsilon}\right)^3n_{\text{imp}}\frac{1}{k_{scr}^6}|\mathbf{E}|^2\times\frac{1}{32}\left(\frac{\omega}{2v_F}\right)^{2}\frac{1}{\hbar(\hbar v)^2}2\pi^3 
    \\
    & \times \bigg\{2\,\real\!\left(e_y e_z^*\right)\int d\theta d\theta'd\theta'' \frac{\sin^3\theta}{(1+\alpha\cos\theta)^3}\sin^3\theta'\cos\theta'(1+\alpha\cos\theta')^4\frac{\sin\theta''\cos\theta''}{(1+\alpha\cos\theta'')^3}
    \\
    &+2\,\imag\!\left(e_z e_x^*\right)\int d\theta d\theta'd\theta'' \frac{\sin^3\theta}{(1+\alpha\cos\theta)^3}\sin^3\theta'(1+\alpha\cos\theta')^4\frac{\sin\theta''\cos\theta''}{(1+\alpha\cos\theta'')^3}\bigg\}
    \\
    =& \frac{1}{(2\pi)^3}\frac{|e|^3}{\hbar^2}\tau_{\text{eff}}^2 \left(\frac{e^2/\epsilon}{\hbar v}\right)^3\frac{n_{\text{imp}}}{k_{scr}^6}\left(\frac{\omega}{2v_F}\right)^{2}v\big(f_1(\alpha)\,\real\!\left(e_y \overline{e}_z\right)-f_2(\alpha)\,\imag\!\left(e_z \overline{e}_x\right)\big),
\end{aligned}
\end{equation}
where $f^{l}(\alpha) = \frac{1}{105}\frac{(\text{arctanh}(\alpha)-\alpha/(1-\alpha^2))(7+3\alpha^2)}{\alpha(1-\alpha^2)^2}$ and $f^{c}(\alpha) = \frac{35+3a^2(14+a^2)}{105}\frac{(\alpha/(1-\alpha^2)-\text{arctanh}(\alpha))}{4\alpha^2(1-\alpha^2)^2}$. 

Similarly, if we look at $j_y$,  only $-2\,\real\!\left(e_x \overline{e}_z\right)\sin\theta'\cos\theta'\cos\phi'$ in $1-|\hat{e}\cdot\hat{q}'|^2$ and $2\imag(e_z\overline{e}_y)\sin\theta'\cos\phi'$ in $i(\hat{e}\times \hat{\overline{e}})\cdot \hat{q}'$  will contribute and we get
\begin{equation}
j_y = -\frac{1}{(2\pi)^3}\frac{|e|^3}{\hbar^2}\tau_{\text{eff}}^2 \left(\frac{e^2/\epsilon}{\hbar v}\right)^3\frac{n_{\text{imp}}}{k_{scr}^6}\left(\frac{\omega}{2v_F}\right)^{2}v\big(f_1(\alpha)\,\real\!\left(e_x \overline{e}_z\right)+f_2(\alpha)\,\imag\!\left(e_z \overline{e}_y\right)\big),
\end{equation}
which matches the symmetry analysis.

Evaluation of the second term is more tricky, but fortunately it is always zero. To see this,  we need to first calculate $[|\hat{e}\cdot \mathbf{A}_{cv,\mathbf{q}^{\prime}}|^2\delta(
2\hbar v_Fq'-\hbar\omega)]$
\begin{equation}
\begin{aligned}
&=\frac{\int d\mathbf{k}|\hat{e}\cdot \mathbf{A}_{cv,\mathbf{k}}|^2\delta(
2\hbar v_Fk-\hbar\omega)\delta(E_{c\mathbf{k}}-E_{c\mathbf{q}'})}{\int d\mathbf{k}\delta(E_{c\mathbf{k}}-E_{c\mathbf{q}'})}
\end{aligned}
\end{equation}
The numerator can be calculated through
\begin{equation}
\begin{aligned}
&\int d\mathbf{k}|\hat{e}\cdot \mathbf{A}_{cv,\mathbf{k}}|^2\delta(
2\hbar v_Fk-\hbar\omega)\delta(E_{c\mathbf{k}}-E_{c\mathbf{q}'})
\\
&=\int \sin\theta d\theta dk \frac{2\pi}{4}\left(1-|e_z|^2\cos^2\theta-\frac{|e_x|^2+|e_y|^2}{2}\sin^2\theta-2\imag(e_x\overline{e}_y)\cos\theta\right)\delta(
2\hbar v_Fk-\hbar\omega)\delta(E_{c\mathbf{k}}-E_{c\mathbf{q}'})
\\
&=-\frac{2\pi}{4}\frac{1}{\hbar v_{F}}\frac{1}{2\hbar v_{F}}\int d\cos\theta\left(1-|e_z|^2\cos^2\theta-\frac{|e_x|^2+|e_y|^2}{2}\sin^2\theta-2\imag(e_x\overline{e}_y)\right)\delta(\frac{\omega}{2v_{F}}(1+\alpha\cos\theta)-q'({1+\alpha\cos\theta')})
\end{aligned}
\end{equation}
 which is $\phi'$ independent. Thus, the integral over $\phi$ and $\phi'$ will always lead to zero due to the presence of $(\sin\phi\cos\phi'-\sin\phi'\cos\phi)$.
 \\

\noindent\textit{Anisotropic Weyl with tilting.--}
 Now, let us consider a more general case where
 \begin{equation}
     H = \hbar (f_1 (k_x\sigma_x+k_y\sigma_y)+vk_z\sigma_z)
 \end{equation}
 with $f_1\neq v$.  For simplicity we can define
 \begin{equation}
(\tilde{k}_x,\tilde{k}_y,\tilde{k}_z) = (f_1 k_x, f_1 k_y, v k_z)
 \end{equation}
 With this new coordinate, we go back to the isotropic case, where tilting along any direction should be equivalent. After this rescale, the only thing changed in the current is a factor.
\\

\noindent\textit{n-Weyl fermions with tilting.--}
The Hamiltonian of a n-Weyl fermion is given by 
\begin{equation}
    H=\hbar \mathbf{d}\cdot \boldsymbol{\sigma}, \    \mathbf{d} = (f_n \real(k_+^n), f_n \imag (k_+^n), vk_z).
\end{equation}

Following the procedure discussed above,  let us first consider the tilting along $z$-direction. We look at the term proportional to affinity because the term proportional to the average of the affinity is again $\phi'$ independent and will vanish. Similarly, the imaginary part of the Bargamann invariant only contributes 
$$-\frac{1}{4} \hat{d}_{\mathbf{q}'',z}(\hat{d}_{\mathbf{q}'}\times \hat{d}_{\mathbf{q}})_z =-\frac{1}{4}\frac{vq_z''}{d_{\mathbf{q}}d_{\mathbf{q}'}d_{\mathbf{q}''}}\rho ^n\rho'^n(\cos n\phi'\sin n\phi -\sin n\phi'\cos n\phi). $$
However, remember that in all other terms of the integrand, only the velocity term contributes harmonics of $\phi$ and only the optical affinity term contributes harmonics of $\phi'$. Since the velocity term only contributes $\cos\phi$ and $\sin\phi$, the integral over $\phi$ will always vanish because the $\phi$ from Bargamann invariant arises as higher harmonics when $n>1$. Thus, for tilting along z, the current vanishes. 

Next, let us consider tilting along x, the only different thing happens in the Dirac-delta function -- It now contributes the $\cos\phi$ and $\cos \phi'$. The integral can then be expressed as
\begin{equation}
\begin{aligned}
&\int d\phi d\phi' d\phi'' (\cos n\phi'\sin n\phi -\sin n\phi'\cos n\phi) \int \rho d\rho dq_z\rho'd\rho' d q_z' \rho'' d\rho '' dq_z'' \mathbf{v}_{c} \left(-\frac{1}{4}\frac{vq_z''}{d_{\mathbf{q}}d_{\mathbf{q}'}d_{\mathbf{q}''}}\rho ^n\rho'^n\right)|\hat{e}\cdot \mathbf{A}_{cv,\mathbf{q}'}|^2 
\\
&\times\delta(2\hbar d_{\mathbf{q}'}-\hbar\omega)\delta(\hbar (d_{\mathbf{q}}+v_{xt}\rho\cos\phi)-\hbar (d_{\mathbf{q}''}+v_{xt}\rho''\cos\phi''))\delta(\hbar (d_{\mathbf{q}}+v_{xt}\rho\cos\phi)-\hbar (d_{\mathbf{q}'}+v_{xt}\rho'\cos\phi'))
\end{aligned}
\end{equation}
After integrating out the Dirac- delta function, all $\cos \phi$ and $\cos \phi''$ arise from the process can be replaced by a function of $\cos \phi',\rho, \rho',\rho'', k_z,k_z', k_z''$. As a result, the integral over $\phi$ will always be zero.  Thus, no matter whether for x-tilted or z-tilted n-Weyl fermions, the ballistic photocurrent is zero.

\subsection{Appendix to `Warp-activated $\jball$ for massive $n$-Dirac fermions'}
\la{app:current-ndirac-warped}

In this appendix, we consider the details of the calculations for the warped $n$-Dirac Hamiltonian relevant for $n>1$-layer rhombohedral graphene~\cite{Koshino-McCann-TrigonalWarpingBerrys-2009},
\begin{eqnarray}
H_n &=& \bm{d}\cdot \bm{\sigma} = 
\begin{pmatrix}
\Delta & f_{n}  (k_-)^n 
\\
f_{n}(k_+)^n 
& -\Delta
\end{pmatrix}
+ H^{6',n}_{\bk}, \\
k_{\pm}&=&k_x\pm ik_y, \quad \Delta= m_Dv^2,
\label{13-multiWeyl-nDirac-w-H}
\end{eqnarray}
see also Eq.~\eqref{Hdn-warp}.
The trigonal warping here originates from the skewed interlayer coupling and the next-nearest interlayer coupling.

In what follows, we focus on the effects of warping and assume the simplest monopolar impurity potential. In addition, we work in the low-frequency regime where the warping effect on the energy-momentum dispersion is small. Thus we retain only the terms linear in the warping strength. In this case, we expand $E_{\bk} \approx E_{\bk}^{(0)} + E_{\bk}^{(1)}$ where $E_{\bk}^{(1)}$ quantifies the effects of the warping term.

The imaginary part of the Bargmann invariant for a generic two-band system is
\begin{eqnarray}
\label{13-multiWeyl-nDirac-w-B}
\operatorname{Im} \scrb_{\bk' \bk \bk''} =  -\frac{\left[\hat{\mathbf{d}}_{\bk'}\times \hat{\mathbf{d}}_{\bk}\right]\cdot\hat{\mathbf{d}}_{\bk''}}{4}. 
\end{eqnarray}
For monopolar impurities, the integral over $\bk'$ reduces to
\begin{eqnarray}
\label{13-multiWeyl-nDirac-w-B-int}
&&\int \frac{d^2k''}{(2\pi)^2} \mbox{Im}{\scrb_{\bk' \bk \bk''}} \df{E_{\rm ex} -E_{\bk''}} = - \frac{\left[\hat{\mathbf{d}}_{\bk'}\times \hat{\mathbf{d}}_{\bk}\right]_z}{4}  \int \frac{d^2k'}{(2\pi)^2} \frac{d_{3,\bk''}}{d_{\bk''}} \df{E_{\rm ex} -E_{\bk''}^{(0)} -E_{\bk''}^{(1)}} \nonumber\\
&&\approx - \frac{\left[\hat{\mathbf{d}}_{\bk'}\times \hat{\mathbf{d}}_{\bk}\right]_z}{4}  \int \frac{d^2k'}{(2\pi)^2} \frac{d_{3,\bk''}}{E_{\bk''}^{(0)}} \left[1 + \frac{E_{\bk''}^{(1)}}{E_{\bk''}^{(0)}} \right] \left[\df{E_{\rm ex} -E_{\bk''}^{(0)}} -E_{\bk''}^{(1)} \frac{d}{d E_{\rm ex}} \df{E_{\rm ex} -E_{\bk''}^{(0)}}\right] \nonumber\\
&&\approx - \frac{\left[\mathbf{d}_{\bk'}\times \mathbf{d}_{\bk}\right]_z}{4 E_{\rm ex}^2} \frac{\Delta}{E_{\rm ex}} \nu(E_{\rm ex})
\Bigg\{
1 + \int_{0}^{2\pi} \frac{d\varphi''}{2\pi} \frac{E_{\bk''}^{(1)}}{E_{\rm ex}}\Big|_{k'' \to k_{\rm ex}} -\frac{1}{2\pi \nu(E_{\rm ex})} \frac{1}{|d E_{\bk''}^{(0)}/dk''|} \left(\frac{d}{d k''}k'' \frac{1}{d E_{\bk''}^{(0)}/dk''} \int_{0}^{2\pi} \frac{d\varphi''}{2\pi} E_{\bk''}^{(1)} \right) \Big|_{k'' \to k_{\rm ex}}
\Bigg\}\nonumber\\
\end{eqnarray}
with $\nu(E_{\rm ex}) = E_{\rm ex}/(2n\pi f_n^2 k_{\rm ex}^{(2n-2)})$ being the DOS and the excitation vector $k_{\rm ex} = \left[(E_{\rm ex}^2 -\Delta^2)/f_{n}^2\right]^{1/(2n)}$ calculated without warping. 
For the $n$-Dirac Hamiltonian, we have $d E_{\bk''}^{(0)}/dk''\Big|_{k'' \to k_{\rm ex}} = \frac{n f_{n}^2 k_{\rm ex}^{2n-1}}{E_{\rm ex}}$.

In calculating the warping effects in other terms, we use
\begin{eqnarray}
\label{13-multiWeyl-nDirac-w-delta-int}
&&\int\frac{d^2k'}{(2\pi)^2} F_{\bk \bk'} \df{E_{\bk} -E_{\bk'}}
\approx \int\frac{d^2k'}{(2\pi)^2} F_{\bk \bk'} \left[\df{E_{\bk} -E_{\bk'}^{(0)}} + \left(-E_{\bk'}^{(1)}\right) \frac{d}{d E_{\bk}}\df{E_{\bk} -E_{\bk'}^{(0)}}\right] \nonumber\\
&&=\int\frac{d^2k'}{(2\pi)^2} F_{\bk \bk'} \df{E_{\bk} -E_{\bk'}^{(0)}} 
+\frac{1}{2\pi} \int_0^{\infty} dk' \left[\frac{d}{dk'} \frac{d k'}{d E_{\bk'}^{(0)}} k'\int_0^{2\pi}\frac{d\varphi''}{2\pi} F_{\bk \bk'} \left(-E_{\bk'}^{(1)}\right) \df{E_{\bk} -E_{\bk'}^{(0)}}\right] \nonumber\\
&&=\int\frac{d^2k'}{(2\pi)^2} F_{\bk \bk'}\df{E_{\bk} -E_{\bk'}^{(0)}}
+\frac{1}{2\pi} \left[\frac{d}{dk'} \frac{d k'}{d E_{\bk'}^{(0)}} k'\int_0^{2\pi}\frac{d\varphi''}{2\pi} F_{\bk \bk'} \left(-E_{\bk'}^{(1)}\right)\right] \frac{1}{|dE_{\bk'}^{(0)}/dk'|} \Big|_{k' \to k_{\rm ex}}.
\end{eqnarray}

We use the above expressions in Eq.~\eqref{eq:jball}, which, assuming monopolar disorded $\widetilde{V}_{\bk \bk'} =V_0$ reads as 
\begin{equation}
\label{13-multiWeyl-nDirac-w-j-def}
\begin{aligned}
\mathbf{j}_{\text{ballistic}}&=-2_{\text{e-h}}\frac{4\pi^3|e|^3}{\hbar^2}(\tau^{mr}_{E_{ex}})^2 n_{\text{imp}}|\mathbf{E}|^2\sum_{\mathbf{k},\mathbf{k}^{\prime},\mathbf{k}^{\prime\prime}} \mathbf{v}_{\mathbf{k}}\;\operatorname{Im}\scrb_{\bk' \bk \bk''} \delta(E_{\mathbf{k},\mathbf{k}''})\,\delta(E_{\mathbf{k},\mathbf{k}'})   \; \delta |\be\cdot \mathbf{A}_{cv,\mathbf{k}^{\prime}}|^2 \,\delta(E_{\mathbf{k}'}-E_{ex}).
\end{aligned}
\end{equation}

A particularly simple expression for the ballistic current response tensor is obtained for the following warping term $H^{6',n}_{\bk} = h_{1,n-2,0} k_{-}^{n-2}k_{+} \sigma_{+} +h.c.$,
\begin{eqnarray}
\label{13-multiWeyl-nDirac-w-J}
B^{\text{de}}_{abc} = 2_{\rm p-h} |e|^3 (\tau^{mr}_{E_{\rm ex}})^2 \frac{n_{\rm imp}  V_0^3}{2\pi} 
\left[\frac{E_{\rm ex}^2 -\Delta^2}{f_n^2}\right]^{-2+2/n} \frac{E_{\rm ex}^2}{n^2f_n^2}
\frac{\pi h_{1,n-2,0} \Delta}{16f_n} \widetilde{B}_{abc}^{(n)},
\end{eqnarray}
where the nontrivial coefficients $\widetilde{B}_{abc}^{(n)}$ are listed in Tab.~\ref{tab:13-multiWeyl-nDirac-w-J}.



\begin{table}[ht]
	
\centering
		
\begin{tabular} {|c|c|c|} \hline
			
$n$ &  Response  \\  \hline \hline 
$n=1$ & 
$\widetilde{B}_{xxy}^{(1)} = \widetilde{B}_{xyx}^{(1)} =\widetilde{B}_{yxx}^{(1)} = -\widetilde{B}_{yyy}^{(1)} =\dfrac{3}{8}$ \\  \hline
$n=2$ & 
$\widetilde{B}_{xxy}^{(2)} = \widetilde{B}_{xyx}^{(2)} =\widetilde{B}_{yxx}^{(2)} = -\widetilde{B}_{yyy}^{(2)} =1$ \\  \hline
\end{tabular}	
\caption{
The nontrivial coefficients $\widetilde{B}_{abc}^{(n)}$ in the ballistic current response tensor given in Eq.~\eqref{13-multiWeyl-nDirac-w-J} for the warped $n$-Dirac case. 
} 
\label{tab:13-multiWeyl-nDirac-w-J}
\end{table}

One can understand the vanishing current at $n\neq1$ and $n\neq2$ as the result of the mismatch between angular harmonics in the velocity, Bargmann invariant, and affinity deviant. Indeed, we have the following harmonics
\begin{eqnarray}
\mathbf{v}_{\bk} &\propto& \sum_{l_1=0,\pm1, \ldots} \left(\epsilon_{l_1}'(k)\, \hat{\mathbf{k}}+ \frac{3l_1}{k} \epsilon_{l_1}(k)\, \hat{\mathbf{\varphi}}\right) e^{3il_1 \varphi}
= \sum_{l_1=0,\pm1, \ldots} \sum_{\pm} \left(A_{l,\pm} \hat{\mathbf{x}} + B_{l, \pm} \hat{\mathbf{y}}\right) e^{i(\pm 1 +3l_1) \varphi},\\
\delta |\mathcal{A}_{cv, \bk'}|^2 &\propto& \sum_{l_2=0,\pm1, \ldots} a_{l_2} e^{i(\pm 2 +3l_2) \varphi'},\\
\int d\varphi'' \scrb_{\bk' \bk \bk'} &\propto& \sum_{l_3=0,\pm1, \ldots} b_{l_3} e^{i(\pm n+3l_3) (\varphi' -\varphi)},
\end{eqnarray}
where the first expression in the first equation contains the radial and azimuthal components of the velocity~\footnote{To derive the angular expansion of the velocity, we used the symmetry property of the energy $\epsilon(k,\varphi)=\epsilon(k,\varphi +2\pi/3)$ and expanded $\epsilon(k,\varphi) = \sum_{l_1=0,\pm1, \ldots} \epsilon_{l_1}(k) e^{3il_1 \varphi}$. }. After integrating over $\varphi'$ and $\varphi$, one obtains that all values of $n$, except $n=3l$ are allowed. 

\section{Appendix to `Symmetry theorems for bulk photovoltaic response tensors'}\la{app:symmetrytheorems}

The main results of this appendix are derivations of the symmetry theorems for bulk photovoltaic response tensors in \s{sec:symmetry} [specifically the general symmetry theorems in \s{sec:generalsymmetryconstraints}  and the three Hamiltonian-specific theorems in \s{sec:Hspecificsymmetry}], as well as their applications to Dirac-Weyl fermions in \tab{tab:nweyldiracfermions}, \ref{tab:nweyldiracsymmetryreducedgeneral} and \ref{tab:nweyldiracconstraintspecific}. \\

Before getting into the meat of the aforementioned derivations, a substantial digression is needed to fundamentally establish:

\noindent (\app{app:responsetensorsymmetric}) What it means for a response tensor to be point-group-symmetric (with emphasis on the bulk photovoltaic tensor),

\noindent (\app{app:algorithm}) How to view a response tensor as a calculational algorithm that inputs the one-electron distribution, Hamiltonian and collisional integral,

\noindent (\app{app:gtransformcollisionalintegral}) How the aforementioned collisional integral transforms under point-group symmetry,

\noindent (\app{app:compatibility}) What is a symmetry compatibility index of a response tensor.\\

Then we are finally ready to

\noindent (\app{app:generalconstraints})  Derive the general symmetry  theorems on photovoltaic tensors,

\noindent (\app{sec:equivariance}) Apply the newly-derived theorems to  
to  determine the symmetry-allowed photovoltaic response tensor elements, by viewing a $G$-symmetric response tensor as a $G$-equivariant linear map and applying Schur's lemma,

\noindent (\app{app:specificsymconstraints}) Derive the Hamiltonian-specific symmetry theorems on photovoltaic tensors,

\noindent (\app{app:symconstraintsnweyl}) Apply all the preceding theorems to determine the symmetry-allowed photovoltaic tensor elements of $n$-Weyl fermions,

\noindent (\app{app:symconstraintsndirac}) Apply all the preceding theorems to determine the symmetry-allowed photovoltaic tensor elements  of $n$-Dirac fermions.

\subsection{What it means for a response tensor to be symmetric}\la{app:responsetensorsymmetric}

While experimenting on a crystal, if a mischievous devil were to momentarily cover your eyes and to displace/rotate/reflect your spacetime rulers {relative} to the crystal, you wouldn't be able to tell the difference when your eyes are open. The devil can move the rulers and fix the crystal (intervention I), or move the crystal and fix the rulers (intervention II). If, before the devil's intervention, an electric field $\cale$ in the $y$ direction induced a current density $\bj$ in the $x$ direction, then you would go on believing that $\cale_y$  induces $j_x$, though either your notion of direction has changed (I) or your crystallographic axes have changed (II). \\

Symmetry can be formulated equivalently with either intervention, modulo some changes in interpretation. This Appendix chooses to formulate symmetry based on interventions of the second kind, which we refer to as \textit{active transformations}. What exactly is an active transformation? Let $s^e$ denote a reference closed system and  $s^g$ denote a $g$-active-transformed closed system; $e$ is the identity transformation, so the reference system can be viewed as being trivially transformed. In the bulk photovoltaic context,  a `closed system' is the sum of  subsystems: (i) the electrons (ii) a driving agent/apparatus that produces static/dynamic electric fields, and  (iii)  the environment (e.g., impurities,  
bath of phonons and/or photons) which plays the role of a source/sink of energy or  momentum  or both.\\

We are interested in observables which are fields over spacetime, which we parametrize by the coordinate $c=(\br,t)$. For instance, let $\rho^{g'}$ mean the $g'$-active-transformed charge density associated to $s^{g'}$. $\rho$ exemplifies a
\e{
\text{Time-neutral scalar field:}\as  \rho^g(c)= \rho^e(g^{\mo}\circ c); \as c\equiv(\br,t), \la{tneutralscalar}
}
with the action $g\circ$ defined in \q{gspacetime}. Other types of fields transform differently:
\e{
&\text{Time-oriented scalar field:}\as \partial_t\rho^{g}(c)= i_g\partial_t\rho^{e}(g^{\mo}\circ c); \la{torientedscalar}\\
&\text{Time-neutral vector field}:\as \cale^g(c) =g^s\cdot \cale^e(g^{\mo}\circ c);\la{tneutralvector}\\
&\text{Time-oriented vector field}:\as \bj^g(c) =i_g g^s\cdot \bj^e(g^{\mo}\circ c);\la{torientedvector}\\
&\text{Time-oriented pseudovector field}:\as \text{magnetic field}\as \bB^g(c) =i_g \det\{g^s\} g^s\cdot \bB^e(g^{\mo}\circ c).\la{torientedpvector}
}

All the above fields can be Fourier transformed as
\e{
A(\br t)=\sum_{\bq\omega}A_{\bq\omega}e^{i\bq\cdot\br-i\omega t} \in \R \iff A_{\bq\omega}=\overline{A_{-\bq,-\omega}}. 
}
 In particular, the active transformations in \qq{tneutralvector}{torientedvector}  translate in Fourier space to:
\e{
&\cale^g_{\bq\omega}=g^s\cdot \cale^e_{g^{s,\mo}\cdot \bq,i_g\omega}=g^s\cdot \overline{\cale^e_{g^{\mo}\cdot \bq,\omega}}^g; \la{gcaleFourier}\\
&\bj^g_{\bq\omega}=g\cdot \bj^e_{g^{s,\mo}\cdot \bq,i_g\omega}= g\cdot \overline{\bj^e_{g^{\mo}\cdot \bq,\omega}}^g; \as g\cdot =i_gg^s\cdot, \la{gjFourier}
}
with   $\overline{A}^g$ meaning to apply the complex conjugation on $A$ iff $i_g=-1$.\\

Consider the direct current density ($\bj_{\bq=0,\omega=0}$) response (of a 
 homogeneous medium) to  a dynamic electric field $\cale_{\bq\omega}$ (associated to a light wave) and a static homogeneous electric field $\cale^s$:\cite{sturmanfridkin_book}
\e{
j_{\bze,0,a} =\sigma^d_{ab}\cale^s_b + \sigma^q_{abc}\cale^s_b\cale^s_c + \Theta_{abcd}\cale^s_b \cale_{\bq\omega,c}\overline{\cale_{\bq\omega,d}} + \Gamma_{abc}\cale_{\bq\omega,b}\overline{\cale_{\bq\omega,c}}+\ldots \la{expanddirectcurrent}
}
$\{\sigma^{d/q},\Theta,\Gamma\}$ are examples of response tensors. Assuming that the homogeneous medium has a crystalline lattice period which is small compared to the light wavelength, we further expand
\e{
\Gamma_{ijk}(\bq\omega)\eq \Gamma^{(0)}_{ijk}(\omega) + q_l \Gamma^{(1)}_{iljk}(\omega) + O(q^2),\la{qexpand}
}
with $\Gamma^{(0)}$ ($\equiv$ the \textit{bulk photovoltaic tensor}) derived  by setting $\bq=\bze$ in $\Gamma^{\alpha}_{ijk}(\bq\omega)$ in the dipole approximation. We use $\Sigma$ for the bulk photovoltaic tensor and separate it into \textit{linear} and \textit{circular} components,  which are respectively symmetric and antisymmetric under interchange of the last two spatial indices: 
\e{
\text{Bulk photovoltaic tensor}\as  \Sigma_{abc} = \Gamma^{(0)}_{abc}(\omega)= \Sigma_{abc}^l+\Sigma^c_{abc}; \as  \Sigma^l_{abc}=\Sigma^l_{acb}; \as \Sigma^c_{abc}=-\Sigma^c_{acb}. \la{bPVtensordefine2}
}
Altogether, \qq{expanddirectcurrent}{bPVtensordefine2} give a precise meaning to the shorter definition of the bulk photovoltaic tensor in \q{bPVtensordefine}.\\

Suppose the Devil were to   actively $g$-transform the system $(\rho^e\ri \rho^g, \cale^e \ri \cale^g, \bj^e \ri \bj^g)$, while preserving our spacetime rulers. If $g$ were a symmetry of the closed system, then we (with our short-term memory) would not be able to tell that the Devil has played a trick on us. This means we will go on to report the same response tensors while making current measurements on $s^g$ as we had for current measurements of $s^e$:
\e{
\text{Devil's axiom:}\as \text{$g$-symmetric closed system}\imp  R^g=R^e \as \text{for any response tensor}\;R.\la{RgRe}
} 
This is what it means for a response tensor to be symmetric.

\subsection{Response tensors as algorithms}\la{app:algorithm}

All response tensors are calculated by some type of perturbation theory. The product of this perturbation theory is an algorithm whose final output is $R$. We use the same letter to denote the  algorithm: 
\e{
R=R(f^0,H,I); \as I=I(f^0,H).\la{Ralgorithm}
}
The inputs to the algorithm are:\\

\noi{a} The one-electron distribution function $f^0_{B}$ in the zeroth order of the perturbation theory.\\

\noi{b} The Hamiltonian $H=H^0+V$. $H^0$ is the electron quasiparticle/mean-field Hamiltonian in the zeroth order of the perturbation; $H^0=\cheH^0$ with $\cheH^0$ previously defined in \app{app:crystallineHam}, but we will drop the accent for simplicity in the present appendix. $V=V_{\text{drive}}+V_{\text{env}}$ is  the sum of perturbations due to the drive and the environment. \\

\noi{c} One-electron transition rates (induced by $V$) which can be expressed in terms of a collisional integral $I_{B}(f^0,H)$, as exemplified in \app{app:excitationcollision} and \app{app:Iimp}. The collisional integral is the convective time derivative of the electron distribution, i.e., the rate of change of the electron distribution function $f_{B}$ in the frame of the moving electron. \\

We assume $H^0$ is solvable in the reference system:
\e{
(H^{0,e}-E_B^e)\ket{B,e}=0; \as 
}
and whatever $g$-transformed system
\e{
(H^{0,g}-E_B^g)\ket{B,g}=0; \as H^{0,g}= \hg H^{0,e} \hg^{\mo}.
}
($\ket{B,e}$ here should be understood as $\ket{\psi_{B,e}}$ in \app{app:crystallineHam}; once again, we are simplifying notation.)
In general, we should not expect $\ket{B,e}=\ket{B,g}$; instead,
\e{
\text{General transformation of eigenstates:}\as \hg\ket{B,e} = \ket{gB,g};\as  E^e_{B}=E^g_{gB},
}
with $gB$ meaning the g-transformation of the quantum number $B$, e.g., for crystalline $H^0$, $B=(b\bk)$ and $gB=(b,g\cdot \bk)$, in accordance with \q{expposition}. We define $f^g_{B}$  as the one-electron distribution function in system $s^g$; and $f^{0,g}_{B}$ as the zeroth-order approximant of $f^g_{B}$. If $B$ is occupied in $s^e$, then $gB$ is occupied in $s^g$: 
\e{
\text{General transformation of electron distribution:}\as f^{g}_{gB}=f^e_B.\la{fgfe}
}
If for $s^e$ the occupation is increasing with time (in the moving frame of electron $B$) due to collisions, then for $s^g$ (and time-inverting $g$) the occupation is decreasing in time (in the frame of electron $gB$):
\e{
\text{General transformation of collisional integral:}\as I_{gB}^g = i_g I_B^e; \as i_g=\pm 1.\la{IgIe}
}

For $H$ to be an input in \q{Ralgorithm} means that part of the algorithm (for calculating $R$) involves solving the eigenproblem of $H^0$, and that $R$ depends on the zeroth-order energy eigenvalues and eigenstates; another part of the algorithm is to incorporate $V$ perturbatively. (See \app{app:Iimalgorithm} for examples.) $R=R(f^0,H,I)$ means that  $R$ may depend on $f^0$ and $H$ in a manner extraneous to $R$'s dependence on $I(f^0,H)$.\\

If a response is  understandable as a small deformation of a Fermi-Dirac distribution of electrons which existed in the undriven equilibrium  state, hence the zeroth-order one-electron distribution has a Fermi-Dirac form: $f^0=f^{\text{FD}}$ [\q{fermidirac}].\footnote{
The intrinsic linear Hall conductivity  of a metal exemplifies a response tensor that depends on $f^0=f^{\text{FD}}$, depends on $H$ (through needing to compute the Berry curvature $\Omega_{\bk}$) but not $I$: $\sigma^d_{xy}(f^0,H) \propto \sum_{\bk} f^0_{\bk} \Omega_{xy\bk}$.
}  
In some cases it is useful to have the zeroth-order $f^0$ resemble more closely the non-equilibrium steady-state distribution, obtained by solving a Boltzmann transport equation.\cite{belinicher_kinetictheory,zhuAA_anomalousshift} Whether or not $f^0=f^{\text{FD}}$, we assume in discussions of symmetry constraints that 
\e{
\text{$g$-symmetric zeroth-order distribution:} \as f^{0,g}=f^{0,e} \as \text{is shorthand for} \as f^{0}_{gB}= f^{0}_{B}. \la{gsymf0}
}
This assumes a one-electron level with wavevector $g\cdot \bk$ exists if a one-electron level exists with wavevector $\bk$, as is guaranteed by: 
\e{
\text{$g$-symmetric zeroth-order Hamiltonian:} \as H^{0,g}=H^{0,e}. \la{gsymH}
}

The response tensor $R^g$ for a g-transformed closed system is given by
\e{
R^g=R(f^{0,g},H^g,I^g),\la{Rg} 
}
meaning that the calculational algorithm $R$ is unchanged; what changes is the g-transformed input; this reflects that the same physical laws apply to $s^e$ and $s^g$. \\

How does one calculate $I^g$? We  assume $I$ is a linear functional  of the one-electron distribution and symmetry decomposable as
\e{
I=I^++I^-; \as I^{\nu}_B=\sum_{B'}w^{\nu}_{BB'}f_{B'}; \as w^{\nu}_{BB'}=\nu w^{\nu}_{B'B}; \as \nu=\pm 1.\la{Ilinear2}
}
If the algorithm $I_B^{\nu}(f^{0,e},H^e)$ calculates $I_B^{\nu}$ in system $s^e$:
\e{
(I_B^{\nu})^e=I_B^{\nu}(f^{0,e},H^e)
}
then $I_B^{\nu}$ in system $s^g$ is calculated by
\e{
 (I_B^{\nu})^{g}=(-\nu)^g \,I_B^{\nu}(f^{0,g},H^{g}); \as (-\nu)^g=-\nu \as\text{if $i_g=-1$, otherwise} \as (-\nu)^g=1.
}
as will be proven below in \app{app:gtransformcollisionalintegral}.\\

For the closed system to be $g$-symmetric, we require as an operational definition that 
\e{
&\text{g-symmetric electron distribution:} \as  f^{g}_{B}=f_{B}^{e}, \iand\lin
&\text{g-symmetric  Hamiltonian:} \as  H^{0,g}=H^{0,e} \iand  V^{\text{im},g}=V^{\text{im},e} \imp H^g=H^e, \iand \lin
&\text{g-symmetric collisional integral:} \as  I^g_{B}=I_{B}^e. \la{systemgsym}
}
$f$ is the exact one-electron distribution in the driven state, and $f-f^0$ can be calculated given $H$ and $I$.  The last two conditions in \q{systemgsym}, plus our assumption of $g$-symmetric $f^{0}$,  jointly imply $R^g=R^e$, thus satisfying the Devil's axiom [\q{RgRe}]. Because $f^0$ is independent of $g$-transformation, we will often simplify notation by omitting the $f^0$-dependence in algorithms:
\e{
R^g=R(H^g,I^g), \as  (I_B^{\nu})^{g}=(-\nu)^g \,I_B^{\nu}(H^{g}). \la{RgIg}
}

\subsection{$g$-transformation of collisional integrals} \la{app:gtransformcollisionalintegral}

We largely consider collisional integrals that are linear functionals of the one-electron distribution:
\e{
I_B=\sum_{B'}w_{BB'}f_{B'}; \as w_{BB'}=\as\text{collisional operator}. \la{Ilinear}
}
We calculate $w=w(H)$ by an algorithm (also denoted $w$) that inputs the energies and eigenstates of $H^0$, as well as  the perturbation $V$ that induces the collisions. The matrix elements of $w(H^e)$ and $w(H^g)$ are related  by
\e{
w_{BB'}(H^g)=\{w_{g^{\mo}B,g^{\mo}B'}(H^{e})\}^{t_g},\la{wBBp}
}
with  $A^{t_g}$ meaning to transpose $A$ if $g$ inverts time.
\q{wBBp} includes the case of $H^T=H^e\imp w_{BB'}=w_{TB',TB}$, which is the reciprocity theorem.  By splitting this collisional operator into components which are even and odd under interchange of $B\leftrightarrow B'$ [\q{Ilinear2}],
\q{wBBp} simplifies to
\begin{align}
w^{\nu}_{BB'}(H^{g})=
    \nu^g w^{\nu}_{g^{\mo}B,g^{\mo}B'}(H^e); \as \nu^g=\nu \as\text{if $i_g=-1$, otherwise} \as \nu^g=1. \la{wBBp2} 
\end{align}

If the algorithm $w_{BB'}^{\nu}(H^e)$ calculates $w_{BB'}^{\nu}$ in system $s^e$, then
\e{
\text{Transformation of $w$-algorithms:} \as (w_{BB'}^{\nu})^e=w_{BB'}^{\nu}(H^e) \imp (w_{BB'}^{\nu})^{g}=(-\nu)^g w_{BB'}^{\nu}(H^{g}) \refeq{wBBp2} i_g w^{\nu}_{g^{\mo}B,g^{\mo}B'}(H^e). \la{wBBgalgorithm}
}
How does the above factor of $(-\nu)^g$ arise? This is necessitated by the general transformation law for the collisional integral, applied to each (a)symmetric component:
\e{
\sum_{B'}(w^{\nu}_{gBgB'})^gf_{B'}^e\refeq{fgfe}\sum_{B'}(w^{\nu}_{gBgB'})^gf_{gB'}^g=\sum_{B'}(w^{\nu}_{gBB'})^gf_{B'}^g\refeq{Ilinear} I^{\nu,g}_{gB} \refeq{IgIe} i_g I^{\nu,e}_B \refeq{Ilinear} i_g\sum_{B'}(w^{\nu}_{BB'})^ef^e_{B'},
}
which implies
\begin{align}
(w^{\nu}_{gBgB'})^g = i_g(w^{\nu}_{BB'})^e = i_g w_{BB'}^{\nu}(H^e) \refeq{wBBp2} 
    (-\nu)^g w^{\nu}_{gB,gB'}(H^{g}).\la{wnug}
\end{align}
It follows that if the algorithm $I_B^{\nu}(f^{0,e},H^e)$ calculates $I_B^{\nu}$ in system $s^e$:
\e{
\text{Transformation of $I$-algorithms:}\as I_B^{\nu,e}=I_B^{\nu}(f^{0,e},H^e) \imp I_B^{\nu,g}=(-\nu)^g I_B^{\nu}(f^{0,g},H^{g}).  \la{gtransformIalgorithm} 
}
We exemplify these algorithms with the impurity-mediated  and photo-excitation collisional integrals in \app{app:Iimalgorithm} and \app{app:Iexalgorithm}, respectively.

\subsubsection{Impurity-mediated collisional integral as an algorithm}\la{app:Iimalgorithm}


The impurity-mediated collisional integral $I^{im}$ was introduced in \app{app:Iimp} and hereby symmetry-decomposed as
\e{
I^{im}=I^{im,+}+I^{im,-}; \as I^{im,\nu}_B= \sum_{B'}w^{im,\nu}_{B,B'}f_{B'}; \as w^{im,\nu}_{B,B'}=\nu w^{im,\nu}_{B',B}.
}
A direct comparison with \qq{Wimps}{Wimpa} allows to identify 
\e{
&w^{im,+}_{B,B'}= W^{im,s}_{BB'}-\delta_{BB'}\sum_{B''}W^{im,s}_{BB''}; \as W^{im,s}_{BB'}=W^{im,s}_{B'B}= \tf{2\pi}{\hbar}N_{\text{imp}} |\Vim_{BB'}|^2\delta(E_{BB'});\la{wsimp}\\
&w^{im,-}_{BB'}= \tf{(2\pi)^2}{\hbar}N_{\text{imp}} \im \sum_{B''}\Vim_{BB''}\Vim_{B''B'}\Vim_{B'B}\delta(E_{BB'})\delta(E_{BB''}).\la{waimp}
}
with $\Vim_{BB'}$ being the impurity potential matrix element. Observed in system $s^g$,
$V^{\text{im},g}_{BB'}$ is calculated by an algorithm  that inputs $H^g$:
    \e{
    V^{\text{im},g}_{BB'}=    \Vim_{BB'}(H^g)=\braopket{B,g}{V^{\text{im},g}}{B',g}.
    }
How does $V^{\text{im},g}_{BB'}$ relate to $V^{\text{im},e}_{BB'}= \braopket{B,e}{V^{\text{im},e}}{B',e}$? Depending on whether $\hg$ is antiunitary, we transpose the matrix element:
    \begin{align}
    V^{\text{im},g}_{BB'}=\braopket{\hg (g^{\mo}B,e)}{ \hg V^{\text{im},e}\hg^{\mo}\hg}{g^{\mo}B',e} = \{V^{\text{im},e}_{g^{\mo}B,g^{\mo}B'}\}^{t_g}=\begin{cases}
        V^{\text{im},e}_{g^{\mo}B,g^{\mo}B'},&i_g=1;\\
        V^{\text{im},e}_{g^{\mo}B',g^{\mo}B},&i_g=-1.
    \end{cases}\la{VBBpg}
            \end{align}    
            $A^{t_g}=A^t$ (transpose of $A$)  if $g$ inverts time, and otherwise $A^{t_g}=A$.\\

The general identity in \q{wBBp2} reduces in the present context to
\e{
w^{\text{im},+}_{BB'}(H^g)\eq  w^{\text{im},+}_{g^{\mo}B,g^{\mo}B'}(H^e); \la{wsBB}\\
w^{\text{im},-}_{BB'}(H^g)\eq  i_g w^{\text{im},-}_{g^{\mo}B,g^{\mo}B'}(H^e). \la{wimasBB} 
}
\q{wsBB} is derivable from  \q{wsimp}, \q{VBBpg} and
\e{
 |\{V^{\text{im},e}_{g^{\mo}B,g^{\mo}B'}\}^{t_g}|^2\delta(E_{BB'}^g)= |V^{\text{im},e}_{g^{\mo}B,g^{\mo}B'}|^2\delta(E_{g^{\mo}B,g^{\mo}B'}^e). \la{WsBB}
 }
 \q{wimasBB} is derivable from  \q{waimp}, \q{VBBpg} and
 \e{
 &\im \sum_{B''}\{V^{\text{im},e}_{g^{\mo}B,g^{\mo}B''}\}^{t_g}\{V^{\text{im},e}_{g^{\mo}B'',g^{\mo}B'}\}^{t_g}\{V^{\text{im},e}_{g^{\mo}B',g^{\mo}B}\}^{t_g}\delta(E^e_{g^{\mo}B,g^{\mo}B'})\delta(E^e_{g^{\mo}B,g^{\mo}B''})\lin
  \eq \im \sum_{B''}\overline{V^{\text{im},e}_{g^{\mo}B,B''}V^{\text{im},e}_{B'',g^{\mo}B'}V^{\text{im},e}_{g^{\mo}B',g^{\mo}B}}^g\delta(E^e_{g^{\mo}B,g^{\mo}B'})\delta(E^e_{g^{\mo}B,B''}).\la{essentially}
}
$\overline{A}^g=\overline{A}$  if $g$ inverts time and otherwise $\overline{A}^g={A}$; this arises in \q{essentially} because interchanging $B$ and $B'$ in the third-order Bargmann invariant is equivalent to taking the complex conjugate.

\subsubsection{Photo-excitation collisional integral as an algorithm} \la{app:Iexalgorithm}

The photo-excitation collisional integral was introduced in \app{app:excitationcollision}.
Because $W^{ex}_{BB'}=W^{ex}_{B'B}$ [\q{Iexgeneral}], we identify 
\e{
I^{ex,+}_B = I^{ex,s}_B \eq \sum_{B'} w^{ex,+}_{BB'}f_{B'}; \as w^{ex,+}_{BB'}=w^{ex,+}_{B'B}= W^{ex}_{BB'}-\delta_{BB'}\sum_{B''}W^{ex}_{BB''}.
}
In accordance with the general transformation law for collisional integrals [\q{IgIe}], if system  $s^e$ is transformed to $s^{Tg^s}$, a photo-excitation rate at $B$ [with absorption of incoming $(\bq,\omega)$ photons]  becomes a photo-relaxation rate at $gB$ [with emission of outgoing $(g\cdot \bq,\omega)$ photons]:
\e{
I^{ex,+,g}_{gB}=i_g I^{ex,+,e}_{B},\la{Iexplusg}
}
bearing in mind that the left-hand algorithm inputs $V^{\text{ex},g}$ [\q{dipolecoupling}] with a $g$-transformed electric field [\q{gcaleFourier}].\\

Viewing the photo-excitation collisional operator that inputs the Hamiltonian,
\e{
 w^{ex,+}_{BB'}(H^g)=  w^{ex,+}_{g^{\mo}B,g^{\mo}B'}(H^e),\la{thesame}
}
in accordance with the general identity in \q{wBBp2}. \q{thesame} is derivable from
\q{VBBpg} (which applies not just to $\Vim$ but also to $\Vex$), $\dekkp= \delta_{\gmo \bk,\gmo \bkp}=\delta_{\gmo \bk',\gmo \bk}$, and 
\e{
\delta(E^g_{BB'}-\hbar\omega) +\delta(E^g_{B'B}-\hbar\omega)= \delta(E^e_{\gmo B,\gmo B'}-\hbar\omega) +\delta(E^e_{\gmo B',\gmo B}-\hbar\omega).
}

Let us  decompose the collisional operator  into linear and circular components
\e{
w^{ex,+}_{BB'}=\sum_{\alpha}^{l,c}w^{\text{ex},+,\alpha}_{BB'}; \as w^{\text{ex},+,\alpha}_{BB'}(H^0,V)=w^{\text{ex},+,\alpha}_{BB',bc}(H^0) \cale_{\bq\omega,b}\overline{\cale_{\bq\omega,c}}; \as w^{\text{ex},+,\alpha}_{BB',bc}= s_{\alpha} w^{\text{ex},+,\alpha}_{BB',cb}; \as s_{l}=1; \as s_{c}=-1. \la{lcdecom}
}
Accordingly, the collisional operator also splits as
\e{
I^{ex,+}_{B} \eq \sum_{\alpha}^{l,c} I^{ex,+,\alpha}_{B,bc} \cale_{\bq\omega,b}\overline{\cale_{\bq\omega,c}}; \as I^{ex,+,\alpha}_{B,bc} =s_{\alpha}I^{ex,+,\alpha}_{B,cb}. \la{Iexcalpha}
}
I'll refer to $I^{ex,+,\alpha}_{B,bc}$ as the \textit{photo-excitation tensor}; it has the following transformation law:
\e{
\text{Transformation of photo-excitation tensor:}\as (I^{ex,+,\alpha}_{gB,bc} )^g= (-s_{\alpha})^g g^s_{be}g^s_{cf} (I^{ex,+,\alpha}_{B,ef} )^e.\la{gtransformphototensor}
}
This follows from \q{Iexplusg}, with an additional transformation `$(s_{\alpha})^gg^s_{be}g^s_{cf}$' because the electric fields transform [\q{gcaleFourier}] as
\e{
 \cale^{g}_{\bq\omega,b}\overline{\cale^{g}_{\bq\omega,c}}= g^s_{be}g^s_{cf} \overline{\cale^{e}_{\bq\omega,e}\overline{\cale^{e}_{\bq\omega,f}}}^g;
}
note that complex conjugating the field Fourier coefficients (for $i_g=-1$) amounts to a transposition of tensor indices, hence the factor of `$(s_{\alpha})^g$'.
Similarly, we can derive 
\e{
 I^{ex,+,\alpha}_{gB,bc}(H^{0,g})= (s_{\alpha})^g g^s_{be}g^s_{cf} I^{ex,+,\alpha}_{B,ef}(H^{0,e}).\la{IexcHT}
}

\subsection{Symmetry compatibility index of response tensors}\la{app:compatibility}

For symmetries that square to identity: $g^2=e$ (including symmetries which are projectively represented: $\hg^2= -\hat{e}$), we define 
\e{
\text{$g$-compatibility index of response}  \as \text{Ind}^g_{R} =\pm 1; \as  R(H^g,I(H^g))= \text{Ind}^g_{R}\cdot R(H^e,I(H^e)). \la{compindex}
}
If $R(H^e,I(H^e))$ is not proportional to $ R(H^g,I(H^g))$, then it can be decomposed into two terms with well-defined indices: 
\e{
R(H^e,I(H^e))=R^+ + R^-; \as R^{\pm}= \tf{R(H^e,I(H^e))\pm R(H^g,I(H^g))}{2}.
}
We drop the superscript on $R^{\pm}$ for notational simplicity. The practical value of the compatibility index is that
\begin{align}
 \text{Ind}^g_{R} =\begin{cases}
     1, & \text{$R$ is $(H^e=H^g)$-compatible, i.e., can be nonzero if $H^e=H^g$. }\\
     -1,  & \text{$R$ is $(H^e=H^g)$-incompatible, i.e., must be zero if $H^e=H^g$. }
 \end{cases}
\end{align}\\

If $g$ inverts all (or none) of the spatial coordinates, then the compatibility index for $R$ is independent of which tensor element $R_{ab\ldots}$ is selected.  For illustration, the compatibility indices of the photovoltaic tensor under $P,T,PT$
 are listed in \tab{tab:compindex}. If $g$ inverts a strict subset of the spatial coordinates, then the compatibility index is tensor-element-dependent, as exemplified by \qq{TMxlemma}{TC2zlemma} below.\\

The following lemmas for the compatibility index (of the photovoltaic tensor) will be  proven in \s{sec:proofmagnettheorem}:
\e{
&\text{$PT$-Lemma}: \as \text{Ind}^T_{\Sigma^{\alpha}}=-\text{Ind}^{PT}_{\Sigma^{\alpha}}.\la{PTlemma}\\
&\text{$TM^x$-Lemma}: \as  \text{Ind}^{TM^x}_{\Sigma_{abc}}=(-1)^{\text{How many of `abc' are x indices}} \text{Ind}^{T}_{\Sigma_{abc}}.\la{TMxlemma}\\
&\text{$TC^{2z}$-Lemma}: \as \text{Ind}^{TC^{2z}}_{\Sigma_{abc}}=-(-1)^{\text{How many of `abc' are z indices}} \text{Ind}^{T}_{\Sigma_{abc}}.\la{TC2zlemma}
}

\subsubsection{T-compatibility index for the impurity-mediated ballistic photocurrent}		\la{sec:calculateindex}

In this section, we use $\Sigma(H)$ as a shorthand for $\Sigma(H,I(H))$.
The impurity-mediated ballistic photovoltaic tensor is
\e{
\Sigma_{abc}^{\alpha=l,c}(H)=-\tf{|e|}{\calv}\sum_{BB'B''B'''}v_{B,a} (w^{im,+})^{\mo}_{BB'}w^{im,-}_{B'B''} (w^{im,+})^{\mo}_{B''B'''}  \delta I^{ex,+,\alpha}_{B''',bc}; \as \delta I^{ex,+,\alpha}_{B,bc}(H^0) = I^{ex,+,\alpha}_{B,bc}-[\delta I^{ex,+,\alpha}_{B,bc}],\la{Sigmaimp}
}
with $w^{im,\pm}$ being the impurity-mediated collisional operators defined in \app{app:Iimalgorithm}.   \q{Sigmaimp} follows directly from \q{eq:ballisticcurrent} (identifying $W^{mr}=w^{im,+}$ and $W^{im,a}=w^{im,-}$), the definition of the bulk photovoltaic tensor in \qq{expanddirectcurrent}{bPVtensordefine2}, and the definition of the photo-excitation tensor in \q{Iexcalpha}.\\

From \q{wBBp2},
\e{
w^{a}_{BB'}(H^{T})=-w^{a}_{TB,TB'}(H^e), \iand w^{s}_{BB'}(H^{T})=w^{s}_{TB,TB'}(H^e) \imp (w^{s})^{\mo}_{BB'}(H^{T})=(w^{s})^{\mo}_{TB,TB'}(H^e), \la{allabove}
}
with $TB$ meaning the one-electron label that is the time-reverse of $B$. It also holds that
\e{
I^{ex,+,\alpha}_{B,bc}(H^T)\refeq{IexcHT}&\; s_{\alpha}I^{ex,+,\alpha}_{TB,bc}(H^e)
\imp \delta I^{ex,+,\alpha}_{B,bc}(H^T)= s_{\alpha}\delta I^{ex,+,\alpha}_{TB,bc}(H^e).\la{allabove2}
}
Combining \qq{allabove}{allabove2} with $\bv_{B}(H^T)=-\bv_{TB}(H^e)$ [proven below in \qq{veloBBg}{veloBBg2}],
\e{
\Sigma_{abc}^{\alpha}(H^T)=-\tf{|e|}{\calv}\sum_{BB'B''B'''}(-v_{TB,a}) (w^{ex,+})^{\mo}_{TB,TB'}(-w^{im,-}_{TB',TB''}) (w^{ex,+})^{\mo}_{TB'',TB'''}  (s_{\alpha}\delta I^{ex,+,\alpha}_{TB''',bc})\bigg|_{H^e}= s_{\alpha}\Sigma_{abc}^{\alpha}(H^e),
}
which proves the T-compatibility index to be 
\e{
\text{Ind}^T_{\Sigma^{\alpha}_{\text{ballistic}}}=s_{\alpha}.\la{Tindeximpurity}
}

The relation $\bv_{B}(H^T)=-\bv_{TB}(H^e)$ for the group velocity is a special case ($B=B'$) of the more general relation for the velocity matrix element
\begin{align}
\bv_{BB'}^g = \bv_{BB'}(H^g)=\braopket{B,g}{\hbv^g}{B',g}=i_gg^s\cdot \{\bv^e_{g^{\mo}B,g^{\mo}B'}\}^{t_g},\la{veloBBg}
\end{align}
which follows from the velocity operator $(\hbv^e=i[H^{0,e},\chebr])$ being a time-oriented vector observable:
\e{
 \hbv^g=i[H^{0,g},\chebr] = i_g \hg i[H^{0,e},\hg^{\mo}\chebr\hg] \hg^{\mo}  =i_g \hg g_s\cdot \hbv^e \hg^{\mo}.  \la{veloBBg2}
}

\subsection{General  constraints on photovoltaic tensors from Hamiltonian symmetries}\la{app:generalconstraints}

 A point-group-symmetric [\q{spatialvsmagnetic}] Hamiltonian has the following implications for the photovoltaic tensor:
\e{
\text{Spatial-symmetric photovoltaic theorem}: \as H^{g^s}\eq H^e \imp  \bSigma_{abc}= g^s_{ad} g^s_{be} g^s_{cf} \bSigma_{def}; \la{gsconstrainsSigma}\\
\text{Magnet-symmetric photovoltaic theorem}: \as H^{Tg^s}\eq H^e \imp  \bSigma_{abc}= \text{Ind}^T_{\Sigma} g^s_{ad} g^s_{be} g^s_{cf}\bSigma_{def}. \la{TgsconstrainsSigma}
}
The second theorem applies: (i) whether or not $H^T=H^e$, (ii) whether or not $H^{g^s}=H^e$,  and (iii) to  components of the bulk photovoltaic tensor with a well-defined compatibility index under time reversal:
\e{
\Sigma\big(H^T,I(H^T)\big)= \text{Ind}^T_{\Sigma}\cdot \Sigma\big(H^e,I(H^e)\big); \as \text{Ind}^T_{\Sigma}=\pm 1.\la{compatibilityindexT}
}

\subsubsection{Proof of spatial-symmetric photovoltaic theorem}  \la{sec:proofspatialtheorem}

We begin by deriving a general property about symmetry transformations of the photovoltaic tensor:
\e{
&\bj^e=\Sigma^{\alpha,e} \cale^e\overline{\cale^e}   \lin
& g^s\cdot \bj^e=  \bj^{g^s}=\Sigma^{\alpha,g^s} \cale^{g^s}\overline{\cale^{g^s}}=\Sigma^{\alpha,g^s} (g^s\cdot \cale^e)\overline{g^s\cdot \cale^e};\\
&- g^s\cdot \bj^e=  \bj^{Tg^s}=\Sigma^{\alpha,Tg^s} \cale^{Tg^s}\overline{\cale^{Tg^s}}=\Sigma^{\alpha,Tg^s} \overline{(g^s\cdot \cale^e)\overline{g^s\cdot \cale^e}}=s_{\alpha}\Sigma^{\alpha,Tg^s} {(g^s\cdot \cale^e)\overline{g^s\cdot \cale^e}}.\la{routegs1}
}
Since this holds for any $\bj^e$ and $\cale^e$, and $g^s_{ab}$ is an orthogonal matrix,
\e{
\text{General transformation:}\as &\bSigma^{\alpha,g^s}_{abc}= g^s_{ad} g^s_{be} g^s_{cf} \bSigma^{\alpha,e}_{def}   \iff g^s_{da} \bSigma^{\alpha,e}_{abc}= \bSigma^{\alpha,g^s}_{def} g^s_{eb} g^s_{fc}  \la{gstransformSigma} \\
&\bSigma^{\alpha,Tg^s}_{abc}= -s_{\alpha} g^s_{ad} g^s_{be} g^s_{cf} \bSigma^{\alpha,e}_{def}.\la{TgstransformSigma} 
}
We shall speak of \qq{gstransformSigma}{TgstransformSigma} as \textit{general transformation laws} that apply no matter the symmetries of $H$.\\

According to the symmetry transformation of the collisional integral [\q{RgIg}], a spatial-symmetric Hamiltonian implies a spatial-symmetric collisional integral:
\e{
H^{g^s}\eq H^e \imp I^{g^s}_{\bk}=I^{e}_{\bk}.
}
This implies that all operational conditions are satisfied for the closed system to be $g^s$-symmetric, and hence by the Devil's axiom [\q{RgRe}], $R^{g^s}=R^e$. In particular, substituting $\Sigma^{g^s}=\Sigma^{e}\equiv \Sigma$ into \q{gstransformSigma}, we obtain \q{gsconstrainsSigma}, which completes the proof.

\subsubsection{Proof of magnet-symmetric photovoltaic theorem} \la{sec:proofmagnettheorem}

Owing to the time-orientation of the collisional integral [\q{RgIg}], a magnet-symmetric Hamiltonian does not generally imply a magnet-symmetric collisional integral, and therefore does not generally imply a magnet-symmetric closed system  $\imp R^{Tg^s}=R^e$. In fact, even if the Hamiltonian is magnet-symmetric, most physically-interesting closed systems are not magnet-symmetric, owing to collisions which are time-oriented by the second thermodynamic law. In other words, if the Devil were to $Tg^s$-transform our closed system, we could almost always certainly tell that a trick has been played by measuring a decreasing entropy.\\

Thus with magnetic symmetries, without being able to use the Devil's axiom, we can no longer derive a self-constraint on $\Sigma$ using the same route [\s{sec:proofspatialtheorem}] we had taken for spatial symmetries. An alternative route exists for irreducible $\Sigma$ with a well-defined compatibility index under time reversal [\q{compatibilityindexT}], as we now explicate. Combining the compatibility relation [\q{compatibilityindexT}] with the general transformation law [\q{TgstransformSigma}] for the photovoltaic tensor under time reversal:
\e{
\bSigma^{\alpha}\big(H^T,-s_{\nu}I^{\nu}(H)^T)\big)\refeq{RgIg}\bSigma^{T,\alpha}= -s_{\alpha} \bSigma^{e,\alpha} \refeq{RgIg}   -s_{\alpha} \bSigma^{\alpha}\big(H^e,I^{\nu}(H^e)\big)  \as \text{(matrix indices suppressed)},
}
we obtain 
\e{
\bSigma^{\alpha}\big(H^T,-s_{\nu}I^{\nu}(H)^T)\big) = -s_{\alpha}\text{Ind}^T_{\Sigma^{\alpha}}\cdot \bSigma^{\alpha}\big(H^T,I^{\nu}(H^T)\big),
}
which holds whether or not $H^T=H^e$. But since the above relation holds for arbitrary $H^T$, we may as well drop the superscript on `$H^T$':
\e{
\bSigma^{\alpha}\big(H,-s_{\nu}I^{\nu}(H))\big) = -s_{\alpha}\text{Ind}^T_{\Sigma^{\alpha}}\cdot \bSigma^{\alpha}\big(H,I^{\nu}(H)\big).\la{magicindex}
}

Returning to the general transformation law [\q{TgstransformSigma}]  under magnetic symmetries, 
\e{
-s_{\alpha}\text{Ind}^T_{\Sigma^{\alpha}}\cdot
\bSigma^{\alpha}\big(H^{Tg^s},I^{\nu}(H^{Tg^s})\big)\refeq{magicindex}\bSigma^{\alpha}\big(H^{Tg^s},-s_{\nu}I^{\nu}(H^{Tg^s})\big)\refeq{RgIg}&\;\bSigma^{Tg^s,\alpha} = -s_{\alpha} g^s g^s g^s \bSigma^{e,\alpha}\lin
&\refeq{RgIg}-s_{\alpha} g^s g^s g^s \bSigma^{\alpha}\big(H^{e},I^{\nu}(H^e)\big), \la{toreduce}
}
which holds whether or not $H^{Tg^s}=H^e$. However, if 
\e{
H^{Tg^s}=H^e: \as  \text{Ind}^T_{\Sigma^{\alpha}}\cdot
\bSigma^{\alpha}\big(H,I^{\nu}(H)\big) \refeq{toreduce} g^s g^s g^s \bSigma^{\alpha}\big(H,I^{\nu}(H)\big).
} 
Since the above relation is independent of $s_{\alpha}$ and $s_{\nu}$, we may as well drop the $\alpha$ and $\nu$ superscripts, finally arriving at
 \q{TgsconstrainsSigma} and completing the proof.

 \subsubsection{Symmetry compatibility index from general transformation laws}

 As a bonus, let us return to \q{toreduce} and focus on $Tg^s=TP=PT$ with $g^s_{ab}=P_{ab}=-\delta_{ab}$ being the parity transformation (i.e., spatial inversion):
 \e{
 \text{Ind}^T_{\Sigma}\cdot
\bSigma\big(H^{TP},I(H^{TP})\big) =(-1)^3 \bSigma\big(H^{e},I(H^e)\big).
 }
 From the general definition of the compatibility index [\q{compindex}], we identify the index under $PT$ transformation as in \q{PTlemma}, which is our PT-lemma. \\

 For $Tg^s=TM^x$ with $M^x$ reflecting the x coordinate, \q{toreduce} has different implications for $\Sigma_{abc}$ depending on how many of `abc' are x indices:
  \e{
 \text{Ind}^T_{\Sigma_{abc}}\cdot
&\bSigma_{abc}\big(H^{TM^x},I(H^{TM^x})\big) =(-1)^{N_{abc;x}} \bSigma_{abc}\big(H^{e},I(H^e)\big)\imp  \text{$TM^{x}$-lemma in \q{TMxlemma}}.
 }
The $TC^{2z}$-lemma in \q{TC2zlemma} is similarly proven.

\subsection{Symmetric response tensors as equivariant linear maps}\la{sec:equivariance}

In applying the spatial-symmetric photovoltaic theorem [\q{gsconstrainsSigma}] to concretely determine the symmetry-allowed photovoltaic response tensor elements, it is useful to view a $G$-symmetric response tensor as a $G$-equivariant linear map and apply Schur's lemma.\\

Let us first develop this viewpoint of response tensors as linear maps.
Since each of $\{\bj,\bcale,\overline{\bcale}\}$ transforms in the three-vector representation of the point group, we introduce:\\

\noi{i} $V$ is a 3D vector space spanned by $\ket{x},\ket{y},\ket{z}$, with $\ket{x}$ corresponding to either of $\{j_x,\cale_x,\overline{\cale_x}\}$, depending on the context. \\

\noi{ii} $V^2=V\otimes V$ is   tensor-product vector space spanned by nine basis vectors: $\ket{x,x},\ket{x,y},\ldots$ \\

\noi{iii} $V^3=V\otimes V\otimes V$ is spanned by 27 basis vectors: $\ket{x,x,x},\ldots $ \\

\noindent Reproducing \q{gsconstrainsSigma} and specializing to $g\in G$ (a non-magnetic point group),
\e{
 \Sigma_{abc}= g_{ad} g_{be} g_{cf} \Sigma_{def} \la{invarianttensor}
}
describes an $G$-invariant tensor. This can equivalently seen as a $G$-equivariant linear map (in short, $G$-linear map) between vector spaces ($a$ and $b$) equipped with representations ($\rho_a$ and $\rho_b$) of $G$. \\

\noi{Example 1} Let $\R$ be a 1-dim vector space on which $G$ acts as 
\e{
\text{1D trivial representation:}\as \rho_{\R}(g)=1, \as \forall  g\in G.\la{1Dtrivialrep}
}

\noi{Example 2} Let $V^3$ be a 27-dim vector space on which $G$ acts as $\rho_{V^3}(g)=\rho_V(g)\otimes\rho_V(g)\otimes\rho_V(g)$, with matrix representation 
\e{
g_{ad}\eq \braopket{a}{\rho_V(g)}{d}; \as  
g_{ad} g_{be} g_{cf}=  \braopket{a,b,c}{\rho_{V^3}(g)}{d,e,f}.
}\\

\noindent \q{invarianttensor} can be seen as a $G$-equivariant linear map 
\e{
\Sigma: \R \rightarrow V^3, \as 1 \mapsto  \vec{\Sigma} =\sum_{abc}\Sigma_{abc} \ket{a,b,c}, \la{Sigma1}
}
or equivalently\footnote{Both viewpoints are equivalent if $g$ is orthogonal; otherwise, one should take care with distinguishing co- and contravariant indices.} as the reverse $G$-equivariant linear map 
\e{
\Sigma: V^3 \rightarrow \R, \as \vec{\Sigma} \mapsto 1.
}
In general,  $G$-equivariance of a linear map means that the linear map commutes with (or is compatible with) the $G$-action, 
\e{
\Sigma: A \ri B, \as   \Sigma \,\rho_A(g) =\rho_B(g) \,\Sigma, \as  \forall g \in G, \la{Gequivariance}
}
bearing in mind $G$ acts on different vector spaces on different sides of the above equality. Then it should be seen that invariant-tensor equation [\q{invarianttensor}]  is the matrix representation of  the equivariance condition [\q{Gequivariance}] with $A=\R$ and $B=V^3$. From \q{Sigma1}, $\Sigma\circ 1=\vec{\Sigma} \in V^3$, and \q{Gequivariance} implies
\e{
 \Sigma \circ 1\refeq{1Dtrivialrep} \Sigma \rho_{\R}(g)\circ 1 =\rho_{V^3}(g) \Sigma \circ 1 \imp \vec{\Sigma}= \rho_{V^3}(g)  \vec{\Sigma},
}
meaning that $\vec{\Sigma}$ is a $G$-invariant vector.\\

We can rewrite \q{invarianttensor} as a $G$-linear map from $A=V^2$ to $B=V$
\e{
 g^{\mo}_{da}\Sigma_{abc}=   \Sigma_{def}\gmo_{eb} \gmo_{fc}  \iff \rho_V(\gmo)\; \Sigma  = \Sigma\; \rho_{V^2}(\gmo).
}
In the photovoltaic context, this has the most direct physical interpretation: $V$ is the vector space of the current and $V^2$ the vector space of the squared electric field. In general, an invariant tensor in $V^3$ gives the same data as several types of $G$-linear map via hom-tensor adjunctions, or, in lay-speak, by various ways to split the indices (taking care of the distinction between contra-and covariance if $g$ is not orthogonal).

\subsubsection{Application of Schur's lemma to response tensors}\la{sec:schur}

$G$-linear maps between irreducible representations (irreps) are subject to Schur's lemma. In particular, if $\rho_A$ and $\rho_B$ are inequivalent, $\Sigma=0$; if $\rho_A$ and $\rho_B$ are equivalent, then in the basis in which $\rho_A=\rho_B$, $\Sigma$ is proportional to the identity map. Let us give some examples.\\

\underline{Parity group $G=\Z_2^P=\{e,P\}$,  $A=V^2$ and $B=V$ }\\

Each of $\ket{x},\ket{y},\ket{z}$ transforms in the nontrivial irrep  [namely, $\rho_V(e)=1=-\rho_V(P)$], while each of $\ket{x,x},\ket{x,y},\ldots$ transforms in the trivial irrep [namely, $\rho_{V^2}(e)=1=\rho_{V^2}(P)$]. Schur's lemma then implies $\Sigma_{abc}=0$ for any $a,b,c$.\\

\underline{Parity group $G=\Z_2^P=\{e,P\}$,  $A=\R$ and $B=V^3$ }\\

$\rho_{\R}$ is trivial by definition. There are no $G$-invariant vectors $\vec{\Sigma}\in V^3$; after all, each of $\ket{x,x,x},\ket{x,x,y},\ldots$ transforms in the nontrivial irrep [namely, $\rho_{V^2}(e)=1=-\rho_{V^2}(P)$]. Schur's lemma then implies $\Sigma_{abc}=0$ for any $a,b,c$. This must, of course, be consistent with the previous paragraph.\\

\underline{$G=SO(2)$,  $A=V^3$ and $B=\R$ }\\

\begin{table}[H]
	\centering
\begin{tabular} {|c|c|c|} \hline
			
$m$  &  Irrep in $V^3$ & Irrep in $\R$ \\  \hline 
$-3$& $\ket{x^{-}x^{-}x^{-}}$ & $0$\\ \hline
$- 2$& $\ket{\{zx^{-}x^{-}\}}$ & $0$\\ \hline
$-1$&$\ket{\{x^{\pm}x^{\mp}x^{-}\}}$ & $0$ \\ \hline		  
$0$ & $\ket{zzz/\{zx^{\pm}x^{\mp}\}}$ & $1$\\ \hline
$ 1$& $\ket{\{x^{\pm}x^{\mp}x^{+}\}}$ & $0$\\ \hline
$ 2$& $\ket{\{zx^{+}x^{+}\}}$ & $0$\\ \hline
$ 3$& $\ket{x^{+}x^{+}x^{+}}$ & $0$ \\ \hline
\end{tabular}
\caption{ Irreps of $SO(2)$ in the vector spaces $V^3$ and $\R$. Under rotation by angle $\theta$,  $x^{\pm}=x\pm iy \ri \omega^{\pm 1} x^{\pm}$ with $\omega=e^{i\theta}$. An irrep has angular momentum $m$ if it transforms under $\theta$-rotation by picking up a multiplicative phase factor $e^{im\theta}$. $\{abc\}$ means all permutations of the the tensor indices are included.	\label{tab:so2irrep}}
\end{table}
Only trivial maps are possible between  irreps with different $m$:
\e{
&\ket{\{x^{\pm}x^{\mp}x^{+}\}}\mapsto 0 \cdot 1; \la{Map1}\\
&\ket{\{x^{\pm}x^{\mp}x^{-}\}}\mapsto 0 \cdot 1; \la{Map2}\\
&\ket{\{zzx^{\pm}\}}\mapsto 0 \cdot 1; \la{Map3}\\
& \ket{\{zx^{\pm}x^{\pm}\}} \mapsto 0 \cdot 1; \la{Map4}\\
& \ket{\{x^{\pm}x^{\pm}x^{\pm}\}} \mapsto 0 \cdot 1, \la{Map5}
}
which imply the following constraints:
\e{
&\text{\qq{Map1}{Map2}}: \as \Sigma_{xxx}=-\Sigma_{yyx}=-\Sigma_{xyy}=-\Sigma_{yxy};\as \Sigma_{yyy}=-\Sigma_{xxy}=-\Sigma_{xyx}=-\Sigma_{yxx}; \la{c3constraint}\\
&\text{\q{Map3}}: \as 0=\Sigma_{\{zzx\}}=\Sigma_{\{zzy\}}; \\
&\text{\q{Map4}}: \as 0=\Sigma_{zxx}-\Sigma_{zyy}=\Sigma_{xzx}-\Sigma_{yzy}=\Sigma_{xxz}-\Sigma_{yyz}=\Sigma_{zxy}+\Sigma_{zyx}=\Sigma_{xzy}+\Sigma_{yzx}=\Sigma_{xyz}+\Sigma_{yxz};  \la{c3constraint2}\\
&\text{\q{Map5} and \q{c3constraint}}: \as 0=\Sigma_{xxx}=\Sigma_{yyy}.\la{so2constraint} 
}
Altogether these are 20 constraints leaving 7 independent tensor elements (7=4 linear + 3 circular):
\e{
\text{$SO(2)$-allowed photovoltaic tensor elements:}\as
& \Sigma_{zzz}; \as \Sigma_{zxy}=-\Sigma_{zyx}; \as\Sigma_{xzy}=-\Sigma_{yzx}; \as\Sigma_{xyz}=-\Sigma_{yxz}; \lin
&\Sigma_{zxx}=\Sigma_{zyy}; \as\Sigma_{xzx}=\Sigma_{yzy}; \as\Sigma_{xxz}=\Sigma_{yyz}; \lin
\text{$SO(2)$-allowed linear photovoltaic elements:}\as
  &\Sigma^l_{zzz};\as \Sigma^l_{zxx}=\Sigma^l_{zyy};\as\Sigma^l_{xyz}=-\Sigma^l_{yzx};\as \Sigma^l_{xxz}=\Sigma^l_{yyz}.\la{SO2allowedlinear}\\
 \text{$SO(2)$-allowed circular photovoltaic elements:}\as &\Sigma^c_{zxy};\as \Sigma^c_{xyz}=\Sigma^c_{yzx}; \as \Sigma^c_{xxz}=\Sigma^c_{yyz}.\la{SO2allowedcircular}
}\\

\underline{$G=C_3$,  $A=V^3$ and $B=\R$ }\\

\begin{table}[H]
	\centering
\begin{tabular} {|c|c|c|} \hline
			
$m$  &  Irrep in $V^3$ & Irrep in $\R$ \\  \hline 
		  
$0$ & $\ket{zzz/x^{\pm}x^{\pm}x^{\pm}/\{zx^{\pm}x^{\mp}\}}$ & $1$\\ \hline
$1$& $\ket{\{zzx^{+}\}/\{x^{\pm}x^{\mp}x^{+}\}}$ & $0$\\ \hline
$2$& $\ket{\{zzx^{-}\}/\{zx^{\pm}x^{\pm}\}/\{x^{\pm}x^{\mp}x^{-}\}}$ & $0$\\ \hline
\end{tabular}
\caption{ Irreps of $C_3$ in the vector spaces $V^3$ and $\R$. Under three-fold rotation, $x^{\pm}=x\pm iy \ri \omega^{\pm 1} x^{\pm}$ with $\omega=e^{i2\pi/3}$. An irrep has angular momentum $m$ (defined modulo three) if it transforms under three-fold rotation by picking up a multiplicative phase factor $\omega^m$. $\{abc\}$ means all permutations of the the tensor indices are included.	\label{tab:C3irrep}}
\end{table}
Only trivial maps are possible between  irreps with different $m$; these are the same trivial maps in \qq{Map1}{Map4} but excluding \q{Map5}. It follows that the constraints in \qq{c3constraint}{c3constraint2} apply, but not \q{so2constraint}. Altogether these are 18 constraints leaving 9 independent tensor elements (9=6 linear + 3 circular):
\e{
\text{$C_3$-allowed photovoltaic tensor elements:}\as
&\Sigma_{xxx}=-\Sigma_{yyx}=-\Sigma_{xyy}=-\Sigma_{yxy}; \as \Sigma_{yyy}=-\Sigma_{xxy}=-\Sigma_{xyx}=-\Sigma_{yxx};\as \Sigma_{zzz}\lin
  &\Sigma_{zxy}=-\Sigma_{zyx}; \as\Sigma_{xzy}=-\Sigma_{yzx}; \as\Sigma_{xyz}=-\Sigma_{yxz}; \lin
&\Sigma_{zxx}=\Sigma_{zyy}; \as\Sigma_{xzx}=\Sigma_{yzy}; \as\Sigma_{xxz}=\Sigma_{yyz}; \lin
\text{$C_3$-allowed linear photovoltaic elements:}\as
  &\Sigma^l_{xxx}=-\Sigma^l_{xyy}=-\Sigma^l_{yxy}; \as \Sigma^l_{yyy}=-\Sigma^l_{xxy}=-\Sigma^l_{yxx};\as \Sigma^l_{zzz};\lin
  &\Sigma^l_{zxx}=\Sigma^l_{zyy};\as\Sigma^l_{xyz}=-\Sigma^l_{yzx};\as \Sigma^l_{xxz}=\Sigma^l_{yyz}.\la{C3allowedlinear}\\
 \text{$C_3$-allowed circular photovoltaic elements:}\as &\Sigma^c_{zxy};\as \Sigma^c_{xyz}=\Sigma^c_{yzx}; \as \Sigma^c_{xxz}=\Sigma^c_{yyz}.\la{C3allowedcircular}
}\\

For linearly-polarized light in a 2D system, we can assume the electric field amplitude is real and parametrized by one angle:
\e{
\bcale_{\bq\omega} =  |\cale|\be \in \R; \as \be=(e_x,e_y)=(\cos\theta,\sin\theta).
}
Parametrizing the two independent tensor elements (`two' in 2D) in \q{C3allowedlinear} by a different angle:
\e{
\Sigma^l_{xxx}=|\Sigma^l|\sin \phi;  \as \Sigma^l_{xxy}=|\Sigma^l|\cos \phi, \la{paraEfield}
}
and applying \q{C3allowedlinear}, we can express the photovoltaic current as a function of the two angles:
\e{
j_x \eq  |\cale|^2\big(  \Sigma_{xxx} (e_x^2-e_y^2) +2\Sigma_{xxy} e_x e_y \big)=|\Sigma^l||\cale|^2\sin(2\theta+\phi);\lin
j_y \eq  |\cale|^2\big(  \Sigma_{yxx} (e_x^2-e_y^2) +2\Sigma_{yxy} e_x e_y \big)=|\Sigma^l||\cale|^2\cos(2\theta+\phi).\la{jxjyC3}
}
Thus the electric field transforms like it has spin one, while the current transforms like it has spin two, but $C_3$ symmetry alone does not fix the direction in which $|j_x|$ is maximized. \\

Let us enlarge the group $C_3$ to $C_{3v}$, which has an extra generator $M_j$ that inverts the $j$-coordinate; to remind us of this specific mirror plane, we add a superscript to $C_{3v}^j$. According to the spatial-symmetric photovoltaic theorem [\q{gsconstrainsSigma}], the effect of $M_j$ symmetry is to constrain $N_{abc;j}$ to be even. For illustration, $M_x$ symmetry extinguishes 4 of the 9 independent tensor elements allowed by $C_3$: 
\e{
\text{$C_{3v}^x$-allowed linear photovoltaic elements:}\as
  & \Sigma^l_{yyy}=-\Sigma^l_{xxy}=-\Sigma^l_{yxx};\as \Sigma^l_{zzz};\as \Sigma^l_{zxx}=\Sigma^l_{zyy};\as \Sigma^l_{xxz}=\Sigma^l_{yyz}.\la{C3vallowedlinear}\\
 \text{$C_{3v}^x$-allowed circular photovoltaic elements:}\as   &\Sigma^c_{xxz}=\Sigma^c_{yyz}.\la{C3vallowedcircular}
}
In particular, setting $\Sigma^l_{xxx}=0$ in \q{jxjyC3}   gives the $C_{3v}^x$-allowed current in 2D:
\e{
j_x = \Sigma^l_{xxy}|\cale|^2 \sin(2\theta); \as  j_y = \Sigma^l_{xxy}|\cale|^2\cos(2\theta),
}
with mirror symmetry guaranteeing that $|j_x|$ is maximized if the light polarization vector deviates from the mirror x-axis by $\pi/4$.

\subsection{Hamiltonian-specific symmetry constraints on photovoltaic tensors}\la{app:specificsymconstraints}

The two theorems in \qq{gsconstrainsSigma}{TgsconstrainsSigma} apply to all photovoltaic tensors independent of specific details of the Hamiltonian, beyond the assumption that the Hamiltonian is symmetric: $H^g=H^e$. In contrast, we introduce in this subappendix the notion of \textit{$H$-specific symmetry constraints} (short for Hamiltonian-specific), which apply to specific components of the photovoltaic tensor and depend on specific structural details of the Hamiltonian.\\

We shall further distinguish between $H$-specific symmetry constraints of the first and second kinds, or first- and second-kind constraints, in short. A \textit{first-kind constraint} (resp. \textit{second-kind constraint}) is a $H$-specific symmetry constraint that arises  from the Hamiltonian being $g$-symmetric (resp. $g$-\textbf{\textit{a}}symmetric: $H^g \neq H^e$), as will be discussed in \app{app:symconstraintsfirstkind} (resp. \app{app:symconstraintssecondkind}).

\subsubsection{Hamiltonian-specific symmetry constraints of the first  kind}\la{app:symconstraintsfirstkind}

We package two first-kind constraints into two theorems, the first of which is:

\begin{tcolorbox}[colback=white, sharp corners]
\textbf{$\vec{H}$-polarity theorem:} For any two-band $H^0$ with a point group $G$, if there exists no invariant vector in the Hamiltonian-vector representation of $G$, then the  ballistic photocurrent (mediated by overscreened monopole impurities) vanishes.    
\end{tcolorbox}

The notion of the Hamiltonian-vector representation was introduced in  \app{app:dvectorep} for two-band Hamiltonians of the form $H^0_{\bk}=\bd_{\bk}\cdot \bsigma+\delta \Ek \iden$; the related notion of invariant vectors was defined in \q{invariantvectordrep}. To be `mediated by overscreened monopole impurities' means that the impurity potential responsible for long-wavelength skew scattering is that of the overscreened monopole in \q{overscreenedmonopole3D}.  \\

 To prove this theorem, we will show that the asymmetric transition rate matrix element [\q{skewscattering}] vanishes, which would imply that the ballistic photocurrent (being a linear functional of the asymmetric transition rate [\qq{jballisticinwords}{eq:jball}]) also vanishes. Indeed, plugging \q{overscreenedmonopole3D} into \q{skewscattering},
\e{
(\text{Asymmetric transition rate})_{\bk\lea\bk'}= \tf{(2\pi)^2}{\hbar}\Nimp \big(\tf{e^2}{\calv \eps k^2_{\text{scr}}}\big)^3\delta(E_{\bk\bk'})  \sum_{\bk''}   \imag\braket{u_{\bk}}{u_{\bk''}}   \braket{u_{\bk''}}{u_{\bk'}} \braket{u_{\bk'}}{u_{\bk}}  \delta(E_{\bk\bk''}).
}
Plugging in the $\bd$-vector expression for the imaginary component of the Bargmann invariant [\q{Bargmannrealimag}], we find that the asymmetric transition rate matrix element is proportional to
\e{
 \hat{\mathbf{d}}_{\bk}\times \hat{\mathbf{d}}_{\bk'}  \cdot [{\hbd}]_{\Eex}; \as    [{\hbd}]_{\Eex}=\tf{\int  \hbd_{\bk''} \delta(E_{\bk''}-\Eex)d\bk''}{\int  \delta(E_{\bk''}-\Eex)d\bk''},
}
with $\Eex=E_{\bk}$ the excitation energy and $[\hbd]_E$ meaning the iso-energy average of $\hbd$. Let us change the integration variable as $\bk''=g\cdot \bk$, with $g$ an element in the point group of $G$, and $g\circ$ denoting a transformation by the matrix representation of $g$ in $\bk$-space. (a) Because this matrix is orthogonal, the Jacobian determinant for the transformation from $\bk''$ to $\bk$ is unity. (b) Because $g$ is a symmetry of $H^0$ [\q{twobandsymmetry}], (b) $E_{g\cdot \bk}=E_{\bk}$, which is equivalent to $|\bd_{g\cdot \bk}|=|\bd_{\bk}|$, and  (c) $\bd_{g\cdot \bk}= \overg\cdot \bd_{\bk}$, with $\overg$ the Hamiltonian-vector (i.e. $\bd$-vector) representation of $g$. Combining (a-c),  
\e{
[{\hbd}]_{\Eex}=\tf{\int \overg\cdot  \hbd_{\bk} \delta(E_{\bk}-\Eex)d\bk}{\int \delta(E_{\bk}-\Eex)d\bk} = \overg\cdot [{\hbd}]_{\Eex}. \la{abovecond}
}
Because we can in principle change the integration variable ($\bk''=g\cdot \bk$) for any $g\in G$, \q{abovecond} must hold for all $g\in G$, which identifies $[{\hbd}]_{E}$ as an invariant vector in the Hamiltonian-vector representation of $G$ [\q{invariantvectordrep}]. If there is no invariant vector, then $[{\hbd}]_{E}=\bze$ for any $E$, which extinguishes the ballistic photocurrent, thus proving the theorem. \\

As applications of the $\vec{H}$-polarity theorem, we have shown in \app{app:dvectorep} that there are no invariant vectors (in the Hamiltonian-vector representation of $G$) for all $n$-fold Dirac-Weyl fermions, hence the monopole-impurity-mediated $\jball$ vanishes.\\

We now proceed to the second theorem:\\

\begin{tcolorbox}[colback=white, sharp corners]
\textbf{$TC_{2z}$ theorem:} For 2D systems with $TC_{2z}$-symmetric Hamiltonians, (a) the linear component of the ballistic/shift current vanishes, and (b) both linear and circular components of the  ballistic photocurrent (mediated by centrosymmetric impurities) vanishes.    
\end{tcolorbox}

 To be `mediated by centrosymmetric impurities' means that the impurity potential (responsible for long-wavelength skew scattering) is centrosymmetric: $\Vim_{\br}=\Vim_{-\br}$, implying that the Fourier transform is real: $\widetilde{\Vim}_{\bk}\in \R$. (This includes the case of the monopole impurity.) It follows that the imaginary component of the N-point self-interference amplitude [\q{eq:smallangle}] is proportional to the   imaginary component of the N-point Bargmann invariant [\q{Bargmann}]:
 \e{
\imag \{\text{N-point self-interference amplitude}\} \approx \widetilde{\Vim}_{\bk_1-\bk_2}\ldots \widetilde{\Vim}_{\bk_N-\bk_1}\cdot  \imag \scrb_{\bk_1\ldots \bk_N},\la{Bisreal}
 }
 with $\scrb_{\bk_1\ldots \bk_N}=\braket{u_{\bk_1}}{u_{\bk_2}}\ldots \braket{u_{\bk_N}}{u_{\bk_1}}$.\\

Statement (a) in the $TC_{2z}$ theorem follows from applying the magnet-symmetric photovoltaic theorem [\q{TgsconstrainsSigma}], with the recognition that the linear shift/ballistic photovoltaic tensor is compatible with time-reversal symmetry [\tab{tab:compindex}]. Statement (a) is not a $H$-specific symmetry constraint, but statement (b) is. \\

Let us prove statement (b). By assumption, the zeroth-order Hamiltonian $H^0$ is $TC_{2z}$-symmetric, meaning that
\e{
\widehat{TC_{2z}}\, H^0_{\bk} \, \widehat{TC_{2z}}^{\mo} =  H^0_{\bk}
}
holds for all $\bk$, which is possible in 2D systems because $TC_{2z}\circ (k_x,k_y)=(k_x,k_y)$. 
Because $\widehat{TC_{2z}}$ squares to the identity operator in its action on wave functions (independent of whether $\widehat{T}^2=\pm \iden$), all eigenstates ($\{ \ket{u_{b\bk}}\}_b$) of $H^0_{\bk}$ can be chosen to have real-valued wave functions,\cite{nogo_AAJH} implying that the N-point self-interference  amplitude [\q{Bisreal}] vanishes for any $N$. In particular, our impurity-mediated ballistic photocurrent is proportional to the imaginary component of the 3-point self-interference amplitude [\q{eq:jball}] and therefore also vanishes. This holds for effective few-band $H^0_{\bk}$ as well as for the Schr\"odinger-type, infinite-band $H^0_{\bk}$.

\subsubsection{Hamiltonian-specific symmetry constraints of the second kind}\la{app:symconstraintssecondkind}

Combining the general transformation laws [\q{gstransformSigma},\q{toreduce}] and the T-compatibility index specific to the  impurity-mediated ballistic photovoltaic tensor [\q{Tindeximpurity}],
\e{
 \Sigma^{\alpha}_{abc}(H^{g})\eq (s_{\alpha})^g g^s_{ad} g^s_{be} g^s_{cf} \Sigma^{\alpha}_{def}(H^e); \la{gstransformSigmaimp}
 }
 with $\Sigma(H)$ being a shorthand for the algorithmic function $\Sigma(H,I(H))$.
 If the Hamiltonian is symmetric, the above transformation law implies:
\e{
 H^{g}\eq H^e \imp  \Sigma^{\alpha}_{abc}= (s_{\alpha})^g g^s_{ad} g^s_{be} g^s_{cf} \Sigma^{\alpha}_{def}, \la{gsconstrainsSigmaimp}
 }
in accordance with the symmetry theorems in \qq{gsconstrainsSigma}{TgsconstrainsSigma}.
\q{gsconstrainsSigmaimp} is a symmetry constraint that is specific to the impurity-mediated  ballistic photovoltaic tensor, because the form of the constraint changes between the linear  ($s^l=1$) and circular ($s^c=-1$) components of the tensor. \\

Though dipole impurity potentials are asymmetric with respect to certain $g$,  $g$ may nevertheless constrain the response tensor. Let us give two examples:  \\

\noi{i} Commonly, we may encounter  $g$ that is a symmetry of $H^0$ and $V^{\text{dr}}$ (the driving perturbation) but not necessarily of the impurity potentials; the latter are  classified  by their parity under $g$ transformation:
\e{
\text{$g$-parity even/odd}:\as V^{\text{im},g}=p_g V^{\text{im},e};  \as  p_g=\pm 1. \la{gparity}
}
We say that $V^{\text{im}}$ is $g$-\textit{antisymmetric} if it has odd $g$-parity: 
\e{
\text{$g$-antisymmetric $V^{\text{im}}$}: \as H^g = H^{0,g}+V^{\text{im},g}+V^{\text{dr},g} = H^{0,e}-V^{\text{im},g}+V^{\text{dr},e} \neq H^e=H^{0,e}+V^{\text{im},g}+V^{\text{dr},e},
}
which exemplifies a $g$-asymmetric Hamiltonian $H$. A specific property of the ballistic photovoltaic tensor is that for impurity potentials with a well-defined $g$-parity ($p_g$),
 \e{
H^{0,g}=H^{0,e}, \as V^{\text{im},g}=p_g V^{\text{im},e}: \as \Sigma^{\alpha}(H^g) \approx p_g \Sigma^{\alpha}(H^e),\la{lowordersym}
 } 
with `$\approx$' meaning equal to the lowest nontrivial order in the impurity potential $V^{\text{im}}$. Indeed, inspection of \q{Sigmaimp} reveals that $V^{\text{im}}$ enters the formula for $\Sigma$ only through the transition rate matrix elements $(w^s)^{\mo}$ and $w^a$;  $w^s$ is second order in $V^{\text{im}}$ [\q{wsimp}] and therefore even under $V^{\text{im}}\ri -V^{\text{im}}$; $w^a$ is third order in $V^{\text{im}}$ [\q{waimp}] and therefore odd under $V^{\text{im}}\ri -V^{\text{im}}$; this makes $\Sigma^{\alpha}$ (which is a linear functional of $w^a$) also odd under $V^{\text{im}}\ri -V^{\text{im}}$. Combining \q{lowordersym} with the general transformation law in \q{gstransformSigmaimp},
\e{
H^{0,g}=H^{0,e}, \as V^{\text{im},g}=p_g V^{\text{im},e}:\as  \Sigma^{\alpha}_{abc}\approx  p_{g} (s_{\alpha})^g g^s_{ad} g^s_{be} g^s_{cf} \Sigma^{\alpha}_{def}, \la{gparitySigmaimp}
}
which is (for $p_g=-1$) our first example of a symmetry constraint of the second kind.\\

\noi{ii} For our second example of a symmetry constraint of the second kind, we consider  a hybrid impurity potential $V^{\text{im}}=V^m+V^d$ that is the sum of a monopole and dipole potential, which individually have  distinct $g$-parities. For small momentum scattering within a single valley of a Dirac-Weyl fermion, we assume that matrix elements ($V^d_{b\bk,b\bk'}$) of the dipole potential are suppressed by a multiplicative factor $|\bk-\bk'|/$(Brillouin-zone period) compared to matrix elements of the monopole potential. We therefore approximate $|V^m_{BB'}+V^d_{BB'}|^2\approx |V^m_{BB'}|^2$ within the symmetric transition rate matrix element [\q{wsimp}]:
\e{
&w^s_{BB'}= W^s_{BB'}-\delta_{BB'}\sum_{B''}W^s_{BB''}; \as W^s_{BB'}\approx \tf{2\pi}{\hbar}N_{\text{imp}} |V^m_{BB'}|^2\delta(E_{BB'}).
}
We emphasize this approximation by writing $w^s=w^s(H^0,V^m)$ as an algorithm that inputs $V^m$, not $V^m+V^d$. \\

By splitting  $V^{\text{im}}_{BB'}=V^m_{BB'}+V^d_{BB'}$ in  the asymmetric transition rate matrix element [\q{waimp}], we decompose this matrix element into a sum of eight terms:
\e{
&w^a_{BB'}=\sum_{\ab\delta} dw^{a}_{BB'}(H^0,V^{\alpha},V^{\beta},V^{\delta}); \as dw^a_{BB'}(H^0,V^{\alpha},V^{\beta},V^{\delta})= \tf{(2\pi)^2}{\hbar}N_{\text{imp}} \im \sum_{B''}V^{\alpha}_{BB''}V^{\beta}_{B''B'}V^{\delta}_{B'B}\delta(E_{BB'})\delta(E_{BB''}),\la{decomposewa}
}
with all Greek indices being summed over $\{m,d\}$. Being a linear functional of $w^a$ [\q{Sigmaimp}], the photovoltaic tensor also splits into eight terms:
\e{
\Sigma^{l}=\sum_{\ab\delta}d\Sigma^{l,\ab\delta};\as d\Sigma_{abc}^{l,\ab\delta}\eq d\Sigma_{abc}^{l,\ab\delta}\big(\,H^0,w^s(H^0,V^m), dw^{a}(H^0,V^{\alpha},V^{\beta},V^{\delta})\,\big)\lin
\eq -\tf{|e|}{\calv}\sum_{BB'B''B'''}v_{B,a} (w^{s})^{\mo}_{BB'}dw^{a}_{B'B''} (w^{s})^{\mo}_{B''B'''} \delta I^{ex,s,l}_{B''',bc}. \la{dSigmal2}
}
Four of these eight terms vanish for $n$-Dirac/Weyl fermions  [\qq{nweylham}{ndiracham}] coupled to overscreened hybrid impurities, leading to
\e{
\text{$n$-Dirac/Weyl Hamiltonians $H^{d,n}$ and $H^{w,n}$}: \as \Sigma^{l}=d\Sigma^{l,ddd} +d\Sigma^{l,mdd}+d\Sigma^{l,dmd}+d\Sigma^{l,ddm}. \la{dSigmal}
}
Namely, $d\Sigma^{l,mmm}$ (and also $dw^{a}(H^0,V^{m},V^{m},V^{m})$) vanishes by the $\vec{H}$-polarity theorem,  because $n$-Dirac/Weyl fermions [\qq{nweylham}{ndiracham}] do not admit a $\vec{H}$-vector polarization. Moreover,
\e{
0 \eq d\Sigma^{l,mmd}+ d\Sigma^{l,mdm} +d\Sigma^{l,dmm},
}
because their net contribution to the asymmetric transition rate [obtained by plugging into \q{decomposewa} the explicit impurity matrix elements in \qq{overscreenedmonopole3D}{overscreeneddipole3D} and \qq{overscreenedmonopole2D}{overscreeneddipole2D}] is proportional to
\e{
\boldsymbol{\mathscr{d}}\cdot \sum_{B''}\bigg[(\bk-\bk'')+(\bk''-\bk')+(\bk'-\bk)\bigg]\braket{u_{B}}{u_B''}\braket{u_{B''}}{u_B'}\braket{u_{B'}}{u_B}\delta(E_{BB''})=0.
}\\

By $g$-transforming all the Hamiltonian inputs to $w^s$ and $d\omega^s$ in \q{dSigmal2},
\e{
&w^s_{BB'}(H^{0,g},V^{m,g}) \refeq{wBBp2} w^{s}_{g^{\mo}B,g^{\mo}B'}(H^{0,e},V^{m,e}) \la{ggg1}\\
&dw^{a}_{BB'}(H^{0,g},V^{\alpha,g},V^{\beta,g},V^{\delta,g})\refeq{essentially}i_g dw^a_{g^{\mo}B,g^{\mo}B'}(H^{0,e},V^{\alpha,e},V^{\beta,e},V^{\delta,e}).\la{ggg2} 
}
To simplify notation, we henceforth denote an algorithm $A$ with all its Hamiltonian inputs being $g$-transformed as $A(H^g)$, or sometimes $A|_{H^g}$.
Combining \qq{ggg1}{ggg2} with previously derived relations for the velocity matrix [\q{veloBBg}] and photo-excitation tensor [\q{IexcHT}], we obtain an expression for $d\Sigma^l$ with $g$-transformed Hamiltonian inputs: 
\e{
d\Sigma_{abc}^{\alpha,\ab\delta}(H^g) \eq  -\tf{|e|}{\calv}\sum_{BB'B''B'''}\big(i_g g^s_{ad}v_{\gmo B,d}\big) (w^{s})^{\mo}_{\gmo B,\gmo B'}(i_g dw^{a}_{\gmo B',\gmo B''}) (w^{s})^{\mo}_{\gmo B'',\gmo B'''}\big(g^s_{be}g^s_{cf} \delta I^{ex,s,\alpha}_{\gmo B''',ef}\big)\bigg|_{H^e} \lin
\eq (s_{\alpha})^g g^s_{ad}g^s_{be}g^s_{cf} d\Sigma_{def}^{\alpha,\ab\delta}(H^e). \la{dSigmaltransformH}
}
This relation holds no matter the symmetry of $H$.\\

Let  $g$ be a symmetry of the zeroth-order Hamiltonian ($H^{0,g}=H^{0,e}$), and be either a symmetry or antisymmetry of the impurity potential: $V^g=\pm V^e$. Defining a $g$-parity $p_g^{\alpha}$  [\q{gparity}] for each of the three impurity potentials in $dw^a$, we find that the response tensor $d\Sigma^{\alpha}$ can be assigned a $g$-parity which is the net $g$-parity for all three impurity potentials: 
\e{
H^{0,g}=H^{0,e}, \as V^{\alpha,g}=p_g^{\alpha} V^{\alpha,e}: \as d\Sigma_{abc}^{\alpha,\ab\delta}(H^g) \refeq{dSigmal}  (-1)^{N_{\alpha\beta\delta;g\text{-odd}}}d\Sigma_{abc}^{\alpha,\ab\delta}(H^e); \as (-1)^{N_{\alpha\beta\delta;g\text{-odd}}}=p_g^{\alpha}p_g^{\beta}p_g^{\delta},\la{oddgg}
}
 with $N_{\alpha\beta\delta;g\text{-odd}}$ being the number of potentials in $\{V^{\alpha},V^{\beta},V^{\delta}\}$ that are $g$-antisymmetric. Combining \q{oddgg} with \q{dSigmaltransformH},
\e{
H^{0,g}=H^{0,e}, \as V^{\alpha,g}=p_g^{\alpha} V^{\alpha,e}: \as d\Sigma_{abc}^{\alpha,\ab\delta}=(-1)^{N_{\alpha\beta\delta;g-\text{odd}}}(s_{\alpha})^g g^s_{ad}g^s_{be}g^s_{cf} d\Sigma_{def}^{\alpha,\ab\delta}, \la{secondsymmetryconstraint2nd}
}
which is our second example of a second-kind constraint.

\subsection{Symmetry constraints for the photovoltaic response of $n$-Weyl fermions} \la{app:symconstraintsnweyl}

Our main result is summarized in \tab{tab:nweylconstraint} and derived in the remainder of this section.\\

\begin{table}[H]
\centering
\begin{tabular} {|c|c|c|c|c|c|} \hline
   Fermion & Impurity &  Group &  Shift$^l$/Ballistic$^l$ &   Shift$^c$/Ballistic$^c$ & $\Sigma_{\text{ballistic}}^{\text{im,}l}$ \\  \hline 
 $H^{w,n}$ & $V^m$ & $D_{\infty}\times \Z_2^T$ & $\Sigma^l_{xyz}=-\Sigma^l_{yzx}$ &0&  $\Sigma^l_{xyz}=0$ \\ \cline{2-6}						
 &  $V^{d\para z}$&   $SO(2)\times \Z_2^T$ &$\Sigma^l_{zzz}; \as\Sigma^l_{zxx}=\Sigma^l_{zyy};\as\Sigma^l_{xxz}=\Sigma^l_{yyz};\as \Sigma^l_{xyz}=-\Sigma^l_{yzx}$  &0 & -\\ \cline{2-6}
 &$V^m+V^{d\para z}$&  $SO(2)\times \Z_2^{T}$ &  $\Sigma^l_{zzz}; \as \Sigma^l_{zxx}=\Sigma^l_{zyy};\as\Sigma^l_{xxz}=\Sigma^l_{yyz};\as \Sigma^l_{xyz}=-\Sigma^l_{yzx}$ &0 &$ d\Sigma^{l,ddd}_{zzz/zxx/xxz}; \as d\Sigma^{l,\{mdd\}}_{yzx}$ \\ \cline{2-6}
 & $V^{d\para x}$&   $\Z_2^{C_{2x}}\times \Z_2^T$ & $\Sigma^l_{xxx/\{xyy\}/\{xzz\}/\{xyz\}}$ & 0 & $\Sigma^l_{\{xyz\}}=0$ \\ \cline{2-6}	
 & $V^m+V^{d\para x}$&  $\Z_2^{C_{2x}}\times \Z_2^{T}$  & $\Sigma^l_{xxx/\{xyy\}/\{xzz\}/\{xyz\}}$ &0&  $d\Sigma^{l,ddd}_{xxx/\{xyy\}/\{xzz\}};\as  d\Sigma^{l,\{mdd\}}_{\{xyz\}}$ \\ \hline			 	
\end{tabular}
\caption{For $n$-Weyl fermions ($H^{w,n}$) coupled to monopolar ($V^m$) or x-oriented dipolar ($V^{d\para x}$) impurities or their combination, we list their symmetry groups (third column),  symmetry-allowed linear/circular shift/ballistic photovoltaic tensor elements based on general symmetry constraints (fourth and fifth columns), and $H$-specific symmetry constraints of the impurity-mediated ballistic photovoltaic tensor (sixth column). `-' means there are no nontrivial $H$-specific constraints. $d\Sigma^{\alpha,\{\beta\delta\gamma\}}$ means all permutations of $(\beta\delta\gamma)$ are symmetry-allowed.	\label{tab:nweylconstraint}}
\end{table}

The group of the Hamiltonian of $n$-Weyl fermions [\app{app:symmetryndiracweyl}] coupled to monopolar impurities is
\e{
G(H^{w,n}+V^m)=D_{\infty}^T \equiv D_{\infty}\times \Z_2^T.
}
We have introduced the notation $G(H)$ for a symmetry group $G$ of the Hamiltonian $H$.
By assumption, monopolar impurities are $O(3)\times \Z_2^T$-symmetric. Let us separately discuss the role of $\Z_2^T$, $D_2$ and $SO(2)$ symmetry:\\

\noi{$\Z_2^T$} Time-reversal symmetry ensures that the circular tensor vanishes:
 \e{
 H^{T}\eq H^e \imp \Sigma^c=0,
 }
 according to \q{gsconstrainsSigmaimp} with $g=T, g^s_{ab}=\delta_{ab}$ and $(s_c)^T=-1.$ This conclusion holds not just for $n$-Weyl fermions but also for massless $n$-Dirac fermions (which are $T$-symmetric), and not just for monopole impurities but also for dipole impurities (also $T$-symmetric).  This justifies the zeros in the fifth column of  \tab{tab:nweylconstraint}. 
 Only massive $n$-Dirac fermions break $T$ symmetry and can in principle sustain a circular photovoltaic current. \\

\noi{$D_2$} The point group $D_2$ includes all two-fold rotations. According to to
\tab{tab:compindex}, the symmetry-allowed tensor elements have odd $N_{abc;j}$ for all $j\in \{x,y,z\}$, with $N_{abc;j}$ meaning the number of j indices in the subscript of $\Sigma^l_{abc}$. This limits us to  $\Sigma_{\{xyz\}}$, which includes permutations of `$xyz$'.\\

\noi{$SO(2)$} This symmetry allows for only one independent tensor element in the set $\Sigma_{\{xyz\}}$ [\q{SO2allowedlinear}], which we write in the second row, fourth column of  \tab{tab:nweylconstraint}.\\

\noi{$\vec{H}$-polarization theorem} The $\vec{H}$-vector representation of $D_{\infty}^T \equiv D_{\infty}\times \Z_2^T$ does not contain any invariant vectors, hence the impurity-mediated $\Sigma^{\text{im},l}_{\text{ballistic}}=0$ [cf. second row, last column of \tab{tab:nweylconstraint}].\\

Let $V^{d\para z}$ be the impurity potential for dipole impurity  oriented parallel to the $z$ axis. Such an impurity breaks $C_{2x/2y}$ but preserves $SO(2)$ symmetry:
\e{
G(H^{w,n}+V^{d\para z})=SO(2)\times \Z_2^T.
}
This group restricts us to the linear photovoltaic elements in  \q{SO2allowedlinear}, which are reproduced in the third row of  \tab{tab:nweylconstraint}.\\

\noi{$\cancel{C_{2x/2y}}$}  Because $C_{2x}$ (and $C_{2y}$) is a symmetry of $H^{w,n}$ and an antisymmetry of $V^{d\para z}$, we obtain from \q{gparitySigmaimp} two second-kind constraints:
\e{
\Sigma^l_{abc}=-(C_{2x/2y})_{ad}(C_{2x/2y})_{be}(C_{2x/2y})_{cf}\Sigma^l_{def}= (-1)^{N_{abc;x/y}}\Sigma^l_{abc}, \la{Nxy}
}
which should be viewed 
as separate equations for $x$ and $y$. This implies that $\Sigma^l_{abc}=0$ if the number of x indices is odd, or if the number of y indices is odd. Imposing this restriction on \q{SO2allowedlinear} does not give us new information.\\

Let $V^{d\para x}$ be the impurity potential for dipole impurity  oriented parallel to the $x$ axis. Such an impurity breaks $C_{2y/2z}$ but preserves $C_{2x}$ symmetry:
\e{
G(H^{w,n}+V^{d\para z})=\Z_2^{C_{2x}}\times \Z_2^T,
}
with $\Z_2^{C_{2x}}$ the $\Z_2$ group generated by ${C_{2x}}$.\\

\noi{$C_{2x}$} According to 
\tab{tab:compindex}, the symmetry-allowed tensor elements have odd $N_{abc;x}$ [cf.  fifth of  \tab{tab:nweylconstraint}]. \\

\noi{ $\cancel{C_{2y/2z}}$} Because $C_{2y}$ (and $C_{2z}$) is a symmetry of $H^{w,n}$ and an antisymmetry of $V^{d\para x}$, we obtain from \q{gparitySigmaimp} two second-kind constraints:
\e{
\Sigma^l_{abc}=-(C_{2z/2y})_{ad}(C_{2z/2y})_{be}(C_{2z/2y})_{cf}\Sigma^l_{def}= (-1)^{N_{abc;z/y}}\Sigma^l_{abc}, \la{Nzy}
}
which should be viewed 
as separate equations for $z$ and $y$. This constraints both $N_{abc;y}$ and $N_{abc;z}$ to be even, while $N_{abc;x}$ remains odd from the previously-explained general contraint [cf. fifth row. last column of  \tab{tab:nweylconstraint}].\\

Let us consider the hybrid impurity potential that is a sum of monopolar and dipolar components: 
\e{
G(H^{w,n}+V^m+V^{d\para z})= SO(2)\times \Z_2^{T}.
}
This group restricts us to the linear photovoltaic elements in  \q{SO2allowedlinear} [fourth row of  \tab{tab:nweylconstraint}].\\

\noi{ $\cancel{C_{2x/2y}}$}  Each of these tensor elements can be decomposed according to \q{dSigmal}, with each component satisfying two second-type constraints [\q{secondsymmetryconstraint2nd}]:
\e{
d\Sigma_{abc}^{l,\ab\delta}=(-1)^{N_{abc;x/y}+N_{\alpha\beta\delta;C_{2x/2y}\text{-odd}}+1} d\Sigma_{abc}^{l,\ab\delta},
}
which should be viewed as separate equations for $x$ and $y$. For tensor elements in  \q{SO2allowedlinear}  with an even number of $x$ or $y$ indices, what's symmetry-allowed are $d\Sigma^{l,ddd}_{zzz/zxx/zyy/xxz/yyz}$, For tensor elements in  \q{SO2allowedlinear}  with an odd number of $x$ or $y$ indices, what's symmetry-allowed are $d\Sigma^{l,\{mdd\}}_{xyz/yzx}$,  which include all permutations of $(mdd)$. This justifies the fourth row, last column of  \tab{tab:nweylconstraint}.\\

Let us consider the hybrid impurity potential that is a sum of monopolar and dipolar components: 
\e{
G(H^{w,n}+V^m+V^{d\para x})= \Z_2^{C_{2x}}\times \Z_2^{T}.
}
\noi{$C_{2x}$} The same general constraint [\tab{tab:compindex}] allows for nonzero
$\Sigma^l_{xxx/\{xyy\}/\{xzz\}/\{xyz\}}.$ \\

\noi{ $\cancel{C_{2y/2z}}$} Each of these tensor elements can be decomposed according to \q{dSigmal}, with each component satisfying two second-type constraints [\q{secondsymmetryconstraint2nd}]:
\e{
d\Sigma_{abc}^{l,\ab\delta}=(-1)^{N_{abc;y/z}+N_{\alpha\beta\delta;C_{2y/2z}\text{-odd}}+1} d\Sigma_{abc}^{l,\ab\delta},
}
which should be viewed as separate equations for $y$ and $z$. Thus for tensor elements with an even number of $y$ or $z$ indices, what's symmetry-allowed are $d\Sigma^{l,ddd/}_{xxx,\{xyy\}/\{xzz\}}$; for tensor elements with an odd number of $y$ or $z$ indices, what's symmetry-allowed are $d\Sigma^{l,\{mdd\}}_{\{xyz\}}$, which include all permutations of $(mdd)$.  This justifies the last entry in  \tab{tab:nweylconstraint}.

\subsection{Symmetry constraints for the photovoltaic response of $n$-Dirac fermions}\la{app:symconstraintsndirac}

In a `2D system', we restrict current and electric fields to lie in the xy-plane, so the indices in $\Sigma_{abc}$ can only be x or y indices. Let us define $N_{abc;x}$ (resp. $N_{abc;y}$) as the number of `abc' which are x indices (resp. y indices); because $N_{abc;x}+N_{abc;y}=3$, $N_{abc;x}$ is odd if and only if $N_{abc;y}$ is even. Our main result is summarized in \tab{tab:ndiracconstraint} and derived in the remainder of this section.\\

\begin{table}[H]
\centering
\begin{tabular} {|c|c|c|c|c|c|} \hline
Fermion  & Impurity & Group & Shift$^l$/Ballistic$^l$ &  Shift$^c$/Ballistic$^c$ &  $\Sigma_{\text{ballistic}}^{\text{im},l/c}$ \\  \hline   
$H^{d,n}$ & $V^m$ &$O(2)\times \Z_2^T$ & 0&0 & -\\ \cline{2-6}
 & $V^{d\para x}$& $\Z_2^{M_y}\times \Z_2^{T}$ & $\Sigma^l_{xxx/\{xyy\}}$ & 0 & -\\ \cline{2-6} 
 & $V^m+V^{d\para x}$& $\Z_2^{M_y}\times \Z_2^{T}$ & $\Sigma^l_{xxx/\{xyy\}}$ & 0 & $d\Sigma^{l,ddd}_{xxx/\{xyy\}}$\\ \hline
$H^{md,n}$ & $V^m$ &$SO(2)$ & 0&0 & -\\ \cline{2-6}			 
 & $V^{d\para x}$& $\Z^e_1$ & $\Sigma^l_{\text{All}}$ & $\Sigma^c_{\text{All}}$ & - \\ \cline{2-6}		 
 & $V^m+V^{d\para x}$&   $\Z^e_1$ & $\Sigma^l_{\text{All}}$ & $\Sigma^c_{\text{All}}$ & $d\Sigma^{l/c,ddd}_{\text{All}}$\\ \hline	 		
\end{tabular}		
\caption{For massless ($H^{d,n}$) and massive ($H^{md,n}$) $n$-Dirac fermions, coupled to monopolar ($V^m$) or x-oriented dipolar ($V^{d\para x}$) impurities or their combination, we list their symmetry groups (third column),  symmetry-allowed linear/circular shift/ballistic photovoltaic tensor elements based on general symmetry constraints (fourth and fifth columns), and $H$-specific symmetry constraints of the impurity-mediated ballistic photovoltaic tensor (sixth column). `-' means there are no nontrivial $H$-specific constraints.  $\Sigma_{\text{All}}$ means that all tensor elements of $\Sigma$ are symmetry-allowed. `0' means all tensor elements vanish.	\label{tab:ndiracconstraint}}
\end{table}

$C_{2z}$ [which maps $(x,y)$ to $(-x,-y)$] is an element in both $O(2)\times \Z_2^T$ and $SO(2)$, which are respectively the symmetry groups of massless and massive $n$-Dirac fermions [\tab{tab:nDiracsymmetryreps}]. This remains true if the fermions are coupled to monopole impurities:
\e{
G(H^{d,n}+V^m)=O(2)\times \Z_2^T ; \as   G(H^{md,n}+V^m)=SO(2). \la{GHintro}
}
According to \tab{tab:compindex}, $C_{2z}$ enforces $\Sigma_{abc}=0$ for any $(abc)$ in a 2D system [\q{gsconstrainsSigma}], and for both linear and circular components of $\Sigma$. This justifies the zeros in the second and fifth rows of \tab{tab:ndiracconstraint}.\\

$C_{2z}$ symmetry is broken by any impurity with an  in-plane dipole moment; the corresponding impurity potential is $C_{2z}$-antisymmetric, which implies  a second-kind constraint [\q{gparitySigmaimp}] which is trivial: $\Sigma=\Sigma$. Suppose   the  dipole moment is aligned in the $x$ direction (with corresponding impurity potential $V^{d\para x}$), then the symmetry groups  of massless and massive Dirac fermions are reduced to
\e{
G(H^{d,n}+V^{d\para x})= \Z_2^{M_y}\times \Z_2^{T}; \as G(H^{md,n}+V^{d\para x})= \Z_1^e,
}
with $\Z_2^{M_y}$ the $\Z_2$ group generated by mirror reflection $M_y: (x,y)\mapsto (x,-y)$, and $\Z_1^e$ being the trivial group. We discuss the two fermion classes in turn:\\

\noi{Massless} The photovoltaic theorems [\qq{gsconstrainsSigma}{TgsconstrainsSigma}] imply that (i) $\Sigma^c=0$ (due to $T$ symmetry), and (ii) $\Sigma_{abc}\neq 0$ requires $N_{abc;y}$ to be even (due to $M_y$ symmetry):
\e{
H^{d,n}+V^{d\para x}: \as \text{only $\Sigma^l_{xxx/\{xyy\}}$ are symmetry-allowed; cf. third row of \tab{tab:ndiracconstraint}}. \la{MyTsymm}
}
 Because $H^{d,n}$ is $M_x$-symmetric and $V^{d\para x}$ is $M_x$-antisymmetric, a constraint of the second kind [\q{gparitySigmaimp}] implies that $\Sigma_{abc}\neq 0$ requires $N_{abc;x}$ to be odd. However, we do not gain more information than what is stated in  \q{MyTsymm}.  \\

\noi{Massive} There are no general or $H$-specific first-kind constraints because the Hamiltonian is trivially symmetric. The second-kind constraint (associated to $C_2$ being a symmetry of $H^{md,n}$ and an antisymmetry of $V^{d\para x}$) is trivial, as noted earlier. 
\e{
H^{d,n}+V^{d\para x}: \as \text{all $\Sigma^{l,c}$ elements are symmetry-allowed; cf. sixth row of \tab{tab:ndiracconstraint}}.
}\\

Finally let us consider the hybrid impurity potential that is a sum of monopolar and dipolar components: 
\e{
G(H^{d,n}+V^m+V^{d\para x})= \Z_2^{M_y}\times \Z_2^{T}; \as G(H^{md,n}+V^m+V^{d\para x})= \Z_1^e.
}
\noi{Massless} The same symmetry group and same general constraints apply as in \q{MyTsymm}. Combining these general constraints with the second-kind constraint [\q{secondsymmetryconstraint2nd}, with either $g=C_2$ or $g=M_x$, applied to \q{dSigmal}],
\e{
H^{d,n}+V^m+V^{d\para x}: \as \text{$\Sigma^l_{xxx/\{xyy\}}$ and $d\Sigma^{l,ddd}_{xxx/\{xyy\}}$ are symmetry-allowed; cf. fourth row of \tab{tab:ndiracconstraint}}. \la{MyTsymm2}
}\\

\noi{Massive} Applying \q{secondsymmetryconstraint2nd} (with $g=C_2$) to \q{dSigmal} gives a second-kind constraint:
\e{
H^{md,n}+V^m+V^{d\para x}: \as \text{$\Sigma^{l,c}_{abc}$ and $d\Sigma^{l/c,ddd}_{abc}$  are symmetry-allowed, for any $(abc)$; cf. seventh row of \tab{tab:ndiracconstraint}.}
\la{secondsymmetryconstraint2ndC2}
}

\bibliography{bib_Apr2018, Addon}

\end{document}